\documentclass{article}

\usepackage{epubtk}
\usepackage{graphicx}
\usepackage{amsfonts}
\usepackage{amsmath}
\usepackage{amssymb}
\usepackage{bm}
\usepackage{dcolumn}
\usepackage{latexsym}
\usepackage{rotating}
\usepackage{eucal}
\usepackage{booktabs}
\usepackage{xspace} 
\usepackage{ragged2e}
\usepackage{multirow}

\RequirePackage[normalem]{ulem} 
\RequirePackage{color}\definecolor{RED}{rgb}{1,0,0}\definecolor{BLUE}{rgb}{0,0,1} 

\newcommand\be{\begin{equation}}
\newcommand\ba{\begin{eqnarray}}
\newcommand\bal{\begin{align}}
\newcommand\eal{\end{align}}
\newcommand\ee{\end{equation}}
\newcommand\ea{\end{eqnarray}}

\begin{document}

\title{Gravitational-Wave Tests of General Relativity with
  Ground-Based Detectors and Pulsar-Timing Arrays}

\def\headertitle{Gravitational-Wave Tests of GR with
  Ground-Based Detectors and Pulsar-Timing Arrays}

\author{%
\epubtkAuthorData{Nicol\'as Yunes}{%
Illinois Center for Advanced Studies of the Universe, \\
Department of Physics, \\
University of Illinois Urbana-Champaign, \\
Urbana, IL 61801, USA}
{nyunes@illinois.edu}
{https://yunes-gravity-theory-group.physics.illinois.edu}
\and
\epubtkAuthorData{Xavier Siemens}{%
Department of Physics,  \\
Oregon State University,\\
Corvallis, OR 97330, USA}
{xavier.siemens@oregonstate.edu}{https://science.oregonstate.edu/directory/xavier-siemens}
\and
\epubtkAuthorData{Kent Yagi}{%
Department of Physics, \\ 
University of Virginia, \\
Charlottesville, VA 22904, USA}
{ky5t@virginia.edu}
{https://www.phys.virginia.edu/People/personal.asp?UID=ky5t}
}

\date{}
\maketitle

\begin{abstract}
This review is focused on tests of Einstein's theory of general
relativity with gravitational waves that are detectable by
ground-based interferometers and pulsar-timing experiments. Einstein's
theory has been greatly constrained in the quasi-linear,
quasi-stationary regime, where gravity is weak and velocities are
small. Gravitational waves are allowing us to probe a complimentary, yet
previously unexplored regime: the non-linear and dynamical
\emph{extreme gravity regime}. Such a regime is, for example, applicable
to compact binaries coalescing, where characteristic velocities can
reach fifty percent the speed of light and gravitational fields are 
large and dynamical. This review begins with the theoretical basis and 
the predicted gravitational-wave observables of modified gravity 
theories. The review continues with a brief description of the 
detectors, including both gravitational-wave interferometers and 
pulsar-timing arrays, leading to a discussion of the data analysis 
formalism that is applicable for such tests. The review then discusses 
gravitational-wave tests using compact binary systems, and ends
with a description of the first gravitational wave observations by advanced 
LIGO, the stochastic gravitational wave background observations by pulsar timing arrays, and the tests that can be performed with them.
\end{abstract}

\epubtkKeywords{General relativity, Gravitational waves, Pulsar
  timing, Experimental tests, Observational tests, Alternative
  theories, Compact binaries}

\newpage
\tableofcontents

\newpage
\section{Introduction}
\label{section:introduction}
%
\subsection{The importance of testing}

The era of precision gravitational wave astrophysics commenced with the first direct gravitational wave observations by the advanced Laser Interferometer Gravitational Observatory (aLIGO) \cite{TheLIGOScientific:2016pea} and all the other events that were discovered by aLIGO, Virgo and KAGRA~\cite{LIGOScientific:2018mvr,LIGOScientific:2020ibl,LIGOScientific:2021djp}. With it, a plethora of previously unavailable information has flooded in, allowing for unprecedented astrophysical measurements and tests of fundamental theories~\cite{TheLIGOScientific:2016src,Yunes:2016jcc,Berti:2018cxi,Berti:2018vdi,LIGOScientific:2018dkp,LIGOScientific:2019fpa,LIGOScientific:2020tif,LIGOScientific:2021sio}. Nobody would question the importance of more precise astrophysical measurements, but one may wonder whether fundamental tests are truly necessary, considering the many successes of Einstein's theory of General Relativity (GR). GR has passed many tests through Solar System, binary pulsar and cosmological observations (see e.g.~\cite{Stairs:2003eg,lrr-2006-3,lrr-2008-9}), but what many of these have in common is that they sample the quasi-stationary, quasi-linear {\emph{weak field}}. That is, they sample the regime of spacetime where the gravitational field is weak relative to the mass-energy of the system, the characteristic velocities of gravitating bodies are typically small relative to the speed of light, and the gravitational field is stationary or quasi-stationary relative to the characteristic size of the system. Direct electromagnetic imaging of supermassive black holes by the Event Horizon Telescope (EHT)~\cite{EventHorizonTelescope:2019dse,EventHorizonTelescope:2022xnr} has the potential to probe gravity in the strong gravitational potential regime~\cite{EventHorizonTelescope:2022xqj}, provided astrophysical mismodeling and statistical uncertainties can be controlled~\cite{Gralla:2020pra,Gralla:2020srx,Volkel:2020xlc,Lara:2021zth}. The systems EHT observes, however, are quasi-stationary and the curvature of spacetime around supermassive black holes is not as high as that for stellar-mass black holes and neutron stars. On the other hand, the \emph{extreme gravity} regime\epubtkFootnote{Notice that ``extreme gravity'' is not synonymous with Planck scale physics in this context. In fact, a stationary black hole would not serve as a probe of extreme gravity, even if one were to somehow acquire information about the gravitational potential close to the singularity. This is because any such observation would necessarily be lacking information about the dynamical sector of the gravitational interaction. Planck scale physics is perhaps more closely related to strong-curvature physics.}, where spacetime is highly dynamical, gravity is strong and self-gravitating bodies have non-negligible velocities, is precisely the area gravitational waves open for exploration. 

To make this more concrete, let us define the gravitational compactness via  ${\cal{C}} = {{{\cal{M}}}/{\cal{R}}}$, where ${\cal{M}}$ and ${\cal{R}}$ are the characteristic mass and length scale of the system (henceforth, we set $G = c = 1$). Let us also define the characteristic velocities ${\cal{V}}$ of the system as a measure of the rate of change of the gravitational field in the center of mass frame. The characteristic velocity can be related to the timescale on which the gravitational energy changes significantly, ${\cal{T}} = E/\dot{E}$, with $E$ the characteristic gravitational energy and $\dot{E}$ its rate of change, through ${\cal{V}} = {\cal{R}}/{\cal{T}}$.  While the compactness ${\cal{C}}$ is a measure of how strong gravity is in the system, the velocity ${\cal{V}}$ is a measure of how dynamical the spacetime is.  
%
%
With these definitions at hand, we define the weak field, the strong field and the extreme gravity regimes as follows:
\begin{align}
\label{eq:weak-field-def}
{\bf{Weak}}\,{\bf{Field}}\,{\bf{Regime:}} &\quad {\cal{C}} \ll 1 \rightarrow {\cal{M}} \ll {\cal{R}} \quad {\bf{and}} 
\quad {\cal{V}} \ll 1 \rightarrow {\cal{M}} \ll {\cal{T}}\,,
\\
{\bf{Strong}}\,{\bf{Field}}\,{\bf{Regime:}} &\quad {\cal{C}} \lesssim {\cal{O}}(1) \rightarrow {\cal{M}} \lesssim {\cal{R}} \quad {\bf{and}} \quad {\cal{V}} \ll 1 \rightarrow {\cal{M}} \ll {\cal{T}}\,,
\\
{\bf{Extreme}}\,{\bf{Gravity}}\,{\bf{Regime:}} &\quad {\cal{C}} \lesssim {\cal{O}}(1) \rightarrow {\cal{M}} \lesssim {\cal{R}} \quad {\bf{and}}  \quad {\cal{V}} \lesssim {\cal{O}}(1) \rightarrow {\cal{M}} \lesssim {\cal{T}}\,.
\end{align}
In spite of the naming convention, there is much information that can be gained from weak-field and strong-field tests. In fact, entire classes of modified gravity theories have been effectively ruled out (or at the very least stringently constrained) by Solar System observations alone~\cite{lrr-2006-3}, as we discuss below. 

Let us provide some examples of these regimes. The prototypical example for weak-field tests are observations in the Solar System. For the Earth-Sun system, ${\cal{M}}$ is essentially the mass of the Sun, while ${\cal{R}}$ is the Earth-Sun orbital separation, which leads to ${\cal{C}} = {\cal{O}}(10^{-8})$, ${\cal{V}} = {\cal{O}}(10^{-4})$ and ${\cal{T}} = {\cal{O}}(10^{6})$ seconds. The characteristic compactness and velocity are clearly very small, and this is true for any planet or satellite in the Solar System. Even if an object were in a circular orbit at the surface of the Sun, its gravitational compactness would be 
${\cal{O}}(10^{-6})$, its characteristic velocity ${\cal{O}}(10^{-3})$ and the characteristic timescale ${\cal{O}}(10^{3})$ seconds. A 
particularly important Solar System test that has been used to stringently constrain many modified theories of gravity is  
the observation of the Shapiro time delay, i.e.~the delay of photons as they traverse a regime of curved spacetime. In 2003, the Cassini Probe~\cite{Bertotti:2003rm} sent radio signals to Earth while on its way to Saturn while Earth was being eclipsed by the Sun, thus sensitively probing the Shapiro time delay effect. The characteristic radius in this experiment is the radius of the Sun, which gives ${\cal{C}} = {\cal{O}}(10^{-6})$ and ${\cal{V}} = 0$, since this observation does not sample the dynamical nature of spacetime. 

The classic example for strong-field tests are observations of binary pulsars. Neutron stars are strong-field sources of gravity because the ratio of their mass to their radius (their matter compactness) is of ${\cal{O}}(10^{-1})$. However, many observables are not sensitive to the gravitational field at the surface of the pulsar. For example, observations of the orbital decay rate with the double binary pulsar J0737-3039~\cite{Lyne:2004cj,Kramer:2006nb} have a characteristic compactness of ${\cal{C}} = {\cal{O}}(10^{-5})$, velocities of ${\cal{V}} = {\cal{O}}(10^{-3})$ and timescales of ${\cal{T}} = {\cal{O}}(10^{3})$ seconds, where ${\cal{R}}$ is the orbital separation ${\cal{R}} = {\cal{O}}(10^{6} \, {\rm{km}})$ and $M \approx 3 M_{\odot}$ is the total mass. Observations of the Shapiro time delay with the same system, where photons from one neutron star pass close to its binary companion\epubtkFootnote{Contrary to popular belief, this Shapiro time delay observation does not probe distances comparable to the surface of neutron stars. Photons from pulsar A do graze pulsar B, but the photon's distance of closest approach depends sensitively on the inclination angle, which is not exactly $90$ degrees for J0737-3039~\cite{Lyne:2004cj,Kramer:2006nb}.}, have a characteristic compactness of ${\cal{C}} = {\cal{O}}(10^{-4})$, where ${\cal{R}}={\cal{O}}(10^{4} \, {\rm{km}})$ is the photon's distance of closest approach and ${\cal{M}} = {\cal{O}}(1.34 M_{\odot})$ is the mass of one of the pulsars. Shapiro time delay observations, however, are not probes of the dynamics of the spacetime, so ${\cal{V}} = 0$ and ${\cal{T}} = \infty$.

A typical example of an extreme gravity scenario is the merger of compact objects, like black holes or neutron stars. In such scenarios, the characteristic compactness and velocity can reach ${\cal{C}} = {\cal{O}}(1) = {\cal{V}}$ during merger. Moreover, the late inspiral proceeds so fast that the characteristic timescale can reach ${\cal{T}}  = {\cal{O}}(10^{-4})$ seconds. We see then that direct gravitational wave observations sample gravitational compactnesses and velocities much larger than those weak-field and strong-field observations can probe. Of course, this had to be the case, since the inspiral that aLIGO observed is completely driven by gravitational wave emission, where the latter cannot be treated as a small perturbation to linear order. However, given the aLIGO observations to date, which have small to moderate signal-to-noise ratios, the constraints on modified gravity effects derived from these cannot always compete with those obtained with weak-field and strong-field observations yet. 

Even though Solar System and binary pulsar observations do not give us access to the extreme gravity regime, they have indeed served (and will continue to serve) as invaluable tools to learn about gravity. Solar system tests effectively cured an outbreak of modified gravity theories in the 1970s and 1980s, as summarized for example in~\cite{lrr-2006-3}. Binary pulsars were crucial as the first indirect detectors of gravitational waves, and later to kill certain theories, like Rosen's bimetric gravity~\cite{Rosen:1974ua}, and heavily constrain others that predict dipolar energy loss, as we will see in Sections~\ref{section:alt-theories} and~\ref{section:binary-sys-tests}. Similarly,  electromagnetic observations of black hole accretion disks probe GR in another strong-field sector: the non-linear but fully stationary regime, verifying that black holes are described by the Kerr metric~\cite{EventHorizonTelescope:2022xqj}.

Given these successes of Einstein's theory, one may wonder why one should bother with testing GR in the extreme gravity regime. Needless to say, the role of science is to {\emph{predict and verify}} and not to assume without proof. Moreover, the incompleteness of GR in the quantum regime, together with the somewhat unsatisfactory requirement of the dark sector of cosmology (including dark energy and dark matter) have prompted many studies of modifications to GR. Gravitational waves have now begun to verify Einstein's theory in a regime previously inaccessible to us, and as such, these tests are invaluable.  In many areas of physics, however, GR is so ingrained that questioning its validity is synonymous with heresy. Dimensional arguments are usually employed to argue that any quantum gravitational correction will necessarily and unavoidably be unobservable, as the former are expected at a (Planck) scale that is inaccessible to detectors. This rationalization is dangerous, as it introduces a theoretical bias in the analysis of new observations of the Universe, thus quenching the potential for new discoveries. For example, if astrophysicists had followed such a rationalization when studying supernova data, they would not have discovered that the Universe is expanding. Dimensional arguments suggest that the cosmological constant is over 100 orders of magnitude larger than the value required to agree with observations. When observing the Universe for the first time in a completely new way, it seems more conservative to remain {\emph{agnostic}} about what is expected and what is not, thus allowing the data itself to guide our efforts toward theoretically understanding the gravitational interaction. 

\subsection{Testing general relativity versus testing alternative theories}

When {\emph{testing GR}}, one considers Einstein's theory as a {\emph{null hypothesis}} and searches for generic deviations. On the other hand, when {\emph{testing modified gravity}} one starts from a particular modified gravity model, develops its equations and solutions and then predicts certain observables that then might or might not agree with experiment. Similarly, one may define a {\emph{bottom-up}} approach versus a {\emph{top-down}} approach. In the former, one starts from some observables in an attempt to discover fundamental symmetries that may lead to a more complete theory, as was done when constructing the standard model of elementary particles. On the other hand, a top-down approach starts from some fundamental theory and then derives its consequence.

Both approaches possess strengths and weaknesses. In the top-down approach one has complete control over the theory under study, being able to write down the full equations of motion, answer questions about well-posedness and stability of solutions, and predict observables. But, as we see in Section~\ref{section:alt-theories}, carrying out such an approach can be quixotic within any one model. What is worse, the lack of a complete and compelling alternative to GR makes choosing a particular modified theory difficult.

Given this, one might wish to attempt a bottom-up approach, where one considers a set of principles one wishes to test without explicit mention of any particular theory. One usually starts by assuming GR as a null-hypothesis and then considers deformations away from GR. The hope is that experiments will be sensitive to such deformations, thus either constraining the size of the deformations or pointing toward a possible inconsistency. But if experiments do confirm a GR deviation, a bottom-up approach fails at providing a given particular action from which to derive such a deformation. In fact, there can be several actions that lead to similar deformations, all of which can be consistent with the data within its experimental uncertainties. 

Nonetheless, both approaches are complementary. The bottom-up approach draws inspiration from particular examples carried out in the top-down approach, therefore allowing for a \emph{map} between deformations of GR and theoretical physics. Given a verification of GR through a bottom-up approach, one can then use this map to place specific constraints on physical interactions that cannot present in the observations carried out. This mapping, of course, is critically important, since without it one would not know what physics one is constraining with the given observations. This is indeed the route most commonly taken in observations of the extreme gravity regime, as we will see in this review. 

\subsection{Gravitational-wave tests versus other tests of general relativity}

Gravitational wave tests differ from other tests of GR in many ways. Perhaps one of the most important differences is the spacetime regime gravitational waves sample. Indeed, as already mentioned, gravitational waves have access to the most extreme gravitational environments in Nature. Moreover, gravitational waves travel essentially unimpeded from their source to Earth, and thus, they do not suffer from issues associated with obscuration. Gravitational waves also exist in the absence of luminous matter, thus allowing us to observe electromagnetically dark objects, such as black hole inspirals.

This last point is particularly important as gravitational waves from inspiraling black-hole binaries are one of the cleanest astrophysical systems in Nature. In the last stages of inspiral, when such gravitational waves are detectable by ground-based interferometers, the evolution of binary black holes is essentially unaffected by any other matter or electromagnetic fields present in the system. As such, one does not need to deal with uncertainties associated with astrophysical matter. Unlike tests of Einstein's theory with accretion disk observations, binary black hole gravitational wave tests may well be the cleanest probes of GR.

Of course, what is an advantage here, can be also a huge disadvantage in another context. Gravitational waves from compact binaries are intrinsically transient, i.e.~they turn on for a certain amount of time and then shut off. This is unlike binary pulsar systems, for which astrophysicists have already collected tens of years of data. Moreover, gravitational wave tests rely on specific detections that cannot be anticipated beforehand. This is in contrast to Earth-based laboratory experiments, where one has complete control over the experimental setup. Finally, the intrinsic weakness of gravitational waves makes detection a very difficult task that requires complex data-analysis algorithms to extract signals from the noise. As such, gravitational wave tests are limited by the signal-to-noise ratio and affected by systematics associated with the modeling of the waves, issues that are not as important in other loud astrophysical systems.  

\subsection[Ground-based vs space-based detectors and interferometers
  vs pulsar timing]{Ground-based vs space-based detectors and 
    interferometers vs \linebreak pulsar timing}

This review article focuses only on ground-based detectors, by which we mean both gravitational wave interferometers, such as aLIGO~\cite{Abramovici:1992ah,Abbott:2007kv,Harry:2010zz}, advanced Virgo~\cite{Acernese:2005yh,Acernese:2007zze} and the Einstein Telescope (ET)~\cite{Punturo:2010zz,Sathyaprakash:2012jk}, as well as pulsar timing arrays (for a review of gravitational wave tests of GR with space-based detectors, see~\cite{Gair:2012nm,Yagi:2013du}). Ground-based detectors have the limitation of being contaminated by man-made and Nature-made noise, such as ground and air traffic, logging, earthquakes, ocean tides and waves, which are clearly absent in space-based detectors. Ground-based detectors, however, have the clear benefit that they can be continuously upgraded and repaired in case of malfunction, which is obviously not possible with space-based detectors. 

As far as tests of GR are concerned, there is a drastic difference in space-based and ground-based detectors: the gravitational wave frequencies these detectors are sensitive to. For various reasons that we will not go into, space-based interferometers are likely to have a million kilometer long arms, and thus, be sensitive in the milli-Hz band. On the other hand, ground-based interferometers are bound to the surface and curvature of the Earth, and thus, they have kilometer-long arms and are sensitive in the deca- and hecta-Hz band. Different types of interferometers are then sensitive to different types of gravitational wave sources. For example, when considering binary coalescences, ground-based interferometers are sensitive to late inspirals and mergers of neutron stars and stellar-mass black holes, while space-based detectors will be sensitive to supermassive black hole binaries with masses around $10^{5}\,M_{\odot}$.

The impact of a different population of sources in tests of GR depends on the particular modified gravity theory considered. When studying quadratic gravity theories, as we see in Section~\ref{section:alt-theories}, the Einstein-Hilbert action is modified by introducing higher order curvature operators, which are naturally suppressed by powers of the inverse of the radius of curvature of the system. Thus, space-based detectors will not be ideal at constraining these theories with supermassive black holes, as their radius of curvature is much larger than that of the stellar-mass black holes at merger that ground-based detectors observe. However, although space-based detectors will not be sensitive to neutron-star--binary coalescences, they are expected to detect the inspiral of a supermassive black hole with a neutron star or a stellar-mass black hole. In these extreme mass-ratio inspirals, quadratic gravity modifications sourced by the smaller object will be dominant (since recall these modifications scale \textit{inversely} with the radius of curvature), and thus, may be constrained with space-based detectors.

Space-based detectors are also unique in their potential to probe the spacetime geometry of supermassive black holes through gravitational waves emitted during extreme--mass-ratio inspirals. In these inspirals, the stellar-mass compact object is on a generic decaying orbit around the supermassive black hole, generating millions of cycles of gravitational waves in the sensitivity band of space-based detectors (in fact, they can easily out-live the detector itself!). Therefore, even small changes to the radiation-reaction force, or to the background geometry, can lead to noticeable effects in the waveform observable and thus to stringent tests of GR~\cite{AmaroSeoane:2007aw,Ryan:1995wh,Ryan:1997hg,Kesden:2004qx,Glampedakis:2005cf,Barack:2006pq,Li:2007qu,Gair:2007kr,Sopuerta:2009iy,Yunes:2009ry,Apostolatos:2009vu,LukesGerakopoulos:2010rc,Gair:2011ym,Contopoulos:2011dz,Canizares:2012ji,Gair:2012nm,Perkins:2020tra,Maselli:2020zgv,Maselli:2021men,Barausse:2020rsu,LISA:2022kgy}. 

Space-based detectors also have the advantage of range, which is particularly important when considering modified gravity effects that accumulate with the distance traveled by the gravitational wave, e.g.~theories in which gravitons do not travel at light speed~\cite{Mirshekari:2011yq}. Space-based detectors have a horizon distance much larger than second-generation ground-based detectors; the former can see black-hole mergers to redshifts of order 10 if there are any at such early times in the universe, while the latter are confined to events within redshift 1. Gravitational waves emitted from distant regions in spacetime need a longer time to propagate from the source to the detectors. Thus, theories that modify the propagation of gravitational waves will be better constrained by space-based detectors than second-generation ground-based detectors~\cite{Will:1997bb,Berti:2004bd,Stavridis:2009mb,Yagi:2009zm,Arun:2009pq,Keppel:2010qu,Mirshekari:2011yq,Berti:2011jz,Perkins:2020tra}.

Another important difference between detectors is their response to an impinging gravitational wave. Ground-based detectors, as we see in Section~\ref{section:detectorsandtechniques}, cannot separate between the two possible scalar modes (the longitudinal and the breathing modes) of metric theories of gravity, due to an intrinsic degeneracy in the response functions. Space-based detectors in principle also possess this degeneracy, but they may be able to break it through Doppler modulation if the interferometer orbits the Sun. Pulsar timing arrays, on the other hand, lack this degeneracy altogether, and thus, they can in principle constrain the existence of these different  polarizations independently. 

One way in which pulsar-timing arrays differ from interferometers in their potential to test GR is the frequency space they are most sensitive to. Interferometers can observe the late inspiral and merger of compact binaries, while pulsar timing arrays are restricted to gravitational waves emitted in the very early inspiral. This is why the latter do not need very accurate waveform templates that account for the highly-dynamical and non-linear nature of gravity; leading-order quadrupole waveforms are sufficient~\cite{Corbin:2010kt}. In turn, this implies that pulsar timing arrays cannot constrain theories that only deviate significantly from GR in the late inspiral or merger, while they are exceptionally well-suited for constraining low-frequency deviations.  

We therefore see a complementarity emerging: different detectors can test GR in different complementary regimes:
\begin{itemize}
\item Ground-based detectors are best at constraining non-GR effect that are largest in the late inspiral and merger of stellar-mass compact binaries, including both conservative and dissipative modifications. 
\item Space-based detectors are best at constraining modifications to the propagation of gravitational waves, the geometry of supermassive black holes and corrections to the radiation-reaction force in extreme mass-ratio inspirals.
\item Pulsar-timing arrays are best at constraining the polarization content of gravitational radiation and any deviation from GR that dominates at low orbital frequencies. 
\end{itemize}
Through the simultaneous implementation of all these tests, GR can be put on a much firmer footing in all parts of the extreme gravity regime.   

\subsection{Notation and conventions}

We follow mainly the notation of~\cite{Misner:1973cw}, where Greek indices stand for spacetime coordinates and spatial indices in the middle of the alphabet $(i,j,k,\ldots)$ for spatial indices. Parenthesis and square brackets in index lists stand for symmetrization and anti-symmetrization respectively, e.g.~$A_{(\mu \nu)} = (A_{\mu \nu} + A_{\nu \mu})/2$ and  $A_{[\mu \nu]} = (A_{\mu \nu} - A_{\nu \mu})/2$. Partial derivatives with respect to spacetime and spatial coordinates are denoted $\partial_{\mu} A = A_{,\mu}$ and $\partial_{i} A = A_{,i}$ respectively. Covariant differentiation is denoted $\nabla_{\mu} A = A_{;\mu}$, multiple covariant derivatives $\nabla^{\mu \nu \ldots} = \nabla^{\mu} \nabla^{\nu} \ldots$, and the curved spacetime D'Alembertian $\square A = \nabla_{\mu} \nabla^{\mu} A$. The determinant of the metric $g_{\mu \nu}$ is $g$, $R_{\mu \nu \delta \sigma}$ is the Riemann tensor, $R_{\mu \nu}$ is the Ricci tensor, $R$ is the Ricci scalar and $G_{\mu \nu}$ is the Einstein tensor. The Levi-Civita tensor and symbol are $\epsilon^{\mu \nu \delta \sigma}$ and $\bar{\epsilon}^{\mu \nu \delta \sigma}$ respectively, with $\bar{\epsilon}^{0123} = +1$ in an orthonormal, positively oriented frame. We use geometric units ($G = c = 1$) and the Einstein summation convention is implied. 

We will be mostly concerned with {\emph{metric theories}}, where gravitational radiation is only defined much farther than a gravitational wave wavelength from the source. In this far- or radiation-zone, the metric tensor can be decomposed as
\begin{equation}
g_{\mu \nu} = \eta_{\mu \nu} + h_{\mu \nu}\,,
\label{eq:hab-def}
\end{equation}
with $\eta_{\mu \nu}$ the Minkowski metric and $h_{\mu \nu}$ the metric perturbation. If the theory considered has additional fields $\phi$, these can also be decomposed in the far-zone as
\begin{equation}
\phi = \phi_{0} + \psi\,,
\label{eq:psi-def}
\end{equation}
with $\phi_{0}$ the background value of the field and $\psi$ a perturbation. With such a decomposition, the field equations for the metric will usually be wave equations for the metric perturbation and for the field perturbation, in a suitable gauge.

\newpage
\section{Modified Gravity Theories}
\label{section:alt-theories}
%

In this section, we discuss some of the many possible modified gravity theories that have been studied in the context of gravitational-wave tests. We begin with a description of the 
theoretically desirable properties that such theories must have. We then proceed with a review of the theories so far explored as far as gravitational waves are concerned. We will 
leave out the description of many theories in this chapter, especially those which currently lack a gravitational-wave analysis; we refer the interested reader to~\cite{Berti:2015itd,Barack:2018yly}. We 
will conclude with a brief description of unexplored theories as possible avenues for future research.

\subsection{Desirable theoretical properties}
\label{sec:properties}

The space of possible theories is infinite, and thus, one is tempted to reduce it by considering a subspace that satisfies a certain number of properties. Although the number and 
details of such properties depend on the theorist's taste, there is at least one {\emph{fundamental property}} that all scientists would agree on:
\begin{enumerate}
\item {\textbf{Precision Tests}}. The theory must produce predictions that pass all solar system, binary pulsar, cosmological, gravitational-wave and experimental tests that have been 
carried out. 
\end{enumerate}
This requirement can be further divided into the following:
\begin{enumerate}
\item[]
\begin{enumerate}
\item[1.a] {\textbf{General Relativity Limit}}. There must exist some limit, continuous or discontinuous, such as the weak-field one, in which the predictions of the theory are consistent with 
those of GR within experimental precision. 
\item[1.b] {\textbf{Existence of Known Solutions}}~\cite{walds-presentation}. The theory must admit solutions that correspond to observed phenomena, including but not limited to (nearly) 
flat spacetime, (nearly) Newtonian stars, and cosmological solutions. 
\item[1.c] {\textbf{Stability of Solutions}}~\cite{walds-presentation}. The special solutions described in Property~(1.b) must be stable to small perturbations on timescales smaller than the 
age of the Universe. For example, perturbations to (nearly) Newtonian stars, such as impact by asteroids, should not render such solutions unstable.
\end{enumerate}
\end{enumerate}
Of course, these properties  are not all necessarily independent, as the existence of a weak-field limit usually also implies the existence of known solutions. On the other hand, the 
mere existence of solutions does not necessarily imply that these are stable. 

In addition to these fundamental requirements, one might also wish to require that any new modified gravity theory possesses certain {\emph{theoretical properties}}. These properties 
will vary depending on the theorist, but the two most common ones are listed below:
\begin{enumerate}
\item[2.] {\textbf{Well-motivated from Fundamental Physics}}. There must be some fundamental theory or principle from which the modified theory (effective or not) derives. This 
fundamental theory would solve some fundamental problem in physics, such as late time acceleration or the incompatibility between Quantum Mechanics and GR. 
\item[3.] {\textbf{Well-posed Initial-Value Formulation}}~\cite{walds-presentation}. A wide class of freely specifiable initial data must exist, such that there is a uniquely determined solution 
to the modified field equations that depends continuously on this data.
\end{enumerate} 
The second property goes without saying at some level, as one expects modified-gravity--theory constructions to be motivated from some (perhaps yet incomplete) quantum-
gravitational description of nature. As for the third property, the continuity requirement is necessary because otherwise the theory would lose predictive power, given that initial 
conditions can only be measured to a finite accuracy. Moreover, small changes in the initial data should not lead to solutions outside the causal future of the data; that is, causality 
must be preserved. Section~\ref{well-posed} expands on this well-posedness property further. 

One might be concerned that Property~(2) automatically implies that any predicted deviation from astrophysical observables will be too small to be detectable. This argument usually 
goes as follows. Any quantum gravitational correction to the action will ``naturally'' introduce at least one new scale, and this, by dimensional analysis, must be the Planck scale. Since 
this scale is usually assumed to be larger than 1~TeV in natural units (or $10^{-35}\mathrm{\ m}$ in geometric units), gravitational-wave observations will never be able to observe 
quantum-gravitational modifications (see, e.g.,~\cite{Dubovsky:2007zi} for a similar argument). In our view, such arguments can be extremely dangerous, since they induce a certain 
theoretical bias in the search for new phenomena. For example, let us consider the supernova observations of the late time expansion of the universe that led to the discovery of the 
cosmological constant. The above argument certainly fails for the cosmological constant, which on dimensional arguments is over 100 orders of magnitude too small. If the supernova 
teams had respected this argument, they would not have searched for a cosmological constant in their data. Today, we try to explain our way out of the failure of such dimensional 
arguments by claiming that there must be some exquisite cancellation that renders the cosmological constant small; but this, of course, came only {\emph{after}} the constant had been 
measured. One is not trying to argue here that cancellations of this type are common and that quantum gravitational modifications are necessarily expected in gravitational-wave 
observations. Rather, we are arguing that one should remain {\emph{agnostic}} about what is expected and what is not, and allow oneself to be surprised without quenching the 
potential for new discoveries that accompanies the era of gravitational-wave astrophysics. 

One last property that we wish to consider for the purposes of this review is the following:
\begin{enumerate}
\item[4.] {\textbf{Extreme Gravity Inconsistency}}. The theory must lead to observable deviations from GR in extreme gravity.
\end{enumerate}
Many modified gravity models have been proposed that pose {\emph{infrared}} or cosmological modifications to GR, aimed at explaining certain astrophysical or cosmological 
observables, like the late expansion of the Universe. Such modified models usually reduce to GR in the strong-field, and particularly, in the extreme-gravity regime, for example via a 
Vainshtein like mechanism~\cite{Vainshtein:1972sx,Deffayet:2001uk,Babichev:2013usa} in a static spherically-symmetric context. Extending this mechanism to highly-dynamical 
extreme gravity scenarios has not been fully worked out yet~\cite{deRham:2012fg,deRham:2012fw}. Gravitational-wave tests of GR, however, are concerned with modified theories 
that predict deviations in extreme gravity, precisely where cosmological modified models do not. Clearly, Property~(4) is not necessary for a theory to be a valid description of nature. 
This is because a theory might be identical to GR in the weak, strong and extreme gravity regimes, yet different at the Planck scale, where it would be unified with quantum 
mechanics. However, Property~(4) is a desirable feature if one is to test this theory with gravitational-wave observations.

\subsection{Well-posedness and effective theories}
\label{well-posed}

Property~(3) not only requires the existence of an initial-value
formulation, but also that it be well-posed, which is not necessarily
guaranteed. For example, the Cauchy--Kowalewski theorem states that a
system of $n$ partial differential equations for $n$ unknown functions
$\phi_{i}$ of the form $\phi_{i,tt} =
F_{i}(x^{\mu};\phi_{j,\mu};\phi_{j,ti};\phi_{j,ik})$, with $F_{i}$
analytic functions has an initial-value formulation (see,
e.g.,~\cite{Wald:1984rg}). This theorem, however, does not guarantee
continuity or the causal conditions described above. For this, one has
to rely on more general energy arguments, for example constructing a
suitable energy measure that obeys the dominant energy condition and
using it to show well-posedness (see,
e.g.,~\cite{Hawking:1973uf,Wald:1984rg}). One can show that
second-order, hyperbolic partial differential equations, i.e., equations of the form
\be
\nabla^{\mu} \nabla_{\mu}\phi + A^{\mu} \nabla_{\mu} \phi + B \phi + C = 0\,,
\ee
where $A^{\mu}$ is an arbitrary vector field and $(B,C)$ are smooth functions, have a well-posed initial-value formulation. Moreover, the Leray theorem proves that any quasilinear, 
diagonal, second-order hyperbolic system also has a well-posed initial-value formulation~\cite{Wald:1984rg}. 

Proving the well-posedness of an initial-value formulation for systems of higher-than-second-order, partial differential equations is much more difficult. In fact, to our knowledge, no 
general theorems exist of the type described above that apply to third, fourth or higher-order, partial, non-linear and coupled differential equations. Usually, one resorts to the 
Ostrogradski theorem~\cite{Ostro} to rule out (or at the very least cast serious doubt on) theories that lead to such higher-order field equations.  Ostrogradski's theorem states that 
Lagrangians that contain terms with higher than first time-derivatives possess a linear instability in the Hamiltonian (see, e.g.,~\cite{Woodard:2006nt} for a nice review)
\epubtkFootnote{Stability and well-posedness are not the same concepts and they do not necessarily imply each other. For example, a well-posed theory might have stable and 
unstable solutions. For ill-posed theories, it does not make sense to talk about stability of solutions.}. As an example, consider the Lagrangian density
\be
{\cal{L}} = \frac{m}{2} \dot{q}^{2} - \frac{m \omega^{2}}{2} q^{2} - \frac{g m}{2 \omega^{2}} \ddot{q}^{2},
\label{eq:L}
\ee
whose equations of motion,
\be
\ddot{q} + \omega^{2} q = - \frac{g}{\omega^{2}} \ddddot{q}\,,
\label{eq:sho-ho}
\ee
obviously contain higher derivatives. The exact solution to this differential equation is
\be
q = A_{1} \cos{k_{1} t} + B_{1} \sin{k_{1} t} + A_{2} \cos{k_{2} t} + B_{2} \sin{k_{2} t}\,, 
\label{eq:exact-sol}
\ee
where $(A_{i},B_{i})$ are constants and $k_{1,2}^{2}/\omega^{2} = (1 \mp \sqrt{1 - 4 g})/(2 g)$. The on-shell Hamiltonian is then
\be
H = \frac{m}{2} \sqrt{1 - 4 g} k_{1}^{2} \left(A_{1}^{2} + B_{1}^{2}\right) - \frac{m}{2} \sqrt{1 - 4 g} k_{2}^{2} \left(A_{2}^{2} + B_{2}^{2}\right)\,,
\ee
from which it is clear that mode~1 carries positive energy, while mode~2 carries negative energy and forces the Hamiltonian to be unbounded from below. The latter implies that 
dynamical degrees of freedom can reach arbitrarily negative energy states. If interactions are present, then an ``empty'' state would instantaneously decay into a collection of 
positive and negative energy particles, which cannot describe the Universe we live in~\cite{Woodard:2006nt}.  

The Ostrogradski theorem~\cite{Ostro}, however, can be evaded if the Lagrangian in Eq.~\eqref{eq:L} describes an {\emph{effective theory}}, i.e.~a theory that is a truncation of a 
more general or complete theory. Let us reconsider the particular example above, assuming now that the coupling constant $g$ is an effective theory parameter and Eq.~\eqref{eq:L} 
is only valid to linear order in $g$. One approach is to search for perturbative solutions of the form $q_{\mathrm{pert}} = x_{0} + g x_{1} + \ldots$, which leads to the system of 
differential equations
\be
\ddot{x}_{n} + \omega^{2} x_{n} = -\frac{1}{\omega^{2}} \ddddot{x}_{n-1}\,,
\ee
with $x_{-1} = 0$. Solving this set of $n$ differential equations and resumming, one finds
\be
q_{\mathrm{pert}} = A_{1} \cos{k_{1} t} + B_{1} \sin{k_{1} t}\,. 
\ee
Notice that $q_{\mathrm{pert}}$ contains only the positive (well-behaved) energy solution of Eq.~\eqref{eq:exact-sol}, i.e.~perturbation theory acts to retain only the well-behaved, 
stable solution of the full theory in the $g\to0$ limit. One can also think of the perturbative theory as the full theory with additional constraints, i.e.~the removal of unstable modes, 
which is why such an analysis is sometimes called {\emph{perturbative constraints}}~\cite{Cooney:2008wk,Cooney:2009rr,Yunes:2009hc}. 

Another way to approach effective field theories that lead to equations of motion with higher-order derivatives is to apply the method of {\emph{order-reduction}}. In this method, one 
substitutes the low-order derivatives of the field equations into the high-order derivative part, thus rendering the resulting new theory usually well-posed. One can think of this as a 
series resummation, where one changes the non-linear behavior of a function by adding uncontrolled, higher-order terms. Let us provide an explicit example by reconsidering the 
theory in Eq.~\eqref{eq:L}. To lowest order in $g$, the equation of motion is that of a simple harmonic oscillator, 
\be
\ddot{q} + \omega^{2} q = {\cal{O}}(g)\,,
\label{eq:sho-lo}
\ee
which is obviously well posed. One can then order-reduce the full equation of motion, Eq.~\eqref{eq:sho-ho}, by substituting Eq.~\eqref{eq:sho-lo} into the right-hand side of 
Eq.~\eqref{eq:sho-ho}. Doing so, one obtains the order-reduced equation of motion 
\be
\ddot{q} + \omega^{2} q =  g \ddot{q} + {\cal{O}}(g^{2})\,,
\ee
which clearly now has no high-order derivatives and is well posed provided $g \ll 1$. The solution to this order-reduced differential equation is $q_{\mathrm{pert}}$ once more, but with 
$k_{1}$ linearized in $g \ll 1$. Therefore, the solutions obtained with a perturbative decomposition and with the order-reduced equation of motion are the same to linear order  in $g$. 
Of course, since an effective field theory is only defined to a certain order in its perturbative parameter, both treatments are equally valid, with the unstable mode effectively removed in 
both cases.

Such a perturbative analysis, however, can say nothing about the well-posedness of the full theory from which the effective theory derives, or of the effective theory if treated as an 
exact one (i.e.~not as a perturbative expansion). In fact, a well-posed full theory may have both stable and unstable solutions. The arguments presented above only discuss the 
stability of solutions in an effective theory, and thus, they are self-consistent only within their perturbative scheme. A full theory may have non-perturbative instabilities, but these can 
only be studied once one has a full (non-truncated in $g$) theory, from which Eq.~\eqref{eq:L} derives as a truncated expansion. Lacking a full quantum theory of nature, quantum 
gravitational models are usually studied in a truncated low-energy expansion, where the leading-order piece is GR and higher order pieces are multiplied by a small coupling constant. 
One can perturbatively explore the well-behaved sector of the truncated theory about solutions to the leading-order theory. Such an analysis, however, is incapable of answering 
questions about well-posedness or non-linear stability of the full theory.

\subsection{Explored theories}

In this subsection we briefly describe the theories that have so far been studied in some depth as far as gravitational waves are concerned. In particular, we focus only on those 
theories that have been sufficiently studied so that predictions of the expected gravitational waveforms (the observables of gravitational-wave detectors) have been obtained for at 
least a typical source, such as the quasi-circular inspiral of a compact binary.

\subsubsection{Scalar-tensor theories}
\label{sec:ST}

The classical type of scalar-tensor theory \cite{PhysRev.124.925,Damour:1992we,Faraoni:1998qx,Faraoni:1999hp,Fujii:2003pa,Goenner:2012cq} is defined by the Einstein-frame action (where 
we will restore Newton's gravitational constant $G$ in this section)
\begin{equation}
S_{\mathrm{ST}}^{\mathrm{(E)}} = \frac{1}{16 \pi G} \int d^{4}x \sqrt{-g} \left[ R - 2 g^{\mu \nu} \left(\partial_{\mu}\varphi\right) \left(\partial_{\nu} \varphi\right) - V(\varphi)\right]
+ S_{\mathrm{mat}}[\psi_{\mathrm{mat}},A^{2}(\varphi) g_{\mu \nu}],
\label{ST-gen-action}
\end{equation}
where $\varphi$ is a scalar field, $A(\varphi)$ is a coupling function, $V(\varphi)$ is a potential function, $\psi_{\mathrm{mat}}$ represents matter degrees of freedom and $G$ is 
Newton's constant in the Einstein frame. For more details on this theory, we refer the interested reader to the reviews~\cite{lrr-2006-3,Will:1993ns}. 

The Einstein frame is not the frame where the metric governs clocks and rods, and thus, it is convenient to recast the theory in the Jordan frame through the conformal transformation 
$\tilde{g}_{\mu \nu} = A^{2}(\varphi) g_{\mu \nu}$:
\begin{equation}
S_{\mathrm{ST}}^{\mathrm{(J)}} = \frac{1}{16 \pi G} \int d^{4}x \sqrt{-\tilde{g}} \left[ \phi \; \tilde{R} - \frac{\omega(\phi)}{\phi} \tilde{g}^{\mu \nu} \left(\partial_{\mu} \phi\right) 
\left(\partial_{\nu} \phi\right) - \phi^{2} V \right]
+ S_{\mathrm{mat}}[\psi_{\mathrm{mat}},\tilde{g}_{\mu \nu}],
\label{ST-action}
\end{equation}
where $\tilde{g}_{\mu \nu}$ is the physical metric, the new scalar field $\phi$ is defined via $\phi \equiv A^{-2}$, the coupling field is $\omega(\phi) \equiv (\alpha^{-2} - 3)/2$ and $
\alpha \equiv A_{,\varphi}/A$. When cast in the Jordan frame, it is clear that scalar-tensor theories are metric theories (see~\cite{lrr-2006-3} for a definition), since the matter sector 
depends only on matter degrees of freedom and the physical metric (without a direct coupling of the scalar field). When the coupling $\omega(\phi) = \omega_{\mathrm{BD}}$ is 
constant, then Eq.~\eqref{ST-action} reduces to the massless version of Jordan--Fierz--Brans--Dicke theory~\cite{PhysRev.124.925}.  

The modified field equations in this classical scalar-tensor theory in the Einstein frame are
\begin{align}
\square \varphi &= \frac{1}{4} \frac{dV}{d\varphi} - 4 \pi G  \frac{\delta S_{\mathrm{mat}}}{\delta \varphi}\,, 
\nonumber \\
G_{\mu \nu} &= 8 \pi G \left( T_{\mu \nu}^{\mathrm{mat}} + T_{\mu \nu}^{(\varphi)}\right)\,,
\end{align}
where
\begin{equation}
T_{\mu \nu}^{(\varphi)} = \frac{1}{4 \pi} \left[\varphi_{,\mu} \varphi_{,\nu} - \frac{1}{2} g_{\mu \nu} \varphi_{,\delta} \varphi^{,\delta} - \frac{1}{4} g_{\mu \nu} V(\varphi)\right]
\end{equation}
is a stress-energy tensor for the scalar field. The matter stress--energy tensor is not constructed from the Einstein-frame metric alone, but by the combination $A(\varphi)^{2} g_{\mu 
\nu}$. In the Jordan frame and neglecting the potential, the modified field equations are~\cite{Will:1993ns}
\begin{align}
\tilde{\square} \phi &= \frac{1}{3 + 2 \omega(\phi)} \left( 8 \pi T^{\mathrm{mat}} - \frac{d \omega}{d\phi} \tilde{g}^{\mu \nu} \phi_{,\mu} \phi_{,\nu} \right),
\nonumber \\
\tilde{G}_{\mu \nu} &= \frac{8 \pi G}{\phi} T_{\mu \nu}^{\mathrm{mat}} + \frac{\omega}{\phi^{2}} \left(\phi_{,\mu} \phi_{,\nu} 
- \frac{1}{2} \tilde{g}_{\mu \nu} \tilde{g}^{\sigma \rho} \phi_{,\sigma} \phi_{,\rho} \right)
+ \frac{1}{\phi} \left( \phi_{,\mu \nu} - \tilde{g}_{\mu \nu} \tilde{\square} \phi \right),
\end{align}
where $T^{\mathrm{mat}}$ is the trace of the matter stress-energy tensor $T_{\mu \nu}^{\mathrm{mat}}$ constructed from the physical metric $\tilde{g}_{\mu \nu}$. The form of the 
modified field equations in the Jordan frame suggest that in the weak-field limit one may consider scalar-tensor theories as modifying Newton's gravitational constant via $G \to G(\phi) 
= G/\phi$. 

Using the decompositions of Eqs.~\eqref{eq:hab-def}-\eqref{eq:psi-def}, the field equations of massless Jordan--Fierz--Brans--Dicke theory can be linearized in the Jordan frame to 
find (see, e.g.,~\cite{Will:1989sk})
\begin{equation}
\square_{\eta} \theta^{\mu \nu} = -16 \pi \tau^{\mu \nu}\,,
\qquad
\square_{\eta} \psi = -16 \pi S\,,
\label{eq:ST-EOM}
\end{equation}
where $\square_{\eta}$ is the D'Alembertian operator of flat spacetime, we have defined a new metric perturbation
\begin{equation}
\theta^{\mu \nu} = h^{\mu \nu} - \frac{1}{2} \eta^{\mu \nu} h - \frac{\psi}{\phi_{0}} \eta^{\mu \nu}\,,
\end{equation}
with $h$ the trace of the metric perturbation and
\begin{align}
\tau^{\mu \nu} &= \phi_{0}^{-1} T^{\mu \nu}_{\mathrm{mat}} + t^{\mu \nu}\,,
\\
S &= - \frac{1}{6 + 4 \omega_{BD}} \left(T^{\mathrm{mat}} - 3 \phi  \frac{\partial T^{\mathrm{mat}}}{\partial \phi}\right) \left(1 - \frac{\theta}{2} - \frac{\psi}{\phi_{0}} \right) - \frac{1}{16 \pi} 
\left(\psi_{,\mu \nu} \theta^{\mu \nu} + \frac{1}{\phi_{0}} \phi_{,\mu} \psi^{,\mu} \right)\,,
\end{align}
with cubic remainders in either the metric perturbation or the scalar perturbation. The quantity $\partial T^{\mathrm{mat}} / \partial \phi$ arises in an effective point-particle theory, 
where the matter action is a functional of both the Jordan-frame metric and the scalar field. The quantity $t^{\mu \nu}$ is a function of quadratic or higher order in $\theta^{\mu \nu}$ or 
$\psi$. These equations can now be solved given a particular physical system, as done for quasi-circular binaries in~\cite{Will:1989sk,Saijo:1997wu,Ohashi:1996uz}. Given the above 
evolution equations, Jordan--Fierz--Brans--Dicke theory possesses a scalar (spin-0) mode, in addition to the two transverse-traceless (spin-2) modes of GR, i.e.~Jordan--Fierz--Brans--Dicke theory is of Type $N_{3}$ in the $E(2)$ classification~\cite{Eardleyprd,lrr-2006-3}. 

Let us now discuss whether these scalar-tensor theories satisfy the properties discussed in Section~\ref{sec:properties}, starting with Property 1. 
Massless Jordan--Fierz--Brans--Dicke theory agrees with all known experimental tests provided $\omega_{\mathrm{BD}} > 4 \times 10^{4}$, 
a bound imposed by the tracking of the Cassini spacecraft through observations of the Shapiro time 
delay~\cite{Bertotti:2003rm}. Massive Jordan--Fierz--Brans--Dicke theory has been constrained to $\omega_{\mathrm{BD}} > 4 \times 10^{4}$ and $m_{\mathrm{s}} < 2.5 
\times 10^{-20} {\mathrm{\ eV}}$, with $m_{\mathrm{s}}$ the mass of the scalar field~\cite{Perivolaropoulos:2009ak,Alsing:2011er}. Of course, these bounds are not independent, as 
when $m_{\mathrm{s}} \to 0$ one recovers the standard massless constraint, while when $m_{\mathrm{s}} \to \infty$, $\omega_{\mathrm{BD}}$ cannot be bounded as the scalar 
becomes non-dynamical. Observations of the Nordtvedt effect with Lunar Laser Ranging observations, as well as observations of the orbital period derivative of white-dwarf/neutron-
star binaries, yield similar or slightly stronger constraints~\cite{Damour:1996ke,Damour:1998jk,Alsing:2011er,Freire:2012mg,Voisin:2020lqi,Kramer:2021jcw}. For example, the Cassini bound was recently updated by the observation of the Nordtvedt effect in a pulsar triple system J0337+1715~\cite{Ransom:2014xla,Archibald:2018oxs} to $\omega_\mathrm{BD} > 1.4\times 10^5$~\cite{Voisin:2020lqi}.  Neglecting any homogeneous, cosmological solutions to the scalar-field 
evolution equation, it is clear that in the limit $\omega \to \infty$ one recovers GR, i.e.~scalar-tensor theories have a continuous limit to Einstein's theory, but see~\cite{Faraoni:1999yp} 
for caveats for certain spacetimes. Moreover, \cite{Salgado:2008xh,2007CQGra..24.5667L,Wald:1984rg} have verified that scalar-tensor theories with minimal or non-minimal coupling 
in the Jordan frame can be cast in a strongly-hyperbolic form, and thus, they possess a well-posed initial-value formulation. Therefore, scalar-tensor theories possess both Properties 
(1) and (3). 

These classical scalar-tensor theories also possess Property~(2), since they can be derived from the low-energy limit of certain string theories. The integration of string quantum fluctuations leads to 
a higher-dimensional string theoretical action that reduces locally to a field theory similar to a scalar-tensor one~\cite{Garay:1992ej,1985NuPhB.261....1F}, the mapping being $\phi = 
e^{-2 \psi}$, with $\psi$ one of the string moduli fields~\cite{Damour:1994zq,Damour:1994ya}. Moreover, scalar-tensor theories can be mapped to $f(R)$ theories, where one replaces 
the Ricci scalar by some functional of $R$. In particular, one can show that $f(R)$ theories are equivalent to Jordan--Fierz--Brans--Dicke theory with $\omega_{\mathrm{BD}} = 0$, via the mapping $
\phi = df(R)/dR$ and $V(\phi) = R \; df(R)/dR - f(R)$~\cite{Chiba:2003ir,Sotiriou:2006hs}. For a review on this topic, see~\cite{lrr-2010-3}.

As for Property~(4), these classical scalar-tensor theories are not typically built with the aim to introduce extreme gravity corrections to GR\epubtkFootnote{The process of spontaneous scalarization 
in a particular type of scalar-tensor theory~\cite{Damour:1992we,Damour:1993hw} does introduce strong-field modifications because it induces non-perturbative corrections that can 
affect the structure of neutron stars. These subclass of scalar-tensor theories would satisfy Property~(4).}. Instead, they typically lead to modifications of Einstein's theory in the weak-
field, i.e.~modifications that dominate in scenarios with sufficiently weak gravitational interactions. Although this might seem strange, it is natural if one considers, for example, one of 
the key modifications introduced by scalar-tensor theories: the emission of dipolar gravitational radiation. Such dipolar emission dominates over the General Relativistic quadrupolar 
emission for systems at large separation, where the gravitational compactness is small, such as in binary pulsars or in the very early inspiral of compact binaries. Therefore, one would 
expect scalar-tensor theories to be best constrained by experiments or observations of compact binaries with large separations, as it has been explicitly shown in~\cite{Yunes:2011aa}.  

Black holes and stars continue to exist in these scalar-tensor theories. Stellar configurations are modified from their GR profile~\cite{Will:1989sk,Damour:1996ke,Harada:1997mr,Harada:1998ge,Tsuchida:1998jw,Damour:1998jk,Sotani:2004rq,DeDeo:2003ju,Sotani:2012eb,Horbatsch:2010hj}, while black holes are not. 
Indeed, Hawking~\cite{Hawking:1972qk,Dykla:1972zz,Hawking:1971tu,Carter:1971zc,Israel:1967za,Robinson:1975bv} has 
proved that Jordan--Fierz--Brans--Dicke black holes that are stationary and the endpoint of gravitational collapse are identical to those of GR. This proof has been extended to a general 
class of scalar-tensor models~\cite{Sotiriou:2011dz}. That is, stationary black holes radiate any excess ``hair'', i.e., any additional degrees of freedom, after gravitational 
collapse, a result sometimes referred to as the {\emph{no-hair}} theorem for black holes in scalar-tensor theories. This result has been extended even further to allow for 
quasi-stationary scenarios in generic scalar-tensor theories through the study of extreme-mass ratio inspirals~\cite{Yunes:2011aa} (small black hole in orbit around a much larger one), 
post-Newtonian comparable-mass inspirals~\cite{Mirshekari:2013vb,Lang:2013fna,Lang:2014osa} and numerical simulations of comparable-mass black hole mergers with a non-trivial initial scalar field 
profile~\cite{Healy:2011ef,Berti:2013gfa}. The only way for black holes to grow hair in scalar tensor theories is if one allows for non-trivial boundary conditions 
for the scalar field~\cite{Healy:2011ef,Horbatsch:2011ye} or if one allows for the presence of matter around black holes~\cite{Cardoso:2013opa,Cardoso:2013fwa}.

Let us now discuss going beyond these classical scalar-tensor theories. 
Damour and Esposito-Far\`{e}se~\cite{Damour:1992we,Damour:1993hw} proposed a theory defined by the action in Eq.~\eqref{ST-action} 
but with the conformal factor $A(\varphi) = e^{\alpha \varphi + \beta \varphi^{2}/2}$ or the coupling function $\omega(\phi) = -3/2 - 2\pi G/(\beta \log{\phi})$, where $\alpha$ and 
$\beta$ are constants. When $\beta = 0$, one recovers standard Jordan--Fierz--Brans--Dicke theory. When $\beta \lesssim -4$ (the precise value of $\beta$ depending on the equation of state), non-
perturbative effects that develop if the gravitational energy is large enough can force neutron stars to spontaneously acquire a non-trivial scalar field profile, to {\emph{spontaneously scalarize}}~\cite{Damour:1992we,Damour:1993hw}. Scalar charges of such neutron stars have been tabulated in~\cite{Anderson:2019hio}, while analytic expressions were found in~\cite{Yagi:2021loe} through resummation or in~\cite{Zhao:2019suc,Guo:2021leu} through surrogate modeling.
Moreover, binary neutron stars that initially had no scalar hair in their early inspiral can also acquire it before they merge, either when their binding energy exceeds some 
threshold (\emph{dynamical scalarization}) or due to the presence of an external scalar field (\emph{induced scalarization})~\cite{Barausse:2012da,Palenzuela:2013hsa}. In this way, 
this new class of scalar-tensor theories, as well as the screened scalar tensor theories discussed in~\cite{Zhang:2017srh}, 
were thought to be consistent with Solar System experiments, yet predict modifications in strong gravity scenarios.
Binary pulsar observations have constrained this theory in the $(\alpha,\beta)$ space; very roughly speaking, $\beta > -4$ and $\alpha < 
10^{-3}$~\cite{Damour:1996ke,Damour:1998jk,Freire:2012mg,Shao:2017gwu,Anderson:2019eay,Voisin:2020lqi,Kramer:2021jcw,Zhao:2022vig}.

Scalarization in the standard Damour-Esposito-Far\`ese scalar-tensor theory, however, has been shown to be inconsistent with solar system constraints upon accounting for 
the cosmological evolution of the scalar field~\cite{Damour:1992kf,Damour:1993id,Sampson:2014qqa,Anderson:2016aoi}. Back in the early 1990s, Damour and 
Nordtvedt~\cite{Damour:1992kf,Damour:1993id} showed that for $\beta > 0$, GR is an attractor in cosmological phase space. This means that the scalar field damps out upon 
cosmological evolution, becoming very small at small cosmological redshifts, and thus, passing solar system tests today. When $\beta < 0$, however, \cite{Sampson:2014qqa} 
showed that the cosmological evolution of the field leads to a runaway solution that maximally violates solar system tests today. One is then forced to consider variations of the 
standard theory (e.g.,~changing the conformal coupling function $A(\varphi)$~\cite{Anderson:2016aoi,Mendes:2016fby}, giving the scalar a mass~\cite{Ramazanoglu:2016kul}, etc.) 
or the standard theory with $\beta > 0$~\cite{Palenzuela:2015ima} if one wishes both solar system tests to be passed and scalarization to occur in neutron stars. For example,~\cite{dePireySaintAlby:2017lwc} showed that endowing the scalar field with a mass introduces oscillatory behavior in the scalar field upon cosmological evolution, suppressing the runaway solution and allowing the theory to pass solar system constraints, while still allowing for spontaneous scalarization~\cite{Ramazanoglu:2016kul}. 

Another type of scalar-tensor theory is Horndeski gravity~\cite{Horndeski:1974wa} (see e.g.~\cite{Kobayashi:2019hrl} for a recent review). This theory is the most general model with a single scalar field that leads to field equations with at most second-order derivatives~\cite{Horndeski:1974wa}. This theory has become popular because it was shown to be equivalent to a curved-spacetime, generalized scalar-tensor theory with Galilean shift symmetry~\cite{Deffayet:2009mn}, which is related to theories that aim at explaining the late-time acceleration of the Universe with a modified theory of gravity~\cite{Ratra:1987rm,Caldwell:1997ii,ArmendarizPicon:2000ah,Dvali:2000hr,Nicolis:2008in,Sotiriou:2008rp,Tsujikawa:2010zza,Clifton:2011jh,Nojiri:2010wj}. Horndeski gravity, however, is not a single theory, but rather a class of models, where the action depends on free functional degrees of freedom of the scalar field and its kinetic energy. Its action is given by
\begin{align}
S_{\mathrm{Horndeski}} &= \frac{1}{16 \pi G} \int d^{4}x \sqrt{-g} \big\{  G_{2}(\phi, X)-G_{3}(\phi, X) \square \phi+G_{4}(\phi, X) R+G_{4 X}\left[(\square \phi)^{2}-\phi^{\mu \nu} \phi_{\mu \nu}\right] \nonumber \\
&+G_{5}(\phi, X) G^{\mu \nu} \phi_{\mu \nu}-\frac{G_{5 X}}{6}\left[(\square \phi)^{3}-3 \square \phi \phi^{\mu \nu} \phi_{\mu \nu}+2 \phi_{\mu \nu} \phi^{\nu \lambda} \phi_{\lambda}^{\mu}\right]\big\}
+ S_{\mathrm{mat}}[\psi_{\mathrm{mat}},g_{\mu \nu}], 
\end{align}
where the $G_i(\cdot,\cdot)$ are arbitrary functionals of the scalar field $\phi$ and $X = - \phi_{,\mu} \phi^{,\mu}/2$, with $G_{iX} \equiv \partial G_i/\partial X$.
In an attempt to reduce this class to a smaller subset, one can restrict attention to models that allow for a stable Minkowski background and non-trivial cosmological evolution in the presence of matter with a homogeneous and isotropic spacetime, which defines the Fab Four class of theories~\cite{Charmousis:2011ea,Charmousis:2011bf}. The action of this class of theories is
\begin{align}
S_{\mathrm{Fab-Four}} &= \frac{1}{16 \pi G} \int d^{4}x \sqrt{-g} \left[V_{1}(\phi) R + V_{2}(\phi) R_{\mathrm{GB}} + V_{3}(\phi) G^{\mu \nu} (\partial_{\mu} \phi) (\partial_{\nu} \phi) 
\right. \nonumber \\ 
& \left. \qquad \qquad + V_{4}(\phi) P^{\mu \nu \rho \sigma} (\partial_{\mu} \phi) (\partial_{\nu} \phi)  (\partial_{\rho} \phi) (\partial_{\sigma} \phi) \right] + S_{\mathrm{mat}}[\psi_{\mathrm{mat}},g_{\mu \nu}], 
\label{ST-gen-action-fab-four}
\end{align}
where $R_{\mathrm{GB}}$ is the Gauss-Bonnet invariant, $P^{\mu \nu \rho \sigma}$ is the double-dual Riemann tensor~\cite{Maselli:2016gxk} and the $V_{i}(\phi)$ are free functions of the scalar field. Different choices of the latter define different scalar-tensor theories. For example, when $V_{1}(\phi) = 1$, $V_{2}(\phi) = \phi$ and all other $V_{i}(\phi)$, the theory reduces to Einstein--dilaton--Gauss--Bonnet gravity in the decoupling limit~\cite{Yunes:2011we,Yagi:2011xp}, which we will discuss in more detail in Sec.~\ref{subsec:MQG}. Horndeski theories admit a modified harmonic formulation~\cite{Kovacs:2020pns,Kovacs:2020ywu} that makes the theories well-posed, as long as the coupling parameter in the theories is smaller than all the other length scales in the problem. The well-posedness of scalar-tensor theories was also studied in~\cite{Bezares:2021yek,Lara:2021piy} by applying a formulation inspired by that of M\"uller, Israel and Stewart for relativistic viscous hydrodynamics~\cite{Cayuso:2017iqc,Allwright:2018rut,Cayuso:2020lca,Cayuso:2023aht,Corman:2024cdr}.

Compact black holes and stars have been studied in Horndeski gravity and in the Fab Four subclass. Hui and Nicolis~\cite{Hui:2012qt} have proved that black holes in shift-symmetric Horndeski gravity have no-hair, i.e.~the scalar field does not activate, provided the spacetime is asymptotically flat, static and spherically symmetric, with the scalar field inheriting these symmetries. One can of course break some of the assumptions that Hui and Nicolis used, and thus, construct hairy black hole solutions (see e.g.~\cite{Rinaldi:2012vy,Sotiriou:2013qea,Anabalon:2013oea,Babichev:2013cya,Minamitsuji:2013ura,Sotiriou:2014pfa,Minamitsuji:2014hha,Cisterna:2015uya} or the review by~\cite{Herdeiro:2015waa}). Many of these results were extended in the slow-rotation limit in~\cite{Maselli:2015yva}. Slowly-rotating neutron star solutions in the Fab Four subclass of Horndeski theories were studied in~\cite{Maselli:2016gxk}. This analysis showed that a particular theory (the so-called ``Paul'' member of the Fab Four class) is ruled out due to its inability to allow for neutron star solutions. Sakstein {\it{et al.}}~\cite{Sakstein:2016oel} found very massive neutron star solutions embedded in an asymptotically de Sitter spacetime within a particular type of extended theory. See e.g.~\cite{Chagoya:2018lmv,Kobayashi:2018xvr,Ogawa:2019gjc,Barranco:2021auj,Boumaza:2022abj} for other recent works on relativistic stars in Horndeski and beyond.

\subsubsection{Massive graviton theories}
\label{sec:MG-LV}
Massive graviton theories are those in which the gravitational interaction is propagated by a massive gauge boson, i.e., a graviton with mass $m_{g} \neq 0$ or Compton wavelength $
\lambda_{g} \equiv h/(m_{g} c) < \infty$. Einstein's theory predicts massless gravitons and thus gravitational propagation at light speed, but if this were not the case, then a certain 
delay would develop between electromagnetic and gravitational signals emitted simultaneously at the source. Fierz and Pauli~\cite{Fierz:1939ix} were the first to write down an action 
for a free massive graviton, and ever since then, much work has gone into the construction of such models. For a detailed review, see, e.g.,~\cite{Hinterbichler:2011tt,deRham:2014zqa}.

Gravitational theories with massive gravitons are somewhat well-motivated from a fundamental physics perspective, and thus, one can say they possess Property~(2). Indeed, in loop 
quantum cosmology~\cite{Ashtekar:2003hd,Bojowald:2006da}, the cosmological extension to loop quantum gravity, the graviton dispersion relation acquires holonomy corrections 
during loop quantization that endow the graviton with a mass~\cite{Bojowald:2007cd} $m_{g} = \Delta^{-1/2} \gamma^{-1} (\rho/\rho_{c})$, with $\gamma$ the Barbero--Immirzi 
parameter, $\Delta$ the area operator, and $\rho$ and $\rho_{c}$ the total and critical cosmological energy densities respectively. In string-theory--inspired effective theories, such as 
Dvali's compact, extra-dimensional theory~\cite{Dvali:2000hr} such massive modes also arise. 

Massive graviton modes also occur in many other modified gravity models. In Rosen's bimetric theory~\cite{Rosen:1974ua}, for example, photons and gravitons follow null geodesics 
of different metrics~\cite{lrr-2006-3,Will:1993ns}. In Visser's massive graviton theory~\cite{Visser:1997hd}, the graviton is given a mass at the level of the action through an effective 
perturbative description of gravity, at the cost of introducing a non-dynamical background metric, i.e., a prior geometry. A recent re-incarnation of this model goes by the name 
of new massive gravity, or its generalization bigravity, where again two metric tensors are introduced~\cite{Pilo:2011zz,Paulos:2012xe,Hassan:2011zd,Hassan:2011ea,DeFelice:2013nba,Narikawa:2014fua}. In Bekenstein's Tensor-Vector-Scalar (TeVeS) theory~\cite{Bekenstein:2004ne}, the existence of a scalar and a vector field lead to subluminal GW 
propagation. 

Old massive graviton theories have a theoretical issue, the van Dam---Veltman--Zakharov (vDVZ) discontinuity~\cite{vanDam:1970vg,Zakharov:1970cc}, which is associated with 
Property 1.a, i.e.~a GR limit.  The problem is that certain predictions of massive graviton theories do not reduce to those of GR in the $m_{g} \to 0$ limit. This can be understood 
qualitatively by studying how the $5$ spin states of the graviton behave in this limit. Two of them become the two GR helicity states of the massless graviton. Another two become 
helicity states of a massless vector that decouples from the tensor perturbations in the $m_{g} \to 0$ limit. The last state, the scalar mode, however, retains a finite coupling to the 
trace of the stress-energy tensor in this limit. Therefore, massive graviton theories in the $m_{g} \to 0$ limit do not reduce to GR, since the scalar mode does not decouple. 

Given these difficulties, and lacking a specific action that was not vDVZ discontinuous, the community began to consider certain {\emph{phenomenological}} effects that one would 
expect to be present in massive gravity. If the graviton is truly massive, whatever the action may be, two class of modifications to Einstein's theory are expected to be present:
\begin{itemize}
\item[(i)] Modification to Newton's laws;
\item[(ii)] Modification to gravitational wave propagation.  
\end{itemize}
Typically, modifications of class (i) correspond to the replacement of the Newtonian potential by a Yukawa type potential (in the non-radiative, near-zone of any body of mass $M$): $V 
= (M/r) \to (M/r) \exp(-r/\lambda_{g})$, where $r$ is the distance to the body~\cite{Will:1997bb}. Tests of such a Yukawa interaction have been proposed through observations of bound 
clusters, tidal interactions between galaxies~\cite{Goldhaber:1974wg} and weak gravitational lensing~\cite{Choudhury:2002pu}, but such tests are model dependent. On the other hand, solar system bounds are more robust, as we will discuss later.

Modifications of class (ii) are in the form of a modified gravitational-wave dispersion relation. Such a modification was originally parameterized via~\cite{Will:1997bb}
\begin{equation}
\frac{v_{g}^{2}}{c^{2}} = 1 - \frac{m_{g}^{2} c^{4}}{E^{2}}\,,
\label{eq:vg-standard}
\end{equation}
where $v_{g}$ and $m_{g}$ are the speed and mass of the graviton, while $E$ is its energy, usually associated to its frequency via the quantum mechanical relation $E = h f$. This 
modified dispersion relation is inspired by special relativity,  a more general version of which, inspired by quantum gravitational theories, is~\cite{Mirshekari:2011yq}
\begin{equation}
\frac{v_{g}^{2}}{c^{2}} = 1 - \lambda^{\alpha}\,,
\label{eq:vg-LV}
\end{equation}
where $\alpha$ is now a parameter that depends on the theory and $\lambda$ is a dimensionless quantity that represents deviations from light speed propagation. For example, in 
Rosen's bimetric theory~\cite{Rosen:1974ua}, the graviton does not travel at the speed of light, but at some other speed partially determined by the prior geometry. In many metric 
theories of gravity, $\lambda = A  m_{g}^{2} c^{4}/E^{2}$, where $A$ is some amplitude that depends on the metric theory (see discussion in~\cite{Mirshekari:2011yq}). Either 
modification to the dispersion relation has the net effect of slowing gravitons down, such that there is a difference in the time of arrival of photons and gravitons. Moreover, such an 
energy-dependent dispersion relation would affect the accumulated gravitational wave phase observed at gravitational-wave detectors, as we will discuss in 
Section~\ref{section:binary-sys-tests}. Given these modifications to the dispersion relation, one would expect the generation of gravitational waves to also be greatly affected in such 
theories. We will present general arguments in Section~\ref{section:binary-sys-tests}, however, that point at these generation effects being subdominant relative to modifications that 
arise in the propagation of gravitational waves, since the latter accumulate as they travel and sources of gravitational waves are at cosmological distances~\cite{Yunes:2016jcc}. 

From the structure of the above phenomenological modifications, it is clear that GR can be recovered in the $m_{g} \to 0$ limit, avoiding the vDVZ issue altogether by construction. 
Such phenomenological modifications have been constrained by several types of experiments and observations~\cite{deRham:2016nuf}. Using the modification to Newton's third law 
and precise observations of the motion of the inner planets of the solar system together with Kepler's third law,~\cite{Will:1997bb} found a bound of $\lambda_{g} > 2.8 \times 10^{12} 
{\mathrm{\ km}}$, which was recently updated to $\lambda_g > 3.9 \times 10^{13}$km~\cite{Will:2018gku,Bernus:2020szc} and $1.22 \times 10^{15}$km~\cite{Mariani:2023ubf}. Recent observations of the S2 star orbit around Sgr A$^*$ have placed a new bound of $\lambda_g > 4.3 \times 10^{11}$ km~\cite{Zakharov:2016lzv}. Such constraints are purely static, as they do not probe the radiative sector of the theory. Dynamical constraints, however, do exist: through observations of the decay 
of the orbital period of binary pulsars,~\cite{Finn:2001qi} found a bound\epubtkFootnote{The model considered by~\cite{Finn:2001qi} is not phenomenological, but it contains a ghost 
mode.} of $\lambda_{g} > 1.6 \times 10^{10} {\mathrm{\ km}}$; working within cubic Galileon theory, which is effectively a massive gravity theory,~\cite{Shao:2020fka} found $\lambda_g > 7 \times 10^{18}$km due to the absence of Galileon radiation~\cite{deRham:2012fw} in binary pulsars, taking into account the Vainshtein suppression in radiation; by investigating the stability of Schwarzschild and Kerr black holes, \cite{Brito:2013wya} placed the constraint $
\lambda_{g} > 2.4 \times 10^{13} {\mathrm{\ km}}$ in Fierz-Pauli theory~\cite{Fierz:1939ix}. New constraints that use gravitational waves have been proposed, including measuring a 
difference in time of arrival of electromagnetic and gravitational waves~\cite{Cutler:2002ef,2008ApJ...684..870K}, as well as direct observation of gravitational waves emitted by binary 
pulsars. As we shall see in Section~\ref{generic-tests:MG-LV}, aLIGO has placed stringent constraints on massive graviton effects in the propagation of gravitational 
waves~\cite{TheLIGOScientific:2016src,Yunes:2016jcc,LIGOScientific:2018dkp,LIGOScientific:2019fpa,LIGOScientific:2020tif,LIGOScientific:2021sio}.  

Somewhat recently, however, it has been found that the vDVZ discontinuity can in fact be evaded by carefully including non-linearities in the action. Vainshtein~\cite{Vainshtein:1972sx,Kogan:2000uy,Deffayet:2001uk,Babichev:2013usa} showed that around any spherically-symmetric source of mass $M$, there exists a certain radius $r < r_{V} \equiv (r_{S} 
\lambda_{g}^{4})^{1/5}$, with $r_{S}$ the Schwarzschild radius, where linear theory cannot be trusted. Since $r_{V} \to \infty$ as $m_{g} \to 0$, this implies that there is no radius at all 
in which the linear approximation can be trusted in the massless limit. Of course, to determine then whether massive graviton theories have a continuous limit to GR, one must include 
non-linear corrections to the action (see also an argument by~\cite{ArkaniHamed:2002sp}), which are much more difficult to construct and work with. These considerations gave birth 
to new, non-linear massive gravity theories~\cite{Bergshoeff:2009zz,deRham:2010kj,Gumrukcuoglu:2012wt,Bergshoeff:2013hr,deRham:2012fg,deRham:2012fw}, or new massive 
gravity for short. In this theory, non-linear interactions are added to the action through the inclusion of an auxiliary fixed metric. A generalization of new massive gravity that allows the 
auxiliary metric to be dynamical goes by the name of bigravity.     

There has been much activity in the study of gravitational waves in the ghost-free bigravity extension of new massive gravity theories~\cite{DeFelice:2013nba,Narikawa:2014fua,Max:2017flc,Max:2017kdc}, so let us here present more details of this model. The ghost-free bigravity action is given by 
\be
S_{\mathrm{bg}} = \frac{M_{g}^{2}}{2} \int d^{4}x \sqrt{-g} R + \frac{\kappa_{b} M_{g}^{2}}{2} \int d^{4}x \sqrt{-\tilde{g}} \tilde{R} - m^{2} M^{2}_{g} \int d^{4}x \sqrt{-g} \sum_{n=0}^{4} 
c_{n} V_{n}(Y^{\mu}_{\nu}) + S_{m}[\psi_{\mathrm{mat}},g_{\mu \nu}]\,,
\ee
where $g_{\mu \nu}$ is the metric tensor of our universe, i.e.~that which couples to the matter degrees of freedom $\psi_{\mathrm{mat}}$ with coupling $M_{g} = (8 \pi G)^{-1/2}$, with 
$R$ its associated Ricci scalar and $g$ its determinant. The metric tensor $\tilde{g}_{\mu \nu}$, with its associated Ricci scalar $\tilde{R}$, determinant $\tilde{g}$ and coupling 
constant $\kappa_{b}$, couples non-minimally to $g_{\mu \nu}$ through the third term in the action. The latter depends on the graviton mass $m$, coupling constants $c_{n}$ and 
complicated functionals $V_{n}[\cdot]$ of $Y^{\mu}_{\nu} := \sqrt{g^{\mu \alpha} \tilde{g}_{\alpha \nu}}$. 

The generation of gravitational waves has not been worked out in this theory yet, but their propagation has. In bigravity, the $g_{\mu \nu}$ and $\tilde{g}_{\mu \nu}$ metric tensors are 
massless and massive spin-2 degrees of freedom respectively, which implies they carry $2+5$ modes (2 transverse-traceless ones in $g_{\mu \nu}$, 2 transverse-traceless ones in $
\tilde{g}_{\mu \nu}$, 2 transverse vector ones in $
\tilde{g}_{\mu \nu}$, and a breathing mode in $\tilde{g}_{\mu \nu}$). The evolution equations for the dominant transverse-traceless modes (assuming propagation on the same 
background) are~\cite{Narikawa:2014fua}
\begin{align}
\square_{c} h_{+,\times} + m^{2} \Gamma_{c} \left(h_{+,\times} - \tilde{h}_{+,\times}\right) &= 0\,,
\\
\square_{\tilde{c}} \tilde{h}_{+,\times} + \frac{m^{2} \Gamma_{c}}{\kappa_{b} \; \xi_{c}^{2}} \left(\tilde{h}_{+,\times} - h_{+,\times}\right) &= 0\,,
\end{align}
where $(\Gamma_{c},\xi_{c})$ are constants that depend on the coupling parameters of the theory, while the operator $\square_{c} = \partial_{t}^{2} - v^{2} \nabla^{2}$, with $v$ 
the propagation speed of the mode. One then sees that the perturbation to the physical metric couples to the perturbation of the auxiliary metric, leading to oscillations between the 
two modes. This behavior is analogous to neutrino oscillations in the standard model, where the ``mass matrix'' is non-diagonal.  

Given how relatively new massive gravity and bigravity are, many of the properties listed in Sec.~\ref{sec:properties} have not yet been fully explored. Clearly, these theories are 
constructed to satisfy Property 1.a, the GR limit. Black holes, neutron stars and white dwarfs in ghost-free massive gravity and bigravity have been studied in~\cite{Volkov:2012wp,Volkov:2013roa,Babichev:2014tfa,Enander:2015kda,Katsuragawa:2015lbl,Li:2016fbf,Aoki:2016eov,Sullivan:2017kwo,Hendi:2017ibm,Yamazaki:2018szv,EslamPanah:2018evk,Gervalle:2020mfr}, and their stability has also been analyzed~\cite{Babichev:2015zub} (see~\cite{Babichev:2015xha} for a review on black hole solutions in massive gravity). 
Spherically symmetric black holes were constructed in~\cite{Comelli:2011wq,Volkov:2012wp}, which appear to be linearly stable in a region of parameter space~\cite{Babichev:2013una,Brito:2013wya,Brito:2013yxa,Babichev:2014oua,Kobayashi:2015yda}. Rotating black hole solutions were also found with asymptotically flat boundary conditions in~\cite{Babichev:2013una} and with anti-de Sitter boundary conditions in~\cite{Ayon-Beato:2015qtt}, although it is not clear whether these solutions are stable. A weak field analysis was carried out in~\cite{DeFelice:2013nba}, where the authors found that Solar System tests can be avoided if the mass of the graviton is large enough due to the activation of a form of Vainshtein screening in bigravity.
Similarly, the construction of stars in massive bigravity began around 2012. The first neutron stars in massive bigravity were constructed by Volkov~\cite{Volkov:2012wp}, assuming spherical symmetry and an incompressible fluid equation of state. This analysis was then extended in~\cite{Enander:2015kda}, who also considered whether black holes and neutron stars can arise naturally from gravitational collapse in these theories. Sullivan and Yunes~\cite{Sullivan:2020zpf} constructed the first rotating neutron star solutions in massive bigravity, using the Hartle-Thorne slow-rotation approximation and a wide class of realistic equations of state. The stability of these solutions has not yet been studied, and thus, much more work remains to be done to understand their non-linearly stability. Lacking knowledge of what a bigravity spacetime for binary systems looks like, the study of the generation of gravitational waves in dynamical spacetimes has not yet been tackled.

Whether all of these theories are well-posed as an initial-value problem (Property 3) remains unclear. For ghost-free massive gravity with flat reference metric, however, the theory has indeed been shown to be well-posed by introducing higher order gradient terms~\cite{deRham:2023ngf}. Massive gravity theories are well-motivated extensions of GR from a fundamental physics standpoint, given that many particles in the standard model that are known to have a mass. Finally, it is 
currently unclear whether these theories lead to modifications in the extreme gravity regime, except perhaps for the graviton oscillations discussed above. One might be concerned 
that the mass of the graviton and subsequent modifications to the graviton dispersion relation should be suppressed by the Planck scale. However, Collins, et~al.~\cite{2004PhRvL..93s1301C,2006hep.th....3002C} have suggested that Lorentz violations in perturbative quantum field theories could be dramatically enhanced when one regularizes and renormalizes 
them. This is because terms that vanish upon renormalization due to Lorentz invariance do not vanish in Lorentz-violating theories, thus leading to an 
enhancement~\cite{2011arXiv1106.1417G}. Whether such an enhancement is truly present in massive gravity depends on the particular model considered and has not yet been studied in 
detail.

Let us close this section with a note of caution about Lorentz violations and massive gravity: although massive gravity theories unavoidably lead to a modification to the graviton 
dispersion relation, and the latter are also common in Lorentz-violating theories, the converse is not necessarily true, i.e. a modification to the dispersion relation, for example due to 
Lorentz-violating effects, does not necessarily imply a massive graviton. In fact, modifications to the dispersion relation are usually accompanied by modifications to either the Lorentz 
group or its action in real or momentum space. Such Lorentz-violating effects are commonly found in quantum gravitational theories, including loop quantum 
gravity~\cite{2008PhRvD..77b3508B} and string theory~\cite{2005hep.th....8119C,2010GReGr..42....1S}, as well as other effective models~\cite{Berezhiani:2007zf,Berezhiani:2008nr}.  
In Doubly Special Relativity~\cite{2001PhLB..510..255A,2002PhRvL..88s0403M,AmelinoCamelia:2002wr,2010arXiv1003.3942A}, the graviton dispersion relation is modified at high 
energies by modifying the law of transformation of inertial observers. Modified graviton dispersion relations have also been shown to arise in generic extra-dimensional 
models~\cite{2011PhLB..696..119S}, in Ho\v{r}ava--Lifshitz theory~\cite{Horava:2008ih,Horava:2009uw,2010arXiv1010.5457V,Blas:2011zd} and in theories with non-commutative 
geometries~\cite{2011arXiv1102.0117G,Garattini:2011kp,Garattini:2011hy}. Even though the dispersion relation is modified in these theories, they do not require a massive graviton.

\subsubsection{Einstein-\AE{}ther Theory and Khronometric gravity}
\label{sec:EA-theory}

Violations of Lorentz symmetry are inherent in quantum gravitational models. Particle physics experiments have already placed very stringent constraints on Lorentz violation in the 
matter sector~\cite{Kostelecky:2003fs,Kostelecky:2008ts,Mattingly:2005re,Jacobson:2005bg,Kostelecky:2010ze}. A generic way to represent such constraints is to bound the 
coefficients of the standard model extension (SME)~\cite{Colladay:1998fq,Kostelecky:1998id,Kostelecky:1999rh}, a framework in which one adds all possible Lorentz-violating 
interactions to the action of the Standard Model. These experiments, however, cannot place stringent constraints on Lorentz violation that enters primarily in the gravitational sector, 
inducing violations in the matter sector only as a second-order process~\cite{Pospelov:2010mp,Liberati:2013xla}. 

The most common way to represent violations of Lorentz symmetry in the gravity sector is through a preferred time direction at every spacetime point. This preferred frame is typically 
described by an \AE{}ther vector field $U^{\mu}$ that is timelike and unit-norm, thus breaking boost symmetry, and consequently Lorentz invariance.  The most generic covariant action 
that (i) depends only on the metric tensor and the \AE{}ther vector field, and their first derivatives, and (ii) is quadratic in the latter, is given by~\cite{Jacobson:2000xp,Eling:2004dk,Jacobson:2008aj}
\be 
\label{ae-action} 
S_{\rm EA} = \frac{1}{16\pi G_{\rm EA}}\int  d^{4}x \sqrt{-g} \; (R - M^{\alpha \beta}{}_{\mu\nu} \nabla_\alpha U^\mu \nabla_\beta U^\nu)+S_{m}(\psi_{\mathrm{mat}},g_{\mu \nu}) \,,
\ee 
up to total divergences, where $g$ is the determinant of the metric $g_{\mu \nu}$, $R$ is its associated Ricci scalar,  
\be
M^{\alpha\beta}{}_{\mu\nu} \equiv c_1 g^{\alpha\beta}g_{\mu\nu}+c_2\delta^{\alpha}_{\mu}\delta^{\beta}_{\nu} + c_3 \delta^{\alpha}_{\nu}\delta^{\beta}_{\mu}+c_4 U^\alpha U^\beta 
g_{\mu\nu}\, ,
\ee 
($c_1$, $c_2$, $c_3$ $c_4$) are (dimensionless) coupling constants, and $G_N={G_{\rm EA}} [{1-(c_1+c_4)/2}]^{-1}$, with $G_{\rm EA}$ the ``bare'' gravitational constant and $G_{N}$ the 
``Newtonian'' gravitational constant measured with Cavendish-type experiments~\cite{Carroll:2004ai}. This action defines Einstein-\AE{}ther theory~\cite{Jacobson:2000xp}, which 
ought to be thought of as an effective field theory, i.e., the low-energy description of some as-of-yet-unknown high-energy theory~\cite{ArmendarizPicon:2010mz}.

A simpler way to violate Lorentz symmetry gravitationally is to require that the preferred time direction be present \emph{globally}, rather than locally in spacetime. This amounts to 
restricting Einstein-\AE{}ther theory by requiring the \AE{}ther field to be orthogonal to hypersurfaces of constant preferred time. Defining these hypersurfaces through level surfaces of 
the \emph{khronon}\epubtkFootnote{This word comes from the Greek word $\chi\rho\acute o\nu o \varsigma$ (khronos), meaning time. The letter ``k'' was chosen to avoid confusion 
with other theories that use the prefix ``chrono''~\cite{Levi,Segal}} scalar field $T$, then this implies requiring that 
\be
\label{U_dT}
U_\mu=\frac{\partial_\mu T}{\sqrt{g^{\mu\nu}\partial_\mu T \partial_\nu T}},
\ee
and thus, the \AE{}ther vector field is no longer generic. Choosing the time coordinate to coincide with the khronon field, and reducing the Einstein-\AE{}ther action through the above 
constraint, one finds~\cite{Jacobson:2010mx,Blas:2009ck,Blas:2010hb}
\begin{align}
\label{action-K}
S_{\mathrm{KG}} &=\frac{1-\beta}{16\pi G_{\rm EA}} \int dT d^3x \, N\sqrt{h} \, \left(K_{ij}K^{ij} - \frac{1+\lambda}{1-\beta} K^2
+  \frac{1}{1-\beta}{}^{(3)} R +  \frac{\alpha}{1-\beta}\, a_i a^i\right) +S_{\mathrm{mat}}(\psi_{\mathrm{mat}},g_{\mu\nu})\,,
\end{align}
where $N=(g^{TT})^{-1/2}$ is the lapse function, $K^{ij}$, ${}^{(3)}\!R$ and $h^{ij}$ are the extrinsic curvature, the $3$-Ricci curvature and the 3-metric associated with the constant 
$T$ hypersurfaces respectively, the acceleration of the khronon field is $a_i\equiv\partial_i \ln{N}$, and $(\alpha,\beta,\lambda)$ are coupling constants related to the Einstein-\AE{}ther couplings via $(\alpha,\beta,\lambda) = (c_{1}+c_{4},c_{1}+c_{3},c_2)$. This action defines khronometric gravity~\cite{Blas:2009ck,Blas:2010hb}, and it also ought to be thought 
of as an effective theory. In fact, khronometric gravity is the low-energy limit of Ho\v rava gravity~\cite{Horava:2009uw}, a power-counting renormalizable theory of 
gravity~\cite{Horava:2009uw,Blas:2009qj}.

Variation of the action with respect to all fields yields the field equations of the theory. The field equations of Einstein-\AE{}ther theory are
\begin{align}
\label{E_def}
&G_{\alpha\beta} - T^{\rm EA}_{\alpha\beta} = 8\pi G_{\rm EA} T^{\mathrm{mat}}_{\alpha\beta}\,, 
\\
\label{AE_def} 
&\AE_{\mu} \equiv \left (\nabla_\alpha J^{\alpha\nu} - c_4 \dot{U}_\alpha \nabla^\nu U^\alpha \right) \left( g_{\mu\nu} -U_\mu
U_\nu\right) = 0\,,  
\end{align}
where $G_{\alpha\beta}$ is the Einstein tensor, and the \AE{}ther stress-energy tensor is 
\ba
\label{Tae}
T^{\rm EA}_{\alpha\beta}&=&\nabla_\mu\left(J^{\phantom{(\alpha}\mu}_{(\alpha}U_{\beta)}-J^\mu_{\phantom{\mu}(\alpha}U_{\beta)}-J_{(\alpha\beta)}U^\mu\right)+c_1\,
\left[
  (\nabla_\mu U_\alpha)(\nabla^\mu U_\beta)-(\nabla_\alpha U_\mu)(\nabla_\beta U^\mu)
  \right]\nonumber\\ &&+\left[ U_\nu(\nabla_\mu J^{\mu\nu})-c_4 \dot{U}^2
  \right] U_\alpha U_\beta +c_4 \dot{U}_\alpha \dot{U}_\beta+\frac{1}{2} M^{\sigma\rho}{}_{\mu\nu} \nabla_\sigma 
  U^\mu \nabla_r U^\nu g_{\alpha\beta}\,,
\ea 
with the shorthands $J^\alpha_{\phantom{a}\mu} \equiv M^{\alpha\beta}_{\phantom{ab}\mu\nu} \nabla_\beta U^\nu$ and $\dot{U}_\nu \equiv U^\mu\nabla_\mu U_\nu$. The field 
equations of khronometric gravity are
\begin{align}
\label{hl1}
G_{\alpha\beta} - T^{\rm EA}_{\alpha\beta}  -2\AE_{(\alpha}U_{\beta)} & = 8\pi G_{\rm EA} T^{\mathrm{mat}}_{\alpha\beta}\,,
\\
 \label{hleq}
\nabla_\mu \left(\frac{\AE^\mu}{\sqrt{\nabla^\alpha T \nabla_\alpha T}}  \right)& =0\,,
\end{align}
where $\AE_\alpha$ was defined in Eq.~\eqref{AE_def}.

The field equations in these two Lorentz-violating gravity theories have been solved in a variety of scenarios. For example, spherically symmetric, static, and flat 
 vacuum solutions (non-rotating black holes) have been found, and in fact discovered to be the same in the two theories~\cite{Jacobson:2010mx,Blas:2011ni,Blas:2010hb,Barausse:2012ny}.  In 
fact, hypersurface-orthogonal solutions to Einstein-\AE ther theory will also be solutions in khronometric gravity.  Slowly-rotating black hole solutions have also been found, but they are not the same~\cite{Wang:2012nv,Barausse:2012ny,Barausse:2012qh,Barausse:2013nwa}. Recently, rotating black hole solutions without the slow-rotation approximation were constructed numerically~\cite{Adam:2021vsk}. Slowly-moving black hole solutions were studied in~\cite{Ramos:2018oku,Kovachik:2023fos} with which one can extract sensitivities that are responsible for scalar and vector radiation in a binary~\cite{Foster:2007gr,Yagi:2013qpa,Yagi:2013ava}. Quasi-normal modes of black holes were studied in~\cite{Konoplya:2006rv,Konoplya:2006ar,Ding:2017gfw,Ding:2018nhz,Churilova:2020bql,Franchini:2021bpt}. These black holes have been shown numerically to be the end state of 
gravitational collapse for values of the coupling constants that are consistent with observational constraints~\cite{Eling:2006ec,Garfinkle:2007bk,Barausse:2011pu,Akhoury:2016mrc}. 
Regarding neutron stars, non-rotating configurations were studied in~\cite{Eling:2007xh}, while slowly-moving solutions were constructed in~\cite{Yagi:2013qpa,Yagi:2013ava,Barausse:2019yuk,Gupta:2021vdj}. Recently,~\cite{Ajith:2022uaw} studied
slowly-rotating or tidally-deformed neutron stars and found that there exist new types of Love numbers.
Moreover, the motion of test-particles has been found to be geodesic in both theories, which implies the absence of  a ``fifth force'' and violations of the WEP. This comes about 
because both theories are diffeomorphism invariant, which then implies that the matter stress-energy tensor associated with $S_{m}$ is covariantly conserved, $\nabla^{\mu} T_{\mu 
\nu}^{\mathrm{mat}} = 0$. This is not, however, the case for strongly self-gravitating bodies (like neutron stars and black holes), which follow non-geodesic paths that depend on the 
bodies' internal structure, a violation of the strong equivalence principle that is sometimes referred to as the N\"ordvedt effect~\cite{nordvedt,nordvedt_dickes_interpretation}.  

Lorentz-violating gravity has been constrained with solar system observations that check for the existence of preferred-frame effects. Constraints derived with such observations are 
parameterized in terms of the $(\alpha_{1}^{\mathrm{ppN}},\alpha_{2}^{\mathrm{ppN}})$ preferred-frame parameters of the ppN framework. In particular, lunar laser observations and 
observations of the solar alignment with the ecliptic plane have placed the constraints $|\alpha_1^{\mathrm{ppN}}|\lesssim 10^{-4}$ and $|\alpha_2^{\mathrm{ppN}}| \lesssim 
10^{-7}$~\cite{lrr-2006-3} (binary pulsar observations have also placed stringent constraints on the strongly self-gravitating counterparts to $\alpha_{1,2}$~\cite{Shao:2013wga,Shao:2012eg}). In Einstein-\AE{}ther theory, the ppN parameters $(\alpha_{1},\alpha_{2})$ are functions of the coupling constants $c_{i}$, namely  
\begin{align}
	\label{eq:alpha1-2-EA}
	\alpha_1^{\mathrm{ppN}, \rm EA} &= -\frac{8(c_3^2+c_1c_4)}{2c_1-c_+c_-}\,,
	\qquad
	\alpha_2^{\mathrm{ppN}, \rm EA} = \frac{\alpha_1^{\mathrm{ppN}}}{2}-\frac{(c_1+2c_3-c_4)(2c_1+3c_2+c_3+c_4)}{(2-c_{14}) (c_{1}+c_{2}+c_{3})}\,,
\end{align}
while in khronometric gravity, the ppN parameters are functions of the $(\alpha,\beta,\lambda)$, namely
\begin{align}
	\label{eq:alpha1-2-KG}
	\alpha_1^{\mathrm{ppN},\mathrm{KG}} &= 4 \frac{\alpha - 2\beta}{\beta-1}\,, 
	\\
	\alpha_2^{\mathrm{ppN},\mathrm{KG}} &= \frac{(\alpha - 2 \beta)}{(\beta-1)(\lambda+\beta)(\alpha-2)}
	[-\beta^2+\beta (\alpha-3)+\alpha  + \lambda(-1-3\beta+2\alpha)]\,.
\end{align}
Notice that $\alpha_{2}^{\mathrm{ppN}, \mathrm{KG}}$ can be written in terms of $\alpha_1^{\mathrm{ppN},\mathrm{KG}}$, so requiring that the latter be small automatically ensures 
the former is also small. 

Because of the many coupling constants in these theories, solar system constraints on only two combinations ($\alpha_{1}^{\mathrm{ppN}}$ and $\alpha_{2}^{\mathrm{ppN}}$) are 
not enough to break all degeneracies and constrain the individual coupling parameters; for this, one requires new constraints, such as those obtained from binary pulsar 
observations. The latter are typically studied by assuming $\alpha_{1}^{\mathrm{ppN}} = 0 = \alpha_{2}^{\mathrm{ppN}}$ to reduce the $4$- and $3$-dimensional coupling 
parameter spaces to a $2$-dimensional one. In Einstein-\AE{}ther theory, this implies~\cite{Foster:2005dk,Jacobson:2008aj} 
\begin{align}
\label{eq:Aecons}
c_2&=\frac{-2 c_1^2-c_1 c_3+c_3^2}{3 c_1}\,,
\qquad
c_4=-\frac{c_3^2}{c_1}\,,
\end{align}
leaving $c_{\pm} \equiv c_{1} \pm c_{3}$ as the only two free coupling parameters, while in khronometric gravity the condition implies $\alpha = 2 \beta$, leaving $(\beta,\lambda)$ as 
the only two free coupling parameters.  The strongest constraints on $c_{\pm}$ and $(\beta,\lambda)$ come from observations of the orbital decay rate of PSR 
J1141-6545~\cite{bhat}, PSR J0348+0432~\cite{2.01NS}, PSR J0737-3039~\cite{kramer-double-pulsar} and PSR J1738+0333~\cite{freire}, which require $(c_{+},c_{-}) \lesssim 
(0.03,0.003)$ and $(\beta,\lambda) \lesssim (0.005,0.1)$ ~\cite{Yagi:2013qpa,Yagi:2013ava}. The khronometric constraint also makes use of Big-Bang Nucleosynthesis 
constraints~\cite{Audren:2013dwa,Carroll:2004ai,Zuntz:2008zz,Jacobson:2008aj}, derived from the agreement between observed and predicted metal abundances in the early 
universe, which require that the ratio of Newton's gravitational constant in that cosmological era to that measured today be close to unity.  Weaker constraints on $c_{\pm}$ have also 
been obtained by requiring the stability of perturbations about a Minkowski background~\cite{Jacobson:2004ts} and the absence of gravitational Cherenkov radiation~\cite{Elliott:2005va}. Combining all of these constraints together, one arrives at $c_{i} \lesssim 10^{-2}$, while $(\alpha,\beta,\lambda) \lesssim (0.01,0.005,0.1)$.

As we will discuss later, the binary neutron star merger event GW170817~\cite{LIGOScientific:2017vwq}, together with its electromagnetic counterparts GRB 170817A~\cite{LIGOScientific:2017ync}, has constrained deviations in the propagation speed of gravitational waves to less than one part in $10^{15}$ relative to the speed of light~\cite{LIGOScientific:2017zic}. In Einstein-\AE{}ther and khronometric gravity, the propagation speed of the tensor mode depends only on $c_1+c_3$ or $\beta$. This means that this combination of parameters is constrained to be smaller than $10^{-15}$. Imposing this new bound and all the other bounds mentioned earlier, two viable regions remain in the Einstein-\AE ther parameter space~\cite{Sarbach:2019yso,Oost:2018tcv}. The first region can be found by saturating the solar system bound on $\alpha_1$ (namely, assuming $|\alpha_1| \lesssim 10^{-4}$ but not $|\alpha_1| \ll 10^{-4}$). Doing so implies that $c_1 \approx - c_3 + \mathcal{O}(10^{-15})$, $c_4 \approx c_3+\mathcal{O}(10^{-4})$, and $c_2 \approx (c_4-c_3)[1+\mathcal{O}(10^{-3})]$, which then means that $c_1+c_3 \approx 0$, $c_4-c_3 \approx c_2 \approx 0$, while $c_1-c_3$ remains unconstrained. The second region can be found when one does \textit{not} saturate the bound on $\alpha_1$, but instead requires that its magnitude be much smaller than $10^{-14}$. In doing so, the $\alpha_2$ bound is automatically satisfied when $c_1 + c_3 \approx 0$ and one is left with a two-dimensional parameter space $(c_2, c_1-c_3)$ with the only constraint $|c_2| \lesssim 0.1$ coming from Big Bang Nucleosynthesis bounds. Gupta \textit{et al.}~\cite{Gupta:2021vdj} reanalyzed the binary pulsar bounds on Einstein-\AE{}ther theory within these viable parameter spaces and derived a new bound on $\alpha_1$, namely $|\alpha_1| \lesssim 10^{-5}$, which is 10 times stronger than the bound from solar system experiments. Regarding khronometric gravity, the new bounds on the coupling constants after GW170817 are given by $|\alpha| \lesssim 10^{-7}$, $|\beta| \lesssim 10^{-15}$ and $|\lambda| \lesssim 10^{-1}$~\cite{EmirGumrukcuoglu:2017cfa}.

Let us conclude this overview with a discussion of whether Einstein-\AE{}ther theory and khronometric gravity satisfy the criteria laid out in Sec.~\ref{sec:properties}. As discussed 
above, both theories satisfy Property 1, since they pass all constraints for sufficiently small coupling constants and solutions that could present astrophysical systems have been found 
and shown to be stable. Both theories also possess Property 2, since they are very well-motivated from fundamental physics. In fact, it is difficult to find a quantum gravitational model 
that does not violate Lorentz symmetry. Einstein-\AE{}ther theory also satisfies Property 3, at least in part, since it has been shown to have an initial-value formulation that is also well-
posed~\cite{Coley:2015qqa,Sarbach:2019yso}. Khronometric gravity, on the other hand, has not been studied in sufficient detail to determine whether the theory is well-posed, although given the 
results from Einstein-\AE{}ther theory, one would expect it to be (at least in some limit). Finally, property 4 is also satisfied, since extreme gravity will be altered in Lorentz-violating 
theories. These modifications, however, are not just confined to the extreme gravity regime, thus allowing for rather stringent bounds from solar system observations and from strong-
field binary pulsar tests. 

\subsubsection{Modified quadratic gravity}
\label{subsec:MQG}

Modified quadratic gravity is a family of models first discussed in the context of black holes and gravitational waves in~\cite{Yunes:2011we,Yagi:2011xp}. The 4-dimensional action is 
given by  
\begin{align} 
S &\equiv \int d^4x \sqrt{-g} \Big\{ \kappa R + \alpha_{1}
f_{1}(\vartheta) R^{2} + \alpha_{2} f_{2}(\vartheta) R_{\mu \nu} R^{\mu \nu}
+ \alpha_{3} f_{3}(\vartheta) R_{\mu \nu \delta \sigma}
R^{\mu \nu \delta \sigma} 
\nonumber \\
&+ \alpha_{4} f_{4}(\vartheta) 
R_{\mu \nu \delta \sigma} \!{}^{*}R^{\mu \nu \delta \sigma}
-  \frac{\beta}{2} \left[\left(\nabla_{\mu} \vartheta\right) \left(\nabla^{\mu} \vartheta\right) 
+ 2 V(\vartheta) \right] +
\mathcal{L}_{\mathrm{mat}} \Big\}\,.
\label{exactaction} 
\end{align}
The quantity $^{*}R^{\mu}{}_{\nu\delta\sigma} = (1/2) \epsilon_{\delta \sigma}{}^{\alpha \beta} R^{\mu}{}_{\nu \alpha \beta}$ is the dual to the Riemann tensor. The quantity $
\mathcal{L}_{\mathrm{mat}}$ is the external matter Lagrangian, while $f_{i}(\cdot)$ are functionals of the field $\vartheta$, with $(\alpha_{i},\beta)$ coupling constants and $\kappa = 
(16 \pi G)^{-1}$. Clearly, the two terms second to last in Eq.~\eqref{exactaction} represent a canonical kinetic energy term and a potential. At this stage, one might be tempted to set $
\beta=1$ or the $\alpha_{i} = 1$ via a rescaling of the scalar field functional, but we shall not do so here.

The action in Eq.~\eqref{exactaction} is well-motivated from
fundamental theories, as it contains all possible quadratic, algebraic
curvature scalars with running (i.e., non-constant) couplings. The only restriction here is that all quadratic terms are assumed to couple to the {\emph{same}} field, which need not be 
the case. For example, in string theory some terms might couple to the dilaton (a scalar field), while other couple to the axion (a pseudo scalar field)~\cite{Kanti:1995cp,Cano:2021rey}. Nevertheless, one can recover 
well-known and motivated modified gravity theories in simple cases. For example, dynamical Chern--Simons modified gravity~\cite{Alexander:2009tp} is recovered when $\alpha_{4} 
=-\alpha_{\mathrm{CS}}/4$ and all other $\alpha_{i} = 0$. Einstein--dilaton--Gauss--Bonnet gravity~\cite{Pani:2009wy} is obtained when $\alpha_{4} = 0$ and $
(\alpha_1,\alpha_2,\alpha_3)=(1,-4,1)\alpha_{\mathrm{EDGB}}$\epubtkFootnote{Technically, Einstein--Dilaton--Gauss--Bonnet gravity has a very particular set of coupling functions 
$f_{1}(\vartheta) = f_{2}(\vartheta) = f_{3}(\vartheta) \propto e^{\gamma \vartheta}$, where $\gamma$ is a constant. In most cases, however, one can expand about $\gamma \vartheta 
\ll 1$, so that the functions become linear in the scalar field.}. Both theories unavoidably arise as low-energy expansions of heterotic string theory~\cite{Green:1987sp,Green:1987mn,Kanti:1995cp,Alexander:2004xd,lrr-2004-5,Cano:2021rey}. As such, modified quadratic gravity theories should be treated as a class of effective field theories. Moreover, dynamical Chern--Simons 
Gravity also arises in loop quantum gravity~\cite{Ashtekar:2004eh,Rovelli:2004tv} when the Barbero--Immirzi parameter is promoted to a field in the presence of 
fermions~\cite{Ashtekar:1988sw,Alexander:2008wi,Taveras:2008yf,Mercuri:2009zt,Gates:2009pt}.

One should make a clean and clear distinction between the theory defined by the action of Eq.~\eqref{exactaction} and that of $f(R)$ theories. The latter are defined as functionals of 
the Ricci scalar only, while Eq.~\eqref{exactaction} contains terms proportional to the Ricci tensor and Riemann tensor squared. One could think of the subclass of $f(R)$ theories with 
$f(R) = R^{2}$ as the limit of modified quadratic gravity with only $\alpha_{1} \neq 0$ and $f_{1}(\vartheta) = 1$. In that very special case, one can map quadratic gravity theories and 
$f(R)$ gravity to a scalar-tensor theory. Another important distinction is that $f(R)$ theories are usually treated as exact, while the action presented above is to be interpreted as an 
{\emph{effective theory}}~\cite{lrr-2004-5} truncated to quadratic order in the curvature in a low-energy expansion of a more fundamental theory. This implies that there are cubic, 
quartic, etc.~terms in the Riemann tensor that are not included in Eq.~\eqref{exactaction} and that presumably depend on higher powers of $\alpha_{i}$. Thus, when studying such an 
effective theory one should also order-reduce the field equations and treat all quantities that depend on $\alpha_{i}$ perturbatively, the so-called {\emph{{small-coupling 
approximation}}. One can show that such an order reduction removes any additional polarization modes in propagating metric perturbations~\cite{Sopuerta:2009iy,Stein:2010pn} that 
naturally arise in $f(R)$ theories. In analogy to the treatment of the Ostrogradski instability in Section~\ref{sec:properties}, order-reduction also lead to a theory with a well-posed initial 
value formulation~\cite{Delsate:2014hba}. 

This family of theories is usually simplified by making the assumption that the coupling functions $f_{i}(\cdot)$ admit a Taylor expansion: $f_{i}(\vartheta) = f_{i}(0)+ f_{i}'(0)\vartheta + 
\mathcal{O}(\vartheta^{2})$ for small $\vartheta$, where $f_{i}(0)$ and $f_{i}'(0)$ are constants and $\vartheta$ is assumed to vanish at asymptotic spatial infinity. Reabsorbing $f_{i}
(0)$ into the coupling constants $\alpha_{i}^{(0)} \equiv \alpha_{i} f_{i}(0)$ and $f_{i}^{\prime}(0)$ into the constants $\alpha_{i}^{(1)} \equiv \alpha_{i} f_{i}^{\prime}(0)$, 
Eq.~\eqref{exactaction} becomes $S = S_{\mathrm{GR}} + S_{0} + S_{1}$ with
\begin{subequations}
\label{eq:quad-action-simped}
\begin{align}
S_{\mathrm{GR}} &\equiv \int d^4x \sqrt{-g} \left\{ \kappa R + \mathcal{L}_{\mathrm{mat}} \right\}\,,
\\
S_{0} &\equiv  \int d^4x \sqrt{-g} \left\{ 
\alpha_{1}^{(0)} R^{2} 
+ \alpha_{2}^{(0)} R_{\mu \nu} R^{\mu \nu}
+ \alpha_{3}^{(0)} R_{\mu \nu \delta \sigma} R^{\mu \nu \delta \sigma} 
\right\}\,,
\\
S_{1} &\equiv  \int d^4x \sqrt{-g} \left\{ 
\alpha_{1}^{(1)} \vartheta R^{2} + \alpha_{2}^{(1)} \vartheta R_{\mu \nu} R^{\mu \nu}
+ \alpha_{3}^{(1)} \vartheta R_{\mu \nu \delta \sigma} R^{\mu \nu \delta \sigma} 
\right. \nonumber \\
&+ \left. \alpha_{4}^{(1)} \vartheta  \; R_{\mu \nu \delta \sigma} \; \!{}^{*}R^{\mu \nu \delta \sigma}
- \frac{\beta}{2} \left[ \left(\nabla_{\mu} \vartheta\right) \left(\nabla^{\mu} \vartheta\right) + 2 V(\vartheta) \right] \right\}\,.
\end{align}
\label{action}
\end{subequations}
Here, $S_{\mathrm{GR}}$ is the Einstein--Hilbert plus matter action, while $S_{0}$ and $S_{1}$ are corrections. The former is decoupled from $\vartheta$, where the omitted term 
proportional to $\alpha_4^{(0)}$ does not affect the classical field equations since it is topological, i.e.~it can be rewritten as the total $4$-divergence of some $4$-current. Similarly, if 
$\alpha_{i}^{(0)}$ were chosen to reconstruct the Gauss--Bonnet invariant, $(\alpha_1^{(0)},\alpha_2^{(0)},\alpha_3^{(0)})=(1,-4,1)\alpha_{\mathrm{GB}}$, then this combination 
would also be topological and not affect the classical field equations. On the other hand, $S_{1}$ is a modification to GR with a direct (non-minimal) coupling to $\vartheta$, such that 
as the field goes to zero, the modified theory reduces to GR. 

Another restriction one usually makes to simplify quadratic gravity theories is to neglect the $\alpha_{i}^{(0)}$ terms and only consider the $S_{1}$ modification, which defines 
{\emph{restricted}} quadratic gravity. The $\alpha_{i}^{(0)}$ terms represent corrections that are non-dynamical. The term proportional to $\alpha_{1}^{(0)}$ resembles a certain class 
of $f(R)$ theories. As such, it can be mapped to a scalar tensor theory with a complicated potential, which has been heavily constrained by torsion-balance E\"ot-Wash experiments to 
$\alpha_{1}^{(0)} < 2 \times 10^{-8}\mathrm{\ m}^{2}$~\cite{Hoyle:2004cw,Kapner:2006si,Berry:2011pb}. Moreover, these theories have a fixed coupling constant that does not run with energy 
or scale. In restricted quadratic gravity, the scalar field effectively forces the running of the coupling. Nevertheless, gravitational waves in quadratic gravity with $\alpha_i^{(1)}=0$ have been studied in~\cite{Naf:2011za,Kim:2019sqk,Tachinami:2021jnf,Alves:2022yea}.

Let us then concentrate on restricted quadratic gravity and drop the superscript in $\alpha_i^{(1)}$. 
The modified field equations are
\begin{align}
G_{\mu \nu} &+ \frac{\alpha_1 \vartheta}{\kappa} \mathcal{H}_{\mu \nu}^{(0)} +
\frac{\alpha_2 \vartheta }{\kappa} \mathcal{I}_{\mu \nu}^{(0)} 
+ \frac{\alpha_3 \vartheta}{\kappa} \mathcal{J}_{\mu \nu}^{(0)} 
+ \frac{\alpha_1}{\kappa}
\mathcal{H}_{\mu \nu}^{(1)} + 
\frac{\alpha_2}{\kappa} \mathcal{I}_{\mu \nu}^{(1)} 
+ \frac{\alpha_3}{\kappa} \mathcal{J}_{\mu \nu}^{(1)} +
\frac{\alpha_4}{\kappa} \mathcal{K}_{\mu \nu}^{(1)} 
= \frac{1}{2\kappa} \left( T_{\mu \nu}^{\mathrm{mat}} + T_{\mu \nu}^{(\vartheta)} \right)\,,
\label{FEs} 
\end{align}
where we have defined 
\begin{subequations}
\allowdisplaybreaks[1]
\begin{align}
\mathcal{H}_{\mu \nu}^{(0)} \equiv & 2 R R_{\mu \nu}  -
\frac{1}{2} g_{\mu \nu} R^{2} - 2 \nabla_{\mu \nu} R + 2 g_{\mu \nu} \square R\,,\\
\mathcal{I}_{\mu \nu}^{(0)} \equiv & \square R_{\mu \nu} + 2
R_{\mu \delta \nu \sigma} R^{\delta \sigma} - \frac{1}{2} g_{\mu \nu} 
R^{\delta \sigma} R_{\delta \sigma}
+ \frac{1}{2} g_{\mu \nu} \square R - \nabla_{\mu \nu} R\,,\\
\mathcal{J}_{\mu \nu}^{(0)}
\equiv & 8 R^{\delta \sigma} R_{\mu \delta \nu \sigma} - 
2 g_{\mu \nu} R^{\delta \sigma} R_{\delta \sigma} + 4 \square R_{\mu \nu}
- 2 R \, R_{\mu \nu} + \frac{1}{2} g_{\mu \nu} R^{2} - 2 \nabla_{\mu \nu} R\,,\\
\mathcal{H}_{\mu \nu}^{(1)} \equiv & -4
(\nabla_{(\mu} \vartheta) \nabla_{\nu)} R - 2 R \nabla_{\mu\nu} \vartheta 
+ g_{\mu \nu} \left[2 R \square \vartheta + 4 (\nabla^{\delta}\vartheta) \nabla_{\delta}R \right]\,,\\
\mathcal{I}_{\mu \nu}^{(1)} \equiv & -(\nabla_{(\mu}\vartheta) \nabla_{\nu)} R - 2\nabla^\delta\vartheta
  \nabla_{(\mu} R_{\nu)\delta} 
+ 2 \nabla^\delta\vartheta \nabla_{\delta} R_{\mu \nu}
+R_{\mu\nu}\square \vartheta
\nonumber \\*
&- 2 R_{\delta(\mu}\nabla^{\delta} \nabla_{\nu)}\vartheta + g_{\mu \nu} \left( \nabla^\delta \vartheta
  \nabla_{\delta} R + R^{\delta \sigma} \nabla_{\delta\sigma} \vartheta \right)\,,\\
\mathcal{J}_{\mu \nu}^{(1)}
\equiv & - 8 \left(\nabla^\delta \vartheta \right) \left( \nabla_{(\mu} R_{\nu)\delta} - \nabla_{\delta} R_{\mu \nu}\right) + 4
R_{\mu \delta \nu \sigma} \nabla^{\delta\sigma} \vartheta\,, \\
\mathcal{K}_{\mu \nu}^{(1)}
\equiv & -4 \left( \nabla^\delta\vartheta \right) \epsilon_{\delta \sigma \chi(\mu} \nabla^{\chi} R_{\nu)}^{~\sigma} + 4
(\nabla_{\delta\sigma} \vartheta) {}^{*}\!R_{(\mu}{}^{\delta}{}_{\nu)}{}^{\sigma}\,.
\end{align}
\end{subequations}
The $\vartheta$ stress-energy tensor is
\be
T_{\mu \nu}^{(\vartheta)} = \beta \left\{ (\nabla_{\mu}\vartheta) (\nabla_{\nu}\vartheta)
 - \frac{1}{2}g_{\mu \nu} \left[\left( \nabla_{\delta}\vartheta\right) 
 \left(\nabla^{\delta}\vartheta \right) - 2
V(\vartheta) \right] \right\}\,.
\label{theta-Tab}
\ee
The field equations for the scalar field are
\begin{align}
\beta \square \vartheta - \beta \frac{dV}{d\vartheta}
=&\, -\alpha_1 R^{2} - \alpha_2 R_{\mu \nu} R^{\mu \nu} 
-\alpha_3 R_{\mu \nu \delta \sigma} R^{\mu \nu \delta \sigma} 
- \alpha_4 R_{\mu \nu \delta \sigma} \!{}^{*}R^{\mu \nu \delta \sigma}\,.
\label{EOM} 
\end{align}
Notice that unlike traditional scalar-tensor theories, the scalar field is here sourced by the geometry and not by the matter distribution directly. This implies that black holes in such theories are hairy, and thus, their inertial mass does not coincide with their gravitational mass~\cite{Yunes:2011we,Benkel:2016rlz,Benkel:2016kcq,Berti:2018cxi,Prabhu:2018aun,Julie:2019sab,Julie:2022huo,HegadeKR:2022xij}, violating the strong equivalence principle (a violation typically quantified through certain sensitivity parameters -- see also Sec.~\ref{section:binary-sys-tests}). Other compact stars, such as neutron stars, however, can be shown to not be hairy\epubtkFootnote{To be more specific, ``hair'' here refers to the monopole scalar charge. Neutron stars can have higher order hair, like dipole scalar hair in dynamical Chern-Simons gravity~\cite{Yagi:2013mbt}.}, and thus have vanishing sensitivities, in the above restricted quadratic gravity theory with couplings to the Gauss-Bonnet invariant or the Pontryagin density~\cite{Yagi:2011xp,Yagi:2015oca}, as well as in more generic shift-symmetric Horndenski theories~\cite{Barausse:2015wia}. This fact is particularly important since dipole radiation, which has been stringently constrained with binary pulsars, is proportional to the square of the difference in the sensitivities of the stars. Thus, a suppression in the sensitivities implies that these theories automatically pass binary pulsar constraints, while still allowing for extreme gravity modifications in systems with black holes~\cite{Yagi:2015oca,Barausse:2015wia}.  

Let us review compact objects in quadratic gravity in more detail.
In non-dynamical theories (when $\beta = 0$ and the scalar-fields are constant, refer to Eq.~\eqref{exactaction}), Stein and Yunes~\cite{Yunes:2011we} have shown that all metrics that are Ricci tensor flat are also solutions of the modified field equations (see also~\cite{Psaltis:2007cw}). This is not so for dynamical theories, since then the $\vartheta$ field is sourced by curvature, leading to corrections to the field equations proportional to the Riemann tensor and its dual. 

In dynamical Chern--Simons gravity, stationary and spherically-symmetric spacetimes are still described by GR solutions, but stationary and axisymmetric spacetimes are not. Instead, they are represented by~\cite{Yunes:2009hc,Konno:2009kg} 
\begin{equation}
ds^{2}_{\mathrm{CS}} = ds^{2}_{\mathrm{Kerr}} + \frac{5}{4} \frac{\alpha_{\mathrm{CS}}^{2}}{\beta \kappa} \frac{a}{r^{4}} \left(1 + \frac{12}{7} \frac{M}{r} + \frac{27}{10}\frac{M^{2}}{r^{2}} \right) \sin^{2}{\theta} \, d\theta \, dt + {\cal{O}}(a^{2}/M^{2})\,, 
\end{equation}
with the scalar field
\begin{equation}
\vartheta_{\mathrm{CS}} = \frac{5}{8} \frac{\alpha_{\mathrm{CS}}}{\beta} \frac{a}{M}  \frac{\cos{\theta}}{r^{2}} \left(1 + \frac{2 M}{r} + \frac{18 M^{2}}{5 r^{2}} \right) + {\cal{O}}(a^{3}/M^{3})\,,
\end{equation}
where $ds^{2}_{\mathrm{Kerr}}$ is the line element of the Kerr metric, and we recall that $\alpha_{\mathrm{CS}} = -4 \alpha_{4}$ in the notation of Section~\ref{subsec:MQG}. These expressions are obtained in Boyer--Lindquist coordinates and in the small-rotation/small-coupling limit to ${\cal{O}}(a/M)$ in~\cite{Yunes:2009hc,Konno:2009kg}, to ${\cal{O}}(a^{2}/M^{2})$ in~\cite{Yagi:2012ya}, and solutions to higher order in spin in~\cite{Maselli:2015tta}. The linear-in-spin corrections modify the frame-dragging effect, and they are of 3.5 post-Newtonian order. The quadratic-in-spin corrections modify the quadrupole moment, which induces 2 post-Newtonian-order corrections to the binding energy. However, the stability of these black holes has only been studied linearly~\cite{Garfinkle:2010zx}. 

In Einstein--Dilaton--Gauss--Bonnet gravity, stationary and spherically-symmetric spacetimes are described, in the small-coupling approximation, by the line element~\cite{Yunes:2011we}
\begin{equation}
ds^{2}_{\mathrm{EDGB}} = -f_{\mathrm{Schw}} \left(1 + h\right) dt^{2} + f^{-1}_{\mathrm{Schw}} \left(1 + k \right)  dr^{2} + r^{2} d\Omega^{2}\,,
\end{equation}
in Schwarzschild coordinates, where $d \Omega^{2}$ is the line element on the two-sphere, $f_{\mathrm{Schw}}=1 - 2 M/r$ is the Schwarzschild factor, and we have defined
\begin{align}
h &= \frac{\alpha_{3}^{2}}{\beta \kappa M^{4}} \frac{1}{3 f_{\mathrm{Schw}}} \frac{M^{3}}{r^{3}} \left(1  + 26 \frac{M}{r} + \frac{66}{5} \frac{M^{2}}{r^{2}} + \frac{96}{5} \frac{M^{3}}{r^{3}} -  80 \frac{M^{4}}{r^{4}}\right)\,,
\\
k &= -\frac{\alpha_{3}^{2}}{\beta \kappa M^{4}} \frac{1}{f_{\mathrm{Schw}}} \frac{M^{2}}{r^{2}} \left[ 1 +  \frac{M}{r} + \frac{52}{3} \frac{M^{2}}{r^{2}} + 2 \frac{M^{3}}{r^{3}} + \frac{16}{5} \frac{M^{4}}{r^{4}} - \frac{368}{3} \frac{M^{5}}{r^{5}} \right]\,,
\end{align}
while the corresponding scalar field is 
\begin{equation}
\vartheta_{\mathrm{EDGB}} = \frac{\alpha_{3}}{\beta} \frac{2}{M r} \left(1 + \frac{M}{r} + \frac{4}{3} \frac{M^{2}}{r^{2}}\right)\,.
\label{eq:theta-EDGB}
\end{equation}
This solution is not restricted just to Einstein--Dilaton--Gauss--Bonnet gravity, but it is also the most general, stationary and spherically-symmetric solution in quadratic gravity. This is because all terms proportional to $\alpha_{1,2}$ are proportional to the Ricci tensor, which vanishes in vacuum GR, while the $\alpha_{4}$ term does not contribute in spherical symmetry (see~\cite{Yunes:2011we} for more details). 
Linear slow-rotation corrections to this solution have been found in~\cite{Pani:2011gy}, and analytic solutions to higher order in spin were found in~\cite{Maselli:2015tta}. Although the stability of these black holes has only been studied linearly as well~\cite{Ayzenberg:2013wua}, other dilatonic black hole solutions obtained numerically  (equivalent to those in Einstein--Dilaton-Gauss--Bonnet theory in the limit of small fields)~\cite{Kanti:1995vq} have been found to be stable under axial perturbations~\cite{Kanti:1997br,Torii:1998gm,Pani:2009wy,Blazquez-Salcedo:2016enn}.

Neutron stars also exist in quadratic modified gravity. In dynamical Chern--Simons gravity, the mass-radius relation remains unmodified to first order in the slow-rotation expansion, but the moment of inertia changes to this order~\cite{Yunes:2009ch,AliHaimoud:2011fw}, while the quadrupole moment and the mass measured at spatial infinity change to quadratic order in spin~\cite{Yagi:2013mbt}. This is because the mass-radius relation, to first order in slow-rotation, depends on the spherically-symmetric part of the metric, which is unmodified in dynamical Chern--Simons gravity. In Einstein--Dilaton--Gauss--Bonnet gravity, the mass-radius relation is modified~\cite{Pani:2011xm}. As in GR, these functions must be solved for numerically, and they depend on the equation of state.

The above restricted quadratic gravity model is not a good representative member of the wider class of quadratic gravity theories when $f_i(\vartheta)$ cannot be approximated as a linear function. Examples include $f_i(\vartheta) =\vartheta^2$ and $f_i(\vartheta) = [1-\exp(-6 \vartheta^2)]/12$. For these cases and with a Gauss-Bonnet combination of $(\alpha_1,\alpha_2,\alpha_3,\alpha_4) = (1,-4,1,0)\alpha$, recent analyses have shown that black holes can spontaneously scalarize~\cite{Doneva:2017bvd,Silva:2017uqg,Macedo:2019sem,Cunha:2019dwb,East:2021bqk,Doneva:2022yqu}, similar to spontaneous scalarization for neutron stars in scalar-tensor theories. The scalar charges can also be induced by spins~\cite{Collodel:2019kkx,Herdeiro:2020wei,Berti:2020kgk,Dima:2020yac,Elley:2022ept}. In a binary, dynamical scalarization can occur in analogy to scalar-tensor theories~\cite{Silva:2020omi,Doneva:2022byd,Julie:2023ncq,Kuan:2023trn,Annulli:2023ydz,Lara:2024rwa}, but a crucial difference is that in scalar-Gauss-Bonnet gravity, black holes in a binary can also \emph{de-scalarize}~\cite{Silva:2020omi,Doneva:2022byd}.

From the structure of the above equations, it should be clear that the dynamics of $\vartheta$ guarantee that the modified field equations are covariantly conserved exactly. That is, 
one can easily verify that the covariant divergence of Eq.~\eqref{FEs} identically vanishes upon imposition of Eq.~\eqref{EOM}. Such a result had to be so, as the action is 
diffeomorphism invariant. If one neglected the kinetic and potential energies of $\vartheta$ in the action, as was originally done in~\cite{Jackiw:2003pm}, the theory would possess 
preferred-frame effects and would not be covariantly conserved A manifestation of the latter is the fact that such a theory requires an additional constraint, i.e., the right-hand side of \eqref{EOM} would have to 
vanish, which is an unphysical consequence of treating $\vartheta$ as prior structure~\cite{Yunes:2007ss,Grumiller:2007rv}.

One last simplification that is usually made when studying modified quadratic gravity theories is to ignore the potential $V(\vartheta)$, i.e., set $V(\vartheta) = 0$. This potential can in 
principle be non-zero, for example if one wishes to endow $\vartheta$ with a mass or if one wishes to introduce a cosine driving term, like that for axions in field and string theory (see e.g. Nashed and Nojiri~\cite{Nashed:2022kes} for slowly-rotating black holes in dynamical Chern-Simons gravity with non-vanishing potentials). 
However, reasons exist to restrict the functional form of such a potential. First, a mass for $\vartheta$ will modify the evolution of any gravitational degree of freedom only if this mass 
is comparable to the inverse length scale of the problem under consideration (such as a binary system). This could be possible if there is an incredibly large number of fields with 
different masses in the theory, such as perhaps in the string axiverse picture~\cite{PhysRevD.83.044026,Kodama:2011zc,PhysRevD.85.103514}. In that picture, however, the moduli 
fields are endowed with a mass due to shift-symmetry breaking by non-perturbative effects; such masses are not expected to be comparable to the inverse length scale of binary 
systems. Second, no mass term may appear in a theory with a shift symmetry, i.e.~invariance under the transformation $\vartheta \to \vartheta + {\mathrm{const}}$. Such symmetries 
are common in four-dimensional, low-energy, effective string theories~\cite{Boulware:1985wk,Green:1987mn,Green:1987sp,1992PhLB..285..199C,lrr-2004-5}, such as dynamical 
Chern--Simons and Einstein--Dilaton--Gauss--Bonnet theory. Similar considerations apply to other, more complicated potentials, such as a cosine term. 

Given these field equations, one can linearize them about Minkowski space to find evolution equations for the perturbation in the small-coupling approximation. Doing so, one 
finds~\cite{Yagi:2011xp}
\begin{align}
\square_{\eta} \vartheta =&
- \frac{\alpha_1}{\beta} \left( \frac{1}{2\kappa} \right)^{2} T_{\mathrm{mat}}^2
- \frac{\alpha_2}{\beta} \left( \frac{1}{2\kappa} \right)^{2} T_{\mathrm{mat}}^{\mu\nu} T^{\mathrm{mat}}_{\mu\nu}
\nonumber \\
& -  \frac{2\alpha_3}{\beta} (h_{\alpha \beta ,\mu \nu} h^{\alpha [\beta ,\mu] \nu} + h_{\alpha \beta ,\mu \nu} h^{\mu [\nu ,\alpha] \beta} ) 
\nonumber \\
& -  \frac{2 \alpha_4}{\beta} \bar{\epsilon}^{\alpha \beta \mu \nu} h_{\alpha \delta,\gamma \beta} h_{\nu}{}^{[\gamma,\delta]}{}_{\mu}\,,
\end{align}
where we have order-reduced the theory where possible and used the harmonic gauge condition (which is preserved in this class of theories~\cite{Sopuerta:2009iy,Stein:2010pn}). 
The corresponding equation for the metric perturbation is rather lengthy and can be found in Eqs.~(17)\,--\,(24) in~\cite{Yagi:2011xp}. Since these theories are to be considered 
effective, working always to leading order in $\alpha_{i}$, one can show that they are perturbatively of type $N_{2}$ in the $E(2)$ classification~\cite{Eardleyprd}, i.e.~in the far zone, 
the only propagating modes that survive are the two transverse-traceless (spin-2) metric perturbations~\cite{Sopuerta:2009iy,Wagle:2019mdq}. In the strong-field region, however, it is possible that 
additional modes are excited, although they decay rapidly as they propagate to future null infinity. 

Lastly, let us discuss what is known about whether modified quadratic gravity satisfies the requirements discussed in Section~\ref{sec:properties}. As it should be clear from the action 
itself, this modified gravity theory satisfies the fundamental requirement, i.e.~passing all precision tests, provided the couplings $\alpha_{i}$ are sufficiently small. This is because such 
theories have a continuous limit to GR as $\alpha_{i} \to 0$.\epubtkFootnote{Formally, as $\alpha_{i} \to 0$, one recovers GR with a dynamical scalar field. The latter, however, does not 
couple to the metric or the matter sector, so it does not lead to any observable effects that distinguish it from GR.}  Dynamical Chern--Simons gravity is constrained only weakly from solar system experiments, $\xi_{4}^{1/4} < 10^{8} {\mathrm{\ km}}$, where $\xi_{4} \equiv \alpha_{4}^{2}/(\beta \kappa)$, through observations of Lense--Thirring precession~\cite{AliHaimoud:2011fw,Nakamura:2018yaw}. However, much stronger bounds have been obtained through gravitational wave observations. Ringdown observations place a new bound of $\xi_{4}^{1/4} < 103$ km~\cite{Silva:2022srr} by assuming a small-spin expansion (which may not be valid for rapidly-spinning black hole remnants), while a multi-messenger observations of gravitational wave and X-ray observations place a bound of $\xi_{4}^{1/4} < 22.6$ km~\cite{Silva:2020acr} as a null test of GR . The coupling constant of Einstein--Dilaton--Gauss--Bonnet gravity, $\xi_{3} \equiv \alpha_{3}^{2}/(\beta \kappa)$, on the other hand, has been 
constrained by several experiments: solar system observations of the Shapiro time delay with the Cassini spacecraft placed the bound $\xi_{3}^{1/4} < 1.3 \times 10^{7} 
{\mathrm{\ km}}$~\cite{Bertotti:2003rm,Amendola:2007ni}; the requirement that neutron stars still exist in this theory placed the constraint $\xi_{3}^{1/4} \lesssim 26  {\mathrm{\ km}}
$~\cite{Pani:2011xm}, with the details depending somewhat on the central density of the neutron star; the requirement that the neutron star maximum mass exceeds $2M_\odot$ also places the bound $\xi_{3}^{1/4} \lesssim 3.43$ km~\cite{Saffer:2021gak}, though this bound depends on the choice of the equation of state of nuclear matter;
observations of the rate of change of the orbital period in the low-mass X-ray binary $A0620-00$~\cite{Psaltis:2005ai,Johannsen:2008tm} has led to the $1\sigma$ constraint $\xi_{3}
^{1/4} < 4.3 {\mathrm{\ km}}$~\cite{Yagi:2012gp}; finally, recent gravitational wave observations place a 90\%-credible constraint $\xi_{3}
^{1/4} < 3.1$ km~\cite{Nair:2019iur,Perkins:2021mhb,Wang:2021jfc,Lyu:2022gdr}, which provides the strongest bound (see also~\cite{Shao:2023yjx,Wang:2023wgv,Gao:2024rel} for more recent works).

Not all sub-properties of the fundamental requirement, however, are known to be satisfied. One can show that certain members of modified quadratic gravity possess known solutions 
and these are stable, at least linearly and in the small-coupling approximation. For example, in dynamical Chern--Simons gravity, spherically symmetric vacuum solutions are given by 
the Schwarzschild metric with constant $\vartheta$ to all orders in $\alpha_{i}$~\cite{Jackiw:2003pm,Yunes:2007ss,Rogatko:2013cma}. Moreover, one can show that such a solution, 
as well as non-spinning black holes and branes in anti-de~Sitter space~\cite{Delsate:2011qp}, are linearly stable to small perturbations~\cite{Molina:2010fb,Garfinkle:2010zx}. 
Spinning solutions, on the other hand, continue to be elusive, with linearly-stable~\cite{Ayzenberg:2013wua,Wagle:2021tam}, approximate solutions in the slow-rotation/small-coupling limit known 
both for black holes~\cite{Yunes:2009hc,Konno:2009kg,Pani:2011gy,Yagi:2012ya,Maselli:2017kic,Alexander:2021ssr} and stars~\cite{Yunes:2009ch,AliHaimoud:2011fw,Pani:2011xm,Yagi:2013mbt} and solutions in the fast-rotation/small-coupling 
limit known for black holes only~\cite{Konno:2014qua,Stein:2014xba,McNees:2015srl,Delsate:2018ome}. In Einstein--Dilaton--Gauss--Bonnet theory, both non-spinning~\cite{Yunes:2011we} and 
spinning~\cite{Pani:2011gy,Ayzenberg:2014aka,Maselli:2015tta,Kleihaus:2015aje,Kleihaus:2011tg} black hole solutions are known, and they have been found to be linearly stable~\cite{Pani:2009wy,Blazquez-Salcedo:2016enn,Okounkova:2019zep,Pierini:2021jxd,Pierini:2022eim}. Neutron stars solutions have been constructed for non-spinning~\cite{Pani:2011xm,Saffer:2019hqn}, spinning~\cite{Kleihaus:2014lba,Kleihaus:2016dui} and tidally-deformed~\cite{Saffer:2021gak} configurations. Ripley and Pretorius~\cite{Ripley:2019aqj} found evidence that scalarized black holes are non-linearly stable when the coupling constant is sufficiently small.

The study of modified quadratic gravity theories as effective theories is valid provided one is sufficiently far from its cut-off scale, i.e.~the scale beyond which higher-order curvature 
terms cannot be neglected any longer. One can estimate the magnitude of this scale by studying the size of loop corrections to the quadratic curvature terms in the action due to $n$-
point interactions~\cite{Yagi:2012ya}. Simple counting requires that the number of scalar and graviton propagators, $P_{s}$ and $P_{g}$, satisfy the following relation in terms of the 
number of vertices $V$:
\begin{equation}
P_s = \frac{V}{2}, \quad P_g = (n-1)\frac{V}{2}\,.
\end{equation}
Loop corrections are thus suppressed by factors of $\alpha_{i}^{V} M_\mathrm{pl}^{(2-n)V} \Lambda^{nV}$, with $M_\mathrm{pl}$ the Planck mass and $\Lambda$ an energy scale 
introduced by dimensional arguments. The cut-off scale above which the theory cannot be treated as an effective one can be approximated as the value of $\Lambda$ at which the 
suppression factor becomes equal to unity:
\begin{equation}
\Lambda_c \equiv M_\mathrm{pl}^{1-2/n} \alpha_{i}^{1/n}\,,
\label{lambdac}
\end{equation}
This cut-off scale automatically places a constraint on the magnitude of $\alpha_{i}$ above which higher-curvature corrections must be included. Setting the largest value of $
\Lambda_{c}$ to be equal to $\mathcal{O}(10\mu \mathrm{\ m}$), thus saturating bounds from table-top experiments~\cite{Kapner:2006si}, and solving for $\alpha_{i}$, we find
\begin{equation}
\alpha_{i}^{1/2} < \mathcal{O}(10^8 \mathrm{\ km}).
\label{alpha-ineq}
\end{equation}
Current bounds on $\alpha_{i}$ require the coupling constant to be much smaller than $10^{8} {\mathrm{\ km}}$~\cite{AliHaimoud:2011fw,Yagi:2012gp,Nakamura:2018yaw,Nair:2019iur,Silva:2020acr,Perkins:2021mhb,Wang:2021jfc,Lyu:2022gdr,Silva:2022srr}, thus 
justifying the treatment of these theories as effective models. 

As for the other requirements discussed in Section~\ref{sec:properties}, it is clear that modified quadratic gravity is well-motivated from fundamental theory, but it is not clear  
whether it has a well-posed initial-value formulation. From an effective point of view, a perturbative treatment in $\alpha_{i}$ naturally leads to stable solutions and a well-posed initial 
value problem, but this is probably not the case when the theory is treated as exact~\cite{Delsate:2014hba}. In fact, if one were to treat such a theory as exact (to all orders in $
\alpha_{i}$), then the evolution system is not hyperbolic in general, as higher than second time derivatives now drive the evolution~\cite{Delsate:2014hba}. Notice, however, that this says 
nothing about the fundamental theories that modified quadratic gravity derives from. This is because even if the truncated theory were ill-posed, higher order corrections that are 
neglected in the truncated version could restore well-posedness. 
For the Einstein--dilaton--Gauss--Bonnet case, the field equations remain second order in derivatives. Ripley, Pretorius and East~\cite{Ripley:2019irj,Ripley:2019aqj,East:2020hgw} showed that, under spherical symmetry, the field equations remain hyperbolic and are well-posed when the coupling constant is small, On the other hand, when the coupling constant is large, there arises elliptic regions in spacetime where hyperbolicity of the equations is broken (see also~\cite{R:2022hlf}).

As for the last requirement (that the theory modifies extreme gravity), modified quadratic theories are ideal in this respect. This is because they introduce corrections to the action that 
depend on higher powers of the curvature. In the extreme gravity regime, such higher powers could potentially become non-negligible relative to the Einstein--Hilbert action. Moreover, 
since the curvature scales inversely with the mass of the black holes under consideration, one expects the largest deviations in systems with small mass, such as stellar-mass black hole 
mergers or extreme--mass-ratio inspirals~\cite{Sopuerta:2009iy,Maselli:2020zgv,Maselli:2021men,Barsanti:2022ana,Tan:2024utr}. 

\subsubsection{Variable \textit{G} theories and large extra dimensions}
\label{sec:Variable-G}

Variable $G$ theories are defined as those where Newton's gravitational constant is promoted to a spacetime function. Such a modification breaks the principle of equivalence 
(see~\cite{lrr-2006-3}) because the laws of physics now become local position dependent. In turn, this implies that experimental results now depend on the spacetime position of the 
laboratory frame at the time of the experiment. 

Many known modified gravity theories that violate the principle of equivalence, and in particular, the strong equivalence principle, predict a varying gravitational constant. A classic 
example is scalar-tensor theory~\cite{Will:1993ns}, which, as explained in Section~\ref{sec:ST}, modifies the gravitational sector of the action by multiplying the Ricci scalar by a scalar 
field (in the Jordan frame). In such theories, one can effectively think of the scalar as promoting the coupling between gravity and matter to a field-dependent quantity $G \to G(\phi)$, 
thus violating local position invariance when $\phi$ varies. Another example are bimetric theories, such as that of Lightman--Lee~\cite{1973PhRvD...8.3293L}, where the gravitational 
constant becomes time-dependent even in the absence of matter, due to possibly time-dependent cosmological evolution of the prior geometry. A final example is higher-
dimensional, brane-world scenarios, where enhanced Hawking radiation~\cite{Emparan:2002px,Tanaka:2002rb} may lead to a time-varying effective 4D gravitational constant~\cite{Deffayet:2007kf}, whose rate of change 
depends on the curvature radius of extra dimensions~\cite{Johannsen:2008tm,McWilliams:2009ym,Yagi:2011yu}. 

One can also construct $f(R)$-type actions that introduce variability to Newton's constant. For example, consider the $f(R)$ model~\cite{Frolov:2011ys}
\be
S = \int d^{4}x \sqrt{-g} \; \kappa \; R \left[1 + \alpha_{0} \ln \left(\frac{R}{R_{0}}\right) \right] + S_{\mathrm{mat}}\,,
\ee
where $\kappa = (16 \pi G)^{-1}$, $\alpha_{0}$ is a coupling constant and $R_{0}$ is a curvature scale. This action is motivated by certain renormalization group flow 
arguments~\cite{Frolov:2011ys}. The field equations are
\begin{align}
G_{\mu \nu} = \frac{1}{2 \bar{\kappa}} T_{\mu \nu}^{\mathrm{mat}} - \frac{\alpha_{0}}{\bar{\kappa}} R_{\mu \nu} - 2 \frac{\kappa}{\bar{\kappa}} \frac{\alpha_{0}}{R^{2}} \nabla_{(\mu} R 
\nabla_{\nu)} R - \frac{1}{2} \frac{\alpha_{0} \kappa}{\bar{\kappa}} g_{\mu \nu} \square R\,,
\end{align}
where we have defined the new constant
\begin{equation}
\bar{\kappa} := \kappa \left[1 + \frac{\alpha_{0}}{\kappa} \ln \left(\frac{R}{R_{0}}\right)\right]\,.
\end{equation}
Clearly, the new coupling constant $\bar{\kappa}$ depends on the curvature scale involved in the problem, and thus, on the geometry, forcing $G$ to run to zero in the ultraviolet limit. 

Another example is a scalar-tensor theory with a non-minimal coupling to a topological invariant. For example, consider the model
\be
S = \int d^{4}x \sqrt{-g} \; \left[\kappa \; \vartheta \; R + \vartheta \; {\cal{T}} - \frac{1}{2} \left(\nabla_{\mu}
\vartheta \right) \left( \nabla^{\mu} \vartheta \right) \right] + S_{\mathrm{mat}}\,,
\ee
where $\kappa = (16 \pi G)^{-1}$ and ${\cal{T}}$ is a topological invariant constructed from the curvature tensor, like the Gauss-Bonnet invariant or the Pontryagin density. Such a 
model, for example, arises in the axiverse scenario~\cite{PhysRevD.83.044026,Kodama:2011zc}, where here we have redefined the field so that its kinetic energy is standard, and we 
have neglected any potential. The field equations are
\begin{align}
G_{\mu \nu} &= \frac{1}{2 \bar{\kappa}} T_{\mu \nu}^{\mathrm{mat}} + \frac{1}{2 \bar{\kappa}} T_{\mu \nu}^{(\vartheta)}  -  \frac{1}{\kappa \sqrt{-g}} \; \frac{\delta }{\delta g^{\mu \nu}} 
\left(\sqrt{-g} \; {\cal{T}} \right)\,,
\\
\square \vartheta &= -\kappa \;R - {\cal{T}}
\end{align}
where $T_{\mu \nu}^{(\vartheta)}$ is the stress-energy tensor of the $\vartheta$ scalar field [see Eq.~\eqref{theta-Tab}], and we have defined the new constant $\bar{\kappa} := \kappa 
\; \vartheta$. We see then that $G \to G(\vartheta) = G/\vartheta$, where $\vartheta$ is sourced by a topological invariant constructed from the curvature tensor and its trace. 

An important point to address is whether variable $G$ theories can lead to modifications to a {\emph{vacuum}} spacetime, such as a black hole binary inspiral. In Einstein's theory, 
$G$ appears as the coupling constant between geometry, encoded by the Einstein tensor $G_{\mu\nu}$, and matter, encoded by the stress energy tensor $T_{\mu\nu}
^{\mathrm{mat}}$. When considering vacuum spacetimes, $T_{\mu \nu}^{\mathrm{mat}} = 0$ and one might naively conclude that a variable $G$ would not introduce any modification 
to such spacetimes. In fact, this is the case in scalar-tensor theories (without homogeneous, cosmological solutions to the scalar field equation), where the no-hair theorem establishes 
that black hole solutions are not modified~\cite{Hawking:1972qk}. On the other hand, scalar-tensor theories with a non-trivial boundary conditions 
for the scalar field~\cite{Healy:2011ef,Jacobson:1999vr,Horbatsch:2011ye} or in non-vacuum spacetimes~\cite{Cardoso:2013fwa,Cardoso:2013opa} can evade the no-hair theorem, 
endowing black holes with time-dependent hair, which in turn would introduce variability into $G$ even in vacuum spacetimes~\cite{Berti:2013gfa}. 

In general, Newton's constant plays a much more fundamental role than merely a coupling constant: it defines the relationship between energy and length. For example, for the 
\emph{vacuum} Schwarzschild solution, $G$ establishes the relationship between the radius $R$ of the black hole and the rest-mass energy $E$ of the spacetime via $R=2 G E/ 
c^4$. Similarly, in a black-hole--binary spacetime, each black hole introduces an energy scale into the problem that is quantified by a specification of Newton's constant. Therefore, one 
can think of variable $G$ modifications as induced by some effective theory that modifies the mapping between the curvature scale and the energy scale of the problem, as is done for 
example in scalar-tensor theories with a non-minimal coupling to a topological invariant (as shown above) or as done in theories with extra dimensions. 

An explicit example of the latter is realized in braneworld models. Superstring theory suggests that physics should be described by 4 large dimensions, plus another 6 that are 
compactified and very small~\cite{1998stth.book.....P,Polchinski:1998rr}. The size of these extra dimensions is greatly constrained by particle theory experiments. Braneworld models, 
where a certain higher-dimensional membrane is embedded in a higher dimensional bulk spacetime, can however evade this constraint as only gravitons can interact with the bulk. 
The ADD model~\cite{ArkaniHamed:1998rs,ArkaniHamed:1998nn} is a particular example of such a braneworld, where the bulk is flat and compact and the brane is tensionless with 
ordinary fields localized on it. The size of these extra dimensions is constrained to micrometer scales by table-top experiments~\cite{Kapner:2006si,Adelberger:2006dh}. 

What is relevant to gravitational-wave experiments is that in many of these braneworld models, black holes may not remain static~\cite{Emparan:2002px,Tanaka:2002rb}. The 
argument goes roughly as follows: a five-dimensional black hole is dual to a four-dimensional one with conformal fields on it by the ADS/CFT conjecture~\cite{Maldacena:1997re,Aharony:1999ti}, but since the latter must evolve via Hawking radiation, the black hole may lose mass. The Hawking mass loss rate is here enhanced by the large number of 
degrees of freedom in the conformal field theory, leading to an effective modification to Newton's laws and to the emission of gravitational radiation. Effectively, one can think of the 
black hole mass loss as due to the black hole being stretched away from the brane into the bulk, reducing the size of the brane-localized black hole. For black-hole binaries, one can 
then draw an analogy between this induced time-dependence in the black hole mass and a variable $G$ theory, where Newton's constant becomes time-dependent~\cite{Yunes:2009bv}. However, Figueras et al.~\cite{Figueras:2011va,Figueras:2011gd,Figueras:2013jja} numerically found stable static solutions that do not require a radiation component and this was recently extended for rotating black holes~\cite{Biggs:2021iqw}. If 
such solutions were the ones realized in nature as a result of gravitational collapse on the brane, then the black hole mass would be time-independent, up to quantum correction due 
to Hawking evaporation, a negligible effect for realistic astrophysical systems. This is likely to be the case based on numerical simulations of the dynamics of gravitational collapse in such 
scenarios~\cite{Wang:2016nqi} that are consistent with the static black hole solutions found in~\cite{Figueras:2011gd}.  

Many experiments have been carried out to measure possible deviations from a constant $G$ value, and they can broadly be classified into two groups: (a) those that search for the 
present or nearly present rate of variation (at redshifts close to zero); (b) those that search for {\emph{secular}} variations over long time periods (at very large redshifts). Examples of 
experiments or observations of the first class include planetary radar-ranging~\cite{2005AstL...31..340P,2018NatCo...9..289G}, surface temperature observations of low-redshift millisecond 
pulsars~\cite{Jofre:2006ug,Reisenegger:2009cq}, lunar ranging observations~\cite{Williams:2004qba} and pulsar timing observations~\cite{Kaspi:1994hp,Deller:2008jx,Zhu:2015mdo,Zhu:2018etc}, the latter two 
being the most stringent. Examples of experiments of the second class include the evolution of the Sun~\cite{1998ApJ...498..871G} and Big-Bang Nucleosynthesis (BBN) 
calculations~\cite{Copi:2003xd,Bambi:2005fi,Alvey:2019ctk}, again with the latter being more stringent. For either class, the strongest constraints are about $\dot{G}/G \lesssim 10^{-13} \; \mathrm{ 
yr}^{-1}$, varying somewhat from experiment to experiment. 

Lacking a particularly compelling action to describe variable $G$ theories, one is usually left with a phenomenological model of how such a modification to Einstein's theory would 
impact gravitational waves. Given that the part of the waveform that detectors are most sensitive to is the gravitational wave phase, one can model the effect of variable $G$ theories 
by studying how the rate of change of its frequency would be modified. Assuming a Taylor expansion for Newton's constant, one can derive the modification to the evolution equation 
for the gravitational wave frequency, given whichever physical scenario one is considering. Solving such an evolution equation then leads to a modification in the accumulated 
gravitational wave phase observed at detectors on Earth. In Section~\ref{section:binary-sys-tests} we will provide an explicit example of this for a compact binary system. 

Let us now discuss whether such theories satisfy the criteria defined in Section~\ref{sec:properties}. The fundamental property can be satisfied if the rate of change of Newton's 
constant is small enough, as variable $G$ theories usually have a continuous limit to GR (as all derivatives of $G$ go to zero). Whether variable $G$ theories are well-motivated from 
fundamental physics (Property 2) depends somewhat on the particular effective model or action that one considers. But in general, Property 2 is usually satisfied, considering that such 
variability naturally arises in theories with extra dimensions, and the latter are also natural in all string theories. Variable $G$ theories, however, usually fail at introducing modifications 
in the extreme gravity regime. Usually, such variability is parameterized as a Taylor expansion about some initial point with constant coefficients. That is, the variability of $G$ is not 
usually constructed to become stronger closer to merger.  The well-posed property and the sub-properties of the fundamental property depend somewhat on the particular effective 
theory used to describe varying $G$ modifications. In the $f(R)$ case, one can impose restrictions on the functional form $f(\cdot)$ such that no ghosts ($f' >0$) or instabilities ($f'' > 
0$) arise~\cite{Frolov:2011ys}. This, of course, does not guarantee that this (or any other such) theory is well-posed. A much more detailed analysis would be required to prove well-posedness of the class of theories that lead to a variable Newton's constant, but such is currently lacking.  

\subsubsection{Non-commutative geometry}
\label{sec:NCG}

Non-commutative geometry is a gravitational theory that generalizes the continuum Riemannian manifold of Einstein's theory with the product of it with a tiny, discrete, finite non-
commutative space, composed of only two points. Although the non-commutative space has zero spacetime dimension, as the product manifold remains four dimensional, its internal 
dimensions are $6$ to account for Weyl and chiral fermions. This space is discrete to avoid the infinite tower of massive particles that would otherwise be generated, as in string 
theory. Through this construction, one can recover the Standard Model of elementary particles, while accounting for all (elementary particle) experimental data to date. Of course, the 
simple non-commutative space described above is expected to be replaced by a more complex model at Planckian energies. Thus, one is expected to treat such non-commutative 
geometry models as effective theories. Essentially nothing is currently known about the full non-commutative theory, of which the theories described in this section are an effective low-
energy limit.

Before proceeding with an action-principle description of non-commutative geometry theories, we must distinguish between the spectral geometry approach championed by 
Connes~\cite{Connes:1996gi}, and Moyal-type non-commutative geometries~\cite{Snyder:1946qz,Groenewold:1946kp,Moyal:1949sk}. In the former, the manifold is promoted to a 
non-commutative object through the product of a Riemann manifold with a non-commutative space. In the latter, instead, a non-trivial set of commutation relations is imposed between 
operators corresponding to position. These two theories are in principle unrelated. In this review, we mostly concentrate on the former, though we will comment on the latter too. 

The effective action for spectral non-commutative geometry theories (henceforth, non-commutative geometries for short) is
\begin{align}
S = \int d^{4}x \sqrt{-g} \left( \kappa R + \alpha_{0} C_{\mu \nu \delta \sigma} C^{\mu \nu \delta \sigma} + \tau_{0} R^{*} R^{*} - \xi_{0} R \; |H|^{2}\right) + S_{\mathrm{mat}}\,,
\label{eq:NCG-action}
\end{align}
where $H$ is related to the Higgs field, $C_{\mu \nu \delta \sigma}$ is the Weyl tensor, $(\alpha_{0}, \tau_{0}, \xi_{0})$ are couplings constants and we have defined the quantity
\begin{equation}
R^{*} R^{*} :=\frac{1}{4} \epsilon^{\mu \nu \rho \sigma} \epsilon_{\alpha \beta \gamma \delta} R_{\mu \nu}{}^{\alpha \beta} R_{\rho \sigma}{}^{\gamma \delta}\,.
\end{equation}
Notice that this term integrates to the Euler characteristic, and since $\tau_{0}$ is a constant, it is topological and does not affect the classical field equations. The last term of 
Eq.~\eqref{eq:NCG-action} is usually ignored, as $H$ is assumed to be relevant only in the early universe. Finally, the second term can be rewritten in terms of the Riemann and Ricci 
tensors as
\begin{equation}
C_{\mu \nu \delta \sigma} C^{\mu \nu \delta \sigma} = \frac{1}{3} R^{2} -2 R_{\mu \nu} R^{\mu \nu} + R_{\mu \nu \delta \sigma} R^{\mu \nu \delta \sigma}\,.
\end{equation}
Notice that this corresponds to the modified quadratic gravity action of Eq.~\eqref{action} with all $\alpha_{i}^{(1)} = 0$ and $(\alpha_{1}^{(0)},\alpha_{2}^{(0)},\alpha_{3}^{(0)}) = 
(1/3,-2,1)$, which is not the Gauss--Bonnet invariant. Notice also that this model is not usually studied in modified quadratic gravity theory, as one usually concentrates on the terms 
that have an explicit scalar field coupling. 

The field equations of this theory can be read directly from Eq.~\eqref{FEs}, but we repeat them here for completeness: 
\begin{align}
G_{\mu \nu} - \frac{2 \alpha_{0}}{\kappa} \left[2 \nabla^{\kappa \lambda } + R^{\lambda \kappa} \right] C_{\mu \lambda \nu \kappa} = \frac{1}{2 \kappa} T_{\mu \nu}^{\mathrm{mat}}\,.
\label{eq:NCG-FEs}
\end{align}
One could in principle rewrite this in terms of the Riemann and Ricci tensors, but the expressions become quite complicated, as calculated explicitly in Eqs.~(2) and (3) of~\cite{Yunes:2011we}. Due to the absence of a dynamical degree of freedom coupling to the modifications to the Einstein--Hilbert action, this theory is not covariantly conserved in vacuum. By this, 
we mean that the covariant divergence of Eq.~\eqref{eq:NCG-FEs} does not vanish in vacuum, thus violating the weak-equivalence principle and leading to additional equations that 
might over-constrain the system. In the presence of matter, the equations of motion will not be given by the vanishing of the covariant divergence of the matter stress-energy alone, 
but now there will be additional geometric terms. 

Given these field equations, one can linearize them about a flat background to find the evolution equations for the metric perturbation~\cite{Nelson:2010rt,Nelson:2010ru}
\begin{equation}
\left(1 - \beta^{-2} \square_{\eta}\right) \square_{\eta} h_{\mu \nu} = -16 \pi T_{\mu \nu}^{\mathrm{mat}}\,,
\label{eq:NCG-h-eq}
\end{equation}
where the term proportional to $\beta^{2} = (-32 \pi \alpha_{0})^{-1}$ acts like a mass term. Here, one has imposed the transverse-traceless gauge (a refinement of Lorenz gauge), 
which can be shown to exist~\cite{Nelson:2010rt,Nelson:2010ru}. Clearly, even though the full non-linear equations are not covariantly conserved, its linearized version is, as one can 
easily show that the divergence of the left-hand side of Eq.~\eqref{eq:NCG-h-eq} vanishes. Because of these features, if one works perturbatively in $\beta^{-1}$, then such a theory 
will only possess the two usual transverse-traceless (spin-2) polarization modes, i.e., it is perturbatively of type $N_{2}$ in the $E(2)$ classification~\cite{Eardleyprd}.  

Let us now discuss whether such a theory satisfies the properties discussed in Section~\ref{sec:properties}. Non-commutative geometry theories clearly possess the fundamental 
property, as one can always take $\alpha_{0} \to 0$ (or equivalently $\beta^{-2} \to 0$) to recover GR. Therefore, there must exist a sufficiently small $\alpha_{0}$ such that all 
precision tests carried out to date are satisfied. As for the existence and stability of known solutions, Refs.~\cite{Nelson:2010rt,Nelson:2010ru} have shown that Minkowski spacetime 
is stable only for $\alpha_{0} < 0$, as otherwise a tachyonic term appears in the evolution of the metric perturbation, as can be seen from Eq.~\eqref{eq:NCG-h-eq}. This then 
automatically implies that $\beta$ must be real.

Current constraints on Weyl terms of this form come mostly from solar system experiments. Ni~\cite{Ni:2012sa} recently studied an action of the form of Eq.~\eqref{eq:NCG-action} 
minimally coupled to matter in light of solar system experiments. He calculated the relativistic Shapiro time-delay and light deflection about a massive body in this theory and found 
that observations of the Cassini satellite place constraints on $|\alpha_{0}|^{1/2} < 5.7 {\mathrm{\ km}}$~\cite{Ni:2012sa}. This is currently the strongest bound we are aware of on $
\alpha_{0}$.  

Many solutions of GR are preserved in non-commutative geometries. Regarding black holes, all solutions that are Ricci flat (vacuum solutions of the Einstein equations) are also 
solutions to Eq.~\eqref{eq:NCG-FEs}. This is because by the second Bianchi identity, one can show that
\begin{equation}
\nabla^{\kappa \lambda} R_{\mu \lambda \nu \kappa} = \nabla^{\kappa}{}_{\nu}R_{\mu \kappa} - \square R_{\mu \nu}\,,
\end{equation}
and the right-hand side vanishes in vacuum, forcing the entire left-hand side of Eq.~\eqref{eq:NCG-FEs} to vanish. However, this is not so for neutron stars, where the equations of 
motion are likely to be modified, unless they are static~\cite{2010PhRvD..82j4026N}. Moreover, as of now there has been no stability analysis of black-hole or stellar solutions and no 
study of whether the theory is well-posed as an initial-value problem, even as an effective theory. Thus, except for the fundamental property, it is not clear that non-commutative 
geometries satisfy any of the other criteria listed in Section~\ref{sec:properties}.

Let us now discuss the second approach of non-commutative theories, the Moyal-type. We promote the coordinates $x^\mu$ to an operator $\hat x^\mu$ and impose the following canonical commutation relation:
 \begin{equation}
 [\hat x^\mu, \hat x^\nu] = i \; \theta^{\mu\nu}\,,
 \end{equation}
where $\theta^{\mu\nu}$ represents the amount of violation of commutation (like $\hbar$  in quantum mechanics) that corresponds to the ``quantum fuzziness'' of spacetime. Within this model and working in an effective field theory formulation in which one assumes a black hole is sourced by a massive scalar field~\cite{Bjerrum-Bohr:2002fji,Kobakhidze:2007jn}, the stress-energy tensor for a black hole (assumed to be a point particle) at $\bm y(t)$ is given by~\cite{Kobakhidze:2016cqh}
\begin{equation}
\label{eq:Tmunu_NC}
T^{\mu\nu} (\bm x, t) = m \gamma(t) v^\mu (t) v^\nu (t) \delta^3[\bm x - \bm y (t)] + \frac{m^3 \Lambda^2}{8} v^\mu (t) v^\nu(t) \theta^k \theta^l \partial_k \partial_l \delta^3[\bm x - \bm y (t)]\,,
\end{equation}
Here $\gamma(t)$ is the Lorentz factor, $m$ and $v^\mu$ are the mass and four-velocity of the black hole, while $\Lambda$ and the unit vector $\theta^i$ are defined by
\begin{equation}
\label{eq:Lambda_theta}
\Lambda \; \theta^i = \frac{\theta^{0i}}{l_p t_p}\,,
\end{equation}
where $l_p$ and $t_p$ are the Planck length and time respectively. The above non-commutative correction to the stress-energy tensor leads to modifications in the orbital dynamics at 2nd post-Newtonian order.

We end this section by explaining whether the Moyal-type non-commutative theory satisfies the properties discussed in Sec.~\ref{sec:properties}. The theory has a continuous limit to GR ($\Lambda \to 0$). From the pericenter precession of binary pulsars, $\Lambda$ has been constrained to $\sqrt{\Lambda} \lesssim \mathcal{O}(10)$~\cite{Jenks:2020gbt}, while gravitational wave observations place a slightly stronger bound, as we will discuss in Sec.~\ref{sec:NC_GW}. The theory possesses the property that the non-GR correction grows as one moves to the extreme gravity regime, as the correction enters at 2nd post-Newtonian order. Further analysis is necessary to reveal the stability of compact objects and well-posedness of the theory.

\subsubsection{Gravitational parity violation}
\label{sec:GPV}

Parity, the symmetry transformation that flips the sign of the spatial triad, has been found to be broken in the Standard Model of elementary interactions. Only the combination of a 
parity transformation, time inversion and charge conjugation (CPT) remains still a true symmetry of the Standard Model. Experimentally, it is curious that the weak interaction exhibits 
maximal parity violation, while other fundamental forces seem to not exhibit any. Theoretically, parity violation unavoidably arises in the Standard Model~\cite{Bell:1969ts,1969PhRv..177.2426A,AlvarezGaume:1983ig}, as there exist one-loop chiral anomalies that give rise to parity violating terms coupled to lepton number~\cite{Weinberg:1996kr}. In certain sectors 
of string theory, such as in heterotic and in Type~I superstring theories, parity violation terms are also generated through the Green--Schwarz gauge anomaly-canceling 
mechanism~\cite{Green:1987mn,Polchinski:1998rr,Alexander:2004xd}. Finally, in loop quantum gravity~\cite{Ashtekar:1988sw}, the scalarization of the Barbero--Immirzi parameter 
coupled to fermions leads to an effective action that contains parity-violating terms~\cite{Taveras:2008yf,Calcagni:2009xz,Mercuri:2009zt,Gates:2009pt}. Even without a particular 
theoretical model, one can generically show that effective field theories of inflation generically contain non-vanishing, second-order, parity violating curvature corrections to the 
Einstein--Hilbert action~\cite{Weinberg:2008hq}. Alternatively, phenomenological parity-violating extensions of GR have been proposed through a scalarization of the fundamental 
constants of nature~\cite{Contaldi:2008yz}.

One is then naturally led to ask whether the gravitational interaction is parity invariant in extreme gravity. A violation of parity invariance would occur if the Einstein--Hilbert action were 
modified through a term that involved a Levi-Civita tensor and parity invariant tensors or scalars. Let us try to construct such terms with only single powers of the Riemann tensor and 
a single scalar field $\vartheta$:
\begin{enumerate}
\item[(ia)] $R_{\alpha \beta \gamma \delta} \; \epsilon^{\alpha \beta \gamma \delta}$,
\qquad (ib) $R_{\alpha \beta \gamma \mu} \; \epsilon^{\alpha \beta \gamma \nu} \; \nabla^{\mu}{}_{\nu} \vartheta$,
\item[(ic)] $R_{\alpha \beta \gamma \mu} \; \epsilon^{\alpha \beta \delta \nu} \; \nabla^{\mu\gamma}{}_{\nu\delta} \vartheta$\,,
\qquad (id) $R_{\alpha \zeta \gamma \mu} \; \epsilon^{\alpha \beta \delta \nu} \; \nabla^{\mu\gamma}{}_{\beta \nu\delta}{}^{\zeta} \vartheta$\,.
\end{enumerate}
Option (ia) and (ib) vanish by the Bianchi identities. Options (ic) and (id) include the commutator of covariant derivatives, which can be rewritten in terms of a Riemann tensor, and thus 
it leads to terms that are at least quadratic in the Riemann tensor. Therefore, no scalar can be constructed that includes contractions with the Levi-Civita tensor from a single Riemann 
curvature tensor and a single field. One can try to construct a scalar from the Ricci tensor 
\begin{enumerate}
\item[(iia)] $R_{\alpha \beta} \; \epsilon^{\alpha \beta \gamma \delta} \nabla_{\gamma \delta} \vartheta$,
\qquad (iib) $R_{\alpha \beta} \; \epsilon^{\alpha \mu \gamma \delta} \nabla_{\gamma \delta \mu}{}^{\beta} \vartheta$,
\end{enumerate}
but again (iia) vanishes by the symmetries of the Ricci tensor, while (iib) involves the commutator of covariant derivatives, which introduces another power of the curvature tensor. 
Obviously, the only term one can write with the Ricci scalar would lead to a double commutator of covariant derivatives, leading to extra factors of the curvature tensor. 

One is then forced to consider either theories with two mutually independent fields or theories with quadratic curvature tensors. Of the latter, the only combination that can be 
constructed and that does not vanish by the Bianchi identities is the so called Pontryagin density, i.e.~$R{}^{*}R$, and therefore, the action~\cite{Jackiw:2003pm,Alexander:2009tp}
\begin{equation}
S = \int d^{4}x \sqrt{-g} \left(\kappa \; R + \frac{\alpha}{4} \; \vartheta \; R{}^{*}R\right)\,,
\label{CS-action}
\end{equation}
is the most general, quadratic action with a single scalar field that introduces gravitational parity violation\epubtkFootnote{If we allow for multiple scalar fields or multiple derivatives of the scalar field, one can construct a more general parity-violating gravity theory, including a term with a single curvature tensor in the action~\cite{Crisostomi:2017ugk}. In fact, one can even construct ghost-free, parity-violating theories that evade the Ostrogradskii instability~\cite{Crisostomi:2017ugk}, though such theories predict that gravitational waves propagate at speeds different from $c$ in general~\cite{Nishizawa:2018srh}.}, where we have rescaled the $\alpha$ prefactor to follow historical 
conventions. This action defines non-dynamical Chern--Simons modified gravity, initially proposed by Jackiw and Pi~\cite{Jackiw:2003pm,Alexander:2009tp}. Notice that this is the 
same as the term proportional to $\alpha_{4}$ in the quadratic gravity action of Eq.~\eqref{eq:quad-action-simped}, except that here $\vartheta$ is prior geometry, i.e.~it does not 
possess self-consistent dynamics or an evolution equation. Such a term violates parity invariance because the Pontryagin density is a pseudo-scalar, while $\vartheta$ is assumed to 
be a scalar.

The field equations for this theory are\epubtkFootnote{The tensor ${\cal{K}}^{(1)}_{\mu \nu}$ is sometimes written as $C_{\mu \nu}$ and referred to as the C-tensor.}
\begin{equation}
G_{\mu \nu} + \frac{\alpha}{4 \kappa} {\cal{K}}^{(1)}_{\mu \nu} = \frac{1}{2 \kappa} T_{\mu \nu}^{\mathrm{mat}}\,,
\end{equation}
which is simply Eq.~\eqref{FEs} with $(\alpha_{1},\alpha_{2},\alpha_{3})$ set to zero and no stress-energy for $\vartheta$. Clearly, these field equations are not covariantly conserved 
in vacuum, i.e.~taking the covariant divergence one finds the constraint
\begin{equation}
\alpha R{}^{*}R = 0\,.
\label{eq:RR}
\end{equation}
This constraint restricts the space of allowed solutions, for example disallowing the Kerr metric~\cite{Grumiller:2007rv}. Therefore, it might seem that the evolution equations for the 
metric are now overconstrained, given that the field equations provide 10 differential conditions for the 10 independent components of the metric tensor, while the constraint adds one 
additional, independent differential condition. Moreover, unless the Pontryagin constraint, Eq.~\eqref{eq:RR}, is satisfied, matter fields will not evolve according to $\nabla^{\mu} 
T_{\mu \nu}^{\mathrm{mat}} = 0$, thus violating the equivalence principle. 

From the field equations, we can derive an evolution equation for the metric perturbation when linearizing about a flat background, namely
\begin{equation}
\square_{\eta} h_{\mu \nu} 
+ \frac{\alpha}{\kappa}  \left(\vartheta_{,\gamma} \; \epsilon_{(\mu}{}^{\gamma \delta \chi} \square_{\eta} h_{\nu) \delta,\chi}
-  \vartheta_{,\gamma}{}^{\zeta} \; \epsilon_{(\mu}{}^{\gamma \delta \chi} h_{|\delta \zeta|,\nu)\chi}
+ \vartheta_{,\gamma}{}^{\zeta} \;  \epsilon_{(\mu}{}^{\gamma \delta \chi} h_{\nu) \delta,\chi \zeta} \right) = -\frac{2}{\kappa} T_{\mu \nu}^{\mathrm{mat}}\,
\label{eq:h-EOM-NDCS}
\end{equation}
in a transverse-traceless gauge, which can be shown to exist in this theory~\cite{Alexander:2007kv,Yunes:2008bu}. The constraint of Eq.~\eqref{eq:RR} is identically satisfied to 
second order in the metric perturbation. However, without further information about $\vartheta$, one cannot proceed any further, except for a few general observations. As is clear 
from Eq.~\eqref{eq:h-EOM-NDCS}, the evolution equation for the metric perturbation can contain third time derivatives, which generically will lead to instabilities. In fact, as shown 
in~\cite{Alexander:2004wk} the general solution to these equations will contain exponentially growing and decaying modes. The theory defined by Eq.~\eqref{CS-action}, however, is 
an effective theory, and thus, there can be higher order operators not included in this action that may stabilize the solution. Regardless, when studying this theory, order-reduction is 
necessary if one is to consider it an effective model. 

Let us now discuss the properties of such an effective theory. Because of the structure of the modification to the field equations, one can always choose a sufficiently small value for $
\alpha$ such that all solar system tests are satisfied. In fact, one can see from the equations in this section that in the limit $\alpha \to 0$, one recovers GR. Non-dynamical Chern--Simons gravity leads to modifications to the non-radiative (near-zone) metric in the gravitomagnetic sector, leading to corrections to Lense--Thirring 
precession~\cite{Alexander:2007zg,Alexander:2007vt}. This fact has been used to constrain the theory through observations of the orbital motion of the LAGEOS 
satellites~\cite{Smith:2007jm} to $(\alpha/\kappa) \dot{\vartheta} < 2 \times 10^{4} {\mathrm{\ km}}$, or equivalently $(\kappa/\alpha) \dot{\vartheta}^{-1} \gtrsim 10^{-14} {\mathrm{\ eV}}
$. 
Much better constraints, however, can be placed through observations of the double binary pulsar~\cite{Yunes:2008ua,AliHaimoud:2011bk}: $(\alpha_{4}/\kappa) \dot{\vartheta} < 0.8 
{\mathrm{\ km}}$. 

Some of the sub-properties of the fundamental requirement are satisfied in non-dynamical Chern--Simons Gravity. On the one hand, all spherically symmetric metrics that are 
solutions to the Einstein equations are also solutions in this theory for a ``canonical'' scalar field ($\theta \propto t$)~\cite{Grumiller:2007rv}. On the other hand, axisymmetric solutions 
to the Einstein equations are generically not solutions in this theory. Moreover, although spherically symmetric solutions are preserved, perturbations of such spacetimes that are 
solutions to the Einstein equations are not generically solutions to the modified theory~\cite{Yunes:2007ss}. What is perhaps worse, the evolution of perturbations to non-spinning 
black holes have been found to be generically overconstrained~\cite{Yunes:2007ss}. This is a consequence of the lack of scalar field dynamics in the modified theory, which via 
Eq.~\eqref{eq:RR} tends to overconstrain it. Such a conclusion also suggests that this theory does not possess a well-posed initial-value problem.  One can argue that non-dynamical 
Chern--Simons Gravity is well-motivated from fundamental theories~\cite{Alexander:2009tp}, except that in the latter, the scalar field is always dynamical, instead of having to be 
prescribed {\emph{a priori}}. Thus, perhaps the strongest motivation for such a model is as a phenomenological proxy to test whether the gravitational interaction remains parity 
invariant in extreme gravity, a test that is uniquely suited to this modified model.

\subsection{Currently unexplored theories in the gravitational-wave sector}

The list of theories we have here described is by no means exhaustive. In fact, there are many fascinating theories that we have chosen to leave out  because they have not yet been 
analyzed in the gravitational wave context in detail. We will update this review with a description of these theories, once a detailed gravitational-wave study for compact binaries or 
supernovae sources is carried out and the predictions for the gravitational waveform observables are made for any physical system plausibly detectable by current or near future 
gravitational-wave experiments. Similarly, there are theories that have only recently began to be studied in the gravitational wave context, such as bigravity and Horndeski gravity. 
We have chosen to include these theories in this review, as work has already begun to understand their predictions for gravitational waves. Once more work is done, we will update
this review to splinter such theories into whole sections in their own right.

\newpage
\section{Detectors and Testing Techniques}
\label{section:detectorsandtechniques}

\subsection{Gravitational-wave interferometers}

Kilometer-scale gravitational-wave interferometers have been in operation for almost two decades.  This type of
detectors use laser interferometry to monitor the locations of test masses at the ends of the arms with
exquisite precision. Gravitational waves change the relative length of 
the optical cavities in the interferometer (or equivalently, the proper travel time of photons) resulting in a strain
$$
h=\frac{\Delta L}{L},
$$
where $\Delta L$ is the path length difference between the two arms of the interferometer.

Fractional changes in the difference in path lengths along the two arms can be 
monitored to better than 1 part in $10^{20}$.  It is not hard to understand how such 
precision can be achieved. For a simple Michelson interferometer,
a difference in path length of order the size of a fringe can easily be detected. 
For the typically-used, infrared lasers of wavelength $\lambda \sim 1\,\mu\mathrm{m}$, 
and interferometer arms of length $L=4 \mathrm{\ km}$, the minimum 
detectable strain is 
$$
h \sim \frac{\lambda}{L} \sim 3 \times 10^{-10}.
$$

This is still far off the $10^{-20}$ mark. In principle, however, changes in the length of the cavities corresponding to 
fractions of a single fringe can also be measured provided we have a sensitive photodiode 
at the dark port of the interferometer, and enough photons to perform the 
measurement. This way we can track changes in the amount of light incident on the photodiode
as the lengths of the arms change and we move over a fringe.  The rate at which photons arrive
at the photodiode is a Poisson process and the fluctuations in the number of photons is $\sim N^{1/2}$, 
where $N$ is the number of photons. Therefore we can track changes in the path length difference of order
$$
\Delta L \sim \frac{\lambda}{N^{1/2}}.
$$
The number of photons depends on the laser power $P$, and the amount of time available to 
perform the measurement. For a gravitational wave of frequency $f$, we can collect photons for a time 
$t \sim 1/f$, so the number of photons is
$$
N \sim \frac{P}{f h_p \nu},
$$
where $h_p$ is Planck's constant and $\nu=c/\lambda$ is the laser frequency. For a typical laser 
power $P \sim 1 \mathrm{\ W}$, a gravitational wave frequency $f=100 \mathrm{\ Hz}$, and $\lambda \sim 1\,\mu\mathrm{m}$ the number of photons
is
$$
N \sim 10^{16}, 
$$
so that the strain we are sensitive to becomes
$$
h \sim 10^{-18}.
$$

The sensitivity can be further improved by increasing the effective length of the arms. 
In the LIGO instruments, for example, each of the two arms forms a resonant Fabry--P\'erot cavity. For
gravitational-wave frequencies smaller than the inverse of the light storage time, the light in the 
cavities makes many back and forth trips in the arms while the wave is traversing the instrument. 
For gravitational waves of frequencies around 
100~Hz and below, the light makes about 
a thousand back and forth trips while the gravitational wave is traversing the interferometer, which results in a 
three-orders-of-magnitude improvement in sensitivity,
$$
h \sim 10^{-21}.
$$
For frequencies larger than 100~Hz the number of round trips the light makes in the Fabry--P\'erot cavities
while the gravitational wave is traversing the instrument is reduced and the sensitivity is degraded.

The proper light travel time of photons in interferometers is affected by the metric perturbation, which can be expressed as a sum over polarization modes
\begin{equation}
h_{ij}(t,\vec{x}) = \sum_{A} h^A_{ij}(t,\vec{x}),
\end{equation}
where $A$ labels the six possible polarization modes in metric 
theories of gravity. The metric perturbation for each mode can be written in terms of a plane wave expansion,
\begin{equation}
 h^A_{ij}(t,\vec{x})=\int_{-\infty}^{\infty}df\, \int_{S^2}d\hat{\Omega}\, 
e^{i2\pi f(t-\hat{\Omega}\cdot\vec{x})}
\tilde h^A(f,\hat{\Omega})\epsilon_{ij}^A(\hat{\Omega}).
\label{pwexp2}
\end{equation}
Here $f$ is the frequency of the gravitational waves, 
$\vec k= 2 \pi f \hat \Omega$ is the wave vector, $\hat \Omega$ is a unit vector
that points in the direction of propagation of the gravitational waves, $e_{ij}^{A}$ is the $A$th polarization tensor,
with $i,j=x,y,z$ spatial indices. The
metric perturbation due to mode $A$ from the direction $\hat \Omega$  can be written by
integrating over all frequencies,
\begin{equation}
 h^A_{ij}(t -\hat \Omega \cdot \vec{x})=\int_{-\infty}^{\infty}df\
e^{i2\pi f(t-\hat{\Omega}\cdot\vec{x})}
\tilde h^A(f,\hat{\Omega}) \epsilon_{ij}^A(\hat{\Omega}).
\label{pwexp}
\end{equation}
By integrating Eq.~\eqref{pwexp2} over all frequencies we have an
expression for the metric perturbation from a particular direction
$\hat \Omega$, i.e., only a function of $t -\hat \Omega \cdot \vec{x}$.
The full metric perturbation due to a gravitational wave from a 
direction $\hat \Omega$ can be written as a sum over all polarization modes
\begin{equation}
h_{ij}(t -\hat \Omega \cdot \vec{x})=\sum_A h^A(t-\hat \Omega \cdot \vec{x}).
\end{equation}

The response of an interferometer to gravitational waves is generally referred to as the 
antenna pattern response, and depends on the geometry of the detector and the
direction and polarization of the gravitational wave. To derive the
antenna pattern response 
of an interferometer for all
six polarization modes we follow the discussion in~\cite{Nishizawa:2009bf} closely. 
For a gravitational wave propagating in the $z$ direction, the
polarization tensors are as follows
\begin{eqnarray}
\epsilon_{ij}^{+}&=& 
\left(
\begin{array}{ccc} 
1 & 0 & 0  \\
0 & -1 & 0  \\
0 & 0 & 0 
\end{array}
\right), 
\epsilon_{ij}^{\times}= 
\left(
\begin{array}{ccc} 
0 & 1 & 0  \\
1 & 0 & 0  \\
0 & 0 & 0 
\end{array}
\right),\nonumber \\ 
\epsilon_{ij}^{x} &=&
\left(
\begin{array}{ccc} 
0 & 0 & 1  \\
0 & 0 & 0  \\
1 & 0 & 0 
\end{array}
\right), 
\epsilon_{ij}^{y}= 
\left(
\begin{array}{ccc} 
0 & 0 & 0  \\
0 & 0 & 1  \\
0 & 1 & 0 
\end{array}
\right),  \nonumber \\
\epsilon_{ij}^{b}&=& 
\left(
\begin{array}{ccc} 
1 & 0 & 0  \\
0 & 1 & 0  \\
0 & 0 & 0 
\end{array}
\right), 
\epsilon_{ij}^{\ell}=
\left(
\begin{array}{ccc} 
0 & 0 & 0  \\
0 & 0 & 0  \\
0 & 0 & 1 
\end{array}
\right), 
\label{polarizationsz} 
\end{eqnarray}
where the superscripts $+$, $\times$, $x$, $y$, $b$, and $\ell$
correspond to the plus, cross, vector-x, vector-y,  breathing, and
longitudinal modes.

\epubtkImage{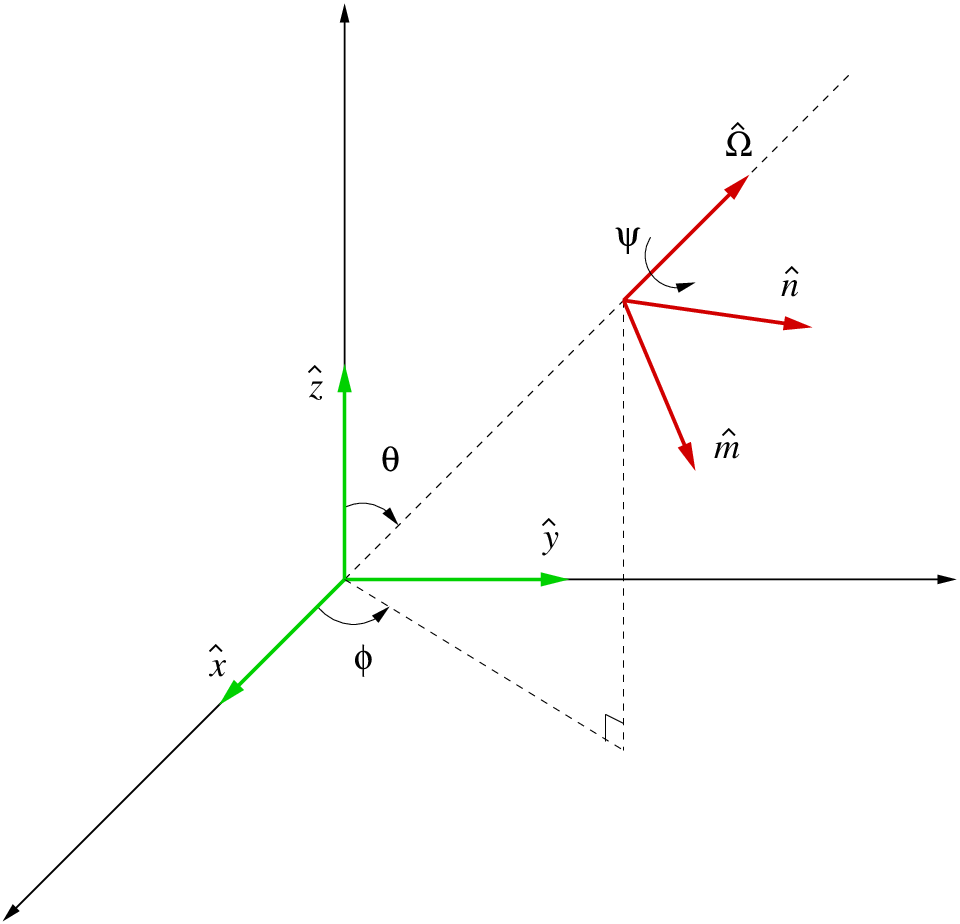}{%
\begin{figure}[htbp]
\centerline{\includegraphics[width=7cm]{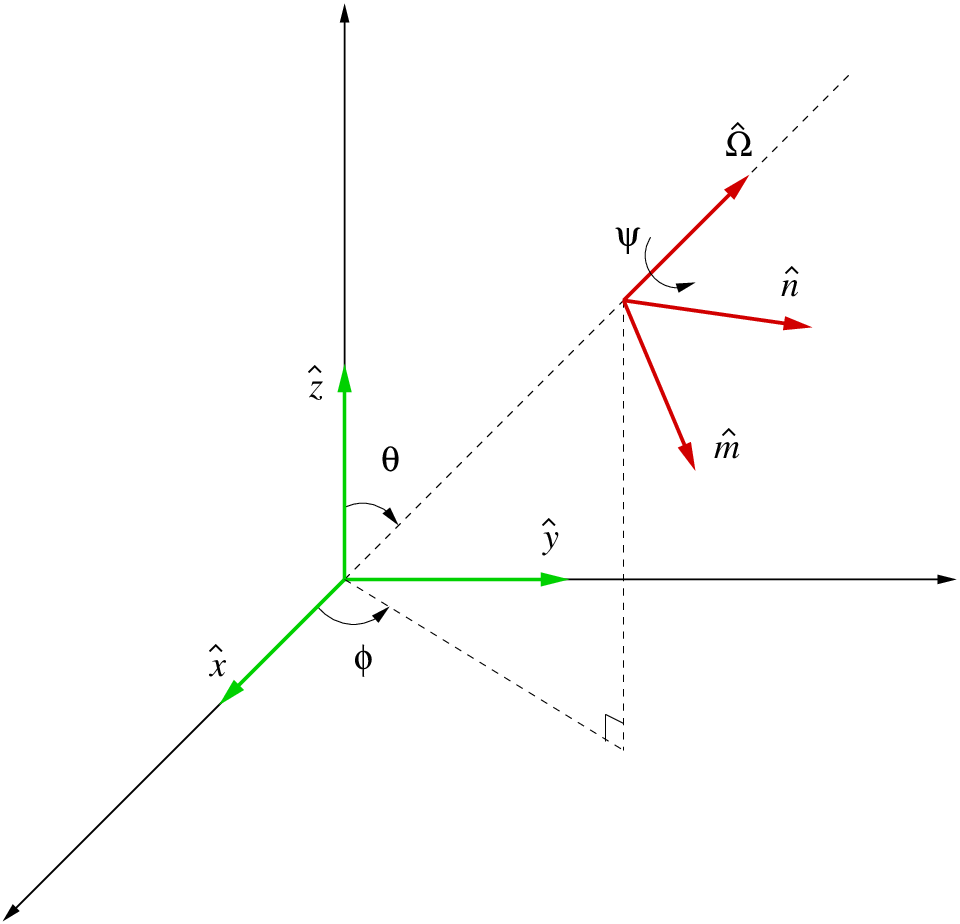}}
\caption{Detector coordinate system and gravitational-wave coordinate
  system.}
\label{coodinatesystems}
\end{figure}}

Suppose that the coordinate system for the detector is $\hat x=
(1,0,0), \hat y = (0,1,0),
\hat z = (0,0,1)$, as in Figure~\ref{coodinatesystems}. Relative to the detector, the gravitational-wave coordinate system is 
rotated by angles $(\theta, \phi)$,
$\hat x ^{\prime} =(\cos \theta \cos \phi , \cos \theta \sin \phi , -\sin \theta)$, 
$\hat y^{\prime} = (- \sin \phi , \cos \phi , 0)$, and 
$\hat z^{\prime} = (\sin \theta \cos \phi , \sin \theta \sin \phi , \cos \theta)$.
We still have the freedom to perform a rotation about the gravitational-wave propagation direction 
which introduces the polarization angle $\psi$, 
\begin{equation}
\begin{array}{lll} 
\displaystyle
\hat m= \hat x^{ \prime} \cos \psi + \hat y^{\prime} \sin \psi\,,\\ 
\displaystyle 
\hat n= - \hat x^{ \prime} \sin \psi + \hat y ^{\prime} \cos \psi\,,
\\ 
\displaystyle 
\hat \Omega = \hat z^{ \prime}\,.
\end{array}
\nonumber
\end{equation}
The coordinate systems $(\hat x,\hat y,\hat z)$ 
and  $(\hat m,\hat n,\hat \Omega)$ 
are also shown in Figure~\ref{coodinatesystems}. 
To generalize the polarization tensors in Eq.~\eqref{polarizationsz} to a wave coming from a 
direction $\hat{{\Omega}}$, we use the unit vectors $\hat{{m}}$, $\hat{{n}}$, 
and $\hat{{\Omega}}$ as follows
\begin{eqnarray}
\epsilon^{+} &=& \hat{{m}} \otimes \hat{{m}} -\hat{{n}} \otimes \hat{{n}} , \nonumber \\
\epsilon^{\times} &=& \hat{{m}} \otimes \hat{{n}} +\hat{{n}} \otimes \hat{{m}} , \nonumber \\
\epsilon^{x} &=& \hat{{m}} \otimes \hat{{\Omega}} +\hat{{\Omega}} \otimes \hat{{m}} , \nonumber \\
\epsilon^{y} &=& \hat{{n}} \otimes \hat{{\Omega}}+\hat{{\Omega}} \otimes \hat{{n}}, \nonumber \\
\epsilon^{b} &=& \hat{{m}} \otimes \hat{{m}} + \hat{{n}} \otimes \hat{{n}} ,  \nonumber \\
\epsilon^{\ell} &=& \hat{{\Omega}} \otimes \hat{{\Omega}}\,.
\end{eqnarray}
For LIGO and VIRGO the arms are perpendicular
so that the antenna pattern response can be written as the difference of projection of the 
polarization tensor onto each of the interferometer arms,
\begin{equation}
F^A (\hat \Omega, \psi) = \frac{1}{2} 
\left(\hat x^i \hat x^j - \hat y^i \hat y^j  \right)\epsilon^A_{ij} (\hat \Omega, \psi).
\end{equation}
This means that the strain measured by an interferometer due to a
gravitational wave from direction $\hat \Omega$ and polarization angle
$\psi$ takes the
form
\begin{equation}
h(t)=\sum_A h_A(t-\hat \Omega \cdot x) F^A (\hat \Omega, \psi).
\label{strain}
\end{equation}
Explicitly, the antenna pattern functions are,
\begin{eqnarray}
F^{+}(\theta, \phi, \psi) &=& \frac{1}{2} (1+ \cos ^2 \theta ) \cos 2\phi \cos 2 \psi
 - \cos \theta \sin 2\phi \sin 2 \psi, \nonumber \\
F^{\times}(\theta, \phi, \psi) &=& -\frac{1}{2} (1+ \cos ^2 \theta ) \cos 2\phi \sin 2 \psi 
 -  \cos \theta \sin 2\phi \cos 2 \psi,\nonumber \\
F^{x}(\theta, \phi, \psi) &=& \sin \theta \,(\cos \theta \cos 2 \phi
\cos \psi -\sin 2\phi \sin \psi) ,\nonumber \\
F^{y}(\theta, \phi, \psi) &=& - \sin \theta \,(\cos \theta \cos 2 \phi
\sin \psi +\sin 2\phi \cos \psi) ,\nonumber \\
F^{b}(\theta, \phi) &=& -\frac{1}{2} \sin^2 \theta \cos 2\phi,
\nonumber \\
F^{\ell}(\theta, \phi) &=& \frac{1}{2} \sin^2 \theta \cos 2\phi. 
\label{IFOresponse} 
\end{eqnarray}
The dependence on the polarization angles $\psi$ reveals that 
the $+$ and $\times$ polarizations are spin-2 tensor modes, 
the $x$ and $y$ polarizations are spin-1 vector modes, and the $b$ 
and $\ell$ polarizations are spin-0 scalar modes. Note that for interferometers
the antenna pattern responses of the scalar modes are degenerate. Figure
\ref{LIGOresponse} shows the antenna patterns for the various
polarizations given in Eq.~\eqref{IFOresponse} with $\psi=0$. The color indicates the
strength of the response with red being the strongest and blue being
the weakest.

\epubtkImage{LIGOAllPolarizations_html.png}{%
\begin{figure}[htbp]
\centerline{\includegraphics[width=15cm]{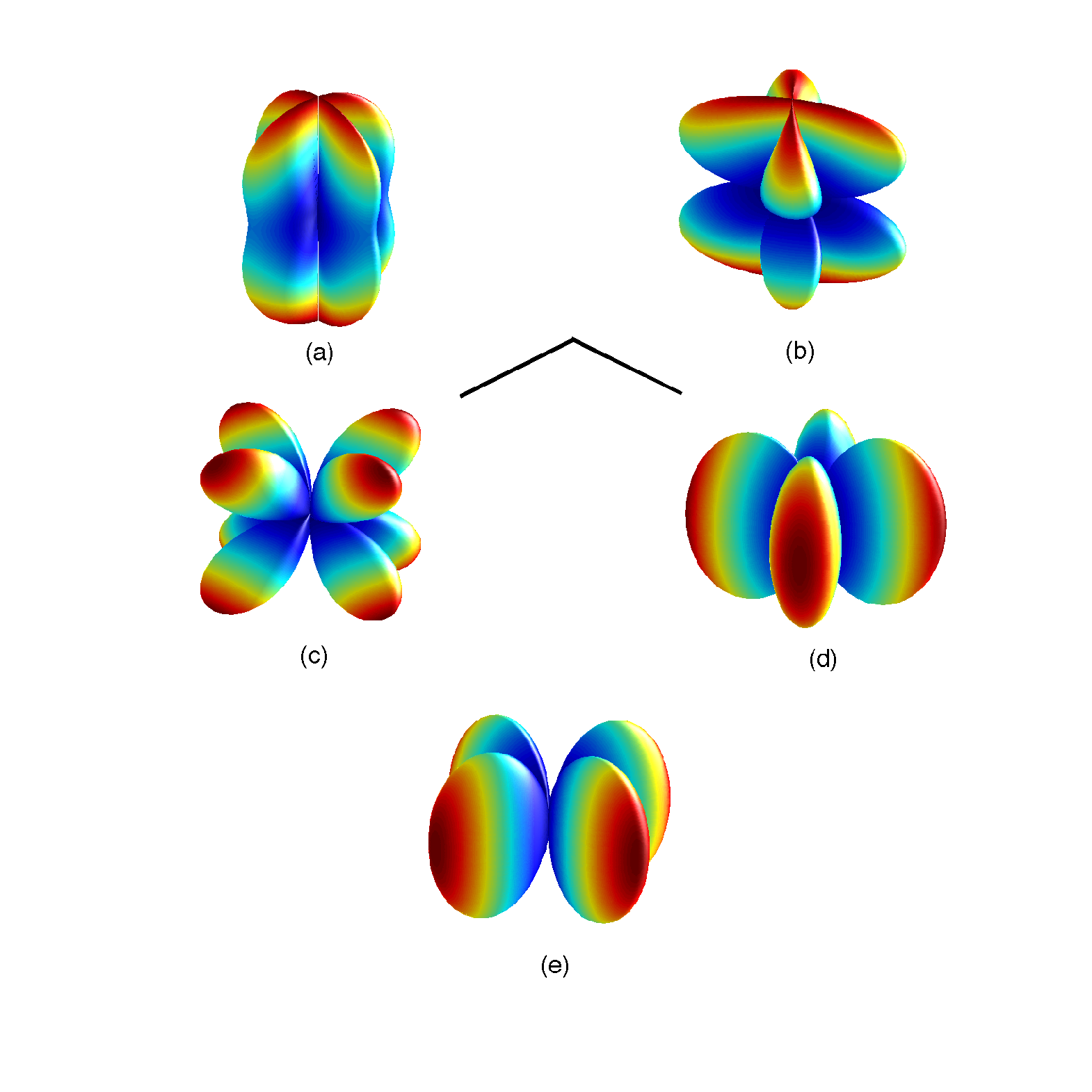}}
\caption{Antenna pattern response functions of an interferometer (see
  Eqs.~\eqref{IFOresponse}) for $\psi=0$. Panels (a) and (b) 
show the plus ($|F_+|$) and cross ($|F_{\times}|$) modes, panels (c) and (d) the vector x and vector y 
modes ($|F_{x}|$ and $|F_{y}|$), and panel (e) shows the scalar modes
(up to a sign, it is the same for both breathing and
longitudinal). Color  indicates the
strength of the response with red being the strongest and blue being
the weakest. The black lines near the center give the orientation of
the interferometer arms.}
\label{LIGOresponse}
\end{figure}}

\subsection{Pulsar timing arrays}

Neutron stars can emit powerful beams of radio waves from their
magnetic poles. If the rotational and magnetic axes are not aligned,
the beams sweep through space like the beacon on a lighthouse.  
If the line of sight is aligned with the magnetic axis at any point 
during the neutron star's rotation the star is observed as a source of
periodic radio-wave bursts.  Such a neutron star is referred to as a pulsar.
Due to their large moment of inertia pulsars are very stable rotators,
and their radio pulses arrive at Earth with extraordinary regularity. Pulsar timing experiments
exploit this regularity; gravitational waves can cause measurable 
deviations in the expected times of arrival of radio pulses from
pulsars. 

The effect of a gravitational wave on the pulses propagating from a
pulsar to Earth was first computed in the late 1970s by Sazhin and Detweiler~\cite{saz78,det79}.
Gravitational waves induce a redshift in the pulse train 
\begin{equation}
z(t,\hat{\Omega}) = 
\frac{1}{2}
\frac{\hat{p}^i\hat{p}^j}{1+\hat{\Omega}\cdot\hat{p}}\Delta h_{ij}\,,
\label{eqzsom}
\end{equation}
where $\hat p$ is a unit vector that points in the direction of the pulsar,
$\hat \Omega$ is a unit vector in the direction of gravitational wave propagation,
and 
\begin{equation}
\Delta h_{ij}
\equiv
h_{ij}(t_{\mathrm{e}},\hat{\Omega}) - 
h_{ij}(t_{\mathrm{p}},\hat{\Omega}),
\label{delhdef}
\end{equation}
is the difference in the metric perturbation at the pulsar when the
pulse was emitted at $t=t_p$ (second term) and the metric perturbation on Earth when the pulse
was received at $t=t_e$ (first term). The inner product in Eq.~\eqref{eqzsom} is computed with 
the Euclidean metric. 

In pulsar timing experiments it is not the redshift, but rather the 
\emph{timing residual} that is measured.  The times of arrival (TOAs) of
pulses are measured, and the timing residual is produced by subtracting
off a model that includes the rotational frequency of the pulsar, the
spin-down (frequency derivative), binary parameters if the pulsar is 
in a binary, sky location and proper motion, etc. The timing
residual induced by a gravitational wave, $R(t)$, is just the integral of the redshift
\begin{equation}
R(t)\equiv \int_0^t dt'\, z(t').
\end{equation}
Times-of-arrival are measured a few times a year over the
course of several years allowing for gravitational waves in the
nano-Hertz band to be detected.  Currently, the best timed pulsars
have residual RMSs of a few tens of ns over a decade or longer. 

The equations above (\eqref{eqzsom}) can be used
to estimate the strain sensitivity of pulsar timing experiments.  
For gravitational waves of frequency $f$ the expected
induced residual is
\begin{equation}
R \sim \frac{h}{f}\,,
\end{equation}
so that for pulsars with RMS residuals $R \sim 100 \mathrm{\ ns}$, and gravitational waves
of frequency $f \sim 10^{-8} \mathrm{\ Hz}$, gravitational waves with strains
\begin{equation}
h \sim Rf \sim 10^{-15}
\end{equation}
would produce a measurable effect.

To find the antenna pattern response of the pulsar-Earth system, we are free to place the pulsar on the $z$-axis. The response to gravitational waves of
different polarizations can then be written as
\begin{equation}
F^A (\hat \Omega, \psi) = \frac{1}{2} \frac{\hat z^i \hat z^j}{1+ \cos
 \theta} \epsilon^A_{ij} (\hat \Omega, \psi)\,,
\end{equation}
which allows us to express the Fourier transform of \eqref{eqzsom} as
\begin{eqnarray}\label{FTantpatts}
	\tilde{z}(f, \hat{\Omega}) = \left(1-e^{-2 \pi i f L 
(1+\hat{\Omega} \cdot \hat{p})} \right) \sum_A \tilde h_A(f,\hat{\Omega})F^A(\hat{\Omega})
\end{eqnarray}
where the sum is over all possible gravitational-wave polarizations: $A=+, 
\times, x, y, b,l$, and $L$ is the distance to the pulsar. 

Explicitly,
\begin{eqnarray}
F^{+}(\theta, \psi) &=& \sin ^2 \frac{\theta}{2}  \cos 2 \psi\\
F^{\times}(\theta,  \psi) &=&  -\sin ^2 \frac{\theta}{2}  \sin 2 \psi\\
F^{x}(\theta, \psi) &=&  -\frac{1}{2} \frac{\sin 2 \theta }{1+\cos \theta}  \cos \psi,\\
F^{y}(\theta, \psi) &=& \frac{1}{2} \frac{\sin 2 \theta }{1+\cos \theta}\sin \psi,\\
F^{b}(\theta) &=&  \sin ^2 \frac{\theta}{2}\\
F^{\ell}(\theta) &=&  \frac{1}{2} \frac{\cos ^2 \theta}{1+\cos \theta}.
\label{PulsarResponse} 
\end{eqnarray}
Just like for
the interferometer case, the dependence on the polarization angle $\psi$, reveals that 
the $+$ and $\times$ polarizations are spin-2 tensor modes, 
the $x$ and $y$ polarizations are spin-1 vector modes, and the $b$ 
and $\ell$ polarizations are spin-0 scalar modes. Unlike
interferometers, the antenna pattern responses of the pulsar-Earth system do not
depend on the azimuthal angle of the gravitational wave, and
the scalar modes are not degenerate. 

\epubtkImage{PTAAllPolarizations2_html.png}{%
  \begin{figure}[htbp]
    \centerline{\includegraphics[width=15cm]{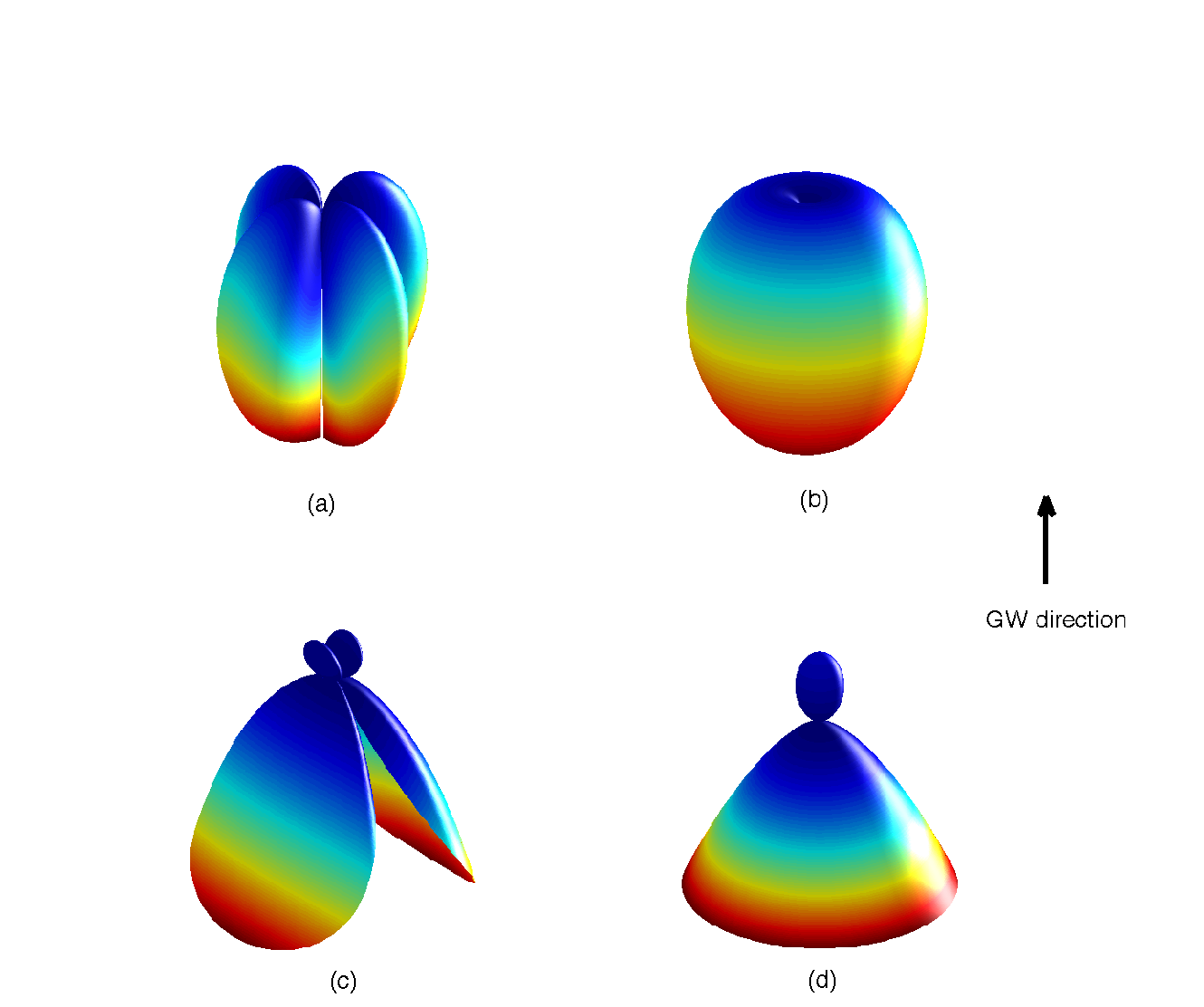}}
    \caption{Antenna patterns for the pulsar-Earth system. The plus
      mode is shown in (a), breathing modes in (b), the vector-x mode
      in (c), and longitudinal modes in (d), as computed from
      Eq.~\eqref{PulsarResponse2}. The cross mode and the vector-y
      mode are rotated versions of the plus mode and the vector-x
      mode, respectively,  so we did not include them here. The
      gravitational wave propagates in the positive $z$-direction with
      the Earth at the origin, and the antenna pattern depends on the
      pulsar's location. The color indicates the strength of the
      response, red being the largest and blue the smallest.}
    \label{PTAantenna}
\end{figure}}

In the literature, it is common to write the antenna pattern response
by fixing the gravitational-wave direction and changing the location of the pulsar. In
this case the antenna pattern responses are~\cite{LeeJenetPrice,daSilvaAlves:2011fp,Chamberlin:2011ev},
\begin{eqnarray}
\tilde F^{+}(\theta_p, \phi_p) &=& \sin ^2 \frac{\theta_p}{2}  \cos 2
\phi_p \nonumber \\
\tilde F^{\times}(\theta_p,  \phi_p) &=& \sin ^2 \frac{\theta_p}{2}  \sin 2 \phi_p \nonumber \\
\tilde F^{x}(\theta_p, \phi_p) &=&  \frac{1}{2} \frac{\sin 2 \theta_p }{1+\cos \theta_p}  \cos \phi_p,\nonumber \\
\tilde F^{y}(\theta_p, \phi_p) &=& \frac{1}{2} \frac{\sin 2 \theta_p }{1+\cos \theta_p}\sin \phi_p,\nonumber \\
\tilde F^{b}(\theta_p) &=&  \sin ^2 \frac{\theta_p}{2}\nonumber \\
\tilde F^{\ell}(\theta_p) &=&  \frac{1}{2} \frac{\cos ^2 \theta_p}{1+\cos \theta_p},
\label{PulsarResponse2} 
\end{eqnarray} 
where $\theta_p$ and $\phi_p$ are the polar and azimuthal angles
of the vector pointing to the pulsar, respectively.  Up to signs, these expressions are the same
as Eq.~\eqref{PulsarResponse} taking $\theta \rightarrow \theta_p$ and
$\psi \rightarrow \phi_p$.
This is because fixing the gravitational-wave propagation direction
while allowing the pulsar location to change 
is analogous to fixing the pulsar position while 
allowing the direction of gravitational-wave propagation to change -- there is degeneracy in the gravitational-wave polarization 
angle and the pulsar's azimuthal angle $\phi_p$. For example, changing
the polarization angle of a
gravitational wave traveling in the $z$-direction is the same as
performing a rotation about the $z$-axis that changes the pulsar's
azimuthal angle. Antenna patterns for the pulsar-Earth system using
Eqs.~\eqref{PulsarResponse2} are shown
in Figure~\ref{PTAantenna}.  The color indicates the strength of the response, red being
the largest and blue the smallest.

\newcommand{\Deqn}[1]{{Eq.~(\ref{#1})}}
\newcommand{\Deqns}[1]{{Eqs.~(\ref{#1})}}
\newcommand{\Ceqn}[1]{{Eq.~(\ref{#1})}}
\newcommand{\Dfig}[1]{{Figure~\ref{#1}}}
\newcommand{\beq}{\begin{equation}}
\newcommand{\eeq}{\end{equation}}
\newcommand{\bea}{\begin{eqnarray}}
\newcommand{\eea}{\end{eqnarray}}
\newcommand{\tr}{{\mathop{\mathrm{tr}}\nolimits}}

\subsection{Coalescence analysis}

Gravitational waves emitted during the inspiral, merger and ringdown of compact binaries are the most studied in the context of data analysis and parameter estimation. In this section, we will review some of the main data analysis techniques employed in the context of parameter estimation and tests of GR. We begin with a discussion of matched filtering and Fisher theory (for a detailed review, see~\cite{Finn:1992xs,Chernoff:1993th,Cutler:1994ys,Finn:2001qi,lrr-2012-4}). We then continue with a discussion of Bayesian parameter estimation and hypothesis testing (for a detailed review, see~\cite{Sivia-Skilling,Gregory,Cornish:2007if,Littenberg:2009bm,Romano:2016dpx}). 

\subsubsection{Matched filtering and Fisher analysis}

When the detector noise $n(t)$ is Gaussian and stationary, and when the signal $s(t)$ is known very well, the optimal detection strategy is matched filtering. For any given realization, such noise can be characterized by its power spectral density $S_{n}(f)$, defined via 
\begin{equation}
\left< \tilde{n}(f) \, \tilde{n}^{*}(f')\right> = \frac{1}{2} S_{n}(f) \delta\left(f - f'\right)\,,
\label{eq:Sn(f)-def}
\end{equation}
where recall that the tilde stands for the Fourier transform, the asterisk for complex conjugation and the brackets for the expectation value.

The detectability of a signal is determined by its {\emph{signal-to-noise ratio}} or SNR, which is defined via
\begin{equation}
\rho^{2} = \frac{\left(s|h\right)}{\sqrt{\left(h|n\right)\left(n|h\right)}}\,,
\end{equation}
where $h$ is a template with parameters $\lambda^{i}$ and we have defined the inner product
\begin{equation}
\left(A|B\right) \equiv 4 \Re \int_{0}^{\infty} \frac{\tilde{A}^{*} \tilde{B}}{S_{n}} df\,.
\label{eq:inner-prod-def}
\end{equation}
If the templates do not exactly match the signal, then the SNR is reduced by a factor of $\bar {\cal{M}}$, called the {\emph{match}}:
\begin{equation}
\bar{{\cal{M}}} \equiv \frac{\left(s|h\right)}{\sqrt{\left(s|s\right)\left(h|h\right)}}\,,
\end{equation}
where $1 - \bar{{\cal{M}}} = {\cal{MM}}$ is the mismatch.

For the noise assumptions made here, the probability of measuring $s(t)$ in the detector output, given a template $h$, is given by
\begin{equation}
p \propto e^{-\left(s - h|s -h\right)/2}\,,
\label{eq:likelihood}
\end{equation}
and thus the waveform $h$ that best fits the signal is that with best-fit parameters such that the argument of the exponential is minimized. For large SNR, the best-fit parameters will have a multivariate Gaussian distribution
centered on the true values of the signal $\hat{\lambda}^{i}$, and thus, the waveform parameters that best fit the signal minimize the argument of the exponential. The parameter errors $\delta \lambda ^{i}$ will be distributed according to
\begin{equation}
p(\delta \lambda^{i}) \propto e^{-\frac{1}{2} \Gamma_{ij} \delta \lambda^{i} \delta \lambda^{j}}\,,
\end{equation}
where $\Gamma_{ij}$ is the {\emph{Fisher matrix}}
\begin{equation}
\Gamma_{ij} \equiv \left.\left(\frac{\partial h}{\partial \lambda^{i}}\right|\frac{\partial h}{\partial \lambda^{j}}\right)\,.
\end{equation}
The root-mean-squared ($1\sigma$) error on a given parameter $\lambda^{\bar{i}}$ is then
\begin{equation}
\label{Fisher-error}
\sqrt{\left<(\delta \lambda^{\bar{i}})^{2}\right>} = \sqrt{\Sigma^{\bar{i} \bar{i}}}\,,
\end{equation}
where $\Sigma^{ij} \equiv (\Gamma_{ij})^{-1}$ is the variance-covariance matrix and summation is not implied in Eq.~\eqref{Fisher-error} ($\lambda^{\bar{i}}$ denotes a particular element of the vector $\lambda^{i}$). This root-mean-squared error is sometimes referred to as the {\emph{statistical}} error in the measurement of $\lambda^{\bar{i}}$. One can use Eq.~\eqref{Fisher-error} to estimate how well modified gravity parameters can be measured assuming an observation consistent with Einstein's theory. Put another way, if a gravitational wave were detected and found consistent with GR, Eq.~\eqref{Fisher-error} would provide an estimate of how close to zero these modified gravity parameters would have to be consistent with statistical fluctuations.

The Fisher method to estimate projected constraints on modified gravity theory parameters is as follows. First, one constructs a waveform model in the particular modified gravity theory one wishes to constrain. Usually, this waveform will be similar to the GR one, but it will contain an additional parameter, $\kappa$, such that the template parameters are now $\lambda^{i}$ plus $\kappa$. Let us assume that as $\kappa \to 0$, the modified gravity waveform reduces to the GR expectation. Then, the accuracy to which $\kappa$ can be measured, or the accuracy to which we can say $\kappa$ is zero given an observation consistent with GR, is approximately $(\Sigma^{\kappa \kappa})^{1/2}$, where the Fisher matrix associated with this variance-covariance matrix must be computed with the non-GR model evaluated at the GR limit ($\kappa \to 0$). Such a method for estimating how well modified gravity theories can be constrained was pioneered by Will in~\cite{Will:1994fb,Poisson:1995ef}, and since then, it has been widely employed as a first-cut estimate of the power of gravitational wave tests. 

The Fisher method described above can dangerously lead to incorrect results if abused~\cite{Vallisneri:2007ev,Vallisneri:2011ts,Rodriguez:2013mla}. One must understand that this method is suitable only if the noise is stationary and Gaussian and if the SNR is sufficiently large. How large a SNR is required for Fisher methods to work depends somewhat on the signals considered, but usually for applications concerning tests of GR, one would be safe with $\rho \gtrsim 30$. In real data analysis, the first two conditions are rarely satisfied, and one must be lucky to make observations at high SNR. Fortuitously, however, the first detection that aLIGO made was at rather high SNR, $\rho \sim 24$~\cite{Abbott:2016blz,TheLIGOScientific:2016wfe}, and as shown in~\cite{Yunes:2016jcc,Cornish:2011ys}, Fisher estimates using the spectral noise of aLIGO during their first observation run are surprisingly close to constraints obtained with a full Bayesian method~\cite{TheLIGOScientific:2016src}. 

\subsubsection{Bayesian theory and model testing}

Bayesian theory is ideal for parameter estimation and model selection. Let us then assume that we have detected a signal and that it can be described by some model ${\cal{M}}$, parameterized by the vector $\lambda^{i}$. Using Bayes' theorem, the posterior distribution function (PDF) or the probability density function for the model parameters, given data $d$ and model ${\cal{M}}$, is 
\begin{equation}
p(\{\lambda^{i}\}|d,{\cal{M}}) = \frac{p(d|\{\lambda^{i}\},{\cal{M}}) p(\{\lambda^{i}\}|{\cal{M}})}{p(d|{\cal{M}})}\,.
\label{eq:PDF}
\end{equation}
Obviously, the global maximum of the PDF in the parameter manifold gives the best fit parameters for that model. The prior probability density $p(\lambda^{i}|{\cal{M}})$ represents our prior beliefs of the parameter range in model ${\cal{M}}$. The marginalized likelihood or {\emph{evidence}}, is the normalization constant
\begin{equation}
p(d|{\cal{M}}) = \int d\lambda^{1} d\lambda^{2} \ldots d\lambda^{i} \; p(d|\{\lambda^{i}\},{\cal{M}}) \; p(\{\lambda^{i}\}|{\cal{M}})\,,
\label{eq:evidence}
\end{equation}
which clearly guarantees that the integral of Eq.~\eqref{eq:PDF} integrates to unity. 
The quantity $p(d|\lambda^{i},{\cal{M}})$ is the likelihood function, which is simply given by Eq.~\eqref{eq:likelihood}, with a given normalization. In that equation we used slightly different notation, with $s$ being the data $d$ and $h$ the template associated with model ${\cal{M}}$ and parameterized by $\lambda^{i}$.  The marginalized PDF, which represents the probability density function for a given parameter $\lambda^{\bar{i}}$ (recall that $\lambda^{\bar{i}}$ is a particular element of $\lambda^{i}$), after marginalizing over all other parameters, is given by
\begin{equation}
p(\lambda^{\bar{i}}|d,{\cal{M}}) = \frac{1}{p(d|{\cal{M}})} \int_{i \neq \bar{i}} 
d\lambda^{1} d\lambda^{2} \ldots d\lambda^{i\neq \bar{i}}
\; p(\{\lambda^{i\neq\bar{i}}\}|{\cal{M}}) \; p(d|\{\lambda^{i\neq\bar{i}}\},{\cal{M}})\,,
\label{eq:marginalizedPDF}
\end{equation}
where the integration is not to be carried out over $\bar{i}$. 
 
Let us now switch gears to model selection. In hypothesis testing, one wishes to determine whether the data is more consistent with {\emph{hypothesis A}} (e.g.~that a GR waveform correctly models the signal) or with {\emph{hypothesis B}} (e.g.~that a non-GR waveform correctly models the signal). Using Bayes' theorem, the PDF for model $A$ given the data is
\begin{equation}
p(A|d) = \frac{p(d|A) p(A)}{p(d)}\,.
\label{eq:PDF2}
\end{equation}
As before, $p(A)$ is the prior probability of hypothesis $A$, namely the strength of our prior belief that hypothesis $A$ is correct. The normalization constant $p(d)$ is given by
\be
p(d) = \int d{\cal{M}} \; p(d|{\cal{M}}) \; p({\cal{M}})\,,
\label{eq:evidence2}
\ee
where the integral is to be taken over all models. Thus, it is clear that this normalization constant does not depend on the model. Similar relations hold for hypothesis $B$ by replacing $A \to B$ in Eq.~\eqref{eq:PDF2}.

When hypothesis A and B refer to fundamental theories of nature we can take different viewpoints regarding the priors. If we argue that we know nothing about whether hypothesis A or B better describes Nature, then we would assign equal priors to both hypotheses. If, on the other hand, we believe GR is the correct theory of Nature, based on all previous experiments performed (including e.g.~those in the Solar System and with binary pulsars), then we would assign $p(A)>p(B)$. This assigning of priors necessarily biases the inferences derived from the calculated posteriors, which is sometimes heavily debated when comparing Bayesian theory to a frequentist approach. However, this ``biasing'' is really unavoidable and merely a reflection of our state of knowledge of Nature (for a more detailed discussion on such issues, please refer to~\cite{Littenberg:2009bm}).      

The integral over all models in Eq.~\eqref{eq:evidence2} can never be calculated in practice, simply because we do not know all models. Thus, one is forced to investigate {\emph{relative}} probabilities between models, so that the normalization constant $p(d)$ cancels out. The so-called {\emph{odds-ratio}} is defined by
\begin{equation}
{\cal{O}}_{A,B} = \frac{p(A|d)}{p(B|d)} = \frac{p(A)}{p(B)} {\cal{B}}_{A,B}\,,
\end{equation}
where ${\cal{B}}_{A,B} \equiv p(d|A)/p(d|B)$ is the {\emph{Bayes Factor}} and the prefactor $p(A)/p(B)$ is the {\emph{prior odds}}. Vallisneri~\cite{Vallisneri:2012qq} investigated the possibility of calculating the odds-ratio using only frequentist tools and without having to compute full evidences. The odds-ratio should be interpreted as the betting-odds of model $A$ over model $B$. For example, an odds-ratio of unity means that both models are equally supported by the data, while an odds-ratio of $10^{2}$ means that there is a 100 to 1 odds that model $A$ better describes the data than model $B$. 

The main difficulty in Bayesian inference (both in parameter
estimation and model selection) is sampling the PDF sufficiently
accurately. Several methods have been developed for this purpose, but
currently the two main workhorses in gravitational wave data analysis
are Markov chain Monte Carlo and Nested Sampling. In the former, one
samples the likelihood through the Metropolis--Hastings
algorithm~\cite{Metropolis:1979dx,Hastings:1970,Cornish:2005qw,Rover:2006ni}. This
is computationally expensive in high-dimensional cases, and thus,
there are several techniques to improve the efficiency of the method,
e.g., parallel tempering~\cite{1986PhRvL..57.2607S}. Once the PDF has been sampled, one can then calculate the evidence integral, for example via thermodynamic integration~\cite{Veitch:2008wd,Feroz:2009de,vanderSluys:2008qx}. In Nested Sampling, the evidence is calculated directly by laying out a fixed number of points in the prior volume, which are then allowed to move and coalesce toward regions of high posterior probability. With the evidence in hand, one can then infer the PDF. As in the previous case, Nested Sampling can be computationally expensive in high-dimensional cases.  

Del Pozzo et al.~\cite{DelPozzo:2011pg} were the first to carry out a Bayesian implementation of model selection in the context of tests of GR. Their analysis focused on tests of a particular massive graviton theory, using the gravitational wave signal from  quasi-circular inspiral of non-spinning black holes.  Cornish, et al.~\cite{Cornish:2011ys,Sampson:2013lpa} extended this analysis by considering model-independent deviations from GR, using the parameterized post-Einsteinian (ppE) approach (Section~\ref{subsubsection:ppE})~\cite{Yunes:2009ke}. This was continued by Li et al.~\cite{Li:2011cg,Li:2011vx}, who carried out a similar analysis to that of Cornish et al.~\cite{Cornish:2011ys,Sampson:2013lpa} on a large statistical sample of aLIGO detections using simulated data and a restricted ppE model. All of these studies suggest that Bayesian tests of GR are possible, given sufficiently high SNR events. And indeed, after the first gravitational wave observations, these tests were carried out in~\cite{TheLIGOScientific:2016src,Yunes:2016jcc}, the signal was shown to be consistent with GR and a plethora of theoretical models were constrained at different levels. 

\subsubsection{Systematics in model selection}

The model selection techniques described above are affected by other systematics present in data analysis. In general, we can classify these into the following~\cite{Vallisneri:2013rc,Gupta:2024gun}: 
\begin{itemize}
\item {\textbf{Mismodeling Systematic}}, caused by inaccurate models of the gravitational-wave template.
\item {\textbf{Instrumental Systematic}}, caused by inaccurate models of the gravitational-wave response.
\item {\textbf{Astrophysical Systematic}}, caused by inaccurate models of the astrophysical environment.
\end{itemize}
Mismodeling systematics are introduced due to the lack of an exact solution to the Einstein equations from which to extract an exact template, given a particular astrophysical scenario. Inspiral templates, for example, are approximated through post-Newtonian theory and become increasingly less accurate as the binary components approach each other. Cutler and Vallisneri~\cite{Cutler:2007mi} were the first to carry out a formal and practical investigation of such a systematic in the context of parameter estimation from a frequentist approach. 

Mismodeling systematics will prevent us from testing GR effectively with signals that we do not understand sufficiently well. For example, when considering signals from black hole coalescences, if the total mass of the binary is sufficiently high, the binary will merge in band. The higher the total mass, the fewer the inspiral cycles that will be in band, until eventually only the merger is in band. Since the merger phase is the least understood phase, it stands to reason that our ability to test GR will deteriorate as the total mass increases. But the situation is not so simple, since the higher the mass of the system, the larger the SNR of the event, and thus, the better we can constrain Einstein's theory. Moreover, we do understand the ringdown phase very well, and tests of the no-hair theorem could be done during this phase, provided a sufficiently large SNR~\cite{Berti:2007zu}. For neutron star binaries or very low-mass black hole binaries, the merger phase is expected to be essentially out of band for aLIGO (above $1$kHz), and thus, the noise spectrum itself may shield us from our ignorance. See~\cite{Moore:2021eok} for recent work on waveform systematics for  tests of GR with gravitational waves.

Instrumental systematics are introduced by our ignorance of the transfer function, which connects the detector response to the incoming gravitational waves. Through sophisticated calibration studies with real data, one can approximate the transfer function very well~\cite{Accadia:2010aa,Abadie:2010px,Tuyenbayev:2016xey,Sun:2020wke,Vitale:2020gvb,Wade:2022utw}. However, this function is not time-independent, because the noise in the instrument is not stationary or Gaussian. Thus, un-modeled drifts in the transfer function can still introduce systematics in parameter estimation that, as of now, are as large as $2\%$ in the amplitude and the phase~\cite{Accadia:2010aa,Tuyenbayev:2016xey,Sun:2020wke,Vitale:2020gvb,Wade:2022utw}. 
In fact, the specific realization of the noise during the first gravitational wave observations did introduce systematics in the first tests of GR carried out~\cite{TheLIGOScientific:2016src}. Some of the impact of these systematics is reduced by cross-correlating the output from multiple detectors; although instrumental systematics are present in each instrument, the noise is mostly uncorrelated between them. Therefore, as more instruments join the gravitational wave effort, cross-correlation should help ameliorate instrumental systematics. 

Astrophysical systematics are induced by our lack of exact {\emph{a priori}} knowledge of the gravitational wave source. As explained above, matched filtering requires knowledge of a waveform template with which to filter the data. Usually, we assume the sources are in a perfect vacuum and isolated. For example, when considering inspiral signals, we ignore any third bodies, electric or magnetic fields in black hole spacetimes, the accelerated expansion of the Universe, etc. Fortunately, however, most of these effects are expected to be small: the probability of finding third bodies sufficiently close to a binary system is very small~\cite{Yunes:2010sm}; for low redshift events, the expansion of the Universe induces an acceleration of the center of mass, which is also very small~\cite{Yunes:2009bv}; electromagnetic fields and neutron star hydrodynamic effects may affect the inspiral of black holes and neutron stars, but not until the very last stages, when most signals will be out of band anyway. For example, tidal deformation effects enter a neutron star binary inspiral waveform at 5 post-Newtonian order, which therefore affects the signal outside the most sensitive part of the Adv.~LIGO sensitivity bucket. See~\cite{Barausse:2014tra} for a systematic study on environmental effects on gravitational wave astrophysics.

Perhaps the most dangerous source of astrophysical systematics is due to the assumptions made about the astrophysical systems we expect to observe. For example, when considering neutron star binary inspirals, one usually assumes the orbit will have circularized by the time it enters the sensitivity band. Moreover, one assumes that any residual spin angular momentum that the neutron stars may possess is very small and aligned or anti-aligned with the orbital angular momentum. These assumptions certainly simplify the construction of waveform templates, but if they happen to be wrong, they would introduce mismodeling systematics that could also affect parameter estimation and tests of GR. See~\cite{Saini:2022igm,Bhat:2022amc} for recent studies on systematic bias on tests of GR with gravitational waves due to neglect of orbital eccentricity.

\subsection{Burst analyses}

In alternative theories of gravity, gravitational-wave sources such as core collapse supernovae may result in the production of gravitational waves in more than just the plus- and cross-polarizations~\cite{Shibata:1994qd,Scheel:1994yr,Harada:1996wt,Novak:1999jg,Novak:1997hw,Ruiz:2012jt}. Indeed, the near-spherical geometry of the collapse can be a source of scalar breathing-mode gravitational waves.  The precise form of the waveform, however, is unknown because it is sensitive to the initial conditions. 

When searching for un-modeled bursts in alternative theories of gravity, a general approach involves using linear combinations of data streams from all available detectors to form maximum likelihood estimators for the waveforms in the various polarizations, and the use of {\emph{null streams}}.  In the context of ground-based detectors and GR, these ideas were first explored by G\"ursel and Tinto~\cite{Guersel:1989th} and later by Chatterji et al.~\cite{Chatterji:2006nh} with the aim of separating false-alarm events from real detections. The main idea was to construct a linear combination of data streams received by a network of detectors, so that the combination contained only noise. In GR, of course, one need only include $h_{+}$ and $h_{\times}$ polarizations, and thus a network of three detectors is sufficient.  This concept can be extended to develop null tests of GR, as originally proposed by Chatziioannou, et al.~\cite{Chatziioannou:2012rf} and later implemented by Hayama, et al.~\cite{Hayama:2012au}. 


Let us consider a network of $D \geq 6$ detectors with uncorrelated noise and a detection by all $D$ detectors.  For a source that emits gravitational waves in the direction $\hat \Omega$, a single data point (either in the time-domain, or a time-frequency pixel) from an array of $D$ detectors (either pulsars or interferometers) can be written as
\begin{equation}
{\boldsymbol{{d}}}={\boldsymbol{F}}{\boldsymbol{{h}}}+{\boldsymbol{{n}}}.
\end{equation}
Here
\begin{equation}
{\boldsymbol{{d}}}\equiv
\left[
        \begin{array}{c}
                {d}_{1}\\ {d}_{2}\\ \vdots\\ {d}_{D}
        \end{array} \right] \, , \quad
         {\boldsymbol{{h}}} \equiv  \left[
        \begin{array}{c}
                        {h}_+ \\ {h}_\times \\ {h}_x \\ {h}_y \\ {h}_b \\ {h}_\ell
        \end{array} \right] \, , \quad
        {\boldsymbol{{n}}} \equiv \left[
        \begin{array}{c}
                {n}_{1}\\ {n}_{1}\\ \vdots\\ {n}_{D}
        \end{array} \right],
\end{equation}
where $ {\boldsymbol{{n}}}$ is a vector with the noise.
The antenna pattern functions are given by the matrix,
\begin{equation}
\left[ \begin{array}{c c c c c c}
                {\boldsymbol{F^+}}& {\boldsymbol{F^\times}}& {\boldsymbol{F^x}}& {\boldsymbol{F^y}} & {\boldsymbol{F^b}}& {\boldsymbol{F^\ell}}
        \end{array} \right]
        \equiv \left[
        \begin{array}{c c c c c c}
                F^+_{1} & F^\times_{1}  & F^x_{1} & F^y_{1}  & F^b_{1} & F^\ell_{1} \\
                F^+_{2} & F^\times_{2}  & F^x_{2} & F^y_{2}  & F^b_{2} & F^\ell_{2}\\
                \vdots & \vdots &\vdots & \vdots& \vdots & \vdots\\
                F^+_{D} & F^\times_{D}  & F^x_{D} & F^y_{D}  & F^b_{D} & F^\ell_{D}
        \end{array} \right].
\end{equation}
For simplicity we have suppressed the sky-location dependence of the antenna pattern functions.  These can be either the interferometric antenna pattern functions in Eqs.~\eqref{IFOresponse}, or the pulsar response functions in Eqs.~\eqref{PulsarResponse}. For interferometers, since the breathing and longitudinal antenna pattern response functions are degenerate,  and even though ${\boldsymbol{F}}$ is a $6\times D$ matrix, there are only $5$ linearly-independent 
vectors~\cite{Boyle:2010gc,Boyle:2010gn,Chatziioannou:2012rf,Hayama:2012au}.
 
If we do not know the form of the signal present in our data, we can obtain maximum likelihood estimators for it. For simplicity, let us assume the data are Gaussian and of unit variance (the latter can be achieved by whitening the data). Just as we did in Eq.~\eqref{eq:likelihood}, we can write the probability of obtaining datum ${\boldsymbol{d}}$, in the presence of a gravitational wave  ${\boldsymbol{h}}$ as
\begin{equation}
P({\boldsymbol{d}} | {\boldsymbol{h}} ) =
                \frac{1}{(2\pi)^{D/2}}
                \exp\left[-\frac12\left|{\boldsymbol{d}} -{\boldsymbol{F}} {\boldsymbol{h}} \right|^2\right].
\end{equation}
The logarithm of the likelihood ratio, i.e., the logarithm of the ratio of the likelihood when a signal is present to that when a signal is absent, can then be written as
\begin{equation}
L\equiv\ln\frac{P( {\boldsymbol{d}} | {\boldsymbol{h}} )}{P({\boldsymbol{d}} | 0 )} = 
                \frac{1}{2} \left[ \left|{\boldsymbol{d}}\right|^2 - \left|{\boldsymbol{d}}-{\boldsymbol{F}}{\boldsymbol{h}}\right|^2 \right] \,.
\end{equation}
If we treat the waveform values for each datum as free parameters, we can maximize the likelihood ratio
\begin{equation}
0 = \left.\frac{\partial L}{\partial {\boldsymbol{h}}}\right|_{{\boldsymbol{h}}={\boldsymbol{h}}_{\mathrm{MAX}}} \,,       
\end{equation}
and obtain maximum likelihood estimators for the gravitational wave,
\begin{equation}
{\boldsymbol{h}}_{\mathrm{MAX}}= ({\boldsymbol{F}}^T {\boldsymbol{F}})^{-1} {\boldsymbol{F}}^T \, {\boldsymbol{d}}.
\end{equation}
We can further substitute this solution into the likelihood, to obtain the value of the likelihood at the maximum,
\begin{equation}
E_{\mathrm{SL}} \equiv 2 L({\boldsymbol{h}}_{\mathrm{MAX}}) 
        = {\boldsymbol{d}}^T \boldsymbol{P}^{\mathrm{GW}} {\boldsymbol{d}},
\end{equation}
where
\begin{equation}
\boldsymbol{P}^{\mathrm{GW}} \equiv {\boldsymbol{F}} \, ({\boldsymbol{F}}^T {\boldsymbol{F}})^{-1} {\boldsymbol{F}}^T.
\end{equation}
The maximized likelihood can be thought of as the power in the signal, and can be used as a detection statistic. $\boldsymbol{P}^{\mathrm{GW}} $ is a projection operator that projects the data into the subspace spanned by ${\boldsymbol{F}}$.  An orthogonal projector can also be constructed,
\begin{equation}
\boldsymbol{P}^{\mathrm{null}}\equiv(\boldsymbol{I}-\boldsymbol{P}^{\mathrm{GW}}),
\end{equation}
which projects the data onto a subspace orthogonal to ${\boldsymbol{F}}$. Thus, one can construct a certain linear combination of data streams that has no component of a certain polarization by projecting them to a direction orthogonal to the direction defined by the beam pattern functions of this polarization mode
\begin{equation}
\boldsymbol{d}^{\mathrm{null}}=\boldsymbol{P}^{\mathrm{null}}\boldsymbol{d}.
\end{equation}
This is called a \textit{null stream} and, in the context of GR, it was introduced as a means of separating false-alarm events due, say, to instrumental glitches from real detections~\cite{Guersel:1989th,Chatterji:2006nh}. 

With just three independent detectors, we can choose to eliminate the two tensor modes (the plus- and cross-polarizations) and construct a \emph{GR null stream}: a linear combination of data streams that contains no signal consistent within GR, but could contain a signal in another gravitational theory, as illustrated in Fig.~\ref{figurenull}. With such a GR null stream, one can carry out null tests of GR and study whether such a stream contains any statistically significant deviations from noise. Notice that this approach does not require a template; if one were parametrically constructed, such as in~\cite{Chatziioannou:2012rf}, more powerful null tests could be applied. As of now, the two aLIGO detectors in the United States, advanced Virgo in Italy, and KAGRA in Japan have started operation, and LIGO-India in India is expected to join in the near future. Given a gravitational wave observation that is detected by all five detectors, one can then construct three GR null streams, each with power in a signal direction. For pulsar timing experiments where one is dealing with the data streams of about a few tens of pulsars, waveform reconstruction for all polarization states, as well as numerous null streams, can be constructed.

\epubtkImage{nullstream-2_html.png}{%
\begin{figure}[htbp]
\centerline{\includegraphics[height=6cm,clip=true]{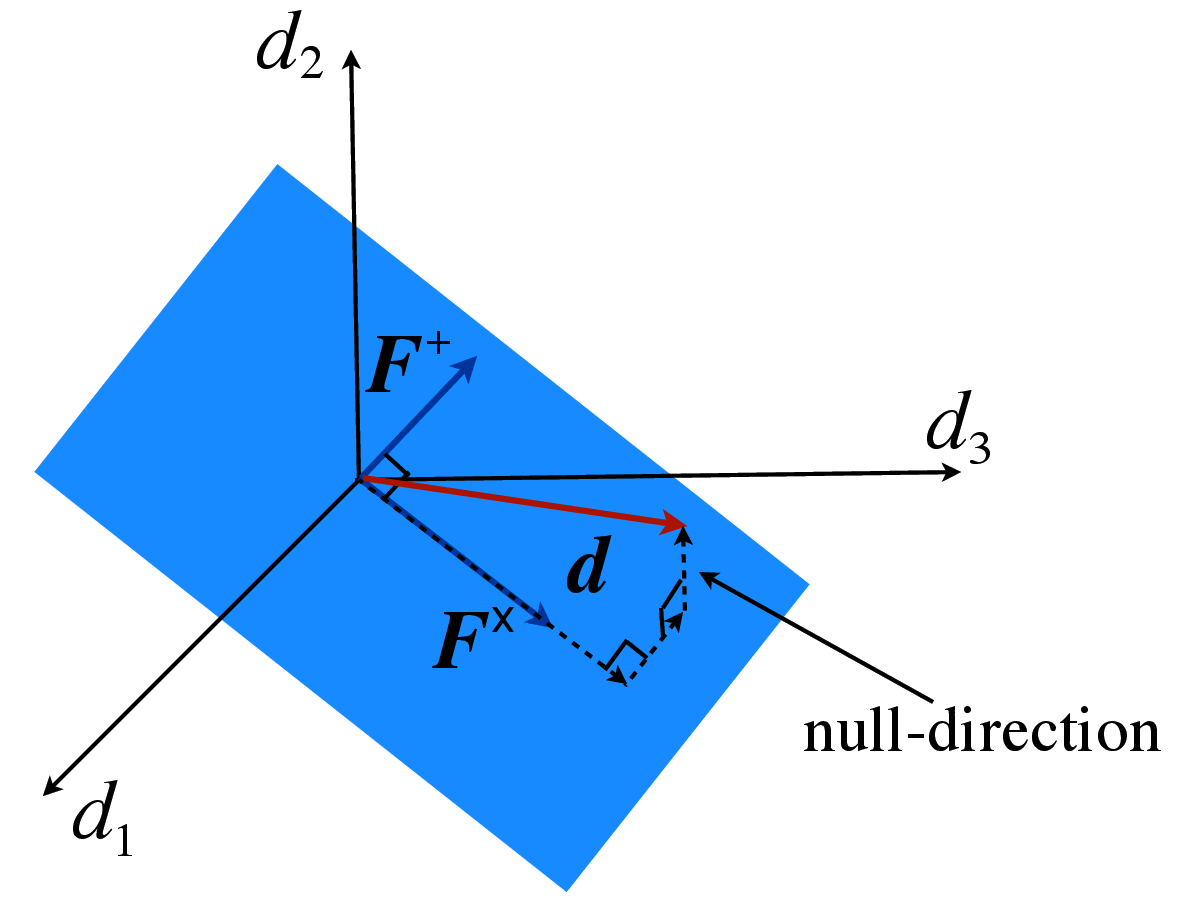}}
\caption{Schematic diagram of the projection of the data stream
  ${\boldsymbol{d}} $ orthogonal to the GR subspace spanned by $F^+$
  and $F^\times$, along with a perpendicular subspace, for 3 detectors
  to build the GR null stream.}
\label{figurenull}
\end{figure}}

\subsection{Stochastic background searches}
\label{sec:Stoch-Anal}

Much work has been done on the response of ground-based interferometers to non-tensorial polarization modes, stochastic background detection prospects, and data analysis techniques~\cite{Maggiore:1999wm,Nakao:2000ug,Gasperini:2001mw,Nishizawa:2009bf,Corda:2010zza}. In the context of pulsar timing, the first work to deal with the detection of such backgrounds in the context of alternative theories of gravity is due to Lee et al.~\cite{LeeJenetPrice}, who used a coherence statistic approach to the detection of non-Einsteinian polarizations. They studied the number of pulsars required to detect the various extra polarization modes, and found that pulsar timing arrays are especially sensitive to the longitudinal mode. Alves and Tinto~\cite{daSilvaAlves:2011fp} also found enhanced sensitivity to longitudinal and vector modes. Here we follow the work in~\cite{Nishizawa:2009bf,Chamberlin:2011ev} that deals with the LIGO and pulsar timing cases using the optimal statistic, a cross-correlation that maximizes the SNR. 

In the context of the optimal statistic, the derivations of the effect of extra polarization states for ground-based instruments and pulsar timing are very similar.  We begin with the metric perturbation written in terms of a plane wave expansion, as in Eq.~\eqref{pwexp2}. If we assume that the background is unpolarized, isotropic, and stationary, we have that
\beq
\langle \tilde h^*_A(f,\hat{\Omega}) \tilde h_{A'}(f',\hat{\Omega}')\rangle = 
\delta^2(\hat{\Omega},\hat{\Omega}')\delta_{AA'}
\delta(f-f')H_A(f),
\label{hev}
\eeq
where $H_A(f)$ is the gravitational-wave power spectrum for polarization $A$.  $H_A(f)$ is related to the energy density in 
gravitational waves per logarithmic frequency interval for that polarization through
\beq
 \Omega_{A}(f)\equiv \frac{1}{\rho_{\mathrm{crit}}}\frac{d\rho_A}{d\ln f},
\label{omdef}
\eeq
where $\rho_{\mathrm{crit}}=3H_0^2/8\pi$ is the closure density of the universe, and 
\beq
\rho_A = \frac{1}{32\pi}\langle \dot{h}_{A\,ij}(t,\vec{x})
\dot{h}_A^{ij}(t,\vec{x})\rangle
\label{eqrho}
\eeq
is the energy density in gravitational waves for polarization $A$.  It follows from the plane
wave expansion in \Deqn{pwexp}, along with \Deqns{hev} and~(\ref{omdef}) in 
\Deqn{eqrho}, that
\beq
H_A(f)=\frac{3H_0^2}{16\pi^3}|f|^{-3}\Omega_A(|f|),
\eeq
and therefore
\bea
\langle \tilde h_A^*(f,\hat{\Omega}) \tilde h_{A'}(f',\hat{\Omega}')\rangle =
\frac{3H_0^2}{16\pi^3}\delta^2(\hat{\Omega},\hat{\Omega}')\delta_{AA'}
\delta(f-f')|f|^{-3}\Omega_A(|f|).
\label{hevom}
\eea

For both ground-based interferometers and pulsar timing experiments, an isotropic stochastic background of gravitational waves appears in the data as correlated noise between measurements from different instruments. The data set from the $a^{\mathrm{th}}$ instrument is of the form 
\begin{eqnarray}
	d_a(t) &=& s_a(t) + n_a(t)\,,
\end{eqnarray}
where $s_a(t)$ corresponds to the gravitational wave signal and $n_a(t)$ to noise. The noise is assumed in this case to be stationary and Gaussian, and uncorrelated between different detectors,
\begin{eqnarray}
	\langle n_a(t) \rangle &=& 0\,, \\
	\langle n_a(t) n_b(t) \rangle &=& 0,
\end{eqnarray}
for $a \neq b$.

Since the gravitational wave signal is correlated, we can use cross-correlations to search for it. The cross-correlation statistic is defined as 
\begin{eqnarray}\label{ccstatQ}
	S_{ab} = \int_{-T/2}^{T/2} dt \int_{-T/2}^{T/2} dt' d_a(t) d_b(t') Q_{ab}(t-t')\,,
\end{eqnarray}
where $Q_{ab}(t-t')$ is a filter function to be determined. Henceforth, no summation is implied on the detector indices $(a,b,\ldots)$.  At this stage it is not clear why  $Q_{ab}(t-t')$ depends on the pair of data sets being correlated. We will show how this comes about later.
 The optimal filter is determined by maximizing the expected SNR
\begin{eqnarray}\label{snr}
	{\mathrm{SNR}} = \frac{\mu_{ab}}{\sigma_{ab}}.
\end{eqnarray}
Here $\mu_{ab}$ is the mean $\langle S_{ab} \rangle$ and $\sigma_{ab}$ is the square root of the variance $\sigma_{ab}^2=\langle S_{ab}^2 \rangle - \langle S_{ab} \rangle^2$. 

The expressions for the mean and variance of the cross-correlation statistic, $\mu_{ab}$ and $\sigma^2_{ab}$ respectively, take the same form for both pulsar timing and ground-based instruments. In the frequency domain, Eq.~\eqref{ccstatQ} becomes
\begin{eqnarray}
	S_{ab} = \int_{-\infty}^{\infty} df \int_{-\infty}^{\infty} df' \delta_T(f-f') \tilde{d}^*_a(f) \tilde{d}_b(f') \tilde{Q}_{ab}(f'),
\label{ccstatQ2}
\end{eqnarray}
by the convolution theorem, and the mean, $\mu_{ab}$, is then
\begin{eqnarray}
	\mu_{ab} \equiv \langle S_{ab} \rangle = \int_{-\infty}^\infty df \int_{-\infty}^\infty df' \, \delta_T(f-f') \langle \tilde{s}_a^{*}(f) \tilde{s}_b(f') \rangle \tilde{Q}_{ab}(f')\,,
\label{mueq}
\end{eqnarray}
where $\delta_T$ is the finite time approximation to the delta function, $\delta_T(f)={\sin{(\pi f t)}}/({\pi f})$. With this in hand, the mean of the cross-correlation statistic is
\begin{eqnarray}
\mu_{ab} = \frac{3H_0^2}{16\pi^3} T \sum_{A} \int_{-\infty}^{\infty} df |f|^{-3} \tilde{Q}_{ab}(f) \Omega_A (f) \Gamma^A_{ab} (f),
\end{eqnarray}
and the variance in the weak signal limit  is
\bea
\sigma_{ab}^2
&\equiv&
\langle S_{ab}^2\rangle-\langle S_{ab}\rangle^2\approx \langle S_{ab}^2\rangle
\nonumber
\\
&\approx& \frac{T}{4}\int_{-\infty}^{\infty} df\, P_a(|f|)P_b(|f|)
\left|\tilde{Q}_{ab}(f)\right|^2,
\label{eqsig}
\eea
where $\Gamma^A_{ab} (f)$ is the overlap reduction function to be discussed in detail later, while the one-sided power spectra of the noise are defined by
\beq\label{e:psd}
\langle \tilde{n}_a^*(f)\tilde{n}_a(f')\rangle = \frac{1}{2}\delta(f-f') P_a(|f|)\,,
\eeq
in analogy to Eq.~\eqref{eq:Sn(f)-def}, where $P_{a}$ plays here the role of $S_{n}(f)$.

The mean and variance can be rewritten more compactly if we define a positive-definite inner 
product using the noise power spectra of the two data streams
\beq
(A,B)_{ab}\equiv\int_{-\infty}^{\infty}df\, A^*(f)B(f)P_a(|f|)P_b(|f|)\,,
\eeq
again in analogy to the inner product in Eq.~\eqref{eq:inner-prod-def}, when considering inspirals.
Using this definition
\bea
\mu_{ab} &=& \frac{3 H_0^2}{16\pi^3}T\,\left(\tilde{Q}_{ab}, 
\frac{\sum_A \Omega_A(|f|)\Gamma^A_{ab}(|f|)}{|f|^3P_a(|f|)P_b(|f|)}
\right)_{ab},\label{optfiltmu}\\
\sigma_{ab}^2 &\approx& \frac{T}{4}\left(\tilde{Q},\tilde{Q}\right)_{ab},
\label{sigip}
\eea
where we recall that the capital Latin indices $(A,B,\ldots)$ stand for the polarization content. From the definition of the SNR and the Schwartz's inequality, it follows that the optimal filter is given by
\beq
\tilde{Q}_{ab}(f)=N\frac{\sum_A\Omega_A(|f|)\Gamma^A_{ab}(|f|)}{|f|^3P_a(|f|)P_b(|f|)},
\label{optfilt}
\eeq
where $N$ is an arbitrary normalization constant, normally chosen so that the mean of the statistic gives the amplitude of the stochastic background.  

The differences in the optimal filter between interferometers and pulsars arise only from differences in the so-called {\emph{overlap reduction functions}}, $\Gamma^A_{ab} (f)$. For ground-based instruments, the signal data $s_a$ are the strains given by Eq.~\eqref{strain}. The overlap reduction functions are then given by
\begin{eqnarray}
\Gamma^A_{ab} (f) = \int_{S^2}  d \hat{\Omega} F_a^A(\hat{\Omega}) F_b^A(\hat{\Omega}) e^{2\pi if \hat{\Omega}\cdot (\vec{x}_a-\vec{x}_b) },
\label{ligoorf}
\end{eqnarray}
where $\vec{x}_a$ and $\vec{x}_b$ are the locations of the two interferometers. The integrals in this case all have solutions in terms of spherical Bessel functions ~\cite{Nishizawa:2009bf}, which we do not summarize here for brevity.

For pulsar timing arrays, the signal data $s_a$ are the redshifts $z_a$, given by Eq.~\eqref{FTantpatts}. The overlap reduction functions are then given by
\bea
\Gamma^A_{ab}(f) = \frac{3}{4\pi} \int_{S^2}d\hat{\Omega}\,
\left(e^{i2\pi f L_a(1+\hat{\Omega}\cdot\hat{p}_a)}-1\right)
\left(e^{-i2\pi f L_b(1+\hat{\Omega}\cdot\hat{p}_b)}-1\right)
F_a^A(\hat{\Omega})F_b^{A}(\hat{\Omega}),
\label{pulsarorf}
\eea
where $L_a$ and $L_b$ are the distances to the two pulsars. For all transverse modes
pulsar timing experiments are in a regime where the exponential
factors in \Deqn{pulsarorf} can be neglected~\cite{Anholm:2008wy,Chamberlin:2011ev}, and the overlap reduction functions 
effectively become frequency independent (see~\cite{Boitier:2020rzg,Boitier:2021rmb,Hu:2022ujx} for recent works on extensions beyond this short wavelength approximation).  For the $+$ and $\times$ mode the
overlap reduction function becomes
\bea
\Gamma^+_{ab} = 3 \left\{ \frac{1}{3} + \frac{1-\cos\xi_{ab}}{2}\left[ \ln\left(
    \frac{1-\cos\xi_{ab}}{2}\right)  -\frac{1}{6}\right]\right\},
\label{orfapprox}
\eea
where $\xi_{ab}=\cos^{-1}(\hat p_a \cdot \hat p_b)$ is the angle between the two 
pulsars.  This quantity is proportional to the Hellings and Downs 
curve~\cite{hd83}. For the breathing mode, the overlap reduction function takes the closed form expression~\cite{LeeJenetPrice,Chamberlin:2011ev}:
\begin{eqnarray}
	\Gamma^b _{ab} = \frac{1}{4} \left( 3+ \cos{\xi_{ab}} \right),
\end{eqnarray} 
and for the vector modes 
\begin{eqnarray}
 \Gamma_{ab}^{v} \approx 3 \log{\left(\frac{2}{1 - \cos{\left(\xi_{ab} \right)}} \right)} - 4 \cos{\left(\xi_{ab} \right)} - 3 \label{VLORF}.
\end{eqnarray}
For the longitudinal mode the overlap reduction functions retains significant frequency dependence and full analytic expressions for all overlap reduction functions including the frequency dependence were found relatively recently~\cite{Hu:2022ujx}.  

The result for the combination of cross-correlation pairs to form an optimal network statistic is also the same in both ground-based interferometer and pulsar timing cases: a sum of the cross-correlations of all detector pairs weighted by their variances. The detector network optimal statistic is,
\beq
S_{\mathrm{opt}} = \frac{\sum_{ab} \sigma^{-2} _{ab} S_{ab} }{\sum_{ab} \sigma^{-2} _{ab}},
\label{optnetwork}
\eeq
 where $\sum_{ab} $ is a sum over all detector pairs.  

In order to perform a search for a given polarization mode one first needs to compute the overlap reduction functions (using either Eq.~\eqref{ligoorf} or \eqref{pulsarorf}) for that mode. With that in hand and a form for the stochastic background spectrum $\Omega_A(f)$, one can construct optimal filters for all pairs in the detector network using Eq.~\eqref{optfilt}, and perform the cross-correlations using either Eq.~\eqref{ccstatQ} (or equivalently Eq.~\eqref{ccstatQ2}). Finally, we can calculate the overall network statistic Eq.~\eqref{optnetwork}, by first finding the variances using Eq.~\eqref{eqsig}.  

It is important to point out that the procedure outlined above is straightforward for ground-based interferometers. Pulsar timing data, however, are irregularly sampled, and have a pulsar timing model subtracted out. This needs to be accounted for, and generally, a time-domain approach is more appropriate for these data sets.  The procedure is similar to what we have outlined above, but power spectra and gravitational-wave spectra in the frequency domain need to be replaced by auto-covariance and cross-covariance matrices in the time domain that account for the model fitting (for an example of how to do this see~\cite{Ellis:2013nrb}). A full description of the pulsar timing data analysis (Bayesian and frequentist) methods used in the most recent analyses can be found in Ref.~\cite{NANOGrav:2023icp}.

Let us summarize some recent studies of the capabilities of pulsar timing arrays to probe non-GR, gravitational-wave polarizations. Nishizawa et al.~\cite{Nishizawa:2009bf} showed that with three spatially separated detectors the tensor, vector, and  scalar contributions to the energy density in gravitational waves can be measured independently. Lee et al.~\cite{LeeJenetPrice} and Alves and Tinto~\cite{daSilvaAlves:2011fp} extended this analysis to show that pulsar timing experiments are especially sensitive to the longitudinal mode, and, to a lesser extent, to the vector modes. Recently, O'Beirne et al.~\cite{OBeirne:2019lwp} proved that the modeling of possible additional polarizations together with modifications to the gravitational wave phase evolution~\cite{Cornish:2017oic} can be used to separate polarization modes if more than tensor ones are present. Cornish et al.~\cite{Cornish:2017oic}, however, demonstrated that longitudinal modes could in practice be difficult to detect due to very large variances in their pulsar-pulsar correlation patterns. Chamberlin and Siemens~\cite{Chamberlin:2011ev} explained that the sensitivity of the cross-correlation to the longitudinal mode using nearby pulsar pairs can be enhanced significantly compared to that of the transverse modes. For example, for the NANOGrav pulsar timing array, two pulsar pairs separated by $3^\circ$ result in an enhancement of 4 orders of magnitude in sensitivity to the longitudinal mode relative to the transverse modes. The main contribution to this effect is due to gravitational waves that are coming from roughly the same direction as the pulses from the pulsars. In this case, the induced redshift for any gravitational-wave polarization mode is proportional to $f \, L$, the product of the gravitational-wave frequency and the distance to the pulsar, which can be large. When the gravitational waves and the pulse direction are exactly parallel, the redshift for the transverse and vector modes vanishes, but it is proportional to $f \, L$ for the scalar-longitudinal mode. Another interesting polarization mode one may wish to detect are circularly polarized tensor modes,  which would be useful for probing e.g. parity violation through amplitude birefringence~\cite{Alexander:2007kv,Alexander:2009tp,Yunes:2010yf,Alexander:2017jmt}. Pulsar timing arrays are not sensitive to circular polarization if the background is isotropic, although they have some sensitivity to this mode for anisotropic backgrounds~\cite{Kato:2015bye,Belgacem:2020nda,Sato-Polito:2021efu}. 

Let us now discuss probing specific fundamental pillars in GR or  modified theories of gravity with pulsar timing arrays.
Lee et al.~\cite{2010ApJ...722.1589L,Lee:2013awh} studied the detectability of massive gravitons in pulsar timing arrays through stochastic background searches. They introduced a modification to Eq.~\eqref{eqzsom} to account for graviton dispersion, and found the modified overlap reduction functions (i.e., modifications to the Hellings--Downs curves Eq.~\eqref{orfapprox}) for various values of the graviton mass for both tensorial~\cite{2010ApJ...722.1589L} and non-tensorial~\cite{Lee:2013awh} polarization modes. They conclude that many stable pulsars ($\geq 60$) are required to distinguish between the massive and massless cases, and that future pulsar timing experiments could be sensitive to graviton masses of about $10^{-22}$~eV ($\sim 10^{13}$~km). This method is competitive with some of the compact binary tests described later in Section~\ref{generic-tests:MG-LV} (see Table~\ref{table:comparison-MG}). In addition, since the method of Lee et al.~\cite{2010ApJ...722.1589L} only depends on the form of the overlap reduction functions, it is advantageous in that it does not involve matched filtering (and therefore prior knowledge of the waveforms), and generally makes few assumptions about the gravitational-wave source properties.  See Liang \textit{et al}.~\cite{Liang:2021bct} for a recent follow-up work on computing overlap reduction functions and modified Hellings--Downs curves in a complete analytical form in massive gravity. Qin \textit{et al}.~\cite{Qin:2020hfy} studied subluminal stochastic gravitational wave background in pulsar timing arrays and its application to $f(R)$ gravity that contains a massive scalar degree of freedom. Gong \textit{et al}.~\cite{Gong:2018cgj} derived the modified Hellings--Downs curves in Einstein-\AE{}ther theory. 
Full analytic expressions for all overlap reduction functions were found relatively recently by Hu, et al.~\cite{Hu:2022ujx}, and can be elegantly cast in the language of spherical harmonics (see the work of Kumar and Kamionkowski~\cite{AnilKumar:2023yfw}).

Gravitational waves have recently been discovered in the nanohertz band using pulsar timing arrays (see, for example, \cite{NANOGrav:2023gor}), and some studies have be implemented to place new constraints on these modified theories~\cite{NANOGrav:2023ygs}. We summarize these results in Chapter 5.

\newpage
\section{Bestiary of Gravitational Wave Tests}
\label{section:binary-sys-tests}

In this section, we present a descriptive treatise on various kinds of tests of GR with gravitational waves. Due to the uneven proportion of studies that deal with compact binary systems, the latter will be emphasized in this section. We begin by explaining the difference between direct and generic tests. We then proceed to describe the many direct or top-down tests and generic or bottom-up tests that have been proposed once gravitational waves are detected, including tests of the no-hair theorems. We concentrate here only on binaries composed of compact objects, such as neutron stars, black holes or other compact exotica. We will not discuss tests one could carry out with electromagnetic information from binary (or double) pulsars, as these are already described in~\cite{lrr-2006-3}. We will also not review tests of GR with accretion disk observations, for which we refer the interested reader to~\cite{lrr-2008-9}.

\subsection{Direct versus generic tests and propagation versus generation}
\label{sec:prop-gen-direc-generic}

Gravitational-wave tests of Einstein's theory can be classed into two distinct subgroups: direct tests
and generic tests. Direct tests employ a top-down approach, where one starts from a particular modified 
gravity theory with a known action, derives the modified field equations and solves them for a particular
gravitational wave-emitting system. On the other hand, generic tests adopt a bottom-up approach, where
one takes a particular feature of GR and asks what type of signature its absence would leave 
on the gravitational-wave observable; one then asks whether the data presents a statistically-significant anomaly 
pointing to that particular signature. 

Direct tests have by far been the traditional approach to testing GR with gravitational waves. 
The prototypical examples here are tests of Jordan--Fierz--Brans--Dicke theory. As described in Section~\ref{section:alt-theories}, one can solve the modified field equations for a binary system in the post-Newtonian approximation to find a prediction for the gravitational-wave observable, as we will see in more detail later in this section. Other examples of direct tests include those concerning modified quadratic gravity models and non-commutative geometry theories. 

The main advantage of such direct tests is also its main disadvantage: one has to pick a particular modified gravity
theory. Because of this, one has a well-defined set of field equations that one can solve, but at the same time, 
one can only make predictions about that modified gravity model. Unfortunately, we currently lack a particular
modified gravity theory that is particularly compelling; many modified gravity theories exist,
but none possess all the criteria described in Section~\ref{section:alt-theories}, except perhaps for the subclass
of scalar-tensor theories with spontaneous scalarization. Lacking a clear alternative to 
GR, it is not obvious which theory one should pick. Given that the full development (from the action 
to the gravitational wave observable) of any particular theory can be incredibly difficult, time and computationally consuming, 
carrying out direct tests of all possible modified gravity models now that gravitational waves have been detected is clearly
unfeasible. 

Given this, one is led to generic tests of GR, where one asks how the absence of specific features
contained in GR could impact the gravitational wave observable. For example, one can ask how such an
observable would be modified if the graviton had a mass, if the gravitational interaction were Lorentz or parity violating, or if
there existed large extra dimensions. From these general considerations, one can then construct a ``meta''-observable, 
i.e., one that does not belong to a particular theory, but that interpolates over all known possibilities in a well-defined way.
This model has come to be known as the parameterized post-Einsteinian framework~\cite{Yunes:2009ke}, in analogy to the parameterized
post-Newtonian scheme used to test GR in the solar system~\cite{lrr-2006-3}. Given such a construction, one can 
then ask whether the data points to a statistically-significant GR deviation. 

The main advantage of generic tests is precisely that one does not have to specify a particular model, but instead
one lets the data select whether it contains any statistically-significant deviations from our canonical beliefs. 
Such an approach is, of course, not new to physics, having most recently been successfully employed by the WMAP 
team~\cite{Bennett:2010jb}. The intrinsic disadvantage of this method is that, if a deviation is found, there is no
one-to-one mapping between it and a particular action, but instead one has to point to a class of possible models. 
Of course, such a disadvantage is not that limiting, since it would provide strong hints as to what type of symmetries
or properties of GR would have to be violated in an ultraviolet completion of Einstein's theory.   

Whether one is considering direct tests or generic tests, modifications to GR can always be classified into those that affect the generation of gravitational waves and those that affect their propagation. The study of the generation of gravitational waves involves the linearization of the modified field equations about a fixed (Minkowski or cosmological) background, and their solution either through a multipolar post-Newtonian formalism (see~\cite{Blanchet:2006zz} for a review) or through purely numerical methods. Such a task is very difficult and time-consuming, which is why it has only been tackled in a few modified gravity cases. The study of the propagation of gravitational waves involves the linearization of the field equations in flat spacetime to derive the wave's dispersion relation or its propagator. Typically, this is a much more straightforward calculation that has been carried out in many specific modified gravity theories and phenomenological models. 

Given these two generic types of modifications to gravity, which one is more important when carrying out gravitational wave tests? The answer to this question depends strongly on the particular modification to gravity that one is considering. Indeed, there are modified theories, such as scalar-tensor theories, in which the propagation of gravitational waves is not modified at all, while the generation is. In such cases, the modifications to the generation of gravitational waves is all that matters. In general, however, a modified theory of gravity will introduce modifications to both the generation and propagation of gravitational waves.  In this case, typically the modifications to the propagation lead to a stronger constraint than the modifications to the generation simply because the former accumulate with distance traveled, while the latter only accumulate with coalescence time, which is typically proportional to the binary's total mass. Indeed, in simple massive gravity theories the ratio of the modification in the Fourier phase introduced by generation to that introduced by propagation effects is roughly~\cite{Finn:2001qi,Yunes:2016jcc} 
\begin{align}
\frac{\Psi_{\mathrm{prop}}(f)}{\Psi_{\mathrm{gen}}(f)} &=10^{18} \left( \frac{\mathcal{M}}{28 M_\odot} \right)^{5/3} 
\left( \frac{D_L}{380\mathrm{Mpc}} \right) \left( \frac{f}{100\mathrm{Hz}} \right)^{8/3}\,,
\end{align}
where $\mathcal{M}$ is the binary's chirp mass, $D_{L}$ is the luminosity distance and we have evaluated the ratio at $100$ Hz. One sees clearly that the propagation effect is clearly dominant, in spite of the propagation effect entering at higher post-Newtonian order than the generation effect. One is thus drawn to conclude that if both generation and propagation effects are present in a modified gravity theory, then typically propagation effects dominate irrespective of the post-Newtonian order they enter at.

\subsection{Direct tests}
\label{sec:Direct-tests}

\subsubsection{Scalar-tensor theories}
\label{sec:direct-test-BD}

Let us first concentrate on Jordan--Fierz--Brans--Dicke theory, where black holes and neutron stars have been shown to exist. In this theory, the gravitational mass depends on the value of the scalar field, as Newton's constant is effectively promoted to a function, thus leading to violations of the weak-equivalence principle~\cite{1975ApJ...196L..59E,1977ApJ...214..826W,Will:1989sk}. The usual prescription for the modeling of binary systems in this theory is due to Eardley~\cite{1975ApJ...196L..59E}.\epubtkFootnote{A modern interpretation in terms of effective field theory can be found in~\cite{Goldberger:2004jt,Goldberger:2006bd}.} He showed that such a scalar-field effect can be captured by replacing the constant inertial mass by a function of the scalar field in the distributional stress-energy tensor and then Taylor expanding about the cosmological constant value of the scalar field at spatial infinity, i.e.,
\begin{equation}
m_{a} \to m_{a}(\phi) = m_{a}(\phi_{0}) \left\{1 + s_{a} \frac{\psi}{\phi_{0}} - \frac{1}{2} \left(s_{a}' - s_{a}^{2} + s_{a} \right) \left(\frac{\psi}{\phi_{0}}\right)^{2} + {\cal{O}}\left[\left(\frac{\psi}{\phi_{0}}\right)^{3}\right] \right\}\,, 
\end{equation}
where the subscript $a$ stands for $a$ different sources, while $\psi \equiv \phi - \phi_{0} \ll 1$ and the sensitivities $s_{a}$ and $s_{a}'$ are defined by 
\begin{equation}
s_{a} \equiv - \left[\frac{\partial \left(\ln m_{a}\right)}{\partial \left(\ln G\right)}\right]_{0}\,,
\label{eq:sensitivity-def}
\qquad
s_{a}' \equiv - \left[\frac{\partial^{2} \left(\ln m_{a}\right)}{\partial \left(\ln G\right)^{2}}\right]_{0}\,,
\end{equation}
where we remind the reader that $G = 1/\phi$, the derivatives are to be taken with the baryon number held fixed and evaluated at $\phi = \phi_{0}$. These sensitivities encode how the gravitational mass changes due to a non-constant scalar field; one can think of them as measuring the gravitational binding energy per unit mass. The internal gravitational field of each body leads to a non-trivial variation of the scalar field, which then leads to modifications to the gravitational binding energies of the bodies.  In carrying out this expansion, one assumes that the scalar field takes on a constant value at spatial infinity $\phi \to \phi_{0}$, disallowing any homogeneous, cosmological solution to the scalar field evolution equation [Eq.~\eqref{eq:ST-EOM}]. 

With this at hand, one can solve the massless Jordan--Fierz--Brans--Dicke modified field equations [Eq.~\eqref{eq:ST-EOM}] for the non-dynamical, near-zone field of $N$ compact objects to obtain~\cite{Will:1989sk}
\begin{align}
\frac{\psi}{\phi_{0}} &= \frac{1}{2 + \omega_{\mathrm{BD}}} \sum_{a} \left(1 - 2 s_{a} \right) \frac{m_{a}}{r_{a}} + \ldots\,,
\label{eq:psi-sol-BD}
\\
g_{00} &=-1 + \sum_{a} \left( 1 - \frac{s_{a}}{2 + \omega_{\mathrm{BD}}} \right) \frac{2 m_{a}}{r_{a}} + \ldots\,,
\\
g_{0i} &= -2 \left(1 + \gamma \right)\sum_{a} \frac{m_{a}}{r_{a}} v_{a}^{i} + \ldots\,,
\\
g_{ij} &= \delta_{ij} \left[1 + 2 \gamma \sum_{a} \left(1 + \frac{s_{a}}{1 + \omega_{\mathrm{BD}}} \right) \frac{m_{a}}{r_{a}} + \ldots \right]\,,
\end{align}
where $a$ runs from 1 to $N$, we have defined the spatial field point distance $r_{a} \equiv |x^{i} - x_{a}^{i}|$, the parameterized post-Newtonian quantity $\gamma = (1 + \omega_{\mathrm{BD}}) (2 + \omega_{\mathrm{BD}})^{-1}$ and we have chosen units in which $G = c = 1$. This solution is obtained in a post-Newtonian expansion~\cite{Blanchet:2006zz}, where the ellipses represent higher-order terms in $v_{a}/c$ and $m_{a}/r_{a}$. From such an analysis, one can also show that compact objects follow geodesics of such a spacetime, to leading order in the post-Newtonian approximation~\cite{1975ApJ...196L..59E}, except that Newton's constant in the coupling between matter and gravity is replaced by $G \to {\cal{G}}_{12} = 1 - (s_{1} + s_{2} - 2 s_{1} s_{2}) (2 + \omega_{\mathrm{BD}})^{-1}$, in geometric units.

As is clear from the above analysis, black-hole and neutron-star solutions in this theory generically depend on the quantities $\omega_{\mathrm{BD}}$ and $s_{a}$. The former characterizes the coupling between the scalar and matter fields and determines the strength of the correction, with the theory reducing to GR in the $\omega_{\mathrm{BD}} \to \infty$ limit~\cite{Faraoni:1999yp}. The latter depends on the compact object that is being studied. For neutron stars, this quantity can be computed as follows. First, neglecting scalar corrections to neutron-star structure and using the Tolman--Oppenheimer--Volkoff equation, one notes that the mass $m \propto N \propto G^{-3/2}$, for a fixed equation of state and central density, with $N$ the total baryon number. Thus, using Eq.~\eqref{eq:sensitivity-def}, one has that 
\begin{equation}
s_{a} \equiv \frac{3}{2} \left[1 - \left( \frac{\partial \ln m_{a}}{\partial \ln N}\right)_{G} \right]\,,
\end{equation}
where the derivative is to be taken holding $G$ fixed. In this way, given an equation of state and central density, one can compute the gravitational mass as a function of baryon number, and from this, obtain the neutron star sensitivities. Eardley~\cite{1975ApJ...196L..59E}, Will and Zaglauer~\cite{Will:1989sk}, and Zaglauer~\cite{Zaglauer:1992bp} have shown that these sensitivities are always in the range $s_{a} \in (0.19,0.3)$ for a soft equation of state and $s_{a} \in (0.1,0.14)$ for a stiff one, in both cases monotonically increasing with mass in $m_{a} \in (1.1,1.5)\,M_{\odot}$. Gralla~\cite{Gralla:2010cd} has found a more general method to compute sensitivities in generic modified gravity theories.

What is the sensitivity of black holes in generic scalar-tensor theories? Will and Zaglauer~\cite{Zaglauer:1992bp} have argued that the no-hair theorems require $s_{a} = 1/2$ for all black holes, no matter what their mass or spin is. As already explained in Section~\ref{section:alt-theories}, stationary black holes that are the byproduct of gravitational collapse (i.e., with matter that satisfies the energy conditions) in a general class of scalar-tensor theories are identical to their GR counterparts~\cite{Hawking:1972qk,1971ApJ...166L..35T,Dykla:1972zz,Sotiriou:2011dz}.\epubtkFootnote{One should note in passing that more general black-hole solutions in scalar-tensor theories have been found~\cite{Kim:1998hc,Campanelli:1993sm}. However, these usually violate the weak-energy condition, and sometimes they require unreasonably small values of $\omega_{\mathrm{BD}}$ that have already been ruled out by observation.} This is because the scalar field satisfies a free wave equation in vacuum, which forces the scalar field to be constant in the exterior of a stationary, asymptotically-flat spacetime, provided one neglects a homogeneous, cosmological solution. If the scalar field is to be constant, then by Eq.~\eqref{eq:psi-sol-BD}, $s_{a} = 1/2$ for a single black-hole spacetime. 

Such an argument formally applies only to stationary scenarios, so one might wonder whether a similar argument holds for binary systems that are in a quasi-stationary arrangement. Will and Zaglauer~\cite{Zaglauer:1992bp} and Mirshekari and Will~\cite{Mirshekari:2013vb} extended this discussion to quasi-stationary spacetimes describing black-hole binaries to higher post-Newtonian order. They argued that the only possible deviations from $\psi = 0$ are due to tidal deformations of the horizon due to the companion, which are known to arise at very high order in post-Newtonian theory, $\psi = {\cal{O}}[(m_{a}/r_{a})^{5}]$. Yunes, et al.~\cite{Yunes:2011aa} extended this argument further by showing that to all orders in post-Newtonian theory, but in the extreme mass-ratio limit, black holes cannot have scalar hair in generic scalar-tensor theories. Finally, Healy, et al.~\cite{Healy:2011ef} have carried out a full numerical simulation of the non-linear field equations, confirming this argument in the full non-linear regime. 

The activation of dynamics in the scalar field for a vacuum spacetime requires either a non-constant distribution of initial scalar field (violating the constant cosmological scalar field condition at spatial infinity) or a pure geometrical source to the scalar field evolution equation. The latter would lead to the quadratic modified gravity theories discussed in Section~\ref{subsec:MQG}. As for the former, Horbatsch and Burgess~\cite{Horbatsch:2010hj} have argued that if, for example, one lets $\psi = \mu t$, which clearly satisfies $\square \psi = 0$ in a Minkowski background,\epubtkFootnote{The scalar field of Horbatsch and Burgess satisfies $\square \psi = \mu g^{\mu \nu} \Gamma^{t}_{\mu \nu}$, and thus $\square \psi = 0$ for stationary and axisymmetric spacetimes, since the metric is independent of time and azimuthal coordinate. However, notice that is not necessarily needed for Jacobson's construction~\cite{Jacobson:1999vr} to be possible.} then a Schwarzschild black hole will acquire modifications that are proportional to $\mu$. Alternatively, scalar hair could also be induced by spatial gradients in the scalar field~\cite{Berti:2013gfa}, possibly anchored in matter at galactic scales. Such cosmological hair, however, is likely to be suppressed by a long timescale; in the example above $\mu$ must have units of inverse time, and if it is to be associated with the expansion of the universe, then it would be natural to assume $\mu = {\cal{O}}(H)$, where $H$ is the Hubble parameter. Therefore, although such cosmological hair might have an effect on black holes in the early universe, it should not affect black hole observations at moderate to low redshifts. 

Scalar field dynamics can be activated in non-vacuum spacetimes, even if initially the stars are not scalarized provided one considers a more general scalar-tensor theory, like the one introduced by Damour and Esposito-Far\`{e}se~\cite{Damour:1992we,Damour:1993hw}. As discussed in Section~\ref{sec:ST}, when the conformal factor takes on a particular functional form, non-linear effects induced when the gravitational energy exceeds a certain threshold can dynamically scalarize merging neutron stars, as demonstrated  by Barausse, et al~\cite{Barausse:2012da,Palenzuela:2013hsa}. Therefore, neutron stars in binaries are likely to have hair in generic scalar-tensor theories, even if they start their inspiral unscalarized. One must be careful, however, to make sure that the scalarized stars found are actually stable to perturbations. This is the case in the standard model of Damour and Esposito-Far\`{e}se~\cite{Damour:1992we,Damour:1993hw} when $\beta < 0$, but it is no longer true when $\beta > 0$~\cite{Mendes:2015gx,Palenzuela:2015ima,Mendes:2016fby}. 

Putting the issue of stability aside, what do gravitational waves look like in (massless) Jordan--Fierz--Brans--Dicke theory? As described in Section~\ref{sec:ST}, both the scalar field perturbation $\psi$ and the new metric perturbation $\theta^{\mu \nu}$ satisfy a sourced wave equation [Eq.~\eqref{eq:ST-EOM}], whose leading-order solution for a two-body inspiral is~\cite{Will:1994fb}
\begin{align}
\theta^{ij} &= 2 \left(1 + \gamma\right) \frac{\mu}{R} \left(v_{12}^{ij} - {\cal{G}}_{12} m \frac{x^{ij}}{r^{3}} \right)\,,
\\
\frac{\psi}{\phi_{0}} &= \left(1 - \gamma\right) \frac{\mu}{R} \left[ \Gamma \left(n_{i} v_{12}^{i}\right)^{2} - {\cal{G}}_{12} \Gamma \frac{m}{r^{3}} \left(n_{i} x^{i}\right)^{2} - \frac{m}{r}  \left({\cal{G}}_{12} \Gamma + 2 \Lambda\right) - 2 S n_{i} v_{12}^{i} \right]\,,
\end{align}
where $R$ is the distance to the detector, $n^{i}$ is a unit vector pointing toward the detector, $r$ is the magnitude of relative position vector $x^{i} \equiv x_{1}^{i} - x_{2}^{i}$, with $x_{a}^{i}$ the trajectory of body $a$, $\mu = m_{1} m_{2}/m$ is the reduced mass and $m = m_{1} + m_{2}$ is the total mass, $v_{12}^{i} \equiv  v_{1}^{i} - v_{2}^{i}$ is the relative velocity vector, and we have defined the shorthands
\begin{align}
\Gamma &\equiv 1 - 2 \frac{m_{1} s_{2} + m_{2} s_{1}}{m}\,,
\qquad
S \equiv s_{2} - s_{1}\,,
\label{eq:S-def}
\\
\Lambda &\equiv{\cal{G}}_{12} \left(1 - s_{1} - s_{2}\right) - \left(2 + \omega_{\mathrm{BD}}\right)^{-1} \left[\left(1- 2 s_{1}\right) s_{2}' + \left(1 - 2 s_{2}\right)s_{1}'\right]\,.
\end{align}
We have also introduced multi-index notation here, such that $A^{ij\ldots} = A^{i} A^{j} \ldots$. Such a solution is derived using the Lorenz gauge condition $\theta^{\mu \nu}{}_{,\nu} = 0$ and in a post-Newtonian expansion, where we have left out subleading terms of relative order $v_{12}^{2}$ or $m/r$. Lang and others~\cite{Mirshekari:2013vb,Lang:2013fna,Lang:2014osa,Sennett:2016klh} have completed this calculation to second post-Newtonian order, which were further extended by Bernard and others to find the equations of motion up to third post-Newtonian~\cite{Bernard:2018hta,Bernard:2018ivi}, tidal effects~\cite{Bernard:2019yfz,Bernard:2023eul}, and scalar modes and non-linear memory effects to 1.5 post-Newtonian order~\cite{Bernard:2022noq}. Spin-orbit effects on orbital dynamics and gravitational waves in scalar-tensor theories were studied in detail in~\cite{Brax:2021qqo}. Juli\'e, et al.~\cite{Julie:2022qux} constructed the Hamiltonian of a binary system in scalar-tensor theories (and scalar Gauss-Bonnet theory) within the effective-one-body formalism up to third post-Newtonian order. Almeida~\cite{Almeida:2024uph} derived the dynamics of binaries in scalar-tensor theories up to 2 post-Newtonian order using effective field theory. Trestini~\cite{Trestini:2024zpi} studied eccentric compact binaries in these theories at 2nd post-Newtonian order within the post-Keplerian framework. 

Given the new metric perturbation $\theta^{ij}$, one can reconstruct the gravitational wave metric perturbation $h^{ij}$, and from this, the response function, associated with the quasi-circular inspiral of compact binaries. After using Kepler's third law to simplify expressions [$\omega = ({\cal{G}}_{12} m/r^{3})^{1/2}$, where $\omega$ is the orbital angular frequency and $m$ is the total mass and $r$ is the orbital separation], one finds for a ground-based L-shaped detector~\cite{Chatziioannou:2012rf}:
\begin{align}
\label{eq:BD-h(t)}
h(t) &=- \frac{{\cal{M}}}{R} u^{2} e^{-2 i \Phi}
\left\{\left[F_{+} \left(1 + \cos^{2}{\iota} \right) + 2 i F_{\times} \cos{\iota}\right] 
\left[1 - \frac{1 - \gamma}{2} \left(1 + \frac{4}{3} S^{2}\right)\right] 
\right.
\nonumber \\
&- \left.
\frac{1-\gamma}{2} \Gamma F_{\mathrm{b}} \sin^{2}{\iota}\right\}
- \eta^{1/5}  \frac{{\cal{M}}}{R}  u e^{-i \Phi} 
S \left(1 - \gamma\right)F_{\mathrm{b}} \sin{\iota}
\nonumber \\
&- \frac{{\cal{M}}}{R} u^{2} 
\frac{1-\gamma}{2} F_{\mathrm{b}} \left(\Gamma + 2 \Lambda\right) + {\rm {c.c.}}\,,
\end{align} 
where we have defined $u \equiv (2 \pi {\cal{M}} F)^{1/3}$, $\eta \equiv \mu/m$ is the symmetric mass ratio, ${\cal{M}} \equiv \eta^{3/5} m$ is the chirp mass, $\iota$ is the inclination angle, c.c.~stands for the complex conjugate, and where we have used the beam-pattern functions in Eq.~\eqref{IFOresponse}. In Eq.~\eqref{eq:BD-h(t)} and henceforth, we linearize all expressions in $1 - \gamma \ll 1$. Jordan--Fierz--Brans--Dicke theory predicts the generic excitation of three polarizations: the usual plus and cross polarizations, and a breathing, scalar mode. We see that the latter contributes to the response at two, one and zero times the orbital frequency. One should note that all of these corrections arise during the generation of gravitational waves, and not due to a propagation effect. In fact, gravitational waves travel at the speed of light (and the graviton remains massless) in standard Jordan--Fierz--Brans--Dicke theory.  

The quantities $\Phi$ and $F$ are the orbital phase and frequency respectively, which are to be found by solving the differential equation
\begin{equation}
\frac{dF}{dt} = \left(1 - \gamma\right) S^{2} \frac{\eta^{2/5}}{\pi} {\cal{M}}^{-2} u^{9} + \frac{48}{5 \pi} {\cal{M}}^{-2} u^{11}  \left[1 - \frac{1-\gamma}{2} \left(1  - \frac{\Gamma^{2}}{6} + \frac{4}{3} S^{2}\right)\right] \ldots\,,
\label{eq:Fdot-BD}
\end{equation}
where the ellipses stand for higher-order terms in the post-Newtonian approximation. In this expression, and henceforth, we have kept only the leading-order dipole term and all known post-Newtonian, GR terms. If one wished to include higher post-Newtonian--order Jordan--Fierz--Brans--Dicke terms, one would have to include monopole contributions as well as post-Newtonian corrections to the dipole term. The first term in Eq.~\eqref{eq:Fdot-BD} corresponds to dipole radiation, which is activated by the scalar mode. That is, the scalar field carries energy away from the system modifying the energy balance law to~\cite{Will:1994fb,Scharre:2001hn,Will:2004xi}
\begin{equation}
\dot{E}_{\mathrm{BD}} = -\frac{2}{3} {\cal{G}}_{12}^{2} \eta^{2} \frac{m^{4}}{r^{4}} \left(1 - \gamma\right) S^{2} - \frac{32}{5} {\cal{G}}_{12}^{2} \eta^{2} \left(\frac{m}{r}\right)^{5} \left[1 - \frac{1-\gamma}{2} \left(1 - \frac{\Gamma^{2}}{6} \right) \right] + \ldots\,,
\end{equation}
where the ellipses stand again for higher-order terms in the post-Newtonian approximation. Solving the frequency evolution equation perturbatively in $1/\omega_{\mathrm{BD}} \ll 1$, one finds
\begin{align}
\frac{256}{5} \frac{t_{c} - t}{{\cal{M}}} &= u^{-8} \left[ 1 - \frac{1}{12} \left(1 - \gamma\right) S^{2} \eta^{2/5} u^{-2} + \dots\right]\,,
\\
\Phi &= -\frac{1}{64 \pi} \left(\frac{256}{5} \frac{t_{c} - t}{{\cal{M}}}\right)^{5/8} \left[1 - \frac{5}{224} \left(1 - \gamma\right) S^{2} \eta^{2/5} \left(\frac{256}{5} \frac{t_{c} - t}{{\cal{M}}}\right)^{1/4}  + \dots \right]\,.
\end{align}
In deriving these equations, we have neglected the last term in Eq.~\eqref{eq:Fdot-BD}, as this is a constant that can be reabsorbed into the chirp mass. Notice that since the two definitions of chirp mass differ only by a term of ${\cal{O}}(\omega_{\mathrm{BD}}^{-1})$, the first term of Eq.~\eqref{eq:Fdot-BD} is not modified.

One of the main ingredients that goes into parameter estimation is the Fourier transform of the response function. This can be estimated in the \emph{stationary-phase} approximation, for a simple, non-spinning, quasi-circular inspiral. In this approximation, one assumes the phase is changing much more rapidly than the amplitude~\cite{Bender,Cutler:1994ys,Droz:1999qx,Yunes:2009yz}. One finds~\cite{Chatziioannou:2012rf}
\begin{align}
\tilde{h}(f) &=  {\cal{A}}_{\mathrm{BD}}  \left(\pi {\cal{M}} f\right)^{-7/6} 
\left[1 - \frac{5}{96} \frac{S^{2}}{\omega_{\mathrm{BD}}} \eta^{2/5} \left(\pi {\cal{M}} f\right)^{-2/3}\right] e^{-i \Psi_{\mathrm{BD}}^{(2)}}
+ \gamma_{\mathrm{BD}}  \left(\pi {\cal{M}} f\right)^{-3/2} e^{-i \Psi_{\mathrm{BD}}^{(1)}}
\end{align}
where we have defined the amplitudes
\begin{align}
{\cal{A}}_{\mathrm{BD}} &\equiv \left(\frac{5 \pi}{96}\right)^{1/2} \frac{{\cal{M}}^{2}}{R} \left[F_{+}^{2} \left(1 + \cos^{2}\iota\right)^{2} + 4 F_{\times}^{2} \cos^{2}{\iota}
- F_{+} F_{\mathrm{b}} \left(1 - \cos^{4}{\iota}\right) \frac{\Gamma}{\omega_{\mathrm{BD}}}\right]^{1/2},
\\ 
\gamma_{\mathrm{BD}} &\equiv - \left(\frac{5 \pi}{48}\right)^{1/2} \frac{{\cal{M}}^{2}}{R} \eta^{1/5} \frac{S}{\omega_{\mathrm{BD}}} F_{\mathrm{b}} \sin{\iota}\,,
\end{align}
and the Fourier phase
\begin{align}
\Psi_{\mathrm{BD}}^{(\ell)} &= - 2 \pi f t_{c} + \ell \Phi_{c}^{(\ell)} + \frac{\pi}{4} - \frac{3 \ell}{256} \left(\frac{2 \pi {\cal{M}} f}{\ell}\right)^{-5/3} \sum_{n=0}^{7} \left(\frac{2 \pi {\cal{M}} f}{\ell}\right)^{n/3} \left(c_{n}^{\mathrm{PN}} + l_{n}^{\mathrm{PN}} \ln f\right) 
\nonumber \\
&+ \frac{5 \ell}{7168} \frac{S^{2}}{\omega_{\mathrm{BD}}} \eta^{2/5} \left(\frac{2 \pi {\cal{M}} f}{\ell}\right)^{-7/3}\,,
\label{eq:BD-phase}
\end{align}
where the Jordan--Fierz--Brans--Dicke correction is kept only to leading order in $\omega_{\mathrm{BD}}^{-1}$ and $v$, while $(c_{n}^{\mathrm{PN}}, l_{n}^{\mathrm{PN}})$ are post-Newtonian GR coefficients (see, e.g.,~\cite{Klein:2013qda}). In writing the Fourier response in this way, we had to redefine the phase of coalescence via
\begin{align}
\Phi_{c}^{(\ell)} = \Phi_{c} - \delta_{\ell,2} \left\{ \arctan\left[\frac{2 \cos{\iota} \; F_{\times}}{\left(1 + \cos^{2}{\iota}\right) F_{+}}\right] + \frac{\Gamma}{\omega_{\mathrm{BD}}} \frac{\cos{\iota} \left(1 - \cos^{2}{\iota}\right) F_{\times} F_{\mathrm{b}}}{\left(1 + \cos^{2}{\iota}\right)^{2} F_{+}^{2} + 4 \cos^{2}{\iota} F_{\times}^{2}} \right\}\,,
\end{align}
where $\delta_{\ell,m}$ is the Kronecker delta and $\Phi_{c}$ is the GR phase of coalescence (defined as an integration constant when the frequency diverges). Of course, in this calculation we have neglected amplitude corrections that arise purely in GR, if one were to carry out the post-Newtonian approximation to higher order.

Many studies have been carried out to determine the level at which such corrections to the waveform could be measured or constrained once a gravitational wave from a non-vacuum system has been detected. The first such study was carried out by Will~\cite{Will:1994fb}, who determined that given a LIGO detection at SNR $\rho = 10$ of a $(1.4,3)\,M_{\odot}$ black-hole/neutron-star non-spinning, quasi-circular inspiral, one could constrain $\omega_{\mathrm{BD}} > 10^{3}$. Scharre and Will~\cite{Scharre:2001hn} carried out a similar analysis but for a LISA detection with $\rho = 10$ of a $(1.4,10^{3})\,M_{\odot}$ intermediate-mass black-hole/neutron-star, non-spinning, quasi-circular inspiral, and found that one could constrain $\omega_{\mathrm{BD}} > 2.1 \times 10^{4}$. Such an analysis was then repeated by Will and Yunes~\cite{Will:2004xi} but as a function of the classic LISA instrument. They found that the bound is independent of the LISA arm length, but inversely proportional to the LISA position noise error, if the position error noise dominates over laser shot noise. All such studies considered an angle-averaged signal that neglected the spin of either body, assumptions that were relaxed by Berti, et al.~\cite{Berti:2004bd,Berti:2005qd}. They carried out Monte-Carlo simulations over all signal sky positions that included spin-orbit precession to find that the projected bound with LISA deteriorates to $\omega_{\mathrm{BD}} > 0.7 \times 10^{4}$ for the same system and SNR. This was confirmed and extended by Yagi, et al.~\cite{Yagi:2009zm}, who in addition to spin-orbit precession allowed for non-circular (eccentric) inspirals. In fact, when eccentricity is included, the bound deteriorates even further to $\omega_{\mathrm{BD}} > 0.5 \times 10^{4}$. The same authors also found that similar gravitational-wave observations with the next-generation detector DECIGO could constrain $\omega_{\mathrm{BD}} > 1.6 \times 10^{6}$. Similarly, for a non-spinning neutron-star/black-hole binary, the future ground-based detector, the Einstein Telescope (ET)~\cite{Punturo:2010zz}, could place constraints about 5 times stronger than the Cassini bound, as shown in~\cite{Arun:2013bp}, and the bound will further improve by stacking many events~\cite{Zhang:2017sym}.  
Many studies have been carried out to determine the level at which such corrections to the waveform could be measured or constrained once a gravitational wave from a non-vacuum system has been detected. The first such study was carried out by Will~\cite{Will:1994fb}, who determined that given a LIGO detection at SNR $\rho = 10$ of a $(1.4,3)\,M_{\odot}$ black-hole/neutron-star non-spinning, quasi-circular inspiral, one could constrain $\omega_{\mathrm{BD}} > 10^{3}$. Scharre and Will~\cite{Scharre:2001hn} carried out a similar analysis but for a LISA detection with $\rho = 10$ of a $(1.4,10^{3})\,M_{\odot}$ intermediate-mass black-hole/neutron-star, non-spinning, quasi-circular inspiral, and found that one could constrain $\omega_{\mathrm{BD}} > 2.1 \times 10^{4}$. Such an analysis was then repeated by Will and Yunes~\cite{Will:2004xi} but as a function of the classic LISA instrument. They found that the bound is independent of the LISA arm length, but inversely proportional to the LISA position noise error, if the position error noise dominates over laser shot noise. All such studies considered an angle-averaged signal that neglected the spin of either body, assumptions that were relaxed by Berti, et al.~\cite{Berti:2004bd,Berti:2005qd}. They carried out Monte-Carlo simulations over all signal sky positions that included spin-orbit precession to find that the projected bound with LISA deteriorates to $\omega_{\mathrm{BD}} > 0.7 \times 10^{4}$ for the same system and SNR. This was confirmed and extended by Yagi, et al.~\cite{Yagi:2009zm}, who in addition to spin-orbit precession allowed for non-circular (eccentric) inspirals. In fact, when eccentricity is included, the bound deteriorates even further to $\omega_{\mathrm{BD}} > 0.5 \times 10^{4}$. The same authors also found that similar gravitational-wave observations with the next-generation detector DECIGO could constrain $\omega_{\mathrm{BD}} > 1.6 \times 10^{6}$. Similarly, for a non-spinning neutron-star/black-hole binary, the future ground-based detector, the Einstein Telescope (ET)~\cite{Punturo:2010zz}, could place constraints about 5 times stronger than the Cassini bound, as shown in~\cite{Arun:2013bp}, and the bound will further improve by stacking many events~\cite{Zhang:2017sym}.  

All such projected constraints are to be compared with the current solar system bound of $\omega_{\mathrm{BD}} > 4 \times 10^{4}$ placed through the tracking of the Cassini spacecraft~\cite{Bertotti:2003rm}, and current pulsar bound of $\omega_\mathrm{BD} > 1.4\times 10^5$ placed through the absence of the Nordtvedt effect in the pulsar triple system~\cite{Voisin:2020lqi}. Table~\ref{table:comparison-BD} presents all such bounds for ease of comparison, normalized to an SNR of 10. As it should be clear, it is unlikely that LIGO observations will be able to constrain $\omega_{\mathrm{BD}}$ better than current bounds. In fact, even LISA would probably not be able to do better than the Cassini bound. Table~\ref{table:comparison-BD} also shows that the inclusion of more complexity in the waveform seems to dilute the level at which $\omega_{\mathrm{BD}}$ can be constrained. This is because the inclusion of eccentricity and spin forces one to introduce more parameters in the waveform, without these modifications truly adding enough waveform complexity to break the induced degeneracies. One would then expect that the inclusion of amplitude modulation due to precession and higher harmonics should break such degeneracies, at least partially, as was found for massive black-hole binary~\cite{Lang:1900bz,Lang:2011je}. However, even then it seems reasonable to expect that only third-generation detectors will be able to constrain $\omega_{\mathrm{BD}}$ beyond solar-system and pulsar levels, as shown in~\cite{Chamberlain:2017fjl}.

 \begin{table}[htbp]
 \caption{Comparison of proposed tests of scalar-tensor theories\epubtkFootnote{All LISA bounds refer to the classic LISA configuration.}.}
 \label{table:comparison-BD}
  \centering
  {\small
  \begin{tabular}{cccl}
    \toprule
	\textbf{Reference} & \textbf{Binary mass}  & $\omega_{\mathrm{BD}} [10^{4}]$ & \textbf{Properties}\\
	\midrule
 	\cite{Bertotti:2003rm} & x & 4 & Solar system\\
 	\cite{Voisin:2020lqi} & $(1.44,0.198,0.410)$ & 14 & Pulsar triple system\\
	\midrule
	\midrule
	\cite{Will:1994fb} & $(1.4,3)\,M_{\odot}$ & 0.1 & LIGO, Fisher, Ang.\ Ave.\\
	&~&~& circular, non-spinning \\
	\midrule
	\cite{Scharre:2001hn} & $(1.4,10^{3})\,M_{\odot}$ & 24 & LISA, Fisher, Ang.\ Ave.\\
	&~&~& circular, non-spinning \\
	\midrule
	\cite{Will:2004xi} & $(1.4,10^{3})\,M_{\odot}$ & 20 & LISA, Fisher, Ang.\ Ave.\\
	&~&~& circular, non-spinning \\
	\midrule
	\cite{Berti:2004bd} & $(1.4,10^{3})\,M_{\odot}$ & 0.7 & LISA, Fisher, Monte-Carlo\\
	&~&~& circular, w/spin-orbit \\
	\midrule
	\cite{Yagi:2009zm} & $(1.4,10^{3})\,M_{\odot}$ & 0.5 & LISA, Fisher, Monte-Carlo\\
	&~&~& eccentric, spin-orbit \\
	\midrule
	\cite{Yagi:2009zz} & $(1.4,10)\,M_{\odot}$ & 160 & DECIGO, Fisher, Monte-Carlo\\
	&~&~& eccentric, spin-orbit \\
	\midrule
	\cite{Arun:2013bp} & $(1.4,10)\,M_{\odot}$ & 10 & ET, Fisher, Ang.\ Ave.\\
	&~&~& circular, non-spinning \\
	\bottomrule
  \end{tabular}}
\end{table}

The main reason that solar-system and pulsar constraints of Jordan--Fierz--Brans--Dicke theory cannot be beaten with current gravitational-wave observations is that the former are particularly well-suited to constrain weak-field deviations of GR. One might have thought that scalar-tensor theories constitute extreme gravity tests of Einstein's theory, but this is not quite true, as argued in Section~\ref{sec:ST}. One can see this clearly by noting that scalar-tensor theory predicts dipolar radiation, which dominates at low velocities over the GR prediction (precisely the opposite behavior that one would expect from an extreme-gravity modification to Einstein's theory). 

Another interesting aspect of gravitational waves can be studied through gravitational-wave memory, which is a permanent shift in the gravitational-wave strain  after the passage of a burst of gravitational waves. Gravitational memory is related to asymptotic symmetries and conserved quantities. The phenomenology of the memory effects in Jordan--Fierz--Brans--Dicke theory were recently studied in~\cite{Hou:2020tnd,Tahura:2020vsa,Tahura:2021hbk} (memory effects in Jordan--Fierz--Brans--Dicke theory are also discussed in~\cite{Lang:2013fna,Bernard:2022noq}). They found that the asymptotic symmetry group is the same as the Bondi--Metzner--Sachs group
in GR. However, there are new memory effects due to the presence of the scalar polarization mode (the breathing mode), and they are not related to asymptotic symmetries nor conserved quantities~\cite{Hou:2020tnd,Tahura:2020vsa}. Gravitational waveforms due to memory in the tensor mode were derived in~\cite{Tahura:2021hbk} up to the Newtonian order. Due to the scalar dipole radiation, the waveform acquires a correction that has a log dependence. The waveform has a different dependence on the inclination angle from the GR case, which may be used to probe the theory~\cite{Yang:2018ceq}. Heisenberg, et al.~\cite{Heisenberg:2023prj} recently derived the gravitational wave memory in the most general scalar-vector-tensor theory
with second-order equations of motion and vanishing potentials. They also proved a theorem stating that the structure of the memory equation remains unchanged in any
metric theories of gravity in which massless gravitational fields satisfy decoupled wave equations to first order in perturbation theory.

However, one should note that all the above analysis considered only the inspiral phase of coalescence, usually truncating the study at the innermost stable-circular orbit. The merger and ringdown phases, where most of the gravitational wave power resides, have so far been mostly neglected. One might expect that an increase in power will be accompanied by an increase in SNR, thus allowing us to constrain $\omega_{\mathrm{BD}}$ further, as this scales with 1/SNR~\cite{Keppel:2010qu}. Moreover, during merger and ringdown, extreme gravity effects in scalar-tensor theories could affect neutron star parameters and their oscillations~\cite{Sotani:2005qx}, as well as possibly induce dynamical scalarization~\cite{Barausse:2012da,Palenzuela:2013hsa}. Sennett, et al.~\cite{Sennett:2016rwa} has attempted to construct an effective-one-body type waveform model to 2PN order that accounts both for the merger-ringdown, and scalarization. All of these non-linear effects could easily lead to a strengthening of projected bounds. However, to date, no detailed analysis has attempted to determine how well one could constrain scalar-tensor theories using full information about the entire coalescence of a compact binary. 

The subclass of scalar-tensor models described by (massless) Jordan--Fierz--Brans--Dicke theory is not the only type of model that can be constrained with gravitational-wave observations. In the extreme--mass-ratio limit, for binaries consisting of a stellar-mass compact object spiraling into a supermassive black hole, Yunes, et al.~\cite{Yunes:2011aa} have shown that generic scalar-tensor theories reduce to either massless or massive Jordan--Fierz--Brans--Dicke theory. Of course, in this case, the sensitivities need to be calculated from the equations of structure within the full scalar-tensor theory. The inclusion of a scalar field mass leads to an interesting possibility: floating orbits~\cite{Cardoso:2011xi}. Such orbits arise when the small compact object experiences superradiance, leading to resonances in the scalar flux that can momentarily counteract the gravitational-wave flux, leading to a temporarily-stalled orbit that greatly modifies the orbital-phase evolution. These authors showed that if an extreme mass-ratio inspiral is detected with a template consistent with GR, this alone allows us to rule out a large region of $(m_{s},\omega_{\mathrm{BD}})$ phase space, where $m_{s}$ is the mass of the scalar field (see Figure~1 in~\cite{Yunes:2011aa}). This is because if such an inspiral had gone through a resonance, a GR template would be grossly different from the signal. Such bounds are dramatically stronger than one of the current most stringent bounds $\omega_{\mathrm{BD}} > 4 \times 10^{4}$ and $m_{s} < 2.5 \times 10^{-20} \mathrm{\ eV}$ obtained from Cassini measurements of the Shapiro time-delay in the solar system~\cite{Alsing:2011er}. Even if resonances are not hit, Berti, et al.~\cite{Berti:2012bp} have estimated that second-generation ground-based interferometers could constrain the combination $m_{s}/(\omega_{\mathrm{BD}})^{1/2} \lesssim 10^{-15} \mathrm{\ eV}$ with the observation of gravitational waves from neutron-star/binary inspirals at an SNR of $10$. These bounds can also be stronger than current constraints, especially for large scalar mass.
Numerical relativity simulations of binary neutron star mergers in massive scalar-tensor theories have been carried out in~\cite{Kuan:2023hrh,Lam:2024wpq}.

Let us now mention possible gravitational-wave constraints on other types of scalar tensor theories. Let us first consider Jordan--Fierz--Brans--Dicke type scalar-tensor theories, where the coupling constant is allowed to vary. Will~\cite{Will:1994fb} has argued that the constraints described in Table~\ref{table:comparison-BD} go through, with the change
\begin{equation}
\frac{2{\cal{G}}_{1,2}}{2 + \omega_{\mathrm{BD}}} \to \frac{2{\cal{G}}_{1,2}}{2 + \omega_{\mathrm{BD}}} \left[1  + \frac{2 \omega_{\mathrm{BD}}'}{(3 + 2 \omega_{\mathrm{BD}})^{2}} \right]^{2}\,,
\end{equation}
where $\omega'_{\mathrm{BD}} \equiv d\omega_{\mathrm{BD}}/d\phi$. In the $\omega_{\mathrm{BD}} \gg 1$ limit, this implies the replacement $\omega_{\mathrm{BD}} \to \omega_{\mathrm{BD}} [1 + \omega_{\mathrm{BD}}'/(2 \omega_{\mathrm{BD}}^{2})]^{-2}$. Of course, this assumes that there is neither a potential nor a geometric source driving the evolution of the scalar field, and is not applicable for theories where spontaneous scalarization is present~\cite{Damour:1992we}. Regarding Horndeski theory, Figueras and Fran\c{c}a~\cite{Figueras:2021abd} carried out binary black hole merger simulations in cubic Horndeski theory and found that the mismatch in gravitational waveforms in the theory and GR can be $\mathcal{O}(10\%)$ for stellar-mass black hole binaries. Higashino and Tsujikawa~\cite{Higashino:2022izi} carried out a post-Newtonian analysis and derived corrections to the gravitational waveforms in a class of Horndeski theory in which the speed of gravitational waves is the same as that of light (see~\cite{Quartin:2023tpl} for a follow-up analysis). f(R) gravity is another class of modified gravity theories that can be mapped to scalar-tensor theories, as discussed in Sec.~\ref{sec:ST}. Gravitational waves in f(R) gravity were studied, e.g. in~\cite{Vilhena:2021bsx}, where correction terms in the action are proportional to $R^2$ and $R\Box R$.
Takeda, et al.~\cite{Takeda:2023wqn} derived bounds on a subclass of Horndeski theory with luminal GW propagation through the neutron star and black-hole binary merger in the GW200115 event.

Another interesting scalar-tensor theory to consider is that studied by Damour and Esposito-Far{\`{e}}se~\cite{Damour:1992we,Damour:1993hw}. As explained in Section~\ref{sec:ST}, this theory is defined by the action of Eq.~\eqref{ST-gen-action} with the conformal factor $A(\psi) = e^{\beta \psi^{2}}$. In standard Jordan--Fierz--Brans--Dicke theory, only mixed binaries composed of a black hole and a neutron star lead to large deviations from GR due to dipolar emission. This is because dipole emission is proportional to the difference in sensitivities of the binary components. For neutron--star binaries with similar masses, this difference is close to zero, while for black holes it is identically zero (see Eqs.~\eqref{eq:S-def} and~\eqref{eq:BD-phase}). 
Bounds on the theory with GW170817 were studied by Zhao, et al.~\cite{Zhao:2019suc}, though such gravitational-wave bounds are much weaker than those from binary pulsars. This work was later improved by Niu, et al.~\cite{Niu:2021nic} by considering several neutron star binaries and neutron star-black hole binaries found by the LIGO/Virgo Collaborations and applied these events to Jordan--Fierz--Brans--Dicke theory, Damour-Esposito-Far\`ese theory and screened modified gravity (that can evade solar system constraints via various screening mechanisms). The authors found that gravitational-wave observations can place bound comparable to binary pulsar ones in Damour-Esposito-Far\`ese theory, but the constraints are much weaker for Jordan--Fierz--Brans--Dicke and screened modified gravity.
Forecasts on projected bounds on these theories with future gravitational-wave observations were made in~\cite{Zhao:2019suc,Carson:2019fxr}.
In the theory considered by Damour and Esposito-Far\`{e}se, when the gravitational energy is large enough, as in the very late inspiral, non-linear effects can lead to drastic modifications from the GR expectation, such as dynamical and induced scalarization~\cite{Barausse:2012da,Palenzuela:2013hsa}. Unfortunately, most of this happens at rather high frequency, and thus, they become observable only if the activation of the scalar field occurs in the sensitivity band of detectors~\cite{Sampson:2014qqa}. 

Although black holes do not acquire scalar charges in typical scalar-tensor theories under stationary configuration, they can acquire these charges under time-dependent situation. This was first pointed out by Jacobson~\cite{Jacobson:1999vr}. Such a ``miraculous scalar hair growth'' in black holes can be probed with gravitational waves from binary black hole mergers. For example, GW151226 places the bound on the rate of the scalar field evolution as $|\dot \phi| < 1.09 \times 10^4/$sec~\cite{Tahura:2019dgr}.

\subsubsection{Bigravity}
\label{sec:GW_bigravity}

What has been studied in some detail in massive bigravity is the propagation of gravitational waves in a Minkowski background~\cite{DeFelice:2013nba,Narikawa:2014fua}. The propagation field equations in Sec.~\ref{sec:MG-LV} can be solved for the two eigenmodes
\begin{align}
h_{1,+/\times} &= \cos{\theta_{g}} \; h_{+/\times} +  \sin{\theta_{g}} \sqrt{\kappa} \; \xi_{c} \tilde{h}_{+/\times}\,,
\\
h_{2,+/\times} &= -\sin{\theta_{g}} \; h_{+/\times} +  \cos{\theta_{g}} \sqrt{\kappa} \; \xi_{c} \tilde{h}_{+/\times}\,,
\end{align}
where $\kappa$ is the ratio between the two gravitational constants for the two metrics, while
\begin{align}
\theta_{g} &= \frac{1}{2} \cot^{-1}\left(\frac{1 + \kappa \; \xi_{c}^{2}}{2 \sqrt{\kappa} \; \xi_{c}} x + \frac{1 - \kappa \; \xi_{c}^{2}}{2 \sqrt{\kappa} \; \xi_{c}}\right)\,,
\end{align}
and
\begin{align}
x &= \frac{2}{\mu^{2}} (2 \pi f)^{2} (\tilde{c} - 1)\,,
\qquad
\mu^{2} = \lambda_{\mu}^{-2} = \frac{(1 + \kappa \xi_{c}^{2}) \Gamma_{c} m^{2}}{\kappa \xi_{c}^{2}}\,,
\end{align}
with $f$ the gravitational wave frequency and $\tilde{c}$ the speed of light in the auxiliary sector~\cite{DeFelice:2013nba}. The latter is close to unity, with deviations proportional to the matter energy density and pressure
\begin{align}
\tilde{c} \approx 1 + \frac{\kappa \xi_{c}^{2} \left(\rho_{m} + p_{m}\right)}{\Gamma_{c} m^{2} \tilde{M}_{G}^{2}}\,,
\end{align}
where $\rho_{m}$ and $p_{m}$ are the energy density and pressure and $\tilde M_G^2 = M_G^2 (1+\kappa \xi_c^2)$ with $M_G^2 = 1/(8\pi G)$. The constant $\xi_{c}$ is the critical value of the ratio of the scale factor $\xi = \tilde{a}/a$ between the two metrics, which is found by enforcing the conservation of energy momentum~\cite{DeFelice:2013nba} and depends only on the coupling constants of the theory $c_{i}$. In turn, $\Gamma_{c}$ is a function of the coupling constants $c_{i}$ and the critical scale factor ratio $\xi_{c}$, while $m$ is the mass scaling parameter that enters the bigravity action. What is remarkable about the eigenfunctions presented above is that they are a linear combination of the metric perturbations associated with the physical and auxiliary metrics ($h$ and $\tilde h$). Thus, one finds that as gravitational waves propagate in bigravity, they experience oscillations between the physical and auxiliary sectors (similar to neutrino oscillations), which from the standpoint of the physical metric perturbation will look like artificial oscillations in the waveform amplitude.  

The propagation field equations in Sec.~\ref{sec:MG-LV} can also be solved for the two eigenfrequencies~\cite{DeFelice:2013nba}
\begin{align}
k_{1,2}^{2} &= (2 \pi f)^{2} - \frac{\mu^{2}}{2} \left(1 + x \mp \sqrt{1 + 2 x \frac{1 - \kappa \; \xi_{c}^{2}}{1 + \kappa \; \xi_{c}^{2}} + x^{2}}\right)\,.
\end{align}
Such a modification to the dispersion relation will then introduce the phase correction
\begin{align}
\delta \Phi_{1,2} &= - \frac{\mu D \sqrt{\tilde{c}-1}}{2 \sqrt{2} x} \left(1 + x \mp \sqrt{1 + 2 x \frac{1 - \kappa \; \xi_{c}^{2}}{1 + \kappa \; \xi_{c}^{2}} + x^{2}}\right)\,.
\end{align}
Clearly then, the physical metric perturbation $h_{+/\times}$ will not only experience amplitude oscillations, but also phase corrections as the wave propagates a distance $D$. 

But the response function that detectors would observe does not depend on the eigenmodes of the problem, but rather on the reconstructed physical metric perturbation. One can solve for these and construct the Fourier transform of the response function using the stationary phase approximation to find~\cite{DeFelice:2013nba} 
\begin{align}
\tilde{h}(f) = {\cal{A}}(f) e^{i \Psi(f)} \left[B_{1} e^{i \delta \Phi_{1}(f)} + B_{2} e^{i \delta\Phi_{2}(f)}\right]\,,
\end{align}
after averaging over all sky, where the amplitude coefficients are
\begin{align}
{\cal{A}}(f) &= \sqrt{\frac{5 \pi}{24}} \frac{{\cal{M}}^{2}}{(8 \pi M_{G}^{2})^{2} D} y^{-7/6}\,,
\\
B_{1} &= \cos{\theta_{g}} \left(\cos{\theta_{g}} + \sqrt{\kappa} \; \xi_{c} \sin{\theta_{g}}\right)\,,
\\
B_{2} &= \sin{\theta_{g}} \left(\sin{\theta_{g}} - \sqrt{\kappa} \; \xi_{c} \cos{\theta_{g}}\right)\,,
\end{align}
and the phases are
\begin{align}
\Psi(f) &= 2 \pi f t_{c} - \phi_{c} - \frac{\pi}{4} + \frac{3}{128} y^{-5/3} + \frac{5}{96} \left(\frac{743}{336} + \frac{11}{4} \eta\right) \eta^{-2/5} y^{-1} - \frac{3 \pi}{8} \eta^{-3/5} y^{-2/3}\,, 
\end{align}
up to 1.5 post-Newtonian  order, with $y = {\cal{M}} f/(8 \tilde{M}_{G}^{2})$. Such a waveform assumes the generation of gravitational waves is not modified in bigravity, which has been justified based on an order-of-magnitude Vainshtein argument.

The bigravity modification to the waveform can be thought of as oscillations between the physical and auxiliary sectors. Since detectors are only sensitive to the physical metric, however, from our viewpoint we would see an unexplained amplitude modulation, reminiscent to that induced by spin precession. The magnitude of the oscillation depends on $\theta_{g}$, which in turn depends on $x$. When $x \ll 1$ and when $x \gg 1$, the oscillations are suppressed, as one can see by evaluating the $B_{1,2}$ amplitude functions. Only when $x \sim 1$ does one observe noticeable oscillations between modes. The value of $x$, however, does not just depend on the frequency and the mass $\mu$, but also on the speed of gravity in the auxiliary sector, which in turn depends on the matter energy density and pressure. One can then expect $x$ to change significantly as the waves leave the environment in which they were generated, where the density is large, and enter a regime of spacetime where the energy density is much smaller. 

This bigravity modified gravitational waveform can in principle be used to constrain the theory given gravitational wave observations. Narikawa, et al.~\cite{Narikawa:2014fua} predicted that a gravitational wave observations consistent with GR with aLIGO at design sensitivity would lead to a constraint on bigravity of $\mu \lesssim 10^{-17} {\rm{cm}}^{-1}$ when $\tilde{c} -1 \gtrsim 10^{-19}$ when one neglects covariances between the bigravity parameters and $\mu \lesssim 10^{-16.5} {\rm{cm}}^{-1}$ for $\kappa \xi_{c}^{2} \gtrsim \sqrt{10}$ in the full model, which would be stronger than all other bounds imposed on bigravity at the time of~\cite{Narikawa:2014fua}. These projections were obtained by carrying out a Fisher analysis using non-spinning, quasi-circular inspiral waveforms. Possible degeneracies between bigravity effects and amplitude corrections due to post-Newtonian effects or amplitude modulation due to spin-precession could potentially deteriorate these bounds. 

\subsubsection{Einstein-\AE{}ther theory and Khronometric gravity }
\label{sec:EA-KG-waveform}

Black holes and neutron stars exist in both of these theories, as briefly discussed in Sec.~\ref{sec:EA-theory}~\cite{Eling:2007xh,Jacobson:2010mx,Blas:2011ni,Blas:2010hb,Wang:2012nv,Barausse:2012ny,Barausse:2012ny,Barausse:2012qh,Barausse:2013nwa,Adam:2021vsk,Ramos:2018oku,Yagi:2013qpa,Yagi:2013ava,Barausse:2019yuk,Gupta:2021vdj,Ajith:2022uaw}. These bodies have strong self-gravity, and thus, they do not follow a geodesic path in spacetime, but rather their motion is affected by their internal structure. Such a modification is encoded through sensitivity parameters, which control the difference between the bodies center of (baryonic) mass and its center of energy, where the latter has contributions both from baryons and from the energy in the vector fields of the theory. The calculation of the sensitivity of an isolated body is difficult in these Lorentz-violating gravity theories because it requires that one first solve the field equations for a compact object moving at a constant velocity, and then that one match this solution to a post-Newtonian solution for a binary system close to one of the objects. Foster~\cite{Foster:2007gr} calculated the sensitivity $s$ of an isolated body in Einstein-\AE{}ther theory in the weak field limit,
\begin{align}
\label{eq:wf-s}
	s^{\rm EA/KG} = -  \left(\alpha_1^{\rm ppN,\rm EA/KG}-\frac{2}{3}\alpha_2^{\rm ppN,\rm EA/KG}\right)\frac{C_*}{2}\,,
\end{align}
where $C_*$ is the compactness of the star and $\alpha_{1,2}^{\rm ppN,\rm EA}$ are the ppN parameter of Eqs.~\eqref{eq:alpha1-2-EA} and~\eqref{eq:alpha1-2-KG}. The sensitivities of neutron stars, however, cannot be modeled by the above formula, since they are not weak-field objects. Instead, one must compute the sensitivities numerically, as done in~\cite{Yagi:2013ava}. Having said this, analytic expressions for the neutron star sensitivities are now available in~\cite{Gupta:2021vdj} for the Tolman VII neutron star models. The sensitivities for black holes have not yet been computed in Einstein-\AE ther theory. For khronometric theory, Ramos and Barausse~\cite{Ramos:2018oku} found that the black hole sensitivities vanish when two of the coupling constants $(\alpha,\beta)$ (which have been constrained stringently from current experiments and observations) are set to zero. 

Neglecting radiation reaction, the leading-order modification in a post-Newtonian expansion to the motion of compact objects can be captured by a redefinition of Newton's gravitational constant. In Einstein-\AE{}ther theory, Foster~\cite{Foster:2007gr,Yagi:2013ava} showed that the constant ${\cal{G}}$ in Kepler's third law for a binary system is different from the constant $G_{N}$ in Newton's third law for a Cavendish-type experiment, and these two are different from the bare constant $G_{\rm EA/KG}$ that enters the action. One can relate these constants via\epubtkFootnote{These expressions follow e.g.~\cite{Zhang:2019iim} and correct typos in~\cite{Hansen:2014ewa}.} 
\begin{align}
\label{eq:Gs}
	\mathcal{G}_{\rm EA/KG} = G_N \left(1 - s_{1}^{\rm EA/KG}\right)\left(1 - s_{2}^{\rm EA/KG} \right)\,, \qquad G_N = \frac{2 G_{\rm EA}}{2-c_{14}} = \frac{2 G_{\rm KG}}{2-\alpha_{\rm KG}}\,,
\end{align}
where $s_{i}$ are the sensitivities of the bodies. At higher post-Newtonian order, these theories introduce other corrections that cannot probably be absorbed via a redefinition of constants. The conservative dynamics of binaries in these theories, however, has not yet been studied beyond Newtonian order. 

Gravitational waves exist in both of these theories, but they are not the only propagating degrees of freedom. In Einstein-\AE{}ther theory, Jacobson and Mattingly~\cite{Jacobson:2004ts} showed that there are five propagating degrees of freedom: two tensor ones, two vector ones and one scalar mode. In khronometric gravity, there are only three propagating modes (the two tensor ones and the scalar mode), because the hypersurface orthogonality condition eliminates the vector modes. The speed of propagation of these modes is~\cite{Foster:2006az}
\begin{align}
\label{eqn:aespeed}
	w_0^{\rm EA} &= \left[\frac{(2-c_{14}c_{123})}{(2+3c_2+c_+)(1-c_+)c_{14}}\right]^{1/2}\,,
	\\
	w_1^{\rm EA} &= \left[\frac{2c_1-c_+c_-}{2(1-c_+)c_{14}}\right]^{1/2}\,,
	\\
\label{eq:w2_EA}
	w_2^{\rm EA} &= \left(\frac{1}{1-c_+}\right)^{1/2}\,,
\end{align}
in Einstein-\AE{}ther theory and
\begin{align}
	w_{0}^{\rm KG} &= \left[\frac{(\alpha_{\rm KG}-2)(\beta_{\rm KG}+\lambda_{\rm KG})}{\alpha_{\rm KG}(\beta_{\rm KG}-1)(2+\beta_{\rm KG}+3\lambda_{\rm KG})}\right]^{1/2}\,,
	\\
	\label{eq:prop-speed-KG}
	w_2^{\rm KG} &= \left(\frac{1}{1-\beta_{\rm KG}}\right)^{1/2}\,,
\end{align}
in khronometric gravity. Gravitational Cherenkov radiation can be evaded, and energy positivity can be enforced if $w_0^{\rm EA/KG}$, $w_1^{\rm EA/KG}$ and $w_2^{\rm EA/KG}$ are all greater than unity~\cite{Jacobson:2007fh,Elliott:2005va}. 

Let us now consider binary systems in a quasi-circular orbit. In Einstein-\AE{}ther theory, the response function in the time-domain is~\cite{Hansen:2014ewa} 
\begin{align}
	\label{eq:responset}
	h_{\rm EA}(t) &= A^{\rm EA}_2 \frac{\mathcal{M}}{r}u^2\left(e^{-2i\Phi+i\Theta}+ e^{2i\Phi-i\Theta} \right) 
	+ A^{\rm EA}_1 \; \bar{\alpha}^{\rm EA} \;  \frac{\mathcal{M}}{r}\eta^{1/5}u\left(e^{-i\Phi}+ e^{+i\Phi} \right)\,,
\end{align}
where $\Theta = \tan^{-1}[2F_{\times}\cos\iota/F_+(1+\cos^2\iota)]$, and to leading post-Newtonian order, 
\begin{align}
\label{eq:A2}
	A^{\rm EA}_2 &= G_{\rm EA}\left[F^2_+(1+\cos^2\iota)^2+4F^2_{\times}\cos^2\iota\right]^{1/2} \,,
	\\
	A_1^{\rm EA} &= 2G_{\rm EA}(s_1-s_2)\,,
	\\	
	\bar{\alpha}^{\rm EA} &\equiv \left[ \frac{4c_+}{2c_1-c_+c_-}F_{l} - \frac{c_2+1}{c_{123}(c_{14}-2)w_0^{\rm EA}}\right.
		\nonumber \\
		&\left.\times \left(F_b+2F_l \right)
		+\frac{1}{(c_{14}-2)c_+w_0^{\rm EA}}\left(F_b-2F_l \right)\right]\sin\iota
		\nonumber \\
		& + \frac{c_+}{2c_1-c_+c_-}\left[iF_{\rm sn}+\cos\iota F_{\rm se}\right]\,,
			\label{eq:alpha}
\end{align}
and where $(F_{+},F_{\times},F_{b},F_{l},F_{\rm se},F_{\rm sn})$ are beam pattern functions. One can show that the breathing and the longitudinal
modes in the response function above are not linearly independent, leaving only 5 independent degrees of freedom. In khronometric gravity, the time-domain response is
\begin{align}
	h(t)^{\rm KG} &= A_2^{\rm KG}\frac{\mathcal{M}}{r}u\left(e^{-2i\Phi+i\Theta}+e^{+2i\Phi-i\Theta} \right)
	+ A_1^{\rm KG}\bar{\alpha}^{\rm KG}\frac{\mathcal{M}}{r}\eta^{1/5}u^2(e^{-i\Phi}+e^{i\Phi})\,,
\end{align}
where 
\begin{align}
	A_2^{\rm KG} &= A_2^{\rm EA}\,,
	\qquad
A_1^{\rm KG} = 4 \; G_{\rm EA} \; (s_1^{\rm KG}-s_2^{\rm KG})
 \; \,,
	\\
	\label{eqn:alphakg}
	\bar{\alpha}^{\rm KG} &= \left[\frac{\sqrt{\alpha^{\rm KG}}(\lambda^{\rm KG}+1)}{\beta^{\rm KG} +\lambda^{\rm KG}}\left(F_b+2F_l\right)
	+\frac{1}{\sqrt{\alpha^{\rm KG}}}\left(F_b-2F_l\right)\right]\,.
\end{align}
Comparing these response functions, we note that although the two expressions for $A_{1,2}$ are very similar, the expressions for $\bar{\alpha}$ are quite different because of the absence of vector modes in khronometric gravity. 
	
The presence of scalar and vector modes enhances the energy and angular momentum loss. Let us consider then again binary systems in a quasi-circular inspiral. In both Lorentz-violating theories, the total rate of energy carried away by all propagating degrees of freedom can be written as~\cite{Hansen:2014ewa}
\begin{align}
\label{eq:Edot-def}
	\dot{E}_{\rm EA/KG}(u) &= \dot{E}_{\rm GR}(u) \left[1 + \frac{7}{4}\eta^{2/5} u^{-2}_{\rm EA/KG} \dot{E}^{\rm EA/KG}_{-1\rm PN}  + \dot{E}^{\rm EA/KG}_{0\rm PN} \right]\,,
\end{align}
where $\dot{E}_{\rm GR}(u) \equiv - (32/5) u^{10} [1 + {\cal{O}}(c^{-2})]$ is the leading post-Newtonian order prediction, with $u_{\rm EA/KG} = (2 \pi {\cal{G}}_{\rm EA/KG} {\cal{M}} F)^{1/3}$ and $F$ the orbital frequency, and where
\begin{align}
	\label{eqn:bm1pn}
	\dot{E}^{\rm EA}_{-1\rm PN} &= \frac{5}{84}\mathcal{G}(s_1-s_2)^2\frac{(c_{14}-2)(w_0^{\rm EA})^3-(w_1^{\rm EA})^3}{c_{14}(w_0^{\rm EA})^3 (w_1^{\rm EA})^3}\,,
	\\
	\label{eqn:b0pn}
	\dot{E}^{\rm EA}_{0\rm PN} &= \mathcal{G}\left(1-\frac{c_{14}}{2}\right)\left(\frac{1}{w_2^{\rm EA}}+\frac{2c_{14}c_+^2}{(2c_1-c_-c_+)^2 w_1^{\rm EA}}
	+\frac{3c_{14}(Z_{\rm EA}-1)^2}{2w_0^{\rm EA}(2-c_{14})}+S\mathcal{A}_2^{\rm EA}+S^2\mathcal{A}_3^{\rm EA}\right) - 1\,,
\end{align}
in Einstein-\AE{}ther theory and
\begin{align}
	\label{eqn:bm1pnkg}
	\dot{E}_{-1\rm PN}^{\rm KG} &= \frac{5}{84}\mathcal{G}(s_1-s_2)^2\sqrt{\alpha^{\rm KG}}
	\left[\frac{(\beta^{\rm KG}-1)(2+\beta^{\rm KG}+3\lambda^{\rm KG})}{(\alpha^{\rm KG}-2)(\beta^{\rm KG}+\lambda^{\rm KG})}\right]^{3/2}\,,
	\\
	\label{eqn:b0pnkg}
	\dot{E}_{0\rm PN}^{\rm KG} &\equiv \tilde{\beta}_{0\rm PN}^{\rm KG} 
	= \mathcal{G} \left(1-\frac{2}{\beta_{\rm KG}}\right)
	\left(\frac{1}{w_{2}^{\rm KG}}+\frac{3\alpha_{\rm KG}(Z_{\rm KG}-1)^2}{2w_{0}^{\rm KG}(2-\alpha_{\rm KG})} + S\mathcal{A}_2^{\rm KG} + S^2\mathcal{A}_3^{\rm KG}\right)
	-1\,,
\end{align}
in khronometric gravity. In these expressions, $S\equiv (s_1 m_2 + s_2 m_1)/m$ is the sum of the mass-weighted sensitivities, $\mathcal{A}_{2,3}^{\rm EA/KG}$ are functions of the coupling parameters given explicitly in~\cite{Yagi:2013ava}, and
\begin{align}
\label{eq:ZEA}
	Z_{\rm EA}&=\frac{(\alpha_1^{\rm ppN}-2\alpha_2^{\rm ppN})(1-c_+)}{3(2c_+-c14)}\,,
	\qquad
    	Z_{\rm KG} = \frac{(\alpha_1^{\rm ppN}-2\alpha_2^{\rm ppN})(1-\beta_{\rm KG})}{3(2\beta_{\rm KG}-\alpha_{\rm KG})}\,.
\end{align}

With this in hand, we can compute the Fourier transform of the response function in the stationary-phase approximation. Focusing again on quasi-circular binary system, one finds~\cite{Hansen:2014ewa}
\begin{align}
	\label{eq:htwidle}
	\tilde{h}_{\rm EA/KG}(f) = \mathcal{A}_{(1)}^{\rm EA/KG} \; e^{-i\Psi^{(1)}_{\rm EA/KG}} + \mathcal{A}_{(2)}^{\rm EA/KG} \; e^{-i\Psi_{\rm EA/KG}^{(2)}} \,,
\end{align}
where 
\begin{align}
	\label{eqn:eaphase}
	\Psi_{\rm EA}^{(\ell)} &= 2\pi ft_c+\Phi_c-\frac{\pi}{4}  -\frac{3\ell}{256}u_{\ell,\rm EA}^{-5} \left[1 + {\cal{O}}(c^{-2}) \right]
	-\frac{3\ell}{256}u_{\ell,\rm EA}^{-5} \left[\dot{E}_{-1\rm PN}^{\rm EA} \eta^{2/5}u_{\ell,\rm EA}^{-2} +  \dot{E}_{0\rm PN}^{\rm EA} + {\cal{O}}(c^{-2}) \right]\,,
	\\
	\mathcal{A}_{(1)}^{\rm EA} &= -\left(\frac{5\pi}{48}\right)^{1/2}  A^{\rm EA}_1\bar{\alpha}^{\rm EA} \frac{\mathcal{M}^2}{r_{12}}\eta^{1/5} u_{1,\rm EA}^{-9/2}\,,
	\qquad
	\mathcal{A}_{(2)}^{\rm EA} = \left(\frac{5\pi}{96}\right)^{1/2} A^{\rm EA}_2\frac{\mathcal{M}^2}{r_{12}}u_{2,\rm EA}^{-7/2}\,,
\end{align}
in Einstein-\AE{}ther theory and
\begin{align}
	\label{eq:phasekg}
	\Psi^{(\ell)}_{\rm KG} &= 2\pi ft_c+\Phi_c - \frac{\pi}{4}-\frac{3\ell}{256}u^{-5}_{\ell} \left[1 + {\cal{O}}(c^{-2})\right]
	-\frac{3\ell}{256}u^{-5}_{\ell} \left[\dot{E}_{-1\rm PN}^{\rm KG} \eta^{2/5}u_{\ell,\rm KG}^{-2} + \dot{E}_{0\rm PN}^{\rm KG} + {\cal{O}}(c^{-2})\right]\,,
	\\
	\mathcal{A}_{(1)}^{\rm KG} &= \left(\frac{5\pi}{48}\right)^{1/2} A_1^{\rm KG}\bar{\alpha}^{\rm KG}\frac{\mathcal{M}^2}{r_{12}}\eta^{1/5}u_{1,\rm KG}^{-9/2}\,,
	\qquad
	\mathcal{A}_{(2)}^{\rm KG} = \left(\frac{5\pi}{96}\right)^{1/2} A^{\rm KG}_2\frac{\mathcal{M}^2}{r_{12}}u_{2,\rm KG}^{-7/2}\,.
\end{align}
in khronometric gravity. In these expressions, $u_{\ell,\rm EA/KG} \equiv \left(2\pi \mathcal{G}_{\rm EA/KG} \mathcal{M}f/\ell\right)^{1/3}$ and $(t_{c},\phi_{c})$ are constant time and phase offsets. One can of course keep higher order terms in the post-Newtonian expansion in the GR sector easily.

The above calculations in Einstein-\AE ther theory were later updated by a few different groups. Zhang, et al.~\cite{Zhang:2019iim} derived the response function in both the time and frequency domains. For the former, the $\ell=2$ harmonic depends not only on the tensor modes but also on the scalar and vector modes, even to leading post-Newtonian order, though such scalar and vector modes vanish when $c_{14}=0$ and $c_+ =0$, respectively. The expression for the $\ell =1$ harmonic is also corrected. For the response function in the frequency domain, the dominant $\ell=2$ harmonic for the tensor mode is given by
\begin{align}
\tilde h_\mathrm{EA}(f) = &
-\sqrt{\frac{5 \pi}{96}} \frac{(G_N\mathcal{M})^2}{R}\mathcal{U}_{2,\mathrm{EA}}^{-7 / 2} \left[F_{+}\left(1+\cos ^2 \iota\right)+2i F_{\times} \cos \iota\right] \nonumber \\
&\times  \left(1 - \frac{1}{2\sqrt{\kappa_3}}\eta^{2/5} \epsilon_x \mathcal{U}_{2,\mathrm{EA}}^{-2} \right) e^{i \Psi^{(2)}_\mathrm{EA}}\,,
\end{align}
with
\begin{align}
\Psi^{(2)}_\mathrm{EA} =& 2\pi f \left(t_c+\frac{r}{w_2^\mathrm{EA}}\right) - \Phi_c - \frac{\pi}{4} + \frac{3}{128} \mathcal{U}_{2,\mathrm{EA}}^{-5} [1+\mathcal{O}(c^{-2})] \nonumber \\
& - \frac{3}{224} \frac{\eta^{2/5} \epsilon_x}{\kappa_3} \mathcal{U}_{2,\mathrm{EA}}^{-7} - \frac{3}{128} \left[ -\frac{2}{3} (s_1 + s_2) - \frac{1}{2}c_{14}+\kappa_3-1 \right]\mathcal{U}_{2,\mathrm{EA}}^{-5} [1+\mathcal{O}(c^{-2})]\,, \\
\mathcal{U}_{2,\mathrm{EA}} =& (\pi G_N \mathcal M f)^{1/3}\,,
\end{align}
while $\kappa_3$ and $\epsilon_x$ are given in~\cite{Zhang:2019iim}. 
Taherasghari and Will~\cite{Taherasghari:2023rwn} carried out a direct integration of the relaxed field equations up to 2.5 post-Newtonian order, while Hou, et al.~\cite{Hou:2023pfz} studied the GW memory effect in Einstein-\AE ther theory.

As one can see from the description above, the propagation speed of the tensor modes is different from the speed of light by a \emph{constant} factor, which makes this effect difficult to measure with detections by a single instrument. However, the detection of a single event by more than one instrument can be used to place a bound on the speed of gravity. Given $N$ detectors located at different places on Earth, a gravitational wave will hit every detector at different times. The time difference between the detection at each instrument (measured for example as the time at which the amplitude of the wave reaches its maximum at the given instrument) is then related to the speed of the wave and the distance between detectors. A time difference that is consistent with waves traveling at the speed of light can place a constraint on the speed of gravity of ${\cal{O}}(1)$ in units of the speed of light~\cite{Blas:2016qmn}. 

Einstein-\AE{}ther theory and khronometric gravity can also be constrained using information about the precise functional form of the response function presented above. Hansen et al~\cite{Hansen:2014ewa} carried out the first analysis to investigate this possibility with aLIGO at design sensitivity and with the Einstein Telescope. Chamberlain and Yunes~\cite{Chamberlain:2017fjl} expanded this analysis by considering a wider range of third-generation future ground-based detectors, as well as space-based detectors. Both analysis first considered the quasi-circular inspiral of binary neutron stars, because the sensitivities for these systems are known~\cite{Yagi:2013ava}. For such binaries, the $-1$PN order term in the phase of the response function (the one proportional to $\dot{E}_{-1\rm PN}^{\rm EA/KG}$ in Eqs.~\eqref{eqn:eaphase} and~\eqref{eq:phasekg}) is suppressed by the square of the difference of the sensitivities as shown in Eqs.~\eqref{eqn:bm1pn} and~\eqref{eqn:bm1pnkg}. This is both because $s_{1} \sim s_{2}$ for binary neutron stars because their masses are similar and because $s_{1,2} \ll 1$ for neutron stars to begin with. Thus, when considering binary neutron stars, the bounds are weaker than one would have expected, since the phase modification is dominated by a term of Newtonian order in the phase. Hansen, et al.~\cite{Hansen:2014ewa} found that aLIGO could not place constraints on $(c_{+},c_{-})$ or $(\lambda_{\rm KG},\beta_{\rm KG})$ that are better than current binary pulsar constraints. However, both Hansen, et al.~\cite{Hansen:2014ewa} and Chamberlain and Yunes~\cite{Chamberlain:2017fjl} found that third-generation detectors will obtain constraints comparable to binary pulsar ones in the future. Chamberlain and Yunes~\cite{Chamberlain:2017fjl} also found that if binary black hole inspirals are considered and if the sensitivities do not suppress the GR modification, then the bounds become roughly an order of magnitude better, since then the modification enters at $-1$PN order. These analyses assumed that the post-Newtonian parameters characterizing the preferred-frame effect vanish, and probed only $(c_{+},c_{-})$ or $(\lambda_{\rm KG},\beta_{\rm KG})$. However, the multi-messenger observations of GW170817 have constrained the deviation in the propagation speed of gravitational waves away from the speed of light to $\sim 10^{-15}$ or smaller~\cite{LIGOScientific:2017zic}. From Eqs.~\eqref{eq:w2_EA} and~\eqref{eq:prop-speed-KG}, this means $c_+$ and $\beta_\mathrm{KG}$ have been constrained effectively to zero. Therefore, the bounds on these theories with gravitational waves need to be reanalyzed by imposing $c_+ =0$ or $\beta_\mathrm{KG}=0$ and varying over the remaining parameters. Schumacher, et al.~\cite{Schumacher:2023cxh} carried out such an analysis for the GW170817 event in Einstein-AEther theory, and found that current GW observations do not place bounds that are stronger than Solar System experiments, or binary pulsar and cosmological observations.

\subsubsection{Modified quadratic gravity}
\label{sec:direct-test-MQG}

Gravitational waves are modified in quadratic modified gravity. In dynamical Chern--Simons gravity, Garfinkle, et al.~\cite{Garfinkle:2010zx} have shown that the propagation of such waves on a Minkowski background remains unaltered, and thus, all modifications arise during the generation stage. In Einstein--dilaton--Gauss--Bonnet theory, a similar analysis shows that gravitational waves can travel at phase velocities different from that of light, but such effects become negligible in the very far-away radiation zone~\cite{Ayzenberg:2013wua}. Yagi, et al.~\cite{Yagi:2011xp} studied the generation mechanism in both theories during the quasi-circular inspiral of comparable-mass, spinning black holes in the post-Newtonian and small-coupling approximations. They found that a standard post-Newtonian analysis fails for such theories because the assumption that black holes can be described by a distributional stress-energy tensor without any further structure fails. They also found that since black holes acquire scalar hair in these theories, and this scalar field is anchored to the curvature profiles, as black holes move, the scalar fields must follow the singularities, leading to dipole scalar-field emission. Juli\'e and Berti~\cite{Julie:2019sab} studied the post-Newtonian dynamics of a two-body system in scalar Gauss-Bonnet gravity through a ``skeltonization'' of black holes. That is, one approximates a black hole with a point-particle that has a mass that depends on the scalar field~\cite{1975ApJ...196L..59E}. The scalar hair is related to the derivative of this mass through the scalar field.

During a quasi-circular inspiral of spinning black holes in dynamical Chern--Simons gravity, the total gravitational wave energy flux carried out to spatial infinity (equal to minus the rate of change of a binary's binding energy by the balance law) is modified from the GR expectation to leading order by~\cite{Yagi:2011xp}  
\begin{equation}
\frac{\delta \dot{E}^{\mathrm{CS}}_{\mathrm{spin}}}{\dot{E}_{\mathrm{GR}}} =
\frac{25}{1236} \; \zeta_{4} \; \eta^{-2} \left[\bar{\Delta}^{2} +27 \left< \left(\bar{\Delta} \cdot \hat{v}_{12} \right)^{2}\right>_{}\right]
\label{eq:Edot-CS-spin}
\end{equation}
due to scalar field radiation and corrections to the metric perturbation that are of magnetic-type, quadrupole form. In this equation, $\dot{E}_{\mathrm{GR}} = (32/5) \eta^2 m^{2} v^4 \omega^2$ is the leading-order GR prediction for the total energy flux, $\zeta_{4} = \alpha_{4}^{2}/(\beta \kappa m^{4})$ is the dimensionless Chern--Simons coupling parameter, $\hat{v}_{12}^{i}$ is the unit relative velocity vector, $\bar{\Delta}^{i} = (m_{2}/m) (a_{1}/m_{1}) \hat{S}^{i}_{1} - (m_{1}/m) (a_{2}/m_{2}) \hat{S}^{i}_{2}$, where $a_{A}$ is the Kerr spin parameter of the $A$th black hole and $\hat{S}^{i}_{A}$ is the unit vector in the direction of the spin angular momentum, and the angle brackets stand for an average over several gravitational wave wavelengths. If the black holes are not spinning, then the correction to the scalar energy flux is greatly suppressed~\cite{Yagi:2011xp} 
\begin{equation}
\frac{\delta \dot{E}^{\mathrm{CS}}_{\mathrm{no-spin}}}{\dot{E}_{\mathrm{GR}}} = \frac{2}{3} \delta_{m}^{2} \zeta_{4} v_{12}^{14}\,,
\end{equation}
where we have defined the reduced mass difference $\delta_{m} \equiv (m_{1} - m_{2})/m$. Notice that this is a 7 post-Newtonian--order correction, instead of a 2 post-Newtonian correction as in Eq.~\eqref{eq:Edot-CS-spin}. In the non-spinning limit, the dynamical Chern--Simons correction to the metric tensor induces a 6 post-Newtonian--order correction to the gravitational energy flux~\cite{Yagi:2011xp}, which is consistent with the numerical results of~\cite{Pani:2011xj}. 

On the other hand, in Einstein--dilaton--Gauss--Bonnet gravity, the corrections to the energy flux are~\cite{Yagi:2011xp} 
\begin{equation}
\frac{\delta \dot{E}^{\mathrm{EDGB}}_{\mathrm{no-spin}}}{\dot{E}_{\mathrm{GR}}} = \frac{5}{96} \eta^{-4} \delta_{m}^{2} \zeta_{3} v_{12}^{-2}\,,
\end{equation}
which is a $-1$ post-Newtonian correction. This is because the scalar field $\vartheta_{\mathrm{EDGB}}$ behaves like a monopole (see Eq.~\eqref{eq:theta-EDGB}), and when such a scalar monopole is dragged by the black hole, it emits electric-type, dipole scalar radiation. Any hairy black hole with monopole hair will thus emit dipolar radiation, leading to $-1$ post-Newtonian corrections in the energy flux carried to spatial infinity.  

Such modifications to the energy flux modify the rate of change of the binary's binding energy through the balance law, $\dot{E} = -\dot{E}_{\mathrm{b}}$, which in turn modify the rate of change of the gravitational wave frequency and phase, $\dot{F} = - \dot{E} \; (dE_{\mathrm{b}}/dF)^{-1}$. For dynamical Chern--Simons gravity (when the spins are aligned with the orbital angular momentum) and for Einstein--dilaton--Gauss--Bonnet theory (in the non-spinning case), the Fourier transform of the gravitational-wave response function in the stationary phase approximation for binary black holes becomes~\cite{Yagi:2011xp,Yagi:2012vf} 
\begin{equation}
\tilde{h}_{\mathrm{dCS,EDGB}} = \tilde{h}_{\mathrm{GR}} (1+\alpha_\mathrm{dCS,EDGB} u^{a_\mathrm{dCS,EDGB}}) e^{i \beta_{\mathrm{dCS,EDGB}} u^{b_{\mathrm{dCS,EDGB}}}}\,,
\end{equation}
where $\tilde{h}_{\mathrm{GR}}$ is the Fourier transform of the response in GR, $u \equiv (\pi {\cal{M}} f)^{1/3}$ with $f$ the gravitational wave frequency. For dynamical Chern-Simons theory, the phase corrections are given by ~\cite{Yagi:2012vf}   
\begin{align}
\label{eq:beta-dCS}
\beta_{\mathrm{dCS}} &= 
-\frac{507775}{7340032} \frac{\zeta_{4}}{\eta^{4/5}} \frac{m^{2}}{m_{1}^{2}} \frac{a_{1}^{2}}{m_{1}^{2}} 
\left[1 - \frac{58833}{20311} \left( \hat{S}_{1} \cdot \hat{L}\right)^{2} \right]
\\ \nonumber
&+ \frac{63625}{1048576}\frac{\zeta_{4}}{\eta^{9/5}} \frac{a_{1}}{m_{1}} \frac{a_{2}}{m_{2}} 
\left[ \left(\hat{S}_{1} \cdot \hat{S}_{2}\right)
- \frac{1467}{509} \left( \hat{S}_{1} \cdot \hat{L} \right) \left( \hat{S}_{2} \cdot \hat{L} \right) \right] + 1 \to 2\,,
\qquad
b_{\mathrm{dCS}} = -1\,,
\end{align}
where $\hat{S}_{1,2}$ and $\hat{L}$ are the unit spin and orbital angular momenta respectively. For the spin-aligned case, the above expression reduces to~\cite{Tahura:2018zuq}
\begin{equation}
\beta_\mathrm{dCS} = \frac{481525}{3670016} \eta^{-14 / 5} \zeta_{4}\left[-2 \delta_m \chi_a \chi_s+\left(1-\frac{4992 \eta}{19261}\right) \chi_a^2+\left(1-\frac{72052 \eta}{19261}\right) \chi_s^2\right]\,, 
\quad
b_{\mathrm{dCS}} = -1\,,
\end{equation}
where $\chi_s \equiv (\chi_1 + \chi_2)/2$, $\chi_a \equiv (\chi_1 - \chi_2)/2$ with $\chi_A \equiv a_A/m_A$, and the amplitude correction is given by~\cite{Tahura:2018zuq}
\begin{equation}
\alpha_\mathrm{dCS} = \frac{57713}{344064} \eta^{-14 / 5} \zeta_4\left[-2 \delta_{\mathbf{m}} \chi_{\mathbf{a}} \chi_{\mathbf{s}}+\left(1-\frac{14976 \eta}{57713}\right) \chi_{\mathbf{a}}^2+\left(1-\frac{215876 \eta}{57713}\right) \chi_{\mathbf{s}}^2\right]\,, 
\quad
a_\mathrm{dCS} = +4\,.
\end{equation}
For Einstein--dilaton--Gauss--Bonnet theory, the phase and amplitude corrections are given by
\begin{align}
\beta_{\mathrm{EDGB}} &= -\frac{5}{7168} \zeta_{3} \eta^{-18/5} \delta_{m}^{2} \,, 
 \qquad
 b_{\mathrm{EDGB}} = -7\,, \\
\alpha_{\mathrm{EDGB}} &= -\frac{5}{192} \zeta_{3} \eta^{-18/5} \delta_{m}^{2} \,, 
 \qquad
 a_{\mathrm{EDGB}} = -2\,,
\end{align}
We have included both deformations to the binding energy and Kepler's third law, in addition to changes in the energy flux, when computing the phase correction. However, in Einstein--dilaton--Gauss--Bonnet theory the binding energy is modified at higher post-Newtonian order, and thus, corrections to the energy flux control the leading modifications to the gravitational-wave response function. 

Can we go beyond leading corrections to the waveforms in these theories?
Higher post-Newtonian corrections in Einstein--dilaton--Gauss--Bonnet gravity were derived in Shiralilou, et al.~\cite{Shiralilou:2020gah,Shiralilou:2021mfl} to first post-Newtonian order higher than the leading, tensor
non-dipole and scalar dipole emission respectively. This was later corrected and extended to second post-Newtonian order relative to the leading tensor/scalar contribution by Lyu, et al.~\cite{Lyu:2022gdr}, which used the results in Sennett, et al.~\cite{Sennett:2016klh} for scalar-tensor theories, but the corresponding scalar charges for black holes in Einstein--dilaton--Gauss--Bonnet gravity. For spin-precessing binaries, the precession equations are modified. Loutrel, Tanaka and Yunes~\cite{Loutrel:2018ydv,Loutrel:2018rxs,Loutrel:2022tbk} solved such equations and derived gravitational waveforms in dynamical Chern-Simons gravity, including spin precession (see also Li, et al.~\cite{Li:2022grj} for related work). Li, et al.~\cite{Li:2023lqz} studied gravitational waves from eccentric compact binaries in dynamical Chern-Simons gravity. Going beyond the inspiral, numerical relativity simulations have been carried out in quadratic gravity to find gravitational waves and scalar waves during the merger-ringdown stage. Okounkova, et al.~\cite{Okounkova:2017yby,Okounkova:2019dfo,Okounkova:2019zjf} carried out such simulations in dynamical Chern-Simons gravity within the small-coupling approximation. Treating the theory as an effective field theory has the advantage that the principal parts of the modified Einstein equations are the same as in GR, and thus, are well-posed. In Einstein--dilaton--Gauss--Bonnet gravity, Witek, et al.~\cite{Witek:2018dmd} derived the scalar dynamics, while Okounkova~\cite{Okounkova:2020rqw} found merger-ringdown gravitational waveforms within the small coupling approximation. Ripley, et al.~\cite{Ripley:2019aqj,East:2020hgw,Corman:2022xqg} performed similar simulations by fully solving the field equations to find waveforms without using the small-coupling approximation. Watarai, et al.~\cite{Watarai:2023yky} recently constructed a parameterized merger waveform via a principal component analysis, and applied their formulation to Einstein--dilaton--Gauss--Bonnet gravity. The ringdown can also be studied through black hole perturbation theory. Quasi-normal mode frequencies were computed for slowly-rotating black holes in dynamical Chern-Simons gravity to first order in spin in~\cite{Wagle:2021tam,Srivastava:2021imr}, while those in Einstein--dilaton--Gauss--Bonnet gravity were derived to zeroth-order~\cite{Blazquez-Salcedo:2016enn,Bryant:2021xdh,Luna:2024spo}, first-order~\cite{Pierini:2021jxd} and second-order~\cite{Pierini:2022eim} in spin, as well as for rapidly-rotating black holes for the first time in~\cite{Chung:2024vaf}, which was recently extended in~\cite{Blazquez-Salcedo:2024oek}. East and Pretorius~\cite{East:2022rqi} performed numerical relativity simulations for binary neutron star mergers in (shift-symmetric) Einstein--dilaton--Gauss--Bonnet gravity and found that, while the non-GR corrections during the inspiral are small (because neutron stars do not possess scalar charges in this theory), there can be strong scalar effects during the post-merger phase. Kuan, et al.~\cite{Kuan:2023trn} carried out numerical relativity simulations for binary neutron star mergers in scalar-Gauss-Bonnet gravity with dynamical scalarization and found a universal relation between the compactness of an isolated neutron star and the critical coupling strength for scalarization.

From the above analysis, it should be clear that the corrections to the gravitational-wave observable in quadratic modified gravity are always proportional to the quantity $\zeta_{3,4} \equiv \xi_{3,4}/m^{4} = \alpha_{3,4}^{2}/(\beta \kappa m^{4})$. Thus, any measurement that is consistent with GR will allow a constraint of the form $\zeta_{3,4} < N \delta$, where $N$ is a number of order unity, and $\delta$ is the accuracy of the measurement. Solving for the coupling constants of the theory, such a measurement would lead to $\xi_{3,4}^{1/4} < (N \delta)^{1/4} m$~\cite{Sopuerta:2009iy}. Therefore, constraints on quadratic modified gravity will weaken for systems with larger characteristic mass. This can be understood by noticing that the corrections to the action scale with positive powers of the Riemann tensor, while this scales inversely with the mass of the object, i.e., the smaller a compact object is, the larger its curvature. Such an analysis then automatically predicts that LIGO will be able to place stronger constraints than LISA-like missions on such theories, because LIGO operates in the 100~Hz frequency band, allowing for the detection of stellar-mass inspirals, while LISA-like missions operate in the mHz band, and are limited to supermassive black-holes inspirals. This reasoning is valid, except for extreme mass-ratio inspirals detectable with LISA, which consist of a stellar mass black hole inspiraling into a supermassive one; in this case, the constraints achievable with LISA can be comparable or in some cases better than those that can be obtained with ground-based instruments~\cite{Chamberlain:2017fjl,Perkins:2020tra,Maselli:2020zgv,Maselli:2021men,Barsanti:2022ana}. 

How well can these modifications be measured with gravitational-wave observations? Yagi, et al.~\cite{Yagi:2011xp} predicted, based on the results of Cornish, et al.~\cite{Cornish:2011ys}, that a sky-averaged LIGO gravitational-wave observation with SNR of 10 of the quasi-circular inspiral of non-spinning black holes with masses $(6,12)\,M_{\odot}$ would allow a constraint of $\xi_{3}^{1/4} \lesssim 20\mathrm{\ km}$, where we recall that $\xi_{3} = \alpha_{3}^{2}/(\beta \kappa)$. A similar sky-averaged, eLISA observation of a quasi-circular, spin-aligned black-hole inspiral with masses $(10^{6},3\times10^{6}\,M_{\odot})$ would constrain $\xi_{3}^{1/4} < 10^{7}\mathrm{\ km}$~\cite{Yagi:2011xp}. The loss in constraining power comes from the fact that the constraint on $\xi_{3}$ will scale with the total mass of the binary, which is six orders of magnitude larger for equal-mass binaries detectable with space-borne sources. Similarly, the observation of the ringdown part of the signal, instead of the inspiral, can also lead to constraints on the theory, since the ringdown spectrum is shifted from the GR one by $\xi_{3}$. Blazquez-Salcedo, et al.~\cite{Blazquez-Salcedo:2016enn} have argued that a ringdown observation with ringdown SNR of 50 could allow for a constraint of $\xi_{3}^{1/4} \lesssim 30 \mathrm{\ km}$. More recently, Chung and Yunes updated this estimate through the calculation of the ringdown frequencies for $10 M_\odot$ black holes of moderate spins, finding that an SNR 10 event should allow a constraint of $\xi_3^{1/4} \lesssim 30 \mathrm{\ km}$~\cite{Chung:2024ira,Chung:2024vaf}.
As already discussed in Sec.~\ref{subsec:MQG}, existing gravitational-wave events from O1-O3 runs can place bounds that are stronger than the above estimates because of the stacking of multiple events (which leads to an enhancement of roughly $1/N^{1/2}$ after stacking $N$ events of the same SNR). Using inspiral-only waveforms, one arrives at the stacked bound $\xi_3^{1/4} \lesssim 3$km~\cite{Nair:2019iur,Perkins:2021mhb,Wang:2021jfc,Lyu:2022gdr,Wang:2023wgv}. The most up-to-date stacked gravitational-wave bound is $\xi_3^{1/4} \lesssim 0.3$km, which was found by Julie, et al.~\cite{Julie:2024fwy} through the construction of an inspiral-merger-ringdown waveform in scalar Gauss-Bonnet gravity via the effective-one-body framework that was compared to O1-O3 events. This bound stronger than other non-gravitational-wave bounds from the existence of compact objects~\cite{Pani:2011xm} ($\xi_{3}^{1/4} < 26\mathrm{\ km}$) and from the change in the orbital period of the low-mass X-ray binary A0620--00 ($\xi_{3}^{1/4} < 1.9\mathrm{\ km}$)~\cite{Yagi:2012gb}. Future gravitational-wave detectors will be able to probe the theory even more stringently~\cite{Carson:2019rda,Carson:2019kkh,Carson:2019fxr,Perkins:2020tra,Carson:2020cqb,Carson:2020ter}. For example, future multiband observations with ground-based and space-based detectors~\cite{Sesana:2016ljz,Barausse:2016eii,Liu:2020nwz} may probe the theory to $\xi_3^{1/4} \lesssim 10^{-1}$km or better~\cite{Carson:2019fxr,Perkins:2020tra}, while a population of GW events with third-generation gravitational-wave detectors can place the bound  $\xi_3^{1/4} \lesssim 10^{-3}-10^{-2}$km~\cite{Perkins:2020tra}.

In dynamical Chern--Simons gravity, one expects similar projected gravitational-wave constraints on $\xi_{4}$, namely $\xi_{4}^{1/4} < {\cal{O}}(M)$, where $M$ is the total mass of the binary system in kilometers. Therefore, for binaries detectable with ground-based interferometers, one expects constraints of order $\xi_{4}^{1/4} < 10\mathrm{\ km}$. In this case, such a constraint would be roughly six orders of magnitude stronger than current LAGEOS bounds~\cite{AliHaimoud:2011fw}. Dynamical Chern--Simons gravity cannot be constrained with binary pulsar observations, since the theory's corrections to the post-Keplerian observables are too high post-Newtonian order, given the current observational uncertainties~\cite{Yagi:2013mbt}. However, the gravitational wave constraint is more difficult to achieve in the dynamical Chern--Simons case, because the correction to the gravitational wave phase is degenerate with spin. However, Yagi, et al.~\cite{Yagi:2012vf} argued that precession should break this degeneracy, and if a signal with sufficiently high SNR is observed, such bounds would be possible. One must be careful, of course, to check that the small-coupling approximation is still satisfied when saturating such a constraint~\cite{Yagi:2012vf}. When using current gravitational-wave observations, the bounds from the correction to the inspiral are too weak and do not satisfy the small-coupling approximation~\cite{Nair:2019iur,Perkins:2021mhb,Wang:2021jfc}, making such bounds unreliable. When focusing on the correction to the ringdown, Silva, et al.~\cite{Silva:2022srr} derived the bound as $\xi_4^{1/4} \lesssim 100$km, but this relies on a small spin expansion (which may not be accurate for remnant black holes that spin fast). Similar to the Einstein--dilaton--Gauss--Bonnet case, future gravitational-wave observations are expected to improve the bound further~\cite{Yagi:2012vf,Carson:2019rda,Carson:2019kkh,Perkins:2020tra}. For example, a population of GW events with third-generation gravitational-wave detectors can place the bound of  $\xi_4^{1/4} \lesssim 10^{-2}-10^{-1}$km~\cite{Perkins:2020tra}.

Although the main focus of this subsection is quadratic gravity theories, in which the action is corrected at quadratic order in curvature, in principle, there are also higher-order curvature corrections that one could study. Endlich, et al.~\cite{Endlich:2017tqa} constructed an effective field theory extension to GR that can be tested with gravitational waves. This effective field theory does not involve additional fields, so it does not reduce to Einstein--dilaton--Gauss--Bonnet or dynamical Chern-Simons gravity, even if one truncates the effective theory at quadratic order in curvature. Sennett, et al.~\cite{Sennett:2019bpc} included quartic order corrections and derived bounds on the theory with existing gravitational-wave observations. Silva, et al.~\cite{Silva:2022srr} derived constraints on the characteristic length scale of the cubic ($\ell_\mathrm{cEFT}$) and quartic ($\ell_\mathrm{qEFT}$) order effective field theory from ringdown observations of GW150914 and GW200129, namely $\ell_\mathrm{cEFT} \leq 38.2$km and $\ell_\mathrm{qEFT} \leq 51.3$km respectively, again assuming a small spin expansion. Liu and Yunes~\cite{Liu:2024atc} constructed an approximate inspiral-merger-ringdown waveform in cubic effective field theory of gravity and derived a new bound from the GW170608 event that is 3.5 times stronger than previous constraints. The curvature dependence of gravitational-wave tests of GR has recently been carried out by Payne, et al.~\cite{Payne:2024yhk} in a model-independent framework that is similar to the one developed by Stein and Yagi~\cite{Stein:2013wza}, which parameterizes scalar-interactions in the gravitational action.

\subsubsection{Non-commutative geometry}
\label{sec:NC_GW}

As described in Sec.~\ref{sec:NCG}, there are two main approaches of non-commutative geometry. We will review gravitational waves in each of these classes below:

\begin{itemize}
\item [(I)] Spectral Geometry
\end{itemize}

Black holes exist in the spectral non-commutative geometry theories.
What is more, the usual Schwarzschild and Kerr solutions of GR persist in these theories. This is not because such solutions have vanishing Weyl tensor, but because the quantity $\nabla^{\alpha \beta} C_{\mu \alpha \nu \beta}$ happens to vanish for such metrics. Similarly, one would expect that the two-body, post-Newtonian metric that describes a black-hole--binary system should also satisfy the non-commutative geometry field equations, although this has not been proven explicitly. Moreover, although neutron-star spacetimes have not yet been considered in non-commutative geometries, it is likely that if such spacetimes are stationary and satisfy the Einstein equations, they will also satisfy the modified field equations. Much more work on this is still needed to put all of these concepts on a firmer basis.  

Gravitational waves exist in non-commutative gravity. Their generation for a compact binary system in a circular orbit was analyzed by Nelson, et al.~in~\cite{Nelson:2010rt,Nelson:2010ru}. They began by showing that a transverse-traceless gauge exists in this theory, although the transverse-traceless operator is slightly different from that in GR. They then proceeded to solve the modified field equations for the metric perturbation [Eq.~\eqref{eq:NCG-h-eq}] via a Green's function approach:
\begin{equation}
h^{ik} = 2 \beta \int \frac{dt'}{\sqrt{(t - t')^{2} - |r|^{2}}} \ddot{I}^{ik}(t') {\cal{J}}_{1}(\beta \sqrt{(t - t')^{2} - |r|^{2}})\,,
\label{eq:h-int-eq-NCG}
\end{equation}
where recall that $\beta^{2} = (-32 \pi \alpha_{0})^{-1}$ acts like a mass term, the integral is taken over the entire past light cone, ${\cal{J}}_{1}(\cdot)$ is the Bessel function of the first kind, $|r|$ is the distance from the source to the observer and the quadrupole moment is defined as usual:
\begin{equation}
I^{ik} = \int d^{3}x \; T^{00}_{\mathrm{mat}} x^{ik} \,,
\end{equation}
where $T^{00}$ is the time-time component of the matter stress-energy tensor. Of course, this is only the first term in an infinite multipole expansion. 

Although the integral in Eq.~\eqref{eq:h-int-eq-NCG} has not yet been solved in the post-Newtonian approximation, Nelson, et al.~\cite{Nelson:2010rt,Nelson:2010ru} did solve for its time derivative to find
\begin{subequations}
\begin{align}
\dot{h}^{xx} &= -\dot{h}^{yy} = 32 \beta \mu r_{12}^{2} \Omega^{4} \left[\sin{(2 \phi)} f_{c}\left(\beta |r|,\frac{\Omega}{\beta}\right) + \cos{(2 \phi)} f_{s}\left(\beta |r|,\frac{2 \Omega}{\beta}\right)\right]\,,
\\
\dot{h}^{xy} &= - 32 \beta \mu r_{12}^{2} \Omega^{4} \left[\sin{\left(2 \phi - \frac{\pi}{2}\right)} f_{c}\left(\beta |r|,\frac{\Omega}{\beta}\right) + \cos{\left(2 \phi - \frac{\pi}{2}\right)} f_{s}\left(\beta |r|,\frac{2 \Omega}{\beta}\right)\right]\,, 
\end{align}
\label{h-NCG}
\end{subequations}
where $\Omega = 2 \pi F$ is the orbital angular frequency, we have defined
\begin{align}
f_{s}(x,z) &= \int_{0}^{\infty} \frac{ds}{\sqrt{s^{2} + x^{2}}} {\cal{J}}_{1}(s) \sin\left(z \sqrt{s^{2} + x^{2}}\right)\,,
\\
f_{c}(x,z) &= \int_{0}^{\infty} \frac{ds}{\sqrt{s^{2} + x^{2}}} {\cal{J}}_{1}(s) \cos\left(z \sqrt{s^{2} + x^{2}}\right)\,,
\end{align}
and one has assumed that the binary is in the $x$-$y$ plane and the observer is on the $z$-axis. However, if one expands these expressions about $\beta = \infty$, one recovers the GR solution to leading order, plus corrections that decay faster than $1/r$. This then automatically implies that such modifications to the generation mechanism will be difficult to observe for sources at astronomical distances.  

Given such a solution, one can compute the flux of energy carried by gravitational waves to spatial infinity. Stein and Yunes~\cite{Stein:2010pn} have shown that in quadratic gravity theories, this flux is still given by
\begin{equation}
\dot{E} = \frac{\kappa}{2} \int d\Omega r^{2} \left< \dot{\bar{h}}_{\mu \nu} \dot{\bar{h}}^{\mu \nu}\right>\,,
\label{eq:Edot-NCG}
\end{equation}
where $\bar{h}_{\mu \nu}$ is the trace-reversed metric perturbation, the integral is taken over a 2-sphere at spatial infinity, and we recall that the angle brackets stand for an average over several wavelengths. Given the solution in Eq.~\eqref{h-NCG}, one finds that the energy flux is
\begin{equation}
\dot{E} = \frac{9}{20} \mu^{2} r_{12}^{2} \Omega^{4} \beta^{2} 
\left[|r|^{2} f_{c}^{2}\left(\beta |r|, \frac{2 \Omega}{\beta}\right) + |r|^{2} f_{s}^{2}\left(\beta |r|, \frac{2 \Omega}{\beta}\right) \right]\,.
\label{eq:NCG-Edot}
\end{equation}
The asymptotic expansion of the term in between square brackets about $\beta = \infty$ is
\begin{equation}
|r|^{2} \left[ f_{c}^{2}\left(\beta |r|, \frac{2 \Omega}{\beta}\right) + f_{s}^{2}\left(\beta |r|, \frac{2 \Omega}{\beta}\right) \right]\sim
|r|^{2} \left\{ \frac{1}{\beta^{2} |r|^{2}} \left[1 + {\cal{O}}\left(\frac{1}{|r|}\right) \right] \right\}\,,
\end{equation}
which then leads to an energy flux identical to that in GR, as any subdominant term goes to zero when the 2-sphere of integration is taken to spatial infinity. In that case, there are no modifications to the rate of change of the orbital frequency. Of course, if one were not to expand about $\beta = \infty$, then the energy flux would lead to certain resonances at $\beta = 2 \Omega$, but the energy flux is only well-defined at future null infinity. 

The above analysis was used by Nelson, et al.~\cite{Nelson:2010rt,Nelson:2010ru} to compute the rate of change of the orbital period of binary pulsars, in the hopes of using this to constrain $\beta$. Using data from the binary pulsar, they stipulated an order-of-magnitude constraint of $\beta \geq 10^{-13}\mathrm{\ m}^{-1}$. However, such an analysis could be revisited to relax a few assumptions used in~\cite{Nelson:2010rt,Nelson:2010ru}. First, binary pulsar constraints on modified gravity theories require the use of at least three observables. These observables can be, for example, the rate of change of the period $\dot{P}$, the line of nodes $\dot{\Omega}$ and the perihelion shift $\dot{w}$. Any one observable depends on the parameters $(m_{1},m_{2})$ in GR or $(m_{1},m_{2},\beta)$ in non-commutative geometries, where $m_{1,2}$ are the component masses. Therefore, each observable corresponds to a surface of co-dimension one, i.e., a two-dimensional surface or sheet in the three-dimensional space $(m_{1},m_{2},\beta)$. If the binary pulsar observations are consistent with Einstein's theory, then all sheets will intersect at some point, within a certain uncertainty volume given by the observational error. The simultaneous fitting of all these observables is what allows one to place a bound on $\beta$. The analysis of~\cite{Nelson:2010rt,Nelson:2010ru} assumed that all binary pulsar observables were known, except for $\beta$, but degeneracies between $(m_{1},m_{2},\beta)$ could potentially dilute constraints on these quantities. Moreover, this analysis should be generalized to eccentric and inclined binaries, since binary pulsars are known to not be on exactly circular orbits. 

But perhaps the most important modification that ought to be made has to do with the calculation of the energy flux itself. The expression for $\dot{E}$ in Eq.~\eqref{eq:Edot-NCG} in terms of derivatives of the metric perturbation derives from the effective gravitational-wave stress-energy tensor, obtained by perturbatively expanding the action or the field equations and averaging over several wavelengths (the \emph{Isaacson} procedure~\cite{Isaacson:1968ra,Isaacson:1968gw}). In modified gravity theories, the definition of the effective stress-energy tensor in terms of the metric perturbation is usually modified, as found for example in~\cite{Stein:2010pn}. In the case of non-commutative geometries, Stein and Yunes~\cite{Stein:2010pn} showed that Eq.~\eqref{eq:Edot-NCG} still holds, provided one considers fluxes at spatial infinity. However, the analysis of~\cite{Nelson:2010rt,Nelson:2010ru} evaluated this energy flux at a fixed distance, instead of taking the $r \to \infty$ limit. 

The balance law relates the rate of change of a binary's binding energy with the gravitational wave flux emitted by the binary, but for it to hold, one must require the following: (i) that the binary be isolated and possess a well-defined binding energy, and (ii) that the total stress-energy of the spacetime satisfies a local covariant conservation law. If (ii) holds, one can use this conservation law to relate the rate of change of the volume integral of the energy density, i.e., the energy flux, to the volume integral of the current density, which can be rewritten as an integral over the boundary of the volume through Stokes' theorem. Since in principle one can choose any integration volume, any physically-meaningful result should be independent of the surface of that volume. This is indeed the case in GR, provided one takes the integration $2$-sphere to spatial infinity. Presumably, if one included all the relevant terms in $\dot{E}$, without taking the limit to $i^{0}$, one would still find a result that is independent of the surface of this two-sphere. However, this has not yet been verified. Therefore, the analysis of~\cite{Nelson:2010rt,Nelson:2010ru} should be taken as an interesting first step toward understanding possible changes in the gravitational-wave metric perturbation in non-commutative geometries.    

Not much beyond this has been done regarding non-commutative geometries and gravitational waves. In particular, one lacks a study of what the final response function would be if the gravitational-wave propagation were modified, which of course depends on the time-evolution of all propagating gravitational-wave degrees of freedom, and whether there are only the two usual dynamical degrees of freedom in the metric perturbation.

\begin{itemize}
\item [(II)] Moyal-type Non-commutative Geometry
\end{itemize}

Gravitational waves from compact binary inspirals in the Moyal-type non-commutative geometry were first derived in Kobakhidze, et al.~\cite{Kobakhidze:2016cqh}, and were improved by Jenks, et al.~\cite{Jenks:2020gbt} by relaxing some of the assumptions made in~\cite{Kobakhidze:2016cqh}. Using the stress-energy tensor for a black hole in Eq.~\eqref{eq:Tmunu_NC} and orbital averaging, the binding energy of a binary is given by
\begin{equation}
E=E_\mathrm{GR} - \frac{m^4\eta(1-2\eta) \Lambda^2}{16 r^3}[1-3(\hat{\bm L} \cdot \bm \theta)^2]\,,
\end{equation}
where $E_\mathrm{GR}$ is the GR contribution, while $\hat{\bm L}$ is the unit orbital angular momentum. We recall that $\Lambda$ and $\bm \theta$ are defined in Eq.~\eqref{eq:Lambda_theta}. The above correction to the binding energy further modifies Kepler's third law. The gravitational-wave luminosity is still given through the quadrupole formula and is not modified from GR, modulo the modifications coming from Kepler's third law. To leading order in small deviations from GR, the luminosity is given by
\begin{equation}
\dot E = \frac{32}{5} \eta^2 x^5\left[1+\frac{\Lambda^2(1-2 \eta)}{32}\left(23-39(\hat{\boldsymbol{L}} \cdot \boldsymbol{\theta})^2\right) x^2\right]\,,
\end{equation}
with $x \equiv (\pi m f)^{2/3}$. From these expressions, we can compute $\dot f = (df/dE) (dE/dt)$ and the Fourier phase in the frequency domain becomes
\begin{equation}
\Psi(f) = \Psi_\mathrm{GR} -\frac{75}{2048} \frac{1-2\eta}{\eta} \Lambda^2 [7-15(\hat{\bm L} \cdot \bm \theta)^2] x^{-1/2}\,.
\end{equation}
Given that the GR phase $ \Psi_\mathrm{GR} \propto x^{-5/2}$ to leading post-Newtonian order, the correction term enters at second post-Newtonian order relative to GR. When $\hat{\bm L} \cdot \bm \theta = 1$, the waveform in the frequency domain is given by
\begin{equation}
\tilde{h}_{\mathrm{NC}} = \tilde{h}_{\mathrm{GR}} (1+\alpha_\mathrm{NC} u^{a_\mathrm{NC}}) e^{i \beta_{\mathrm{NC}} u^{b_{\mathrm{NC}}}}\,,
\end{equation}
with~\cite{Tahura:2018zuq}
\begin{equation}
\beta_{\mathrm{NC}} = -\frac{75}{256} \eta^{-4 / 5}(2 \eta-1) \Lambda^2\,, \quad b_\mathrm{NC} = -1\,, \\
\alpha_{\mathrm{NC}} = -\frac{3}{8} \eta^{-4 / 5}(2 \eta-1) \Lambda^2\,, \quad a_\mathrm{NC} = +4\,,
\end{equation}
clearly mapping to the ppE framework. 

Let us now discuss the bound on the Moyal-type non-commutative geometry. Kobakhidze, et al.~\cite{Kobakhidze:2016cqh} used the bound on the second post-Newtonian order correction from GW150914, obtained by the LIGO/Virgo Collaboration~\cite{TheLIGOScientific:2016src}, and found $\sqrt{\Lambda} \lesssim 3.5$. Jenks, et al.~\cite{Jenks:2020gbt} rederived the bounds from the posterior samples from several events in the GWTC-1 catalog, including the second post-Newtonian, non-GR correction term, and found a probability distribution on $\Lambda^2 [7-15(\hat{\bm L} \cdot \bm \theta)^2]$. Finding 90\%-credible limits, the authors then derived the bound on $\sqrt{\Lambda}$ as a function of $(\hat{\bm L} \cdot \bm \theta)$. In most cases, the bound on $\sqrt{\Lambda}$ is of order unity and is consistent with~\cite{Kobakhidze:2016cqh}.  Jenks, et al.~\cite{Jenks:2020gbt} also derived the non-commutative correction to the pericenter precession of a binary and applied the result to the double binary pulsar PSR J0737-3039. Using the measurements of pericenter precession together with the measurements of the Shapiro delay, mass ratio and mass functions for determining the masses of the pulsars, the authors derived a bound on $\sqrt{\Lambda}$ that ended up being weaker than the gravitational-wave bounds by roughly a factor of 5. Perkins, et al.~\cite{Perkins:2020tra} gave a future forecast on probing non-commutative gravity with gravitational waves. Through a population of gravitational-wave events, the authors found that the theory can be probed to the $\sqrt{\Lambda} \lesssim 10^{-3}$ level.

\subsection{Generic tests}
\label{generic-tests}
\subsubsection{Massive graviton theories}
\label{generic-tests:MG-LV}

Several massive graviton theories have been proposed to later be discarded due to ghosts, non-linear or radiative instabilities. Thus, little work has gone into studying whether black holes and neutron stars in these theories persist and are stable, and how the generation of gravitational waves is modified. Such questions will depend on the specific massive gravity model considered, and of course, if a Vainshtein mechanism is active and effective, then there will not be any significant modifications. 

However, a few generic properties of such theories can still be stated. One of them is that the non-dynamical (near-zone) gravitational field will be corrected, leading to Yukawa-like modifications to the gravitational potential~\cite{Will:1997bb}
\begin{equation}
V_{\mathrm{MG}}(r) = \frac{M}{r} e^{-r/\lambda_{g}}\,,
\qquad \mathrm{or} \qquad
V_{\mathrm{MG}}(r) = \frac{M}{r} \left(1 + \gamma_{\mathrm{MG}} e^{-r/\lambda_{g}} \right)\,,
\label{eq:Yukawa}
\end{equation}
where $r$ is the distance from the source to a field point. For example, the latter parameterization arises in gravitational theories with compactified extra dimensions~\cite{Kehagias:1999my}. Such corrections lead to a \emph{fifth force}, which then in turn allows us to place constraints on $m_{g}$ through solar system observations~\cite{Talmadge:1988qz}. Nobody has yet considered how such modifications to the near-zone metric could affect the binding energy of compact binaries and their associated gravitational waves. 

Another generic consequence of a graviton mass is the appearance of additional propagating degrees of freedom in the gravitational wave metric perturbation. In particular, one expects scalar, longitudinal modes to be excited (see, e.g., \cite{Dilkes:2001av}). This is, for example, the case if the action is of Pauli--Fierz type~\cite{Fierz:1939ix,Dilkes:2001av}. Such longitudinal modes arise due to the non-vanishing of the $\Psi_{2}$ and $\Psi_{3}$ Newman--Penrose scalars, and can be associated with the presence of spin-0 particles, if the theory is of Type~N in the $E(2)$ classification~\cite{lrr-2006-3}. The specific form of the scalar mode will depend on the structure of the modified field equations, and thus, it is not possible to generically predict its associated contribution to the response function. 

A robust prediction of massive graviton theories relates to how the propagation of gravitational waves is affected. If the graviton has a mass, its velocity of propagation will differ from the speed of light, as given for example in Eq.~\eqref{eq:vg-standard}. Will~\cite{Will:1997bb} showed that such a modification in the dispersion relation leads to a correction in the relation between the difference in time of emission $\Delta t_{e}$ and arrival $\Delta t_{a}$ of two gravitons:
\begin{equation}
\Delta t_{a} = \left(1 + z\right) \left[ \Delta t_{e} + \frac{D}{2 \lambda_{g}^{2}} \left(\frac{1}{f_{e}^{2}} + \frac{1}{f_{e}^{'2}} \right) \right],
\label{eq:t-MG}
\end{equation}
where $z$ is the redshift, $\lambda_{g}$ is the graviton's Compton wavelength, $f_{e}$ and $f_{e}'$ are the emission frequencies of the two gravitons and $D$ is the distance measure 
\begin{equation}
D = \frac{1 + z}{H_{0}} \int_{0}^{z} \frac{dz'}{(1 + z')^{2} [\Omega_{M} (1 + z')^{3} + \Omega_{\Lambda}]^{1/2}}\,,
\end{equation}
where $H_{0}$ is the present value of the Hubble parameter, $\Omega_{M}$ is the matter energy density and $\Omega_{\Lambda}$ is the vacuum energy density (for a zero spatial-curvature universe).  

Even if the gravitational wave at the source is unmodified, the graviton time delay will leave an imprint on the Fourier transform of the response function by the time it reaches the detector~\cite{Will:1997bb}. This is because the Fourier phase is proportional to 
\begin{equation}
\Psi \propto 2 \pi \int^{f}_{f_{c}} [t(f) - t_{c}] df'\,,
\end{equation}
where $t$ is now not a constant but a function of frequency as given by Eq.~\eqref{eq:t-MG}. Carrying out the integration, one finds that the Fourier transform of the response function becomes
\begin{equation}
\tilde{h}_{\mathrm{MG}} = \tilde{h}_{\mathrm{GR}} e^{i \beta_{\mathrm{MG}} u^{b_{\mathrm{MG}}}}\,,
\label{eq:response-MG}
\end{equation}
where $\tilde{h}_{\mathrm{GR}}$ is the Fourier transform of the response function in GR, we recall that $u = (\pi {\cal{M}} f)^{1/3}$ and we have defined 
\begin{equation}
\beta_{\mathrm{MG}} = - \frac{\pi^{2} D {\cal{M}}}{\lambda_{g}^{2} (1 + z)}\,,
\qquad
b_{\mathrm{MG}} = -3\,.
\end{equation}
Such a correction is of 1 post-Newtonian order relative to the leading-order, Newtonian term in the Fourier phase. Notice also that there are no modifications to the amplitude at all.

Numerous studies have considered possible bounds on $\lambda_{g}$. The most stringent solar system constraint is $\lambda_{g} > 3.9 \times 10^{13}\mathrm{\ km}$~\cite{Will:2018gku,Bernus:2020szc} and it comes from observations of Kepler's third law, which if the graviton had a mass would be modified by the Yukawa factor in Eq.~\eqref{eq:Yukawa}. Observations of the rate of decay of the period in binary pulsars~\cite{Finn:2001qi,Baskaran:2008za} can also be used to place the more stringent constraint $\lambda > 1.5 \times 10^{14}\mathrm{\ km}$. Similarly, studies of the stability of Kerr black holes in Pauli--Fierz theory~\cite{Fierz:1939ix} have yielded constraints of $\lambda_{g} > 2.4 \times 10^{13}\mathrm{\ km}$~\cite{Brito:2013wya}. 

Gravitational-wave observations of binary systems can also be used to constrain the mass of the graviton. One possible test is to compare the times of arrival of coincident gravitational wave and electromagnetic signals, for example in white-dwarf binary systems. Larson and Hiscock~\cite{Larson:1999kg} and Cutler, et al.~\cite{Cutler:2002ef} estimated that one could constrain $\lambda_{g} > 3 \times 10^{13}\mathrm{\ km}$ with classic LISA. Will~\cite{Will:1997bb} was the first to consider constraints on $\lambda_{g}$ from gravitational-wave observations only, a test that was indeed carried out with the first aLIGO observations leading to the constraint $\lambda_{g} > 10^{13} \mathrm{\ km}$ at $90\%$ confidence with GW150914~\cite{TheLIGOScientific:2016src}. Using 43 gravitational-wave events in the catalog GWTC-3, the above bound has been updated to $\lambda_g > 9.4 \times 10^{13}$km~\cite{LIGOScientific:2021sio}. Will considered sky-averaged, quasi-circular inspirals and found that LIGO observations of $10\,M_{\odot}$ equal-mass black holes would lead to a constraint of $\lambda_{g} > 6 \times 10^{12}\mathrm{\ km}$ with a Fisher analysis, a projection that was quite close to the bound that aLIGO obtained. 

Constraints on the mass of the graviton can of course improve with future observations. For example, constraints are improved to $\lambda_{g} > 6.9 \times 10^{16}\mathrm{\ km}$ with classic LISA observations of $10^{7}\,M_{\odot}$, equal-mass black holes because the massive graviton correction accumulates with distance traveled (see Eq.~\eqref{eq:response-MG}). Will's study was later generalized by Will and Yunes~\cite{Will:2004xi} and by Chamberlain and Yunes~\cite{Chamberlain:2017fjl}, who considered how different detector characteristics affected the possible bounds on $\lambda_{g}$. Will and Yunes found that the bound scales with the square-root of the LISA arm length and inversely with the square root of the LISA acceleration noise~\cite{Will:2004xi}. Chamberlain and Yunes found that the bound can increase to $\lambda_{g} > 10^{15}\mathrm{\ km}$ with single gravitational-wave observations with third-generation ground-based detectors and up to $\lambda_{g} > 10^{18}\mathrm{\ km}$ with certain gravitational-wave observations with space-based detectors~\cite{Chamberlain:2017fjl}. Carson and Yagi~\cite{Carson:2019kkh} showed that multiband observations of GW150914-like events with ground-based and space-based detectors can constrain the graviton Compton wavelength to $\lambda_g \gtrsim 10^{14}$km, while Perkins, et al.~\cite{Perkins:2020tra} found that the bound can be as strong as $\lambda_g \gtrsim 10^{16}-10^{17}$km with a population of gravitational-wave events detected by a network of third-generation ground-based detectors.

The initial projections of Will have been refined by Berti, et al.~\cite{Berti:2004bd}, Yagi and Tanaka~\cite{Yagi:2009zm}, Arun and Will~\cite{Arun:2009pq}, Stavridis and Will~\cite{Stavridis:2010zz} and  Berti, et al.~\cite{Berti:2011jz} to allow for non--sky-averaged responses, spin-orbit and spin-spin coupling, higher harmonics in the gravitational wave amplitude, eccentricity and multiple detections. Although the bound deteriorates on average for sources that are not optimally oriented relative to the detector, the bound improves when one includes spin couplings, higher harmonics, eccentricity, and multiple detections as the additional information and power encoded in the waveform increases, helping to break parameter degeneracies. However, all of these studies neglected the merger and ringdown phases of the coalescence, an assumption that was relaxed by Keppel and Ajith~\cite{Keppel:2010qu}, leading to the strongest projected bounds $\lambda_{g} > 4 \times 10^{17}\mathrm{\ km}$. Moreover, all studies until then had computed bounds using a Fisher analysis prescription, an assumption relaxed by del Pozzo, et al.~\cite{DelPozzo:2011pg}, who found that a Bayesian analysis with priors consistent with solar system experiments leads to bounds stronger than Fisher ones by roughly a factor of two. Perkins and Yunes~\cite{Perkins:2018tir} took into account the effect of screening inside the Milky Way and the host galaxy of a source binary. They found that future gravitational-wave observations can place constraints on $\lambda_g$ and the screening radius of $\mathcal{O}(10^{13})$--$\mathcal{O}(10^{17})$km and $\mathcal{O}(10^{2})$--$\mathcal{O}(10^{4})$Mpc respectively.

In summary, projected constraints on $\lambda_{g}$ are generically stronger than current solar system or binary pulsar constraints by several orders of magnitude, given a LISA observation of massive black-hole mergers. Even aLIGO observations can and have done better than current solar system constraints by factors between a few~\cite{DelPozzo:2011pg} to an order of magnitude~\cite{Keppel:2010qu,TheLIGOScientific:2016src}, depending on the source. All of these results are summarized in Table~\ref{table:comparison-MG}, normalizing everything to an SNR of 10.  

 \begin{table}[htbp]
 \caption[Comparison of proposed tests of massive graviton theories.]{Comparison of constraints and projected constraints on  massive graviton theories. Entries above the double line correspond to actual bounds, while those below correspond to projected bounds with gravitational-wave observations (normalized to an SNR of 10). Ang.\ Ave.\ stands for an angular average over all sky locations.}
 \label{table:comparison-MG}
  \centering
  {\small
  \begin{tabular}{cccl}
    \toprule
	\textbf{Reference} & \textbf{Binary mass}  & $\lambda_{g} [10^{15}\mathrm{\ km}]$ & \textbf{Properties}\\
	\midrule
 	\cite{Will:2018gku,Bernus:2020szc,Mariani:2023ubf} & x & 0.039, 1.22 & Solar-system dynamics\\
	\midrule
	\cite{Finn:2001qi} & x & $1.6 \times 10^{-5}$ & Binary pulsar orbital period \\
	&~&~& in Visser's theory~\cite{Visser:1997hd} \\	
	\midrule 
	\cite{Brito:2013wya} & x & 0.024 & Stability of black holes \\
	&~&~& in Pauli--Fierz theory~\cite{Fierz:1939ix}\\	
	\midrule 
	\cite{Zakharov:2016lzv} & x & 0.00043 & S2 orbit  \\
	\midrule 
	\cite{TheLIGOScientific:2016src} & $(32,39) M_{\odot}$ & $0.01$ & GW150914 aLIGO observation \\
	\midrule 
	\cite{LIGOScientific:2021sio} &x & $0.094$ & 43 gravitational-wave events from GWTC-3  \\
	\midrule
	\midrule
	\cite{Will:1997bb} & $(10,10)\,M_{\odot}$ & 0.006 & LIGO, Fisher, Ang. Ave.\\
	&~&~& circular, non-spinning \\
	\midrule
	\cite{DelPozzo:2011pg} & $(13,3)\,M_{\odot}$ & 0.006\,--\,0.014 & LIGO, Bayesian, Ang.\ Ave.\\
	&~&~& circular, non-spinning \\	
	\midrule
	\cite{Chamberlain:2017fjl} & $(30,30)\,M_{\odot}$ & 1 & third-generation ground-based, Fisher,\\
	&~&~& Ang.\ Ave., circular, non-spinning \\		
	\midrule
	\cite{Larson:1999kg,Cutler:2002ef} & $(0.5,0.5)\,M_{\odot}$ & 0.03 & LISA, WD-WD, coincident\\
	&~&~& with electromagnetic signal \\
	\midrule
	\cite{Will:1997bb,Will:2004xi,Chamberlain:2017fjl} & $(10^{7},10^{7})\,M_{\odot}$ & 50--70 & LISA, Fisher, Ang.\ Ave. \\
	&~&~& circular, non-spinning \\
	\midrule
	\cite{Berti:2004bd} & $(10^{6},10^{6})\,M_{\odot}$ & 10 & LISA, Fisher, Monte-Carlo\\
	&~&~& circular, w/spin-orbit \\
	\midrule	
	\cite{Arun:2009pq} & $(10^{5},10^{5})\,M_{\odot}$ & 10 & LISA, Fisher, Ang.\ Ave.\\
	&~&~& higher-harmonics, circular, non-spinning \\
	\midrule	
	\cite{Yagi:2009zm} & $(10^{6},10^{7})\,M_{\odot}$ & 22 & LISA, Fisher, Monte-Carlo\\
	&~&~& eccentric, spin-orbit \\
	\midrule
	\cite{Yagi:2009zz} & $(10^{6},10^{7})\,M_{\odot}$ & 2.4 & DECIGO, Fisher, Monte-Carlo\\
	&~&~& eccentric, spin-orbit \\
	\midrule
	\cite{Stavridis:2010zz} & $(10^{6},10^{6})\,M_{\odot}$ & 50 & LISA, Fisher, Monte-Carlo\\
	&~&~& circular, w/spin modulations \\
	\midrule
	\cite{Keppel:2010qu} & $(10^{7},10^{7})\,M_{\odot}$ & 400 & LISA, Fisher, Ang.\ Ave.\\
	&~&~& circular, non-spinning, w/merger \\	
	\midrule
	\cite{Berti:2011jz} & $(13,3)\,M_{\odot}$ & 30 & eLISA, Fisher, Monte-Carlo\\
	&~&~& multiple detections, circular, non-spinning \\	
	\bottomrule
  \end{tabular}}
\end{table}

Before proceeding, we should note that the correction to the propagation of gravitational waves due to a non-zero graviton mass are not exclusive to binary systems. In fact, any gravitational wave that propagates a significant distance from the source will suffer from the time delays described in this section. Binary inspirals are particularly useful as probes of this effect because one knows the functional form of the waveform, and thus, one can employ matched filtering to obtain a strong constraint. But, in principle, one could use gravitational-wave bursts from supernovae or other sources.

\subsubsection{Massive boson fields and superradiance}
\label{generic-tests:massive-boson-fields}

Black holes in the presence of massive boson fields are unstable to the superradiant instability~\cite{Zeldovich:1971mw,Misner:1972kx,Damour:1976kh,Ternov:1978gq} (see~\cite{Brito:2015oca} for a review). When the Compton wavelength of the massive boson is comparable to the size of the black hole's horizon, the boson field gains energy (exponentially populating bound Bohr orbits around the black hole~\cite{Arvanitaki:2010sy,Arvanitaki:2014wva}) at the expense of the black hole's rotational energy, forming a Bose-Einstein condensate cloud around it. For stellar mass black holes, this occurs if the boson field is extremely light, with a mass in the range $(10^{-14},10^{-10})$ eV. Bosons in this mass range include the QCD axion~\cite{Weinberg:1977ma,Wilczek:1977pj,Peccei:1977hh}, string theory axions~\cite{Arvanitaki:2009fg} and dark photons~\cite{Holdom:1985ag,Pani:2012bp}.

The discovery of superradiant instabilities would be a smoking gun for the existence of some type of exotic physics, which has therefore prompted many studies that investigate its signature in astrophysical environments. For example, if superradiance is active, then rapidly rotating black holes should not exist, as the black holes should be spun down and energy is dumped into the boson field. This would appear as a drastic deficit in the spin-population of black holes, which could be observed given a large enough catalog of gravitational wave observations from compact binary coalescence~\cite{Arvanitaki:2009fg,Arvanitaki:2010sy,Arvanitaki:2014wva}. In fact, the measurement of spin from X-ray observations from accretion disks around rotating black holes has been used to place constraints on the mass of such light bosons~\cite{Brito:2013wya,Arvanitaki:2014wva}.  

Superradiance and massive boson fields would also have an effect in the emission of gravitational waves. The latter can produce gravitational waves in several ways. One possibility is for the Bose-Einstein condensate cloud to completely collapse if the attractive boson self-interactions become stronger than the gravitational binding energy, producing a \emph{Bosenova}. The resulting gravitational waves are expected to be burst-like, although a quantitative dynamical analysis has not yet been carried out. Another way for the massive bosons to generate gravitational waves is through transitions between energy levels and via bosonic annihilation. In both cases, the emitted gravitational waves are coherent, monochromatic and long-lasting, producing signals that could be observed by ground-based detectors through continuous searches~\cite{Arvanitaki:2016qwi}. These waves would also produce a stochastic background that could also be detected with ground-based instruments~\cite{Brito:2017wnc}. 
Recent works on probing ultralight bosons with gravitational-wave observations include~\cite{Brito:2017wnc,Brito:2017zvb,Tsukada:2018mbp,East:2018glu,Ikeda:2018nhb,Dergachev:2019wqa,Ng:2019jsx,Chen:2019fsq,Palomba:2019vxe,Brito:2020lup,Dergachev:2020fli,Zhu:2020tht,Tsukada:2020lgt,Ng:2020ruv,Yuan:2021ebu,Guo:2021xao,Kalogera:2021bya,KAGRA:2021tse,Zhang:2021mks,Yuan:2022bem}. For example, Tsukada, et al.~\cite{Tsukada:2018mbp} carried out the first search of a stochastic gravitational-wave background from boson clouds with the first observing run of LIGO/Virgo and excluded scalar bosons with masses $[2.0,3.8] \times 10^{-13}$eV, while Yuan, et al.~\cite{Yuan:2022bem} updated the results to rule out the mass range $[1.5,15] \times 10^{-13}$eV with the first and third observing runs.

\subsubsection{Lorentz-violating gravity}
\label{sec:generic-tests-LV}

We have so far concentrated on massive graviton theories, but, as discussed in Section~\ref{sec:MG-LV}, there is a strong connection between such theories and Lorentz violation. Modifications to the dispersion relation are usually a result of a modification of the Lorentz group or its action in real or momentum space. For this reason, it is interesting to consider generic Lorentz-violating-inspired, modified dispersion relations of the form of Eq.~\eqref{eq:vg-LV}, or more precisely~\cite{Mirshekari:2011yq}
\begin{equation}
\frac{v_{g}^{2}}{c^{2}} = 1 - A E^{\alpha_{\mathrm{LV}} - 2}\,,
\label{eq:gen-disp-rel}
\end{equation}
where $\alpha_{\mathrm{LV}}$ controls the structure of the modification and $A$ its amplitude. 

A plethora of models that violate Lorentz symmetry in the gravitational sector can be modeled with the generic modified dispersion relation presented above. Clearly, when $\alpha_{\mathrm{LV}}=0$ and $A = m_{g}^{2} c^{2}$ one recovers the standard modified dispersion relation of Eq.~\eqref{eq:vg-standard}. On the other hand, one can reproduce the predictions of the SME when $A =- 2 \mathring{k}_{(I)}^{(d)}$ for even $d \geq 4$ and $A = \pm 2 \mathring{k}_{(V)}^{(d)}$ for odd $d \geq 5$ with $\alpha = d-2$ in the rotation-invariant limit to linear order in $\mathring{k}_{(V)}^{(d)}$. Here, $\mathring{k}_{(I)}^{(d)}$ and $\mathring{k}_{(V)}^{(d)}$ are constant coefficients to quantify the degree of Lorentz violation. In the rotational invariant case, the modified dispersion relation in the SME is given in Eq.~(5) of~\cite{Kostelecky:2016kfm}. The mapping between Eq.~\eqref{eq:gen-disp-rel} and other modified gravity models is given in Table~\ref{table:mod-disp-mappings}. Niu, et al.~\cite{Niu:2022yhr} constrained anisotropy, birefringence and dispersion in gravitational-wave propagation within the context of the SME with GWTC-3.
Gong, et al.~\cite{Gong:2023ffb} derived similar constraints on non-birefringent dispersions with GWTC-3.

\begin{table}[htbp]
\caption[]{Mapping between different modified gravity models and the modified dispersion relation of Eq.~\eqref{eq:gen-disp-rel}. In double special relativity, $\eta_{\mathrm dsrt}$ characterizes an observer-independent length scale. In extra-dimensional theories, $\alpha_{\mathrm edt}$ represents the square of the Planck length. In Ho\v{r}ava-Lifshitz gravity, $\kappa_{\rm hl}$ is related to the bare gravitational constant, while $\mu_{\rm hl}$ is related to the deformation in the ``detailed balance'' condition. In multifractional spacetimes, $E_*$ is the characteristic length scale above which spacetime is discrete.

}
\label{table:mod-disp-mappings} 
\centering
{\small
\begin{tabular}%
{p{9.25cm}
>{\Centering}p{3.5cm}
>{\Centering}p{1cm}}
\toprule
Theory  & $A$ & $\alpha_{\mathrm{LV}}$ \\ 
\midrule
Double Special Relativity~\cite{AmelinoCamelia:2000ge,Magueijo:2001cr,AmelinoCamelia:2002wr,AmelinoCamelia:2010pd} & 
$\eta_{\mathrm dsrt}$ &  $3$ \\
Extra-Dimensional Theories~\cite{Sefiedgar:2010we} &
$- \alpha_{\mathrm edt}$ & $4$ \\
Ho\v{r}ava-Lifshitz Gravity~\cite{Horava:2008ih,Horava:2009uw,Vacaru:2010rd,Blas:2011zd} & 
$\kappa^{4}_{\rm hl}  \mu^{2}_{\rm hl}/16$ & $4$ \\
Massive Graviton~\cite{Will:1997bb,Rubakov:2008nh,Hinterbichler:2011tt,deRham:2014zqa} & 
$m_{g}^{2} c^4$ & $0$ \\
Multifractional Spacetime Theory (timelike)~\cite{Calcagni:2009kc,Calcagni:2011kn,Calcagni:2011sz,Calcagni:2016zqv}  & 
$2 E_*^{2-\alpha}/(3-\alpha)$  & $2-3$ \\ 
Multifractional Spacetime Theory (spacelike)~\cite{Calcagni:2009kc,Calcagni:2011kn,Calcagni:2011sz,Calcagni:2016zqv}  & 
$- 2 \cdot 3^{1-\alpha/2} E_*^{2-\alpha}/(3-\alpha)$ & $2-3$ \\
Even SME~\cite{Kostelecky:2016kfm} &  $- 2 \mathring{k}_{(I)}^{(d)}$ & $d-2$ \\
Odd SME~\cite{Kostelecky:2016kfm} &  $\pm 2 \mathring{k}_{(V)}^{(d)}$ & $d-2$ \\
\bottomrule
\end{tabular}}
\end{table}

Such a modification to the propagation of gravitational waves introduces a generalized time delay between subsequent gravitons of the form~\cite{Mirshekari:2011yq}
\begin{equation}
\Delta t_{a} = (1 + z) \left[\Delta t_{e} + \frac{D_{\alpha_{\mathrm{LV}}}}{2 \lambda_{a}^{2-\alpha_{\mathrm{LV}}}} \left(\frac{1}{f_{e}^{2-\alpha_{\mathrm{LV}}}} - \frac{1}{f_{e}'{}^{2 - \alpha_{\mathrm{LV}}}} \right) \right]\,,
\end{equation}
where we have defined $\lambda_{A} \equiv h_{p} A^{1/(\alpha_{\mathrm{LV}}-2)}$, with $h_{p}$ Planck's constant, and the generalized distance measure~\cite{Mirshekari:2011yq}
\begin{equation}
D_{\alpha_{\mathrm{LV}}} = \frac{(1 + z)^{1-\alpha_{\mathrm{LV}}}}{H_{0}} \int_{0}^{z} \frac{(1 + z')^{\alpha_{\mathrm{LV}}-2}}{\left[\Omega_{M} (1 + z')^{3} + \Omega_{\Lambda}\right]^{1/2}} dz'\,.  
\end{equation}
Such a modification then leads to the following correction to the Fourier transform of the response function~\cite{Mirshekari:2011yq}
\begin{equation}
\tilde{h}_{\mathrm{LV}} = \tilde{h}_{\mathrm{GR}} e^{i \beta_{\mathrm{LV}} u^{b_{\mathrm{LV}}}}\,,
\label{eq:response-LV}
\end{equation}
where $\tilde{h}_{\mathrm{GR}}$ is the Fourier transform of the response function in GR, and we have defined~\cite{Mirshekari:2011yq}
\begin{equation}
\beta_{\mathrm{LV}}^{\alpha_{\mathrm{LV}}\neq1} = - \frac{\pi^{2-\alpha_{\mathrm{LV}}}}{1-\alpha_{\mathrm{LV}}} \frac{D_{\alpha_{\mathrm{LV}}}}{\lambda_{A}^{2-\alpha_{\mathrm{LV}}}} \frac{{\cal{M}}^{1-\alpha_{\mathrm{LV}}}}{(1 + z)^{1-\alpha_{\mathrm{LV}}}}\,,
\qquad
b_{\mathrm{LV}}^{\alpha_{\mathrm{LV}}\neq1} = 3(\alpha_{\mathrm{LV}}-1)\,.
\end{equation}
The case $\alpha_{\mathrm{LV}}=1$ is special leading to the Fourier phase correction~\cite{Mirshekari:2011yq}
\begin{equation}
\delta \Psi_{\alpha_{\mathrm{LV}}=1} = \frac{3\pi D_{1}}{\lambda_{A}} \ln{u}\,.
\end{equation}
The reason for this is that when $\alpha_{\mathrm{LV}}=1$ the Fourier phase is proportional to the integral of $1/f$, which then leads to a natural logarithm. 

The constraints one can place on Lorentz-violating gravity depend on the particular value of $\alpha_{\mathrm{LV}}$. Of course, when $\alpha_{\mathrm{LV}} = 0$, one recovers the standard massive graviton result with the mapping $\lambda_{g}^{-2} \to \lambda_{g}^{-2} + \lambda_{A}^{-2}$. When $\alpha_{\mathrm{LV}} = 2$, the dispersion relation is identical to that in Eq.~\eqref{eq:vg-standard}, but with a redefinition of the speed of light, and should thus be unobservable. Indeed, in this limit the correction to the Fourier phase in Eq.~\eqref{eq:response-LV} becomes linear in frequency, and this is 100\% degenerate with the time of coalescence parameter in the standard GR Fourier phase. Finally, relative to the standard GR terms that arise in the post-Newtonian expansion of the Fourier phase, the new corrections are of $(1 + 3\alpha_{\mathrm{LV}}/2)$ post-Newtonian order. Then, if LIGO gravitational-wave observations were incapable of discerning between a 4 post-Newtonian and a 5 post-Newtonian waveform, then such observations would not be able to see the modified dispersion effect if $\alpha_{\mathrm{LV}} > 2$. Mirshekari, et al.~\cite{Mirshekari:2011yq} confirmed this expectation with a Fisher analysis of non-spinning, comparable-mass quasi-circular inspirals. They found that for $\alpha_{\mathrm{LV}} = 3$, one can place very weak bounds on $\lambda_{A}$, namely $A <10^{-7} \mathrm{\ eV}^{-1}$ with a LIGO observation of a $(1.4,1.4)\,M_{\odot}$ neutron star inspiral, $A <  0.2 \mathrm{\ eV}^{-1}$ with an enhanced-LISA or NGO observation of a $(10^{5},10^{5})\,M_{\odot}$ black-hole inspiral, assuming a SNR of 10 and 100 respectively. A word of caution is due here, though, as these analyses neglect any Lorentz-violating correction to the generation of gravitational waves, including the excitation of additional polarization modes. One would expect that the inclusion of such effects would only strengthen the bounds one could place on Lorentz-violating theories, but this must be done on a theory by theory basis. Bounds on the modified dispersion relation parameter $A$ have been derived with GW150914 and GW151226 by Yunes, et al.~\cite{Yunes:2016jcc} and with catalogs of binary black hole merger events by the LIGO/Virgo Collaboration~\cite{LIGOScientific:2019fpa,LIGOScientific:2020tif,LIGOScientific:2021sio}. Regarding the Standard Model Extension, one can consider not only isotropic dispersion, but also anisotropic and birefringent ones. Kostelecky and Mewes~\cite{Kostelecky:2016kfm} derived the first bounds on gravitational Lorentz violation within the Standard Model Extension. This analysis was later improved by~\cite{Wang:2021ctl,Niu:2022yhr,Haegel:2022ymk} for gravitational-wave events in GWTC-1, 2 and 3. Gong, et al.~\cite{Gong:2021jgg} derived bounds on Lorentz (and parity) violation effects in the Standard Model Extension due to fifth and sixth spatial derivatives in the action, motivated by Ho\v{r}ava-Lifshitz gravity, using gravitational-wave events from GWTC-1 and 2. 

The modified dispersion relation in Eq.~\eqref{eq:gen-disp-rel} can be further generalized. For example, a generic evolution equation for a single tensor perturbation under a cosmological background spacetime is given by~\cite{Saltas:2014dha,Nishizawa:2017nef}
\begin{equation}
\label{eq:h_evol_eq}
h_{ij}''+(2+\nu) \mathcal{H} h_{ij}' + (c_T^2 k^2 + a^2 m_g^2) h_{ij} = a^2 \gamma_{ij}\,,
\end{equation}
where $a$ is the scale factor, a prime represents the conformal time derivative, $\nu$ is the Planck mass run rate, and $c_T$ is the gravitational-wave propagation speed. When $c_T$ is a function of $E$, the group velocity of the graviton $v_g$ can be mapped to Eq.~\eqref{eq:gen-disp-rel}, after expanding the former about the GR case. Equation~\eqref{eq:h_evol_eq} can be mapped to various example theories, including theories with extra dimension, Horndeski theory, f(R) gravity and Einstein-\AE ther theory. The parameter $\nu$ modifies the Hubble friction rate and changes the amplitude of gravitational waves. Thus, this parameter effectively modifies the luminosity distance measured by gravitational waves $d_L^\mathrm{(GW)}$ relative to that measured by electromagnetic observations $d_L^\mathrm{(EM)}$. For example, a generic modification to  $d_L^\mathrm{(GW)}$ can be parameterized as~\cite{Belgacem:2018lbp}
\begin{equation}
\frac{d_L^\mathrm{(GW)}}{d_L^\mathrm{(EM)}} = \Xi_0 + \frac{1-\Xi_0}{(1+z)^n}\,,
\end{equation}
where $z$ is the redshift, $\Xi_0$ corresponds to the ratio at $z\to\infty$, and $n$ shows the redshift dependence of the ratio. GR is recovered when $\Xi_0 \to 1$, which is the case when $z \to 0$. This parameterization has been tested with the GW170817 event and studied with future events~\cite{Belgacem:2018lbp,Mukherjee:2020mha,Finke:2021aom,Jiang:2021mpd,Mancarella:2021ecn,Finke:2021eio,Finke:2021znb}. If there are multiple fields that couple to each other, the evolution equation in Eq.~\eqref{eq:h_evol_eq} can be further generalized. For example, when there are two tensor perturbations, $h$ and $s$ (omitting the tensor indices for simplicity), the evolution equations can be expressed as~\cite{BeltranJimenez:2019xxx,Ezquiaga:2021ler}
\begin{equation}
\left[\hat{I} \frac{\mathrm{d}^2}{\mathrm{~d} \eta^2}+\hat{\nu}(\eta) \frac{\mathrm{d}}{\mathrm{d} \eta}+\hat{C}(\eta) k^2+\hat{\Pi}(\eta) k+\hat{M}(\eta)\right]\left(\begin{array}{l}
h \\
s
\end{array}\right)=0\,.
\end{equation}
Here $\hat I$ is the identity matrix while $\hat \nu$, $\hat C$, $\hat \Pi$ and $\hat M$ are the friction, velocity, chiral and mass mixing matrices respectively. This mixing induces echoes, distortions, oscillations and birefringence in gravitational waves. For example, the gravitational wave oscillation in bigravity, described in Sec.~\ref{sec:GW_bigravity}, corresponds to having non-vanishing off-diagonal components in $\hat M$.

\subsubsection{Variable \textit{G} theories and large extra dimensions}
\label{sec:generic-tests-G-ED}

The lack of a particular Lagrangian associated with variable $G$ theories, excluding scalar-tensor theories and theories with large extra dimensions, makes it difficult to ascertain whether black-hole or neutron-star binaries exist in such theories. Whether this is so will depend on the particular variable $G$ model considered. In spite of this, if such binaries do exist, the gravitational waves emitted by such systems will carry some generic modifications relative to the GR expectation. 

Most current tests of the variability of Newton's gravitational constant rely on electromagnetic observations of massive bodies, such as neutron stars. As discussed in Section~\ref{sec:Variable-G}, scalar-tensor theories can be interpreted as variable-$G$ theories, where the variability of $G$ is really a scalar-field-induced variation in the coupling between gravity and matter. However, Newton's constant serves the more fundamental role of defining the relationship between geometry or length and energy, and such a relationship is not altered in most scalar-tensor theories, unless the scalar fields are allowed to vary on a cosmological scale (background, homogeneous scalar solution). 

For this reason, one might wish to consider a possible temporal variation of Newton's constant in pure vacuum spacetimes, such as in black-hole--binary inspirals. Such temporal variation would encode $(\dot{G}/G)$ at the time and location of the merger event. Thus, once a sufficiently large number of gravitational wave events has been observed and found consistent with GR, one could reconstruct a constraint map that bounds $(\dot{G}/G)$ along our past light cone (as a function of redshift and sky position). Since our past-light cone with gravitational waves will extend to roughly redshift $10$ with LISA (limited by the existence of merger events at such high redshifts), such a constraint map will be more complete than what one can achieve with current tests at redshifts of almost zero. Big Bang nucleosynthesis constraints also allow us to bound a linear drift in $(\dot{G}/G)$  from $z \sim 10^{3}$ to zero, but these become degenerate with limits on the number of relativistic species. Moreover, these bounds exploit the huge lever-arm provided by integrating over cosmic time, but they are insensitive to local, oscillatory variations of $G$ with periods much less than the cosmic observation time. Thus, gravitational-wave constraint maps would test one of the pillars of GR: local position invariance. This principle (encoded in the equivalence principle) states that the laws of physics (and thus the fundamental constants of nature) are the same everywhere in the universe. 

Let us then promote $G$ to a function of time of the form~\cite{Yunes:2009bv}
\begin{equation}
\label{eq:Gdot-parametrization}
G(t,x,y,z) \approx G_{\mathrm{c}} + \dot{G}_{\mathrm{c}} \left(t_{c} - t\right)\,, 
\end{equation}
where $G_{\mathrm{c}} = G(t_{c},x_{c},y_{c},z_{c})$ and $\dot{G}_{\mathrm{c}} = (\partial G/\partial t)(t_{c},x_{c},y_{c},z_{c})$ are constants, and the sub-index $c$ means that these quantities are evaluated at coalescence. Clearly, this is a Taylor expansion to first order in time and position about the coalescence event $(t_{c},x^{i}_{c})$, which is valid provided the spatial variation of $G$ is much smaller than its temporal variation, i.e., $|\nabla^{i} G| \ll \dot{G}$, and the characteristic period of the temporal variation is longer than the observation window (at most, $T_{\mathrm{obs}} \leq 3$ years for LISA), so that $\dot{G}_{\mathrm{c}} T_{\mathrm{obs}} \ll G_{\mathrm{c}}$. Similar parameterization of $G(t)$ have been used to study deviations from Newton's second law in the solar system~\cite{Dirac:1937ti,1988NuPhB.302..645W,1989RvMP...61....1W,2003RvMP...75..403U}. Thus, one can think of this modification as the consequence of some effective theory that could represent the predictions of several different alternative theories.  

The promotion of Newton's constant to a function of time changes the rate of change of the orbital frequency, which then directly impacts the gravitational-wave phase evolution. To leading order, Yunes, et al.~\cite{Yunes:2009bv} find
\begin{equation}
\dot{F} = \dot{F}_{\mathrm{GR}} + \frac{195}{256\pi} {\cal{M}}^{-2} x^{3} \eta^{3/5} (\dot{G}_{c} {\cal{M}})\,,
\end{equation}
where $\dot{F}_{\mathrm{GR}}$ is the rate of change of the orbital frequency in GR, due to the emission of gravitational waves and $x = (2 \pi M F)^{1/3}$. Such a modification to the orbital frequency evolution leads to the following modification~\cite{Yunes:2009bv} to the Fourier transform of the response function in the stationary-phase approximation~\cite{Bender,Cutler:1994ys,Droz:1999qx,Yunes:2009yz}
\begin{equation}
\tilde{h} = \tilde{h}_{\mathrm{GR}} \left(1 + \alpha_{\dot{G}} u^{a_{\dot{G}}} \right) e^{i \beta_{\dot{G}} u^{b_{\dot{G}}}}\,,
\label{eq:h-Gdot}
\end{equation}
where we recall again that $u = (\pi {\cal{M}} f)^{1/3}$ and have defined the constant parameters~\cite{Yunes:2009bv,Tahura:2018zuq}
\begin{equation}
\alpha_{\dot{G}} = - \frac{35}{512} \frac{\dot{G}_{c}}{G_{c}} \left(G_{c} {\cal{M}}_{z}\right)\,,
\qquad 
\beta_{\dot{G}} = - \frac{275}{851968} \frac{\dot{G}_{c}}{G_{c}} \left(G_{c} {\cal{M}}_{z}\right)\,,
\qquad 
a = -8\,,
\qquad
b = - 13\,,
\end{equation}
to leading order in the post-Newtonian approximation. Tahura and Yagi~\cite{Tahura:2018zuq} generalized the above corrections by allowing the gravitational constant in the conservative and dissipative sectors to be different (which is the case for example in Jordan--Fierz--Brans--Dicke theory with a time-varying scalar field) and including sensitivities as the gravitational self-energy of a body is a function of the gravitational constant. We note that this corresponds to a correction of $-4$ post-Newtonian order in the phase, relative to the leading-order term, and that the corrections are independent of the symmetric mass ratio, scaling only with the redshifted chirp mass ${\cal{M}}_{z}$. Due to this, one expects the strongest effects to be seen in low-frequency gravitational waves, such as those one could detect with LISA or DECIGO/BBO. 

Given such corrections to the gravitational-wave response function, one can investigate the level to which a gravitational-wave observation consistent with GR would allow us to constrain $\dot{G}_{c}$. Yunes, et al.~\cite{Yunes:2016jcc} and Tahura and Yagi~\cite{Tahura:2019dgr} used the first two gravitational-wave events and derived the constraint of $\dot G_c/G_c \leq 2.2 \times 10^{4}$ yr$^{-1}$. Yunes, et al.~\cite{Yunes:2009bv} carried out a study for future detectors and found that for comparable-mass black-hole inspirals of total redshifted mass $m_{z} = 10^{6}\,M_{\odot}$ with LISA, one could constrain $(\dot{G}_{c}/G_{c}) \lesssim 10^{-9} \mathrm{\ yr}^{-1}$ or better to redshift 10, assuming an SNR of $10^{3}$ (see also Carson and Yagi~\cite{Carson:2019kkh} and Perkins, et al.~\cite{Perkins:2020tra} for similar bounds). The constraint is strengthened when one considers intermediate-mass black-hole inspirals and extreme mass-ratio inspirals, where one would be able to achieve a bound of $(\dot{G}_{c}/G_{c}) \lesssim 10^{-11} \mathrm{\ yr}^{-1}$ and $(\dot{G}_{c}/G_{c}) \lesssim 10^{-16} \mathrm{\ yr}^{-1}$ respectively~\cite{Chamberlain:2017fjl}. Although this is not as stringent as the strongest constraints from other observations (see Section~\ref{sec:Variable-G}), we recall that gravitational-wave constraints would measure local variations at the source, as opposed to local variations at zero redshift or integrated variations from the very early universe.

There are other interesting ways to measure the time variation of $G$ with gravitational waves. For example, Zhao, et al.~\cite{Zhao:2018gwk} proposed a method that combines the gravitational-wave standard siren and supernova standard candle measurements as follows. Through gravitational waves from a binary neutron star with associated electromagnetic counterparts, one can determine the luminosity distance and redshift of the source independently (as was the case for GW170817). If there is a known type Ia supernova with a similar redshift, one can identify the luminosity distance of this supernova to be roughly the same as the binary neutron star, which leads one to determine the peak luminosity of the supernova. This peak luminosity depends on the Chandrasekhar mass, and thus, it is proportional to $G^{-3/2}$, so the combination of the gravitational-wave and supernova observations allows one to measure $G$ at a certain redshift. For example, if a future third-generation gravitational-wave interferometer detects a signal at $z=0.4$ and a supernova is known at a similar redshift, the time variation in $G$ can be constrained to $\dot G/G \lesssim 3 \times 10^{-12}$ yr$^{-1}$. Another interesting method was proposed by Vijaykumar, et al.~\cite{Vijaykumar:2020nzc} and it relies on the measurement of the minimum and maximum mass of neutron stars, which depends on $G$ at a particular cosmic epoch. From GW170817, the authors derived the bound $-7 \times 10^{-9} \mathrm{yr}^{-1} < \dot G/G < 5 \times 10^{-8}\mathrm{yr}^{-1}$.  
References~\cite{An:2023rqz,Sun:2023bvy} further studied the effect of time variation in $G$ in GW propagation.

The effect of promoting Newton's constant to a function of time is degenerate with several different effects. One such effect is a temporal variability of the black hole masses, i.e., if $\dot{m} \neq 0$. Such time-variation could be induced by gravitational leakage into the bulk in certain brane-world scenarios~\cite{Johannsen:2008tm}, as explained in Section~\ref{sec:Variable-G}. For a single black hole of mass $M$, the rate of black hole evaporation is given by
\begin{equation}
\frac{dM}{dt} = -2.8 \times 10^{-7} \left(\frac{1\,M_{\odot}}{M}\right)^{2} \left(\frac{\ell}{10 \, \mu{\mathrm{m}}}\right)^{2}\,M_{\odot}\mathrm{\ yr}^{-1}\,,
\end{equation}
where $\ell$ is the size of the large extra dimension. As expected, such a modification to a black-hole--binary inspiral will lead to a correction to the Fourier transform of the response function that is identical in structure to that of Eq.~\eqref{eq:h-Gdot}, but the parameters $(\beta_{\dot{G}},b_{\dot{G}}) \to (\beta_{\mathrm{ED}},b_{\mathrm{ED}})$ with~\cite{Yagi:2011yu,Chamberlain:2017fjl}
\begin{equation}
\label{eq:beta-ED}
\beta_{\mathrm{ED}} = -1.3 \times 10^{-24} 
\left[\left(\frac{M_{\odot}}{m_{1}}\right)^{2} + \left(\frac{M_{\odot}}{m_{2}}\right)^{2}\right] 
\left(\frac{\ell}{10 \mu{\rm{m}}}\right)^{2}
\left(\frac{3 - 26 \eta + 34 \eta^{2}}{\eta^{2/5}(1 - 2 \eta)}\right)\,,
\qquad
b_{\mathrm{ED}} = -13 \,.
\end{equation}
A similar expression is found for a neutron-star/black-hole inspiral, except that the $\eta$-dependent factor in between parenthesis is corrected. 

Given a gravitational-wave detection consistent with GR, one could then, in principle, place an upper bound on $\ell$. Yunes, et al.~\cite{Yunes:2016jcc} showed that the two first gravitational wave observations by aLIGO require $\ell \lesssim 10^{9} \mu{\mathrm{m}}$, a constraint that is many orders of magnitude weaker than current table-top bounds~\cite{Adelberger:2006dh}. Chamberlain and Yunes~\cite{Chamberlain:2017fjl} predicted that this constraint would only improve by three orders of magnitude with future ground-based detectors. Yagi, et al.~\cite{Yagi:2011yu}, however, predicted that this bound could be improved to $\ell \leq 10^{3} \, \mu{\mathrm{m}}$ with a 1-year LISA observation of a $(10,10^{5})\,M_{\odot}$ binary inspiral at an SNR of 100. A similar observation with the third generation, space-based detector DECIGO/BBO should be able to beat current constraints by roughly one order of magnitude. All of these constraints could be strengthened by roughly one order of magnitude further, if one included the statistical enhancement in parameter estimation due to detection of order $10^{5}$ sources by DECIGO/BBO. Care must be taken, however, since the constraints described above weaken somewhat for more generic inspirals, due to degeneracies between $\ell$ and eccentricity and spin. Carson and Yagi~\cite{Carson:2019kkh} and Perkins, et al.~\cite{Perkins:2020tra} derived projected bounds on the black hole evaporation rate $\dot M$ and found that future space-based detectors can constrain such a rate as $\dot M \lesssim (10^{-8} - 10^{-7})M_\odot$/yr.

Another way to place a constraint on $\ell$ is to consider the effect of mass loss in the orbital dynamics~\cite{McWilliams:2009ym}. When a system loses mass, the evolution of its semi-major axis $a$ will acquire a correction of the form $\dot{a} = -(\dot{M}/M) a$, due to conservation of specific orbital angular momentum. There is then a critical semi-major axis $a_{c}$ at which this correction balances the semi-major decay rate due to gravitational wave emission. McWilliams~\cite{McWilliams:2009ym} argues that systems with $a < a_{c}$ are then gravitational-wave dominated and will thus inspiral, while systems with $a > a_{c}$ will be mass-loss dominated and will thus outspiral. If a gravitational wave arising from an inspiraling binary is detected at a given semi-major axis, then $\ell$ can be automatically constrained. Yagi, et al.~\cite{Yagi:2011yu} extended this analysis to find that such a constraint is weaker than what one could achieve via matched filtering with a waveform in the form of Eq.~\eqref{eq:h-Gdot}, using the DECIGO detector. 

There are other ways to constrain the extra dimension models with gravitational waves. For example, if there are non-compact additional spatial dimensions, gravitational waves can leak into such dimensions, while electromagnetic waves are constrained to four dimensions. Therefore, the amplitude of gravitational waves are smaller than when spacetime is four dimensional, which effectively enhances the luminosity distance compared to that measured from electromagnetic waves. Using the multimessenger observations of GW170817, Pardo, et al.~\cite{Pardo:2018ipy} placed a bound on the number of spacetime dimensions, namely $D = 3.98^{+0.07}_{-0.09}$, which is consistent with a similar analysis by the LIGO/Virgo Collaboration~\cite{LIGOScientific:2018dkp}. Hernandez~\cite{MaganaHernandez:2021zyc} carried out a similar analysis but using binary black hole mergers. Using some theoretical features on the pair instability mass gap, the authors were able to break the degeneracy between the mass and redshift of binary black holes. Using gravitational-wave events in GWTC-3, the authors placed the bound $D = 3.95^{+0.07}_{-0.09}$, which is comparable to the one from GW170817 mentioned above. Corman, et al.~\cite{Corman:2021avn} improved these analyses by taking into account the effect of redshift for cosmological sources. One can also probe extra dimensions by counting the number of gravitational-wave sources and studying the distribution of their SNRs~\cite{Calabrese:2016bnu,Garcia-Bellido:2016zmj}. Du, et al.~\cite{Du:2020rlx} studied gravitational waves in simple compactified extra dimension models and showed that such models are inconsistent with GW150914. For brane-world models,  paths for gravitational waves are considered to be different from those for electromagnetic waves, so one can constrain the models through multimessenger observations of binary neutron stars. For example, assuming the bulk spacetime is anti-de Sitter, Visnelli, et al.~\cite{Visinelli:2017bny} derived a bound on $\ell$ using GW170817, namely $\ell < 0.535$Mpc, which is much weaker than that from table-top experiments (see~\cite{Yu:2019jlb} for a short review on this topic). Extra dimensions can also be probed through quasinormal modes~\cite{Chakraborty:2017qve,Mishra:2021waw}.
Du, et al.~\cite{Du:2023liq} derived the stress-energy tensor for GWs without imposing the standard, Isaacson's short-wavelength approximation. This could be useful for deriving GWs in higher-dimensional theories, where the size of extra dimensions is typically constrained to$\mu$m level from table-top experiments, which is much shorter than the typical GW wavelength.

The $\dot{G}$ correction to the gravitational-wave phase evolution is also degenerate with cosmological acceleration. That is, if a gravitational wave is generated at high-redshift, its phase will be affected by the acceleration of the universe. To zeroth-order, the correction is a simple redshift of all physical scales. However, if one allows the redshift to be a function of time 
\begin{equation}
z \sim z_{c} + \dot{z}_{c} (t - t_{c}) \sim z_{c} + H_{0} \left[ \left(1 + z_{c}\right)^{2} - \left(1 + z_{c}\right)^{5/2} \Omega_{M}^{1/2} \right] (t - t_{c})\,,
\end{equation}
then the observed waveform at the detector becomes structurally identical to Eq.~\eqref{eq:h-Gdot} but with the parameters~\cite{Seto:2001qf,Takahashi:2004yr,Yagi:2011bt,Nishizawa:2011eq,Bonvin:2016qxr,Yunes:2009bv}
\begin{equation}
\beta_{\dot{z}} = \frac{25}{32768} \dot{z}_{c} {\cal{M}}_{z}\,,
\qquad
b_{\dot{z}} = -13\,.
\end{equation}
Using the measured values of the cosmological parameters from the WMAP analysis~\cite{Komatsu:2008hk,Dunkley:2008ie}, one finds that this effect is roughly $10^{-3}$ times smaller than that of a possible $\dot{G}$ correction at the level of the possible bounds quoted above~\cite{Yunes:2009bv}. Of course, as one begins to consider observations with space-based detectors, which allow for fairly stringent constraints on $\dot{G}$, one must then also account for possible degeneracies with $\dot{z}$.

A final possible degeneracy arises with modifications to the gravitational waves due to the presence of a third body~\cite{Yunes:2010sm,Robson:2018svj,Kuntz:2022juv,Xuan:2022qkw}, due to migration if the binary is in an accretion disk~\cite{Kocsis:2011dr,Yunes:2011ws}, and due to the interaction of a binary with a circumbinary accretion disk~\cite{Hayasaki:2012qn}. All of these effects introduce corrections to the gravitational-wave phase at negative post-Newtonian order, just like the effect of a variable gravitational constant. However, degeneracies of this type are only expected to affect a small subset of black-hole--binary observations, namely those with a third body sufficiently close to the binary, or a sufficiently massive accretion disk.

\subsubsection{Parity violation}
\label{sec:generic-tests-PV}

As discussed in Section~\ref{sec:GPV} the simplest action to model parity violation in the gravitational interaction is given in Eq.~\eqref{CS-action}. Black holes and neutron stars exist in this theory. A generic feature of this theory is that parity violation imprints onto the propagation of gravitational waves, an effect that has been dubbed \emph{amplitude birefringence}. Such birefringence is not to be confused with optical or electromagnetic birefringence, in which the gauge boson interacts with a medium and is doubly-refracted into two separate rays. In amplitude birefringence, right- (left)-circularly polarized gravitational waves are enhanced or suppressed (suppressed or enhanced) relative to the GR expectation as they propagate~\cite{Jackiw:2003pm,Lue:1998mq,Alexander:2007kv,Yunes:2008bu,Alexander:2009tp,Yunes:2010yf}.

One can understand amplitude birefringence in gravitational wave propagation as a result of the non-commutativity of the parity operator and the Hamiltonian. The Hamiltonian is the generator of time evolution, and thus, one can write~\cite{Yunes:2010yf}
\begin{equation}
\label{eq:matrix}
\left( \begin{array}{c} {h_{+,k}(t)} \\ h_{\times,k}(t) \end{array} \right)
= e^{-i f t}
\left( \begin{array}{rr} u_{c} & - iv \\ iv & u_{c} \end{array} \right)
\left( \begin{array}{c} h_{+,k}(0) \\ h_{\times,k}(0) \end{array} \right)\,,
\end{equation}
where $f$ is the gravitational-wave angular frequency, $t$ is time, and $h_{+,\times,k}$ are the gravitational wave Fourier components with wavenumber $k$.  The quantity $u_{c}$ models possible background curvature effects, with $u_{c} = 1$ for propagation on a Minkowski metric, and $u_{c}$ proportional to redshift for propagation on a Friedman--Robertson--Walker metric~\cite{Laguna:2009re}. The quantity $v$ models possible parity-violating effects, with $v = 0$ in GR. One can rewrite the above equation in terms of right and left-circular polarizations, $h_{\mathrm{R,L}} = (h_+ \mp ih_\times) /\sqrt{2}$ to find
\begin{equation}
\label{eq:matrix2}
\left( \begin{array}{c} h_{\mathrm{R},k}(t) \\ h_{\mathrm{L},k}(t) \end{array} \right)
= e^{-i f t}
\left( \begin{array}{rr} u_{c} + v & 0 \\ 0 & u_{c}-v \end{array} \right)
\left( \begin{array}{c} h_{\mathrm{R},k}(0) \\ h_{\mathrm{L},k}(0) \end{array} \right).
\end{equation}
Amplitude birefringence has the effect of modifying the eigenvalues of the diagonal propagator matrix for right- and left-polarized waves, with right modes amplified or suppressed and left modes suppressed or amplified relative to GR, depending on the sign of $v$. In addition to these parity-violating propagation effects, parity violation should also leave an imprint in the generation of gravitational waves. However, such effects need to be analyzed on a theory by theory basis. Moreover, the propagation-distance--independent nature of generation effects should make them easily distinguishable from the propagation effects we consider here. 

The degree of parity violation, $v$, can be expressed entirely in terms of the waveform observables via~\cite{Yunes:2010yf}
\begin{equation}
v = \frac{1}{2} \left(\frac{h_{\mathrm{R}}}{h_{\mathrm{R}}^{\mathrm{GR}}} - \frac{h_{\mathrm{L}}}{h_{\mathrm{L}}^{\mathrm{GR}}} \right) = \frac{i}{2} \left( \delta \phi_{\mathrm{L}} - \delta \phi_{\mathrm{R}}\right)\,,
\end{equation}
where $h_{\mathrm{R,L}}^{\mathrm{GR}}$ is the GR expectation for a right or left-polarized gravitational wave. In the last equality we have also introduced the notation $\delta \phi \equiv \phi - \phi^{\mathrm{GR}}$, where $\phi^{\mathrm{GR}}$ is the GR gravitational-wave phase and 
\begin{equation}
h_{\mathrm{R,L}} = h_{0,\mathrm{R,L}} e^{-i \left[\phi(\eta) - \kappa_{i} \chi^{i}\right]}\,,
\end{equation}
where $h_{0, \mathrm{R,L}}$ is a constant factor, $\kappa$ is the conformal wave number and $(\eta,\chi^{i})$ are conformal coordinates for propagation in a Friedmann--Robertson--Walker universe. The precise form of $v$ will depend on the particular theory under consideration. For example, in non-dynamical Chern--Simons gravity with a field $\vartheta = \vartheta(t)$, and in an expansion about $z \ll 1$, one finds\epubtkFootnote{We have here explicitly pulled out a factor of $\alpha/kappa$ to make the continuous GR limit explicit.}~\cite{Yunes:2010yf} 
\begin{equation}
v =  \frac{\alpha}{\kappa} \pi f z \left(\dot{\vartheta}_{0} - \frac{\ddot{\vartheta}_{0}}{H_{0}}\right) 
=  \frac{\alpha}{\kappa}  \pi f D \left(H_{0}\dot{\vartheta}_{0} - \ddot{\vartheta}_{0} \right)\,,
\label{eq:v-def}
\end{equation}
where $\vartheta_{0}$ is the Chern--Simons scalar field at the detector, with $\alpha$ the Chern--Simons coupling constant [see, e.g.,~Eq.~\eqref{CS-action}], $z$ is redshift, $D$ is the comoving distance and $H_{0}$ is the value of the Hubble parameter today and $f$ is the observed gravitational-wave frequency. When considering propagation on a Minkowski background, one obtains the above equation in the limit as $\dot{a} \to 0$, so the second term dominates, where $a$ is the scale factor. To leading-order in a curvature expansion, the parity-violating coefficient $v$ will always be linear in frequency, as shown in Eq.~\eqref{eq:v-def}. For more general parity violation and flat-spacetime propagation, $v$ will be proportional to $(f D) f^{n} \alpha$, where $\alpha$ is a coupling constant of the theory (or a certain derivative of a coupling field) with units of $[\mathrm{Length}]^{n}$ (in the previous case, $n =0$, so the correction was simply proportional to $f D \alpha$, where $\alpha \propto \ddot{\vartheta}$). 

How does such parity violation affect the waveform? By using Eq.~\eqref{eq:matrix2} one can easily show that the Fourier transform of the response function becomes~\cite{Alexander:2007kv,Yunes:2008bu,Yunes:2010yf,Yagi:2017zhb}
\begin{equation}
\tilde{h}_{\mathrm{PV}} = \left(F_{+} + i \; v \;F_{\times} \right) \tilde{h}_{+} + \left(F_{\times} - i \; v \; F_{+} \right) \tilde{h}_{\times}\,.
\end{equation}
Of course, one can rewrite this in terms of a real amplitude correction and a real phase correction. Expanding in $v \ll 1$ to leading order, we find~\cite{Yunes:2010yf,Yagi:2017zhb}
\begin{equation}
\tilde{h}_{\mathrm{PV}} = \tilde{h}^{\mathrm{GR}} \left(1 + v \; \delta Q_{\mathrm{PV}} \right) e^{i v \delta \psi_{\mathrm{PV}}}\,,
\label{eq:hpar-PV}
\end{equation}
where $\tilde{h}_{\mathrm{GR}}$ is the Fourier transform of the response function in GR and we have defined
\begin{align}
Q_{\mathrm{GR}} &= \sqrt{F_{+}^{2} \left(1 + \cos^{2}{\iota}\right)^{2} + 4 \cos^{2}{\iota} F_{\times}^{2}}\,,
\\
\delta Q_{\mathrm{PV}} &= \frac{2 \cos{\iota} \left(1 + \cos^{2}{\iota}\right) \left(F_{+}^{2} + F_{\times}^{2}\right)}{Q_{\mathrm{GR}}^{2}}\,,
\\
\delta \psi_{\mathrm{PV}} &= \frac{\left(1 - \cos^{2}{\iota}\right)^{2} F_{+} F_{\times}}{Q_{\mathrm{GR}}^{2}}\,.
\end{align}
We see then that amplitude birefringence modifies both the amplitude and the phase of the response function. Using the non-dynamical Chern--Simons expression for $v$ in Eq.~\eqref{eq:v-def}, we can rewrite Eq.~\eqref{eq:hpar-PV} as~\cite{Yunes:2010yf}
\begin{equation}
\tilde{h}_{\mathrm{PV}} =  \tilde{h}^{\mathrm{GR}} \left(1 + \alpha_{\mathrm{PV}} u^{a_{\mathrm{PV}}}\right) e^{i \beta_{\mathrm{PV}} u^{b_{\mathrm{PV}}}}\,,
\end{equation}
where we have defined the coefficients
\begin{align}
\label{eq:alpha-PV}
\alpha_{\mathrm{PV}} &= 
\left(\frac{D}{\cal{M}}\right) 
\left[\frac{2 \cos{\iota} \left(1 + \cos^{2}{\iota}\right) \left(F_{+}^{2} + F_{\times}^{2}\right)}{Q_{\mathrm{GR}}^{2}}\right]
\left(\frac{\alpha}{\kappa}\right)  \left(H_{0}\dot{\vartheta}_{0} - \ddot{\vartheta}_{0} \right) 
\,,
\quad
a_{\mathrm{PV}} = 3\,,
\\
\label{eq:beta-PV}
\beta_{\mathrm{PV}} &=  
\left(\frac{D}{\cal{M}}\right)
\left[ \frac{\left(1 - \cos^{2}{\iota}\right)^{2} F_{+} F_{\times}}{Q_{\mathrm{GR}}^{2}}\right]
\left(\frac{\alpha}{\kappa}\right) \left(H_{0}\dot{\vartheta}_{0} - \ddot{\vartheta}_{0} \right)^{2}
\,,
\quad
b_{\mathrm{PV}} = 3\,,
\end{align}
where we recall that $u = (\pi {\cal{M}} f)^{1/3}$. The phase correction corresponds to a term of 4 post-Newtonian order relative to the Newtonian contribution, and it scales quadratically with the Chern--Simons coupling field $\vartheta$, which is why it was left out in~\cite{Yunes:2010yf}.  The amplitude correction, on the other hand, is of 1.5 post-Newtonian order relative to the Newtonian contribution. Since both of these appear as positive-order, post-Newtonian corrections, there is a possibility of degeneracy between them and standard waveform template parameters.  

Given such a modification to the response function, one can ask whether parity violation is observable with current detectors. 
Yagi and Yang~\cite{Yagi:2017zhb} carried out a Fisher analysis on GW150914 and found that the SNR is too small and one cannot place any meaningful bound within the weak parity-violation approximation. 
Okounkova, et al.~\cite{Okounkova:2021xjv} used gravitational-wave events in GWTC-2 and placed a constraint on amplitude birefringence, in particular the opacity parameter $\bar\kappa \lesssim 0.74$Gpc$^{-1}$. Here $\bar\kappa$ is defined through
\begin{equation}
\frac{h_R}{h_L} = \frac{e^{-d_c \bar\kappa} (1+\cos\iota)^2}{e^{d_c \bar\kappa} (1-\cos\iota)^2}\,,
\end{equation}
where $d_c$ is the comoving distance to the source. The bound was derived assuming that all events in the catalog have the same distance and using a phenomenological model for parity violation. 
Ng, et al.~\cite{Ng:2023jjt} further derived the updated constraint on amplitude birefringence with GWTC-3 that is $\sim 25$ times more stringent than the constraint in~\cite{Okounkova:2021xjv}. The effect of parity violation has also been studied in~\cite{CalderonBustillo:2024akj}, where the authors measured the emission of net circular polarization across 47 binary black-hole mergers and found that the average is consistent with zero. A future prospect of how well this can be constrained is given in~\cite{Califano:2023aji}.

References~\cite{Yagi:2017zhb,Callister:2023tws} also considered the possibility of probing the parity-violating polarization mode (V-mode)~\cite{Seto:2006hf,Seto:2006dz,Seto:2007tn,Seto:2008sr,Crowder:2012ik,Smith:2016jqs} in stochastic gravitational-wave background from binary black hole mergers. 
Martinovic, et al.~\cite{Martinovic:2021hzy} proposed a method to search for parity violation in the stochastic gravitational-wave background and applied it to LIGO/Virgo O3 data.

For future prospects on probing amplitude birefringence with gravitational waves,
Alexander, et al.~\cite{Alexander:2007kv,Yunes:2008bu} argued that a gravitational wave observation with LISA would be able to constrain an integrated measure of $v$, because LISA can observe massive--black-hole mergers to cosmological distances, while amplitude birefringence accumulates with distance traveled. For such an analysis, one cannot Taylor expand $\vartheta$ about its present value, and instead, one finds that
\begin{equation}
\frac{1 + v}{1 - v} = e^{2 \pi f \zeta(z)}\,,
\end{equation}
where we have defined
\begin{align}
\zeta(z) &= \frac{\alpha H_{0}}{\kappa} \int_{0}^{z} dz \left(1 + z\right)^{5/2} \left[\frac{7}{2} \frac{d\vartheta}{dz} + \left(1 + z\right) \frac{d^{2}\vartheta}{dz^{2}} \right]\,.
\end{align}
We can solve the above equation to find
\begin{equation}
v = \frac{e^{2 \pi f \zeta(z)}  -1}{1 + e^{2 \pi f \zeta(z)}} \sim  \pi f \zeta(z)\,,
\end{equation}
where in the second equality we have linearized about $v\ll1$ and $f \zeta \ll1$. Alexander, et al.~\cite{Alexander:2007kv,Yunes:2008bu} realized that this induces a time-dependent change in the inclination angle (i.e., the apparent orientation of the binary's orbital angular momentum with respect to the observer's line-of-sight), since the latter can be defined by the ratio $h_{\mathrm{R}}/h_{\mathrm{L}}$. They then carried out a simplified Fisher analysis and found that a LISA observation of the inspiral of two massive black holes with component masses $10^{6}\,M_{\odot} (1 + z)^{-1}$ at redshift $z = 15$  would allow us to constrain the integrated dimensionless measure $\zeta < 10^{-19}$ to $1\sigma$. One might worry that such an effect would be degenerate with other standard GR processes that induce similar time-dependencies, such as spin-orbit coupling. However, this time-dependence is very different from that of the parity-violating effect, and thus, Alexander, et al.~\cite{Alexander:2007kv,Yunes:2008bu} argued that these effects would be weakly correlated. 

Another test of parity violation was proposed by Yunes, et al.~\cite{Yunes:2010yf}, who considered the coincident detection of a gravitational wave and a gamma-ray burst with the SWIFT~\cite{Gehrels:2004am} and GLAST/Fermi~\cite{Carson:2006af} gamma-ray satellites, and the ground-based LIGO~\cite{Abbott:2007kv} and Virgo~\cite{Acernese:2007zze} gravitational wave detectors. If the progenitor of the gamma-ray burst is a neutron-star/neutron-star or neutron-star/black-hole merger, the gamma-ray jet is expected to be collimated. Therefore, an electromagnetic observation of such an event implies that the binary's orbital angular momentum at merger must be pointing along the line of sight to Earth, leading to a strongly--circularly-polarized gravitational-wave signal and to maximal parity violation. If an afterglow from the gamma-ray burst observation were to provide an accurate sky location via galaxy identification, one would be able to obtain an accurate distance measurement from the gravitational wave signal alone. Moreover, since GLAST/Fermi observations of gamma-ray bursts occur at low redshift, one would also possess a purely electromagnetic measurement of the distance to the source. Amplitude birefringence would manifest as a discrepancy between these two distance measurements. Therefore, if no discrepancy is found, the error ellipse on the distance measurement would allow us to place an upper limit on any possible gravitational parity violation. Because of the nature of such a test, one is constraining generic parity violation over distances of hundreds of Mpc, along the light cone on which the gravitational waves propagate. 

The coincident gamma-ray burst/gravitational-wave test compares favorably to the pure LISA test, with the sensitivity to parity violation being about 2\,--\,3 orders of magnitude better in the former case. This is because, although the fractional error in the gravitational-wave distance measurement is much smaller for LISA than for LIGO, since it is inversely proportional to the SNR, the parity violating effect also depends on the gravitational-wave frequency, which is much larger for neutron-star inspirals than massive black-hole coalescences. Mathematically, the simplest models of gravitational parity violation will lead to a signature in the response function that is proportional to the gravitational-wave wavelength\epubtkFootnote{Even if it is not linear, the effect should scale with positive powers of $\lambda_{\mathrm{GW}}$. It is difficult to think of any parity-violating theory that would lead to an inversely proportional relation.} $\lambda_{\mathrm{GW}} \propto D f$. Although the coincident test requires small distances and low SNRs (by roughly 1\,--\,2 orders of magnitude), the frequency is also larger by a factor of 5\,--\,6 orders of magnitude for the LIGO-Virgo network. 

The coincident gamma-ray burst/gravitational-wave test also compares favorably to current solar system constraints. Using the motion of the LAGEOS satellites, Smith, et al.~\cite{Smith:2007jm} have placed the $1\sigma$ bound $\dot{\vartheta}_{0} < 2000 \mathrm{\ km}$ assuming $\ddot{\vartheta}_{0} = 0$. A similar assumption leads to a $2\sigma$ bound of $\dot{\vartheta}_{0} < 200 \mathrm{\ km}$ with a coincident gamma-ray burst/gravitational-wave observation. Moreover, the latter test also allows us to constrain the second time-derivative of the scalar field. Finally, a LISA observation would constrain the integrated history of $\vartheta$ along the past light cone on which the gravitational wave propagated.  However, these tests are not as stringent as the recently proposed test by Dyda, et al.~\cite{Dyda:2012rj}, $\dot{\vartheta}_{0} < 10^{-7} \mathrm{\ km}$, assuming the effective theory cut-off scale is less than 10~eV and obtained by demanding that the energy density in photons created by vacuum decay over the lifetime of the universe not violate observational bounds. 

The coincident test is somewhat idealistic in that there are certain astrophysical uncertainties that could hamper the degree to which we could constrain parity violation. One of the most important uncertainties relates to our knowledge of the inclination angle, as gamma-ray burst jets are not perfectly aligned with the line of sight (which was indeed the case for GW170817). If the inclination angle is not known \emph{a priori}, it will become degenerate with the distance in the waveform template, decreasing the accuracy to which the luminosity could be extracted from a pure gravitational wave observation by at least a factor of two. Even after taking such uncertainties into account, Yunes, et al.~\cite{Yunes:2010yf} found that $\dot{\vartheta}_{0}$ could be constrained much better with gravitational waves than with current solar system observations.

So far, we have focused on amplitude birefringence, though there is another possible type of birefringence: velocity birefringence. In a general parity-violating gravitational theory, circular polarizations of tensor perturbations in a flat Friedmann-Robertson-Walker spacetime satisfy the following equation~\cite{Wang:2012fi,Zhu:2013fja}:
\begin{equation}
\label{eq:evol_eq_parity}
\tilde{h}_{\mathrm{A}}^{\prime \prime}+\left(2+\nu_{\mathrm{A}}\right) \mathcal{H} \tilde{h}_{\mathrm{A}}^{\prime}+\left(1+\mu_{\mathrm{A}}\right) k^2 \tilde{h}_{\mathrm{A}}=0, \quad (\mathrm{A}=\mathrm{R}, \mathrm{L})\,.
\end{equation}
Here primes represent derivatives with respect to conformal time and $\mathcal H = a'/a$. The parameters $\nu_A$ and $\mu_A$ characterize the parity violation in the amplitude and velocity respectively. In particular, the latter can be further parameterized as~\cite{Zhao:2022pun}
\begin{equation}
\mu_A = \bar \alpha \lambda_A \left( \frac{k}{a M_\mathrm{PV}} \right)^{\bar \beta}\,,
\end{equation}
where $\lambda_R = +1$, $\lambda_L = -1$ and the parameters $\bar \alpha$ and $\bar \beta$ depend on specific theories of gravitational parity violation. Bounds on velocity birefringence with current gravitational-wave observations can be found in~\cite{Wang:2020pgu,Wang:2020cub,Wang:2021gqm,Zhao:2022pun}. For example, 
Wang, et al.~\cite{Wang:2021gqm} used the 4th-Open Gravitational-wave Catalog and derived a constraint on the energy scale of parity violation $M_\mathrm{PV} > 0.05$GeV for $\bar \beta=1$ and assuming $\bar \alpha = \mathcal{O}(1)$. This bound improves the previous one by a factor of 5. The authors also found that the two most massive events in the catalog (GW190521 and GW191109) showed some evidence for birefringence, with a false alarm rate of a few per 100 observations. However, further analysis is needed because GW190521 may not be a standard, quasi-circular binary~\cite{Romero-Shaw:2020thy,Gayathri:2020coq,Gamba:2021gap}. Lagos, et al.~\cite{Lagos:2024boe} carried out a multi-messenger tests of birefringence with GW170817 and its associated radio observations. The authors derived a new bound on amplitude birefringence, while velocity birefringence remained unconstrained. They also revealed that such velocity birefringence effects are in general difficult to constrain with dark binary mergers even with future-generation detectors.

Parity violation has also been studied in theory-agnostic, parameterized ways. Jenks, et al.~\cite{Jenks:2023pmk} developed a parameterized framework (similar to the parameterized post-Einsteinian waveform to be discussed in Sec.~\ref{subsubsection:ppE}) to study amplitude and velocity birefringence in GW propagation, which was further extended in~\cite{Daniel:2024lev} to account for the presence of the Gauss-Bonnet term in the action. Califano, et al.~\cite{Califano:2023aji} further applied this framework to study future prospects of probing a few different parity-violating theories of gravity with future GW detectors.
Yamada and Tanaka~\cite{Yamada:2020zvt} constructed a parameterized gravitational-waveform in parity-violating gravity (also similar to the parameterized post-Einsteinian waveform) by deriving gravitational waveforms from an evolution equation of the tensor perturbations similar to Eq.~\eqref{eq:evol_eq_parity}. The authors then derived bounds on the generic parity-violating parameter at different post-Newtonian orders from gravitational-wave events in GWTC-1.

\subsubsection{Parameterized post-Einsteinian framework}
\label{subsubsection:ppE}

One of the biggest disadvantages of a top-down or direct approach toward testing GR is that one must pick a particular theory from the beginning of the analysis. However, given the large number of possible modifications to Einstein's theory and the lack of a particularly compelling alternative, it is entirely possible that none of these will represent the correct gravitational theory in the extreme gravity regime. Thus, if one carries out a top-down approach, one will be forced to make the assumption that we, as physicists, know which modifications of gravity are possible and which are not~\cite{Yunes:2009ke}. The parameterized post-Einsteinian (ppE) approach is a framework developed specifically to alleviate such a bias by allowing the data to select the correct theory of nature through the systematic study of statistically significant anomalies. 

For detection purposes, one usually expects to use match filters that are consistent with GR. But if GR happened to be wrong in the extreme gravity regime, it is possible that a GR template would still extract the signal, but with the wrong parameters. That is, the best fit parameters obtained from a matched filtering analysis with GR templates will be biased by the assumption that GR is sufficiently accurate to model the entire coalescence. This \emph{fundamental bias} could lead to a highly distorted image of the gravitational-wave universe. In fact, recent work by Vallisneri and Yunes~\cite{Vallisneri:2013rc} indicates that such fundamental bias could indeed be present in observations of neutron star inspirals, if GR is not quite the right theory in the extreme gravity regime. 

One of the primary motivations for the development of the ppE scheme was to alleviate fundamental bias, and one of its most dangerous incarnations: \emph{stealth-bias}~\cite{Cornish:2011ys}. If GR is not the right theory of nature, yet all our future detections are of low SNR, we may estimate the wrong parameters from a matched-filtering analysis, yet without being able to identify that there is a non-GR anomaly in the data. Thus, stealth bias is nothing but fundamental bias hidden by our limited SNR observations. Vallisneri and Yunes~\cite{Vallisneri:2013rc} have found that such stealth-bias is indeed possible in a certain sector of parameter space, inducing errors in parameter estimation that could be larger than statistical ones, without us being able to identify the presence of a non-GR anomaly. 

\paragraph{Historical development}\mbox{}\\

\noindent
The ppE scheme was designed in close analogy with the parameterized post-Newtonian (ppN) framework, developed in the 1970s to test GR with solar system observations (see, e.g.,~\cite{lrr-2006-3} for a review). In the solar system, all direct observables depend on a single quantity, the metric, which can be obtained by a small-velocity/weak-field post-Newtonian expansion of the field equations of whatever theory one is considering. Thus, Will and Nordtvedt~\cite{Nordtvedt:1968qs,1971ApJ...163..611W,1972ApJ...177..757W,1972ApJ...177..775N,1973ApJ...185...31W} proposed the generalization of the solar system metric into a \emph{meta-metric} that could effectively interpolate between the predictions of many different alternative theories. This meta-metric depends on the product of certain Green function potentials and ppN parameters. For example, the spatial-spatial components of the meta-metric take the form
\begin{equation}
g_{ij} = \delta_{ij} \left(1 + 2 \gamma U + \dots\right),
\end{equation}
where $\delta_{ij}$ is the Kronecker delta, $U$ is the Newtonian potential and $\gamma$ is one of the ppN parameters, which acquires different values in different theories: $\gamma = 1$ in GR, $\gamma = (1 + \omega_{\mathrm{BD}}) (2 + \omega_{\mathrm{BD}})^{-1} \sim 1 - \omega_{\mathrm{BD}}^{-1} + {\cal{O}}(\omega_{\mathrm{BD}}^{-2})$ in Jordan--Fierz--Brans--Dicke theory, etc. Therefore, any solar system observable could then be written in terms of system parameters, such as the masses of the planets, and the ppN parameters. An observation consistent with GR allows for a bound on these parameters, thus simultaneously constraining a large class of modified gravity theories. 

The idea behind the ppE framework was to develop a formalism that allowed for similar generic tests but with gravitational waves instead of solar system observations. The first pre-ppE attempts were by Arun, et al.~\cite{Arun:2006yw,Mishra:2010tp}, who considered the quasi-circular inspiral of compact objects. They suggested the waveform template family 
\begin{equation}
\tilde{h}_{\mathrm{PNT}} = \tilde{h}^{\mathrm{GR}} e^{i \beta_{\mathrm{PNT}} u^{b_{\mathrm{PN}}}}\,.
\label{eq:rppE}
\end{equation}
This waveform depends on the standard system parameters that are always present in GR waveforms, plus one theory parameter $\beta_{\mathrm{PNT}}$ that is to be constrained. The quantity $b_{\mathrm{PN}}$ is a number chosen by the data analyst and is restricted to be equal to one of the post-Newtonian predictions for the phase frequency exponents, i.e., $b_{\mathrm{PN}}=(-5,-3,-2,-1,\ldots)$.

The template family in Eq.~\eqref{eq:rppE} allows for \emph{post-Newtonian tests of GR}, i.e., consistency checks of the signal with the post-Newtonian expansion. For example, let us imagine that a gravitational wave has been detected with sufficient SNR that the chirp mass and mass ratio have been measured from the Newtonian and 1 post-Newtonian terms in the waveform phase. One can then ask whether the 1.5 post-Newtonian term in the phase is consistent with these values of chirp mass and mass ratio. Put another way, each term in the phase can be thought of as a curve in $({\cal{M}},\eta)$ space. If GR is correct, all these curves should intersect inside some uncertainty box, just like when one tests GR with binary pulsar observations. From that standpoint, these tests can be thought of as null-tests of GR and one can ask: given an event, is the data consistent with the hypothesis $\beta_{\mathrm{PNT}} = 0$ for the restricted set of frequency exponents $b_{\mathrm{PN}}$?

A Fisher and a Bayesian data analysis study of how well $\beta_{\mathrm{PNT}}$ could be constrained given a certain $b_{\mathrm{PN}}$ was carried out in~\cite{Mishra:2010tp,Huwyler:2011iq,Li:2011cg}. Mishra, et al.~\cite{Mishra:2010tp} considered the quasi-circular inspiral of non-spinning compact objects and showed that aLIGO observations would allow one to constrain $\beta_{\mathrm{PNT}}$ to 6\% up to the 1.5 post-Newtonian order correction ($b_{\mathrm{PN}}=-2$). Third-generation detectors, such as ET, should allow for better constraints on all post-Newtonian coefficients to roughly 2\%. Clearly, the higher the value of $b_{\mathrm{PN}}$, the worse the bound on $\beta_{\mathrm{PNT}}$ because the signal power contained in higher frequency exponent terms decreases, i.e., the number of useful additional cycles induced by the $\beta_{\mathrm{PNT}} u^{b_{\mathrm{PN}}}$ term decreases as $b_{\mathrm{PN}}$ increases. Huwyler, et al.~\cite{Huwyler:2011iq} repeated this analysis but for LISA observations of the quasi-circular inspiral of black hole binaries with spin precession. They found that the inclusion of precessing spins forces one to introduce more parameters into the waveform, which dilutes information and weakens constraints on $\beta_{\mathrm{PNT}}$ by as much as a factor of 5. Li, et al.~\cite{Li:2011cg} carried out a Bayesian analysis of the odds-ratio between GR and these templates given a non-spinning, quasi-circular compact binary inspiral observation with aLIGO and advanced Virgo. They calculated the odds ratio for each value of $b_{\mathrm{PN}}$ listed above and then combined all of this into a single probability measure that allows one to quantify how likely the data is to be consistent with GR.  

\paragraph{The simplest ppE model}\mbox{}\\

\noindent
One of the main disadvantages of the post-Newtonian template family in Eq.~\eqref{eq:rppE} is that it is not rooted on a theoretical understanding of modified gravity theories. To alleviate this problem, Yunes and Pretorius~\cite{Yunes:2009ke} re-considered the quasi-circular inspiral of compact objects. They proposed a more general, ppE template family through generic deformations of the $\ell=2$ harmonic of the response function in Fourier space :
\begin{equation}
\tilde{h}_{\mathrm{ppE, insp, 1}}^{(\ell=2)} = \tilde{h}^{\mathrm{GR}}  \left(1 + \alpha_{\mathrm{ppE}} u^{a_{\mathrm{ppE}}}\right) e^{i \beta_{\mathrm{ppE}} u^{b_{\mathrm{ppE}}}}\,, 
\label{eq:fullppE}
\end{equation}
where now $(\alpha_{\mathrm{ppE}},a_{\mathrm{ppE}},\beta_{\mathrm{ppE}},b_{\mathrm{ppE}})$ are all free parameters to be fitted by the data, in addition to the usual system parameters. This waveform family reproduces all predictions from known modified gravity theories: when $(\alpha_{\mathrm{ppE}},\beta_{\mathrm{ppE}}) = (0,0)$, the waveform reduces exactly to GR, while for other parameters one reproduces the modified gravity predictions of Table~\ref{table:ppEpars}.

\begin{table}[htbp]
\caption[Parameters that define the deformation of the response
function in a variety of modified gravity theories.]{Parameters that
define the deformation of the response function in a variety of
modified gravity theories. The notation $\cdot$ means that a value for
this parameter is either irrelevant, as its amplitude is zero, or it has not yet been calculated.}
\label{table:ppEpars} 
\centering
{\small
\begin{tabular}%
{p{2.75cm}
>{\Centering}p{2cm}
>{\Centering}p{0.5cm}
>{\Centering}p{5cm}
>{\Centering}p{1.5cm}}
\toprule
Theory  & $\alpha_{\mathrm{ppE}}$ & $a_{\mathrm{ppE}}$ & $\beta_{\mathrm{ppE}}$ & $b_{\mathrm{ppE}}$ \\ 
\midrule
Jordan--Fierz--Brans--Dicke & $-\frac{5}{96} \frac{S^{2}}{\omega_{\mathrm{BD}}} \eta^{2/5}$ & $-2$ & $-\frac{5}{3584} \frac{S^{2}}{\omega_{\mathrm{BD}}} \eta^{2/5}$ & $-7$ \\ 
\midrule
Dissipative Einstein--dilaton--Gauss--Bonnet Gravity & $-\frac{5}{192}  \zeta_{3} \eta^{-18/5} \delta_{m}^{2} $ & $-2$ & $-\frac{5}{7168}  \zeta_{3} \eta^{-18/5} \delta_{m}^{2} $ & $-7$ \\
\midrule
Massive Graviton & $0$ & $\cdot$ & $- \frac{\pi^{2} D {\cal{M}}}{\lambda_{g}^{2} (1 + z)}$ & $-3$ \\ 
\midrule
Lorentz Violation & $0$ & $\cdot$ & $-\frac{\pi^{2 - \gamma_{\mathrm{LV}}}}{(1 - \gamma_{\mathrm{LV}})} \frac{D_{\gamma_{\mathrm{LV}}}}{\lambda_{\mathrm{LV}}^{2 - \gamma_{\mathrm{LV}}}} \frac{{\cal{M}}^{1- \gamma_{\mathrm{LV}}}}{(1  + z)^{1- \gamma_{\mathrm{LV}}}}$ & $-3 \gamma_{\mathrm{LV}} - 3$ \\  
\midrule
$G(t)$ Theory & $-\frac{35}{512} \dot{G}_c {\cal{M}}$ & $-8 $ & $-\frac{275}{851968} \dot{G}_{c} {\cal{M}}$ & $-13$ \\ 
\midrule
Extra Dimensions & $\cdot$ & $\cdot$ & 
$\beta_{\mathrm{ED}}$
& $-13$ \\
\midrule
Non-Dynamical Chern--Simons Gravity & $\alpha_{\mathrm{PV}}$ & $3$ & $\beta_{\mathrm{PV}}$ & $3$ \\ 
\midrule
Dynamical Chern--Simons Gravity & $\alpha_\mathrm{dCS}$ & $+4$ & $\beta_{\mathrm{dCS}}$ & $-1$ \\
\midrule
Non-commutative Gravity & $\alpha_\mathrm{NC}$ & $+4$ & $-\frac{75}{256} \eta^{-4 / 5}(2 \eta-1) \Lambda^2$ & $-1$ \\
\midrule
Einstein-\AE{}ther Theory  & $- \frac{1}{2\sqrt{\kappa_3}}\eta^{2/5} \epsilon_x$ & $-1$ & $- \frac{3}{224} \frac{\eta^{2/5} \epsilon_x}{\kappa_3}$ & $-7$ \\ 
 \midrule
Khronometric Gravity  & $\cdot$ & $\cdot$ & $\frac{3}{128} \dot{E}^{\mathrm{KG}}_{-1{\mathrm{PN}}} \eta^{2/5}$ & $-7$ \\ 
\bottomrule
\end{tabular}}
\end{table}

In Table~\ref{table:ppEpars}, recall the following definitions of constants: 
$S$ is the difference in the square of the sensitivities, $\omega_{\mathrm{BD}}$ is the Jordan--Fierz--Brans--Dicke coupling parameter (see Section~\ref{sec:direct-test-BD}; we have here neglected the scalar mode) and $\eta$ is the symmetric mass ratio;
$\zeta_{3}$ is the coupling parameter in Einstein--dilaton--Gauss--Bonnet theory (see Section~\ref{sec:direct-test-MQG}), where we have here included both the dissipative and the conservative corrections and $\delta_{m}$ is the normalized mass difference parameter;
$D$ is a certain distance measure, $z$ is the cosmological redshift factor and $\lambda_{g}$ is the Compton wavelength of the graviton (see Section~\ref{generic-tests:MG-LV});
$\lambda_{\mathrm{LV}}$ is a distance scale at which Lorentz-violation becomes important and $\gamma_{\mathrm{LV}}$ is the graviton momentum exponent in the deformation of the dispersion relation (see Section~\ref{generic-tests:MG-LV});
$\dot{G}_{c}$ is the value of the time derivative of Newton's constant at coalescence (see Section~\ref{sec:generic-tests-G-ED});
$\beta_{\mathrm{ED}}$ is given in Eq.~\eqref{eq:beta-ED}; 
$\beta_{\mathrm{dCS}}$ is given in Eq.~\eqref{eq:beta-dCS}; 
$(\alpha_{\mathrm{PV}},\beta_{\mathrm{PV}})$ are given in Eqs.~\eqref{eq:alpha-PV} and~\eqref{eq:beta-PV} of Section~\ref{sec:generic-tests-PV};
$\dot{E}_{0,1{\mathrm{PN}}}^{\mathrm{EA/KG}}$ is given in Eqs.~\eqref{eqn:bm1pn}--\eqref{eqn:b0pnkg} of Section~\ref{sec:EA-KG-waveform}.

Although there are only a few modified gravity theories where the leading-order post-Newtonian correction to the Fourier transform of the response function can be parameterized by post-Newtonian waveforms of Eq.~\eqref{eq:rppE}, all such predictions can be modeled with the ppE templates of Eq.~\eqref{eq:fullppE}. In fact, only massive graviton theories, certain classes of Lorentz-violating theories and dynamical Chern--Simons gravity lead to waveform corrections that can be parameterized via Eq.~\eqref{eq:rppE}. For example, the lack of amplitude corrections in Eq.~\eqref{eq:rppE} does not allow for tests of gravitational parity violation or non-dynamical Chern--Simons gravity. 

However, this does not imply that Eq.~\eqref{eq:fullppE} can parameterize all possible deformations of GR. First, Eq.~\eqref{eq:fullppE} can be understood as a single-parameter deformation away from Einstein's theory. If the correct theory of nature happens to be a deformation of GR with several parameters (e.g., several coupling constants, mass terms, potentials, etc.), then Eq.~\eqref{eq:fullppE} will only be able to parameterize the one that leads to the most useful cycles. This was recently verified by Sampson, et al.~\cite{Sampson:2013lpa}. Second, Eq.~\eqref{eq:fullppE} assumes that the modification can be represented as a power series in velocity, with possibly non-integer values. Such an assumption does not allow for possible logarithmic terms, which are known to arise due to non-linear hereditary interactions at sufficiently-high post-Newtonian order. It also does not allow for interactions that are screened, e.g., in theories with massive degrees of freedom. Nonetheless, the parameterization in Eq.~\eqref{eq:fullppE} will still be able to signal that the detection is not a pure Einstein event, at the cost of biasing their true value~\cite{Sampson:2013jpa}.

The inspiral ppE model of Eq.~\eqref{eq:fullppE} is motivated not only from examples of modified gravity predictions, but from generic modifications to the physical quantities that drive the inspiral: the 
reduced effective potential and the radiation-reaction force or the fluxes of the constants of the motion. Yunes and Pretorius~\cite{Yunes:2009ke} and Chatziioannou, et al.~\cite{Chatziioannou:2012rf} considered generic modifications of the form
\begin{align}
V_\mathrm{eff} = \left(-\frac{m}{r} +\frac{L_z^2}{2r^2} \right)\left[1 + A_{\mathrm{ppE}} \left(\frac{m}{r}\right)^{p_{\mathrm{ppE}}}\right]\,,
\qquad
\dot{E} = \dot{E}_{\mathrm{GR}} \left[1 + B_{\mathrm{ppE}} \left(\frac{m}{r}\right)^{q_{\mathrm{ppE}}}\right]\,,
\end{align}
where $L_z$ is the $z$-component of the orbital angular momentum, $\dot E_\mathrm{GR} \propto v^2 (m/r)^4$ with $v$ and $r$ being the relative velocity and orbital separation, $(p_{\mathrm{ppE}},q_{\mathrm{ppE}}) \in \mathbb{Z}$, since otherwise one would lose analytically in the limit of zero velocities for circular inspirals, and where $(A_{\mathrm{ppE}},B_{\mathrm{ppE}})$ are parameters that depend on the modified gravity theory and, in principle, could depend on dimensionless quantities like the symmetric mass ratio. Such modifications lead to the following corrections to the SPA Fourier transform of the $\ell=2$ time-domain response function for a quasi-circular binary inspiral template (to leading order in the deformations and in post-Newtonian theory) 
\begin{align}
\tilde{h} &= A \left(\pi {\cal{M}} f\right)^{-7/6} e^{-i \Psi_{\mathrm{GR}}}  \left[1 - \frac{B_{\mathrm{ppE}}}{2} \eta^{-2q_{\mathrm{ppE}}/5} \left(\pi {\cal{M}} f\right)^{2q_{\mathrm{ppE}}} 
\right. 
\nonumber \\ 
&+ \left.
 \frac{A_{\mathrm{ppE}}}{6} \left(3 + 4 p_{\mathrm{ppE}} - 2 p_{\mathrm{ppE}}^{2}\right) \eta^{-2p_{\mathrm{ppE}}/5} \left(\pi {\cal{M}} f\right)^{2p_{\mathrm{ppE}}}\right] e^{-i \delta\Psi_{\mathrm{ppE}}}\,,
\label{eq:h-new}
\\
\delta \Psi_{\mathrm{ppE}} &= \frac{5}{32} A_{\mathrm{ppE}} \frac{2 p_{\mathrm{ppE}}^{2} - 2 p_{\mathrm{ppE}} - 3}{(4-p_{\mathrm{ppE}})(5-2p_{\mathrm{ppE}})} \eta^{-2p_{\mathrm{ppE}}/5} \left(\pi {\cal{M}} f\right)^{2p_{\mathrm{ppE}}-5} 
\nonumber \\ 
&+ \frac{15}{32} \frac{B_{\mathrm{ppE}}}{(4-q_{\mathrm{ppE}})(5-2q_{\mathrm{ppE}})} \eta^{-2q_{\mathrm{ppE}}/5} \left(\pi {\cal{M}} f\right)^{2q_{\mathrm{ppE}}-5}\,.
\end{align}
Of course, usually one of these two modifications dominates over the other, depending on whether $q_{\mathrm{ppE}}>p_{\mathrm{ppE}}$ or $p_{\mathrm{ppE}}<q_{\mathrm{ppE}}$. In Jordan--Fierz--Brans--Dicke theory, for example, the radiation-reaction correction dominates as $q_{\mathrm{ppE}}<p$. If, in addition to these modifications in the generation of gravitational waves, one also allows for modifications in the propagation, one is then led to the following template family~\cite{Chatziioannou:2012rf}
\begin{align}
\tilde{h}_{\mathrm{ppE, insp, 2}}^{(\ell=2)} = {\cal{A}} \left(\pi {\cal{M}} f\right)^{-7/6} e^{-i \Psi_{\mathrm{GR}}} \left[1 + c \beta_{\mathrm{ppE}} \left(\pi {\cal{M}} f\right)^{b_{\mathrm{ppE}}/3+5/3} \right] e^{2 i \beta_{\mathrm{ppE}} u^{b_{\mathrm{ppE}}}} e^{i \kappa_{\mathrm{ppE}} u^{k_{\mathrm{ppE}}}}\,.
\label{eq:fullppE-2}
\end{align}
Here $(b_{\mathrm{ppE}},\beta_{\mathrm{ppE}})$ and $(k_{\mathrm{ppE}},\kappa_{\mathrm{ppE}})$ are ppE parameters induced by modifications to the generation and propagation of gravitational waves respectively, where still $(b_{\mathrm{ppE}},k_{\mathrm{ppE}}) \in \mathbb{Z}$, while $c$ is fully determined by the former set via
\be
\label{c-cons}
c_{\mathrm{cons}} = -\frac{8}{15} \frac{b_{\mathrm{ppE}}(3 - b_{\mathrm{ppE}}) (b_{\mathrm{ppE}}^{2} + 6 b_{\mathrm{ppE}} -1)}{b_{\mathrm{ppE}}^{2} + 8 b_{\mathrm{ppE}} + 9}\,,
\ee
if the modifications to the binding energy dominate, 
\be
\label{c-diss}
c_{\mathrm{diss}} = -\frac{16}{15} (3 - b_{\mathrm{ppE}}) b_{\mathrm{ppE}}\,,
\ee
if the modifications to the energy flux dominate, or
\be
c_{\mathrm{both}} = -\frac{16}{15} \frac{b_{\mathrm{ppE}}(3 -b_{\mathrm{ppE}}) (b_{\mathrm{ppE}}^{2}  + 7 b_{\mathrm{ppE}} + 4)}{b_{\mathrm{ppE}}^{2} + 8 b_{\mathrm{ppE}} + 9}\,.
\label{c-both}
\ee
if both corrections enter at the same post-Newtonian order. Noticing again that if only a single term in the phase correction dominates in the post-Newtonian approximation (or both will enter at the same post-Newtonian order), one can map Eq.~\eqref{eq:h-new} to Eq.~\eqref{eq:fullppE} by a suitable redefinition of constants. 

The model presented above contains modifications to the propagation of gravitational waves, which enter through frequency-dependent changes in the dispersion relation. The first generic analysis of such effects was in fact carried out by Mirshekari, Yunes and Will in~\cite{Mirshekari:2011yq}, which was then adopted in the model of Chatziioannou, et al.~\cite{Chatziioannou:2012rf}. One could, however, be more general than this and allow for propagation direction-dependent modifications to the dispersion relation. This was indeed considered by Tso, et al.~\cite{Tso:2016mvv}, who found the modified waveform does indeed take the form of Eq.~\eqref{eq:fullppE-2}, but where the ppE amplitude parameters now also depend on a unit vector that points in the direction of wave propagation. 
See~\cite{Zhu:2023wci,Romano:2023bge} other related works.

\paragraph{More complex ppE models}\mbox{}\\

\noindent
Of course, one can introduce more ppE parameters to increase the complexity of the waveform family, and thus, Eq.~\eqref{eq:fullppE} should be thought of as a minimal choice. In fact, one expects any modified theory of gravity to introduce not just a single parametric modification to the amplitude and the phase of the signal, but two new functional degrees of freedom:
\begin{equation}
\alpha_{\mathrm{ppE}}u^{a_{\mathrm{ppE}}} \to \delta A_{\mathrm{ppE}}(\lambda^{a},\theta^{a};u)\,,
\qquad
\beta_{\mathrm{ppE}} u^{b_{\mathrm{ppE}}} \to \delta \Psi_{\mathrm{ppE}}(\lambda^{a},\theta^{a};u)\,,
\label{eq:gen-ppE}
\end{equation}
where these functions will depend on the frequency $u$, as well as on system parameters $\lambda^{a}$ and theory parameters $\theta^{a}$. In a post-Newtonian expansion, one expects these functions to reduce to leading-order on the left-hand sides of Eqs.~\eqref{eq:gen-ppE}, but also to acquire post-Newtonian corrections of the form 
\begin{align}
\delta A_{\mathrm{ppE}}(\lambda^{a},\theta^{a};u) &= \alpha_{\mathrm{ppE}}(\lambda^{a},\theta^{a})u^{a_{\mathrm{ppE}}} \sum_{n} \alpha_{n,\mathrm{ppE}}(\lambda^{a},\theta^{a}) u^{n}\,,
\\
\delta \Psi_{\mathrm{ppE}}(\lambda^{a},\theta^{a};u) &= \beta_{\mathrm{ppE}}(\lambda^{a},\theta^{a}) u^{b_{\mathrm{ppE}}} \sum_{n} \beta_{n,\mathrm{ppE}}(\lambda^{a},\theta^{a}) u^{n}\,,
\end{align}
where here the structure of the series is assumed to be of the form $u^{n}$ with $u>0$.  Such a model, also suggested by Yunes and Pretorius~\cite{Yunes:2009ke}, would introduce too many new parameters that would dilute the information content of the waveform model. Sampson, et al.~\cite{Sampson:2013lpa} demonstrated that the simplest ppE model of Eq.~\eqref{eq:fullppE} suffices to signal a deviation from GR, even if the injection contains three terms in the phase. Indeed, this simple ppE model was the one used in the first aLIGO detections to test for parametric deformations of GR~\cite{TheLIGOScientific:2016src}, albeit in a restricted regime of ppE space, as Yunes, Yagi and Pretorius recently proved explicitly~\cite{Yunes:2016jcc}.

The number of parameters that can be included in the model is precisely one of the most important differences between the ppE and ppN frameworks. In ppN, it does not matter how many ppN parameters are introduced, because the observations are of very high SNR, and thus, templates are not needed to extract the signal from the noise. On the other hand, in gravitational wave astrophysics, templates are essential to make detections and do parameter estimation. Spurious parameters in these templates that are not needed to match the signal will deteriorate the accuracy to which {\emph{all}} parameters can be measured because of an Occam penalty. Thus, in gravitational wave astrophysics and data analysis one wishes to minimize the number of theory parameters when testing GR~\cite{Cornish:2011ys,Sampson:2013lpa}. One must then find a balance between the number of additional theory parameters to introduce and the amount of bias contained in the templates. 

A curious fact of the generalizations proposed above is that the frequency exponents in the amplitude and phase correction were assumed to be integers, i.e., $(a_{\mathrm{ppE}},b_{\mathrm{ppE}},n) \in \mathbb{Z}$. This must be the case if these corrections arise due to modifications that can be represented as integer powers of the momenta or velocity. We are not aware of any theory that predicts corrections proportional to fractional powers of the velocity for circular inspirals. Moreover, one can show that theories that introduce non-integer powers of the velocity into the equations of motion will lead to issues with analyticity at zero velocity and a breakdown of uniqueness of solutions~\cite{Chatziioannou:2012rf}. In spite of this, modified theories can introduce logarithmic terms, that for example enter at high post-Newtonian order in GR due to non-linear propagation effects (see, e.g., \cite{Blanchet:2006zz} and references therein). Moreover, certain modified gravity theories introduce \emph{screened} modifications that become ``active'' only above a certain frequency due to certain non-linearities. Such effects could be modeled through a Heaviside function, which is for example needed when dealing with massive Jordan--Fierz--Brans--Dicke gravity~\cite{Detweiler:1980uk,Cardoso:2011xi,Alsing:2011er,Yunes:2011aa}. However, even these non-polynomial injections would be detectable with the simplest ppE model. In essence, one finds similar results as if one were trying to fit a 3-parameter injection with the simplest 1-parameter ppE model~\cite{Sampson:2013lpa}.
One weakness of the ppE framework is that one can only test non-GR deviations that follow a post-Newtonian series representation in the inspiral. Moreover, one needs to test modifications at different post-Newtonian orders one at a time, making the analysis inefficient and time-consuming. To overcome these issues, Xie, et al.~\cite{Xie:2024ubm} recently proposed a novel
neural post-Einstein framework. Using deep-learning neural network, the new framework allows one to test many different theories simultaneously and identify the best theory in a single run, allowing for the search of deviations that are not power-series in velocity. 

Of course, one can also generalize the inspiral ppE waveform families to more general orbits, for example through the inclusion of spins aligned or counter-aligned with the orbital angular momentum. More general inspirals would still lead to waveform families of the form of Eq.~\eqref{eq:fullppE} or~\eqref{eq:fullppE-2}, but where the parameters $(\alpha_{\mathrm{ppE}},\beta_{\mathrm{ppE}})$ would now depend on the mass ratio, mass difference, and the spin parameters of the black holes. With a single detection, one cannot break the degeneracy in the ppE parameters and separately fit for its system parameter dependencies. However, given multiple detections one should be able to break such a degeneracy, at least to a certain degree~\cite{Cornish:2011ys}. Such breaking of degeneracies begins to become possible when the number of detections exceeds the number of additional parameters required to capture the physical parameter dependencies of $(\alpha_{\mathrm{ppE}},\beta_{\mathrm{ppE}})$. The ppE framework was also recently extended to include spin precession~\cite{Loutrel:2022xok} and higher harmonics~\cite{Mezzasoma:2022pjb} (see also~\cite{Mehta:2022pcn} for a related work to the latter extension). Bonilla, et al.~\cite{Bonilla:2022dyt} extended the simplest ppE waveform that includes a leading non-GR parameter in the inspiral by introducing an additional non-GR parameter in the post-inspiral phase. Maggio, et al.~\cite{Maggio:2022hre} recently constructed a parameterized waveform during plunge-merger-ringdown based on the effective-one-body formulation.

PpE waveforms can be extended to account for the merger and ringdown phases of coalescence. Yunes and Pretorius have suggested the following template family to account for this as well~\cite{Yunes:2009ke}
\begin{equation}
\tilde{h}_{\mathrm{ppE,full}}^{(\ell=2)} =
 \begin{cases}
   \tilde{h}_{\mathrm{ppE}} & f < f_{\mathrm{IM}} \,, \\
   \gamma u^{c} e^{i (\delta + \epsilon u)} & f_{\mathrm{IM}} < f < f_{\mathrm{MRD}} \,, \\
   \zeta \frac{\tau}{1 + 4 \pi^{2} \tau^{2} \kappa \left(f - f_{\mathrm{RD}}\right)^{d}} & f > f_{\mathrm{MRD}} \,,
 \end{cases}
\label{main-eq}
\end{equation}
where the subscripts IM and MRD stand for inspiral merger and merger ringdown, respectively. The merger phase ($f_{\mathrm{IM}} < f < f_{\mathrm{MRD}}$) is modeled here as an interpolating region between the inspiral and ringdown, where the merger parameters $(\gamma,\delta)$ are set by continuity and differentiability, and the ppE merger parameters $(c, \epsilon)$ should be fit for. In the ringdown phase ($f > f_{\mathrm{MRD}}$), the response function is modeled as a single-mode generalized Lorentzian, with real and imaginary dominant frequencies $f_{\mathrm{RD}}$ and $\tau$, ringdown parameter $\zeta$ also set by continuity and differentiability, and the ppE ringdown parameters $(\kappa,d)$ are to be fit for. The transition frequencies $(f_{\mathrm{IM}},f_{\mathrm{MRD}})$ can either be treated as ppE parameters or set via some physical criteria, such as at the light-ring frequency and the fundamental ringdown frequency, respectively.

Another generalization of the ppE model that includes merger and ringdown is the hybridization of the so-called IMRPhenom models~\cite{Ajith:2007qp,Ajith:2009bn,Santamaria:2010yb,Husa:2015iqa,Khan:2015jqa,Schmidt:2012rh}. The latter is given by  
\be
\label{eq:gIMR}
\tilde{h}_{\rm gIMR}(f) =
\begin{cases} 
      A_{\rm I}(f) e^{i \Phi_{\rm I}(f)} e^{i \delta\Phi_{\rm I,\rm gIMR}}  & f\leq f_{\rm Int}\,, \\
      A_{\rm Int}(f) e^{i \Phi_{\rm Int}(f)} & f_{\rm Int}\leq f\leq f_{\rm MR}\,, \\
      A_{\rm MR}(f) e^{i \Phi_{\rm MR}(f)} & f_{\rm MR}\leq f\,, 
   \end{cases}
\ee
where 
\be
\label{eq:deltaPhi-gIMR}
\delta \Phi_{\rm I,\rm gIMR} = \frac{3}{128 \eta} \sum_{i=0}^{7} \phi_{i} \; \delta \phi_{i} \; (\pi m f)^{(i-5)/3}\,.
\ee
Here $\phi_i$ are the phase coefficients in GR while $\delta \phi_i$ are their fractional non-GR corrections. $f_{\rm Int}$ and $f_{\rm MR}$ are transition frequencies from the inspiral to an intermediate phase, and from the latter to merger and ringdown. The Fourier phases above are continuous and differentiable at the transitions. This generalized ppE model is identical to that of Eq.~\eqref{eq:fullppE} in the inspiral, with modifications in the merger and ringdown phases only~\cite{Yunes:2016jcc}.

There has also been effort to generalize the ppE templates to allow for the excitation of non-GR gravitational-wave polarizations. Modifications to only the two GR polarizations map to corrections to terms in the time-domain Fourier transform that are proportional to the $\ell=2$ harmonic of the orbital phase. However, Arun suggested that if additional polarizations are present, other terms proportional to the $\ell=0$ and $\ell=1$ harmonic will also arise~\cite{Arun:2012hf}. Chatziioannou, Yunes and Cornish~\cite{Chatziioannou:2012rf} have found that the presence of such harmonics can be captured through the more complete single-detector template family
\begin{align}
\label{eq:fullppE-with-amp}
\tilde{h}_{\mathrm{ppE, insp}}^{\mathrm{all} \, \ell}(f) &= {\cal{A}} \; \left(\pi {\cal{M}} f\right)^{-7/6} e^{-i \Psi^{(2)_{\mathrm{GR}}}} \left[1 + c \; \beta_{\mathrm{ppE}} \left(\pi {\cal{M}} f\right)^{b_{\mathrm{ppE}}/3+5/3} \right] 
e^{2 i \beta_{\mathrm{ppE}} u_{2}^{b_{\mathrm{ppE}}}} e^{2 i k_{\mathrm{ppE}} u_{2}^{\kappa_{\mathrm{ppE}}}}
\nonumber \\
&+ \gamma_{\mathrm{ppE}} \; u_{1}^{-9/2} e^{-i \Psi^{(1)_{\mathrm{GR}}}}
e^{i \beta_{\mathrm{ppE}} u_{1}^{b_{\mathrm{ppE}}}} e^{2 i k_{\mathrm{ppE}} u_{1}^{\kappa_{\mathrm{ppE}}}}\,,
\\
\Psi^{(\ell)}_{\mathrm{GR}} &=-2 \pi f t_{c} + \ell \Phi_{c}^{(\ell)} + \frac{\pi}{4} - \frac{3 \ell}{256 u_{\ell}^{5}} \sum_{n=0}^{7} u_{\ell}^{n/3} \left(c_{n}^{\mathrm{PN}} + l_{n}^{\mathrm{PN}} \ln u_{\ell}\right)\,,
\end{align}
where we have defined $u_{\ell} = (2 \pi {\cal{M}} f/\ell)^{1/3}$.  The ppE theory parameters are now \sloppy $\vec{\theta} = (b_{\mathrm{ppE}}, \beta_{\mathrm{ppE}}, k_{\mathrm{ppE}}, \kappa_{\mathrm{ppE}}, \gamma_{\mathrm{ppE}}, \Phi_{c}^{(1)})$. Of course, one may ignore $(k_{\mathrm{ppE}},\kappa_{\mathrm{ppE}})$ altogether, if one wishes to ignore propagation effects. Such a parameterization recovers the predictions of Jordan--Fierz--Brans--Dicke theory for a single-detector response function~\cite{Chatziioannou:2012rf}, as well as Arun's analysis for generic dipole radiation~\cite{Arun:2012hf}. 
The above framework was recently extended by Schumacher, et al.~\cite{Schumacher:2023jxq} to account for gravitational-wave polarizations with different propagation speeds.

One might worry that the corrections introduced by the $\ell=1$ harmonic, i.e., terms proportional to $\gamma_{\mathrm{ppE}}$ in Eq.~\eqref{eq:fullppE-with-amp}, will be degenerate with post-Newtonian corrections to the amplitude of the $\ell=2$ mode (not displayed in Eq.~\eqref{eq:fullppE-with-amp}). However, this is clearly not the case, as the latter scale as $(\pi {\cal{M}} f)^{-7/6 + n/3}$ with $n$ an integer greater than 0, while the $\ell=1$ mode is proportional to $(\pi {\cal{M}} f)^{-3/2}$, which would correspond to a $(-0.5)$ post-Newtonian order correction, i.e., $n=-1$. On the other hand, the ppE amplitude corrections to the $\ell=2$ mode, i.e., terms proportional to $\beta_{\mathrm{ppE}}$ in the amplitude of Eq.~\eqref{eq:fullppE-with-amp}, can be degenerate with such post-Newtonian corrections when $b_{\mathrm{ppE}}$ is an integer greater than $-4$. 

\paragraph{Applications of the ppE formalism}\mbox{}\\

\noindent
The different models presented above answer different questions. For example, the model of Eq.~\eqref{eq:fullppE-2} contains a stronger prior (that ppE frequency exponents be integers) than that of Eq.~\eqref{eq:fullppE}, and thus, it is ideal for fitting a particular set of theoretical models. On the other hand, the model of Eq.~\eqref{eq:fullppE} with continuous ppE frequency exponents allows one to search for \emph{generic} deviations that are statistically significant, without imposing such theoretical priors. That is, if a deviation from GR is present, then Eq.~\eqref{eq:fullppE} is more likely to be able to fit it, than Eq.~\eqref{eq:fullppE-2}. If one prioritizes the introduction of the least number of new parameters, Eq.~\eqref{eq:fullppE} with $(a_{\mathrm{ppE}},b_{\mathrm{ppE}}) \in \mathbb{R}$ can still recover deviations from GR, even if the latter cannot be represented as a correction proportional to an integer power of velocity. 

Given these ppE waveforms, how should they be used in a data analysis pipeline? The main idea behind the ppE framework is to match-filter or perform Bayesian statistics with ppE enhanced template banks to allow the data to select the best-fit values of $\theta^{a}$. As discussed in~\cite{Yunes:2009ke,Cornish:2011ys} and then later in~\cite{Li:2011cg}, one might wish to first run detection searches with GR template banks, and then, once a signal has been found, do a Bayesian model selection analysis with ppE templates.  The first study to carry out such a Bayesian analysis was by Cornish, et al.~\cite{Cornish:2011ys}, who concluded that an aLIGO detection at SNR of 20 for a quasi-circular, non-spinning black-hole inspiral would allow us to constrain $\alpha_{\mathrm{ppE}}$ and $\beta_{\mathrm{ppE}}$ much better than existent constraints for sufficiently strong-field corrections, e.g., $b_{\mathrm{ppE}} > -5$. This is because for lower values of the frequency exponents, the corrections to the waveform are weak-field and better constrained with Solar System~\cite{Sampson:2013wia} and binary pulsar observations~\cite{Yunes:2010qb}. These predictions were shown to be very accurate with the first aLIGO observations~\cite{TheLIGOScientific:2016src,Yunes:2016jcc}. The large statistical study of Li, et al.~\cite{Li:2011cg} used a reduced set of ppE waveforms and investigated our ability to detect deviations of GR when considering a future catalog of aLIGO/advanced Virgo detections. The LIGO/Virgo Collaboration derived bounds on non-GR parameters in a generalized inspiral-merger-ringdown formalism with GW150914~\cite{TheLIGOScientific:2016src}, GW170817~\cite{LIGOScientific:2018dkp}, and gravitational-wave events in the GWTC catalogs~\cite{TheLIGOScientific:2016pea,LIGOScientific:2019fpa,LIGOScientific:2020tif,LIGOScientific:2021sio}, mostly on corrections entering at positive post-Newtonian orders. Yunes, et al.~\cite{Yunes:2016jcc} derived the bounds on the ppE parameters from GW150914 and GW151226 including negative post-Newtonian corrections (see Fig.~\ref{fig:beta-cons}). The bounds from gravitational-wave observations are much stronger than those from solar system or binary pulsar observations for non-GR effects entering first at positive post-Newtonian orders. Perkins, et al.~\cite{Perkins:2020tra} gave a forecast on constraining ppE parameters with a population of events for future observations, while a future forecast with multiband observations is discussed in~\cite{Carson:2019rda,Carson:2019kkh,Perkins:2020tra,Gupta:2020lxa,Datta:2020vcj}.
In general, phase corrections are more important than amplitude corrections~\cite{Tahura:2019dgr}, but the latter can be important e.g. when testing GR with astrophysical stochastic gravitational wave backgrounds~\cite{Maselli:2016ekw,Saffer:2020xsw,Chen:2024pcn}. When searching over multiple ppE parameters simultaneously, it is more efficient to apply a principal component decomposition to break degeneracies among these ppE parameters~\cite{Saleem:2021nsb,Datta:2022izc,Shoom:2021mdj}.

\begin{figure}[t]
\begin{center}
\includegraphics[width=10cm,clip=true]{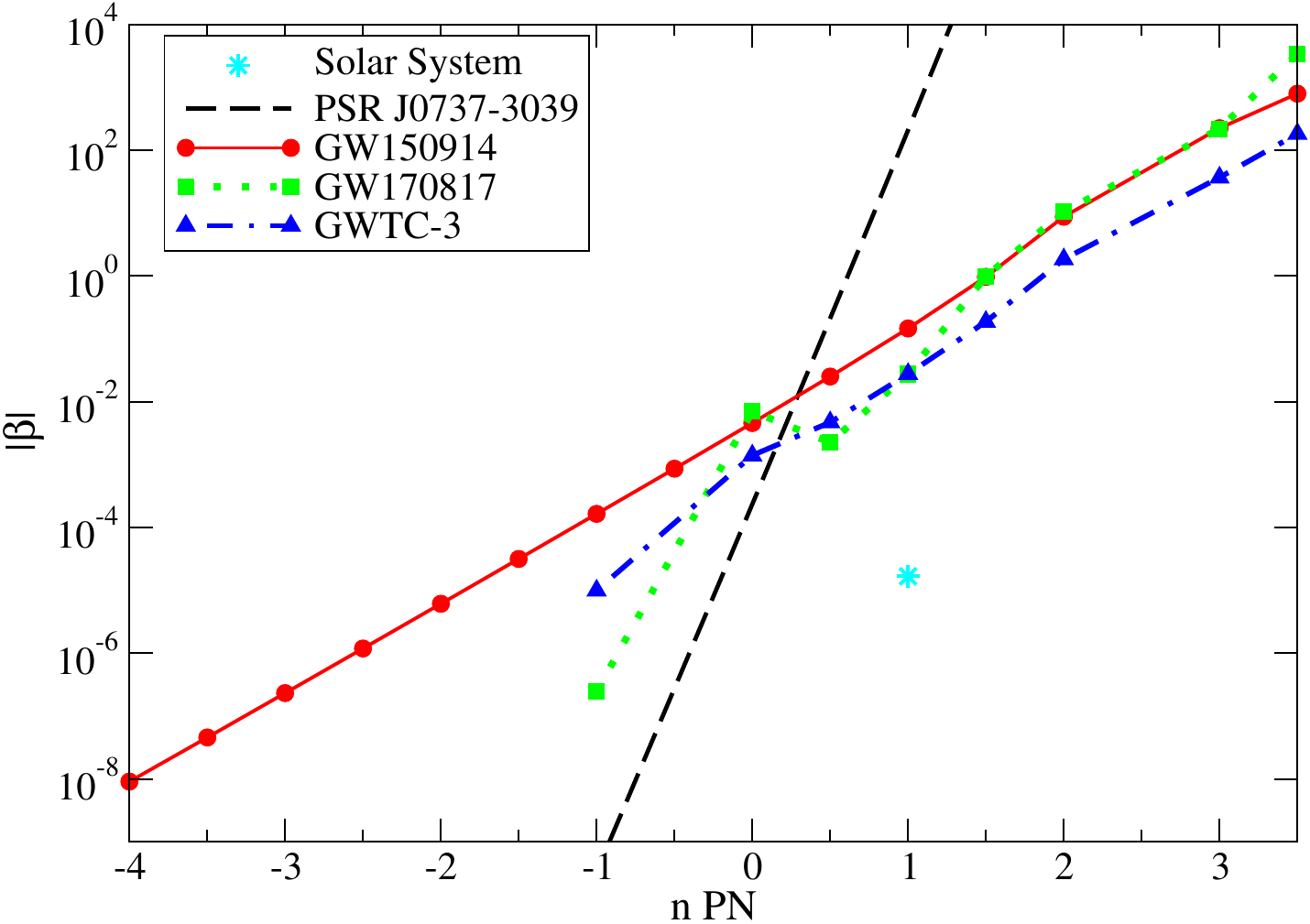} 
\caption{\label{fig:beta-cons} 
90\%-confidence constraints on the ppE parameter $|\beta|$ at $n$th post-Newtonian order. The red solid lines are those from GW150914~\cite{Yunes:2016jcc} (that is consistent with the analysis at positive post-Newtonian orders by the LIGO/Virgo Collaboration~\cite{TheLIGOScientific:2016pea,TheLIGOScientific:2016src}. The green dotted line are bounds from GW170817~\cite{LIGOScientific:2018dkp}, while the blue dotted-dashed line are those obtained by combining binary black hole merger events in the GWTC-3 catalog~\cite{LIGOScientific:2021sio}.
We also present the bounds from Solar system experiments (cyan star)~\cite{Sampson:2013wia} and binary pulsar observations (black line)~\cite{Yunes:2010qb}. 
} 
\end{center}
\end{figure}

Let us now review the parameterized tests of gravity carried out by the LIGO/Virgo Collaboration in more detail\footnote{KAGRA joined the collaboration for tests of GR with gravitational waves with the GWTC-3 data set~\cite{LIGOScientific:2021sio}.}. The first analysis was done on GW150914~\cite{TheLIGOScientific:2016src}. On top of the GR parameters, they added an extra non-GR parameter $\delta \phi_i$ in Eq.~\eqref{eq:deltaPhi-gIMR} in the inspiral waveform to search over and derived posterior distributions on this parameter assuming it entered at positive post-Newtonian orders. 
The collaboration repeated the analysis for similar non-GR parameters entering the intermediate and merger-ringdown portions of the waveform. They also carried out an analysis with multiple $\delta \phi_i$ varied simultaneously. In this case, the constraint was much weaker than when a single $\delta \phi_i$ parameter is included due to the huge degeneracies among different non-GR parameters. Later, the LIGO/Virgo Collaboration derived bounds on $\delta \phi_{-2}$ (the correction at $-1$ post-Newtonian order) with gravitational-wave events in the catalogs~\cite{LIGOScientific:2018dkp,LIGOScientific:2019fpa,LIGOScientific:2020tif,LIGOScientific:2021sio}. The most stringent bound on this parameter comes from GW170817~\cite{LIGOScientific:2018dkp}. This is because the total mass of the binary is smaller than that of binary black holes, leading to smaller relative velocity of binary constituents at a fixed frequency and making the $-1$ post-Newtonian term relatively larger. Figure~\ref{fig:beta-cons} shows the bound on the parameterized post-Einsteinian parameters $\beta$ with GW150914, GW170817, and combined binary black hole merger events in the GWTC-3 catalog\footnote{For the latter two, we assumed all the binaries are equal-mass systems to convert the bounds on $\delta \phi_i$ to $\beta$.}. The Collaboration further derived combined bounds using multiple events in the gravitational-wave catalogs~\cite{LIGOScientific:2018dkp,LIGOScientific:2019fpa,LIGOScientific:2020tif,LIGOScientific:2021sio}. They considered two ways to obtain such combined bounds. The first way was to simply multiply together the posterior distribution from each event. This is meaningful if the $\delta \phi_i$ parameters are common to all sources. The second way was a hierarchical analysis, where one assumes that the $\delta \phi_i$ parameters are drawn from a common underlying distribution and tries to constrain the latter~\cite{Zimmerman:2019wzo,Isi:2019asy}. The LIGO/Virgo Collaboration assumed such a distribution to be Gaussian and derived posterior distributions on the mean and standard deviation of the distribution~\cite{LIGOScientific:2020tif,LIGOScientific:2021sio}. A Gaussian parameterization is reasonable as it is the least informative distribution, and it has been shown to work efficiently, even when the true distribution is non-Gaussian~\cite{Isi:2019asy}. 

A built-in problem with the ppE and the ppN formalisms is that if a non-zero ppE or ppN parameter is detected, then one cannot necessarily map it back to a particular modified gravity action. On the contrary, as suggested in Table~\ref{table:ppEpars}, there can be more than one theory that predicts structurally-similar corrections to the Fourier transform of the response function. For example, both Jordan--Fierz--Brans--Dicke theory and the dissipative sector of Einstein--dilaton--Gauss--Bonnet theory predict the same type of leading-order correction to the waveform phase. However, if a given ppE parameter is measured to be non-zero, this could provide very useful information as to the type of correction that should be investigated further at the level of the action. The information that could be extracted is presented in Table~\ref{table:ppEpars-interpretation}, which is derived from knowledge of the type of corrections that lead to Table~\ref{table:ppEpars}~\cite{Yunes:2016jcc}. Moreover, if a follow-up search is done with the ppE model in Eq.~\eqref{eq:fullppE-2}, one could infer whether the correction is due to modifications to the generation or the propagation of gravitational waves. In this way, a non-zero ppE detection could inform theories of what type of GR modification is preferred by nature. 

\begin{table}[htbp]
\caption{Interpretation of non-zero ppE parameters.}
\label{table:ppEpars-interpretation}
\centering
{\small
\begin{tabular}{>{\Centering}p{1cm}>{\Centering}p{3cm}p{7cm}}
\toprule
$a_{\mathrm{ppE}}$ & $b_{\mathrm{ppE}}$ & Interpretation\\ 
\midrule
$1$ & $-1$ & Gravitational parity violation by activation of pseudo-scalar field, non-commutative geometry \\
\midrule
$-8$ & $-13$ & Anomalous acceleration, large extra dimensions, mass leakage, violation of position invariance. \\
\midrule
$\cdot$ & $-7$ & Dipole gravitational radiation by activation of scalar field, black hole hair. \\
\midrule
$\cdot$ & $(-7,-5)$ & Gravitational Lorentz violation by activation of dynamical vector field. \\
\midrule
$\cdot$ & $(-3,+3,+6,+9)$ & Massive graviton propagation, modified dispersion relations. \\
\bottomrule
\end{tabular}}
\end{table}

\paragraph{Degeneracies}\mbox{}\\

\noindent
Much care must be taken to avoid confusing a ppE theory modification with some other systematic, such as an astrophysical, a mismodeling or an instrumental effect. Instrumental effects can be easily remedied by requiring that several instruments, with presumably unrelated instrumental systematics, independently derive a posterior probability for $(\alpha_{\mathrm{ppE}},\beta_{\mathrm{ppE}})$ that peaks away from zero. Astrophysical uncertainties can also be alleviated by requiring that different astrophysical events lead to the same posteriors for ppE parameters (after breaking degeneracies with system parameters). However, astrophysically there are a limited number of scenarios that could lead to corrections in the waveforms that are large enough to interfere with these tests. For comparable-mass--ratio inspirals, this is usually not a problem as the inertia of each binary component is too large for any astrophysical environment to affect the orbital trajectory~\cite{Hayasaki:2012qn}. Magnetohydrodynamic effects could affect the merger of neutron-star binaries, but this usually does not affect the inspiral. In extreme--mass-ratio inspirals, however, the small compact object can be easily nudged away by astrophysical effects, such as the presence of an accretion disk~\cite{Yunes:2011ws,Kocsis:2011dr} or a third supermassive black hole~\cite{Yunes:2010sm}. These astrophysical effects present the interesting feature that they correct the waveform in a form similar to Eq.~\eqref{eq:fullppE} but with $b_{\mathrm{ppE}} < -5$. This is because the larger the orbital separation, the stronger the perturbations of the astrophysical environment, either because the compact object gets closer to the third body or because it leaves the inner edge of the accretion disk and the disk density increases with separation. Such effects are not likely to be present in all sources observed, as few extreme--mass-ratio inspirals are expected to be embedded in an accretion disk or sufficiently close to a third body ($\lesssim$~0.1~pc) for the latter to have an effect on the waveform. 

Perhaps the most dangerous systematic is mismodeling, which is due to the use of approximation schemes when constructing waveform templates. For example, in the inspiral one builds models using the post-Newtonian approximation, which expands and truncates the waveform at a given power of orbital velocity. These models are typically calibrated to numerical relativity simulations in the late inspiral, so that they can be joined smoothly to ringdown waves. Thus, inspiral-merger-ringdown models can have uncertainties that originate in either the truncation of the post-Newtonian approximation, numerical error or calibration error. Moreover, neutron stars are usually modeled as test-particles (with a Dirac distributional density profile), when in reality they have a finite radius, which will depend on its equation of state. Such finite-size effects enter at 5 post-Newtonian order (due to the \emph{effacement} principle~\cite{Hawking:1987en,Damour:1987pa}), but with a post-Newtonian coefficient that can be rather large~\cite{mora-will,berti-iyer-will,Flanagan:2007ix}. Ignorance of the post-Newtonian series beyond 3 post-Newtonian order can lead to systematics in the determination of physical parameters and possibly also to confusion when carrying out ppE-like tests. Moore, et al.~\cite{Moore:2021eok} estimated the amount of systematic errors in tests of GR due to waveform mismodeling and found that such errors may be evident when stacking as few as 10--30 gravitational events. This result shows how important the accurate modeling of the waveform in GR is.

In spite of these problems with the modeling of signals, recent work indicates that current waveforms are accurate enough to allow for robust tests of GR in a large region of parameter space with aLIGO observations. In particular, if one is interested in constraining GR deformations that enter below 2nd post-Newtonian order, then ppE-type models, which include deformations only in the inspiral, are sufficient to place conservative constraints with events of modest signal-to-noise ratio~\cite{Yunes:2016jcc,Perkins:2022fhr}. Better constraints could of course be obtained if one could also include deformations in the merger and ringdown phases, but doing so would only strengthen ppE-type constraints in all cases investigated thus far. This situation holds for current ground-based observations, since signal-to-noise ratios are not too large. Third generation ground-based detectors, as well as space-based detectors are expected to detect much louder signals. For such events, precise tests of GR will require an improvement in the accuracy of the GR modeling. The importance of including eccentricity in parameterized test of GR has been pointed out in~\cite{Narayan:2023vhm,Saini:2023rto}. Payne, et al.~\cite{Payne:2023kwj} proposed a framework to simultaneously infer non-GR deviations, as well as the astrophysical population, to avoid bias from prior assumptions on the latter, while Magee, et al.~\cite{Magee:2023muf} addressed the issue of selection biases on parameterized tests of GR with gravitational waves.

\subsubsection{Searching for non-tensorial gravitational-wave polarizations}

Another way to search for generic deviations from GR is to ask whether any gravitational-wave signal detected contains more than the two traditional polarizations expected in GR. Indeed, this is expected in theories with degrees of freedom in addition to the metric tensor, since these tend to source non-tensorial modes in the metric perturbation. For example, the merger of compact objects, as well as supernova explosion, tend to activate a scalar mode in the metric perturbation of scalar-tensor theories~\cite{Gerosa:2016fri}. A general approach to answer this question is through null streams, as discussed in Section~\ref{sec:Stoch-Anal}. This concept was first studied by G\"ursel and Tinto~\cite{Guersel:1989th} and later by Chatterji, et al.~\cite{Chatterji:2006nh} with the aim to separate false-alarm events from real detections. Chatziioannou, et al.~\cite{Chatziioannou:2012rf} proposed the extension of the idea of null streams to develop null tests of GR, which had also been studied with stochastic gravitational wave backgrounds in~\cite{Nishizawa:2009bf,Nishizawa:2009jh} and implemented in~\cite{Hayama:2012au} to reconstruct the independent polarization modes in time-series data of a ground-based detector network.

Given a gravitational-wave detection, one can ask whether the data is consistent with two polarizations by constructing a null stream through the combination of data streams from 3 or more detectors. As explained in Section~\ref{sec:Stoch-Anal}, such a null stream should be consistent with noise in GR, while it would present a systematic deviation from noise if the gravitational wave metric perturbation possessed more than two polarizations. Notice that such a test would not require a template; if one were parametrically constructed, such as in~\cite{Chatziioannou:2012rf}, more powerful null tests could be applied to such a null steam. In the future, as several gravitational wave detectors go online (the two aLIGO ones in the United States, advanced Virgo in Italy, LIGO-India in India, and KAGRA in Japan), gravitational-wave observations from multiple detectors could be used to construct three enhanced GR null streams, each with power in a signal null direction. 
Pang, et al.~\cite{Pang:2020pfz} developed two methods for probing additional polarizations based on null streams with electromagnetic counterparts and applied them to GW170817.
The LIGO/Virgo Collaboration has probed the existence of additional polarization modes with GW150914~\cite{TheLIGOScientific:2016src}, GW170814~\cite{LIGOScientific:2017ycc}, GW170817~\cite{LIGOScientific:2018dkp}, and gravitational-wave events in the GWTC catalogs~\cite{TheLIGOScientific:2016pea,LIGOScientific:2019fpa,LIGOScientific:2020tif,LIGOScientific:2021sio}. Similar analyses are done in~\cite{Hagihara:2019ihn,Takeda:2020tjj}, while future prospects are given in~\cite{Takeda:2018uai,Takeda:2019gwk,Philippoz:2018xrm,Hu:2023soi}, including multiband observations with space- and ground-based detectors~\cite{Philippoz:2018xrm}. Chatziioannou, et al.~\cite{Chatziioannou:2021mij} proposed a new method for constraining additional polarization modes on top of the GR tensor modes. The method is model independent (does not rely on waveform templates), phase coherent and no prior-information is needed for the sky location of a transient source. Isi, et al.~\cite{Isi:2015cva,Isi:2017equ} and Kuwahara and Asada~\cite{Kuwahara:2022dyx} proposed the use of continuous gravitational waves to probe extra polarizations. Nishizawa, et al.~\cite{Nishizawa:2009bf} studied the detectability of various polarization modes with observations of stochastic gravitational-wave backgrounds with a network of ground-based interferometers. Omiya and Seto~\cite{Omiya:2021zif,Omiya:2023rhj} studied the overlap reduction functions of even and odd-parity components of the tensor, vector and scalar polarizations for isotropic stochastic gravitational-wave background observations with ground-based detectors. If gravitational-wave signals are lensed, multiple copies of the same signal effectively increase the number of detectors in a network thanks to Earth's rotation~\cite{Goyal:2020bkm}, which can help separate different polarization modes.

Let us review the polarization tests carried out by the LIGO/Virgo Collaboration in more detail. The first test was performed on the GW150914 event~\cite{TheLIGOScientific:2016src}, though the results were inconclusive because the number of GR tensorial (plus and cross) modes are equal to the number of detectors that observed this event (LIGO Hanford and Livingston). The first informative test of polarization asked whether certain loud events supported the hypothesis that gravity contains only tensorial polarizations versus the hypotheses that it contains only scalar or only vectorial polarizations. Such a test should be understood as purely null test because no viable theory currently exists that predicts purely scalar or purely vectorial gravitational waves; such theories existed in the 1970s, but they were stringently constrained by solar system experiments over 50 years ago~\cite{Will:1993ns}. Nonetheless, such a null test was performed on the GW170814 event~\cite{LIGOScientific:2017ycc} and the collaboration found that the tensor modes are preferred over pure-scalar or pure-vector modes with Bayes factors ($B^T_S$ and $B^T_V$) of more than 1000 and 200 respectively. This analysis was improved with the GW170817 event~\cite{LIGOScientific:2018dkp}, where the Bayes factors found were $\log_{10}B^T_S = +23.09\pm 0.08$ and $\log_{10}B^T_V = +20.81\pm 0.08$. The improvement is due to (i) the sky position of the binary neutron star source relative to the detectors, and (ii) the fact that the sky position is measured precisely from the electromagnetic counterparts. The asymmetry in the measurement of $B^T_S$ and $B^T_V$ is due to the intrinsic geometries of the detector antenna patterns, making scalar modes easier to distinguish than vector modes. 

The null-stream technique mentioned earlier was applied to binary black hole merger events in the GWTC-2 catalog~\cite{LIGOScientific:2020tif}. The Collaboration found $\log_{10}B^T_S = \mathcal{O}(1)$, indicating that the pure-scalar hypothesis is disfavored over the pure-tensorial one, while the pure-vector hypothesis could not be ruled out (in fact, some binaries slightly preferred the pure-vector hypothesis over the pure-tensor one). The Bayes factors with the null stream analysis are much smaller than those obtained for the GW170814 and GW170817 events because the former relies on an incoherent stacking of signal power and does not coherently track the phase over time. The LIGO/Virgo Collaboration further applied the null stream analysis to test mixed polarization states of tensor, vector, and scalar modes with binary black hole merger events in the GWTC-3 catalog~\cite{LIGOScientific:2021sio}. The method uses an effective antenna pattern function constructed by choosing a subset $L$ of polarization modes and projecting the relevant polarization state to be tested into the chosen basis in this subspace~\cite{Wong:2021cmp}. Each polarization mode can then be described by a linear combination of the basis modes plus an additional orthogonal component. By choosing the dimension of the subset as $L=1$\footnote{A single basis is sufficient to capture the two tensorial polarizations. This is because, in the quadrupolar approximation, the plus and cross modes only differ by a relative amplitude and phase that can be marginalized over when computing the evidence.}, the collaboration found that the pure-scalar, pure-vector and vector-scalar mixed hypotheses are strongly disfavored, while any mixed hypothesis containing tensor modes cannot be ruled out. On the other hand, by choosing $L=2$, they found that mixed hypotheses can be more strongly disfavored than the pure-vector hypothesis (the pure-scalar hypothesis cannot be tested because the longitudinal and breathing modes for interferometers are linearly dependent). This is because mixed hypotheses involve a larger number of free parameters that leads to a larger Occam penalty. The bottom line is that the population of events in GTWC-3 is consistent with the pure tensorial hypothesis. 

\subsubsection{I-Love-Q tests}
\label{sec:I-Love-Q}

Neutron stars in the slow-rotation limit can be characterized by their mass and radius (to zeroth-order in spin), by their moment of inertia (to first-order in spin), and by their quadrupole moment and Love numbers (to second-order in spin). One may expect these quantities to be quite sensitive to the neutron star's internal structure, which can be parameterized by its equation of state, i.e., the relation between its internal pressure and its internal energy density. Since the equation of state cannot be well-constrained in the laboratory at super-nuclear densities, one is left with a variety of possibilities that predict different neutron-star mass-radius relations. 

Recently, however, Yagi and Yunes~\cite{Yagi:2013bca,Yagi:2013awa,Yagi:2016bkt} have demonstrated that there are relations between the moment of inertia ($I$), the Love numbers ($\lambda)$, and the quadrupole moment ($Q$), the \emph{I-Love-Q relations} that are essentially insensitive to the equation of state. Figure~\ref{fig:I-Love-Q} shows two of these relations (the normalized I-Love and Q-Love relations -- see caption) for a variety of equations of state, including APR~\cite{APR}, SLy~\cite{SLy,shibata-fitting}, Lattimer--Swesty with nuclear incompressibility of 220~MeV (LS220)~\cite{LS, ott-EOS}, Shen~\cite{Shen1,Shen2,ott-EOS}, the latter two with temperature of 0.01~MeV and an electron fraction of 30\%, and polytropic equations of state with indices of $n=0.6$, $0.8$ and $1$\epubtkFootnote{Notice that these relations are independent of the polytropic constant $K$, where $p=K \rho^{(1+1/n)}$, as shown in~\cite{Yagi:2013awa}.}. The bottom panels show the difference between the numerical results and the analytical, fitting curve. Observe that all equations of state lead to the same I-Love and Q-Love relations, with discrepancies smaller than 1\% for realistic neutron-star masses. These results have been verified by many groups: 
in~\cite{Lattimer:2012xj} to a wide range of equations of state,
in~\cite{Yagi:2013awa,Chan:2014tva,Chan:2015iou} via a post-Minkowskian analysis, 
in~\cite{Pappas:2013naa,Stein:2014wpa,Yagi:2014bxa} to higher multiple order,
in~\cite{Haskell:2013vha,Zhu:2020imp} to weakly-magnetized neutron stars,
in~\cite{Doneva:2013rha,Pappas:2013naa,Chakrabarti:2013tca,Yagi:2014bxa} to rapidly rotating neutron stars,
in~\cite{Maselli:2013mva} to dynamical tides during coalescence~\cite{Ferrari:2011as,Maselli:2012zq}, in~\cite{Bretz:2015rna} to differential rotation, in~\cite{Majumder:2015kfa} for different normalizations, in stellar oscillations~\cite{Chan:2014kua,Benitez:2020fup,Sotani:2021kiw}, in hybrid stars~\cite{Paschalidis:2017qmb,Tan:2021nat}. See Yagi and Yunes~\cite{Yagi:2016bkt} for a review on universal relations for neutron stars. The universal relations were revisited and improved in Carson, et al.~\cite{Carson:2019rjx} after GW170817.

\epubtkImage{I-Love-fit-Dup-MaxC-PRD-Q-Love-fit-Dup-MaxC-PRD.png}{%
\begin{figure}[htbp]
\centerline{
 \includegraphics[width=7.25cm,clip=true]{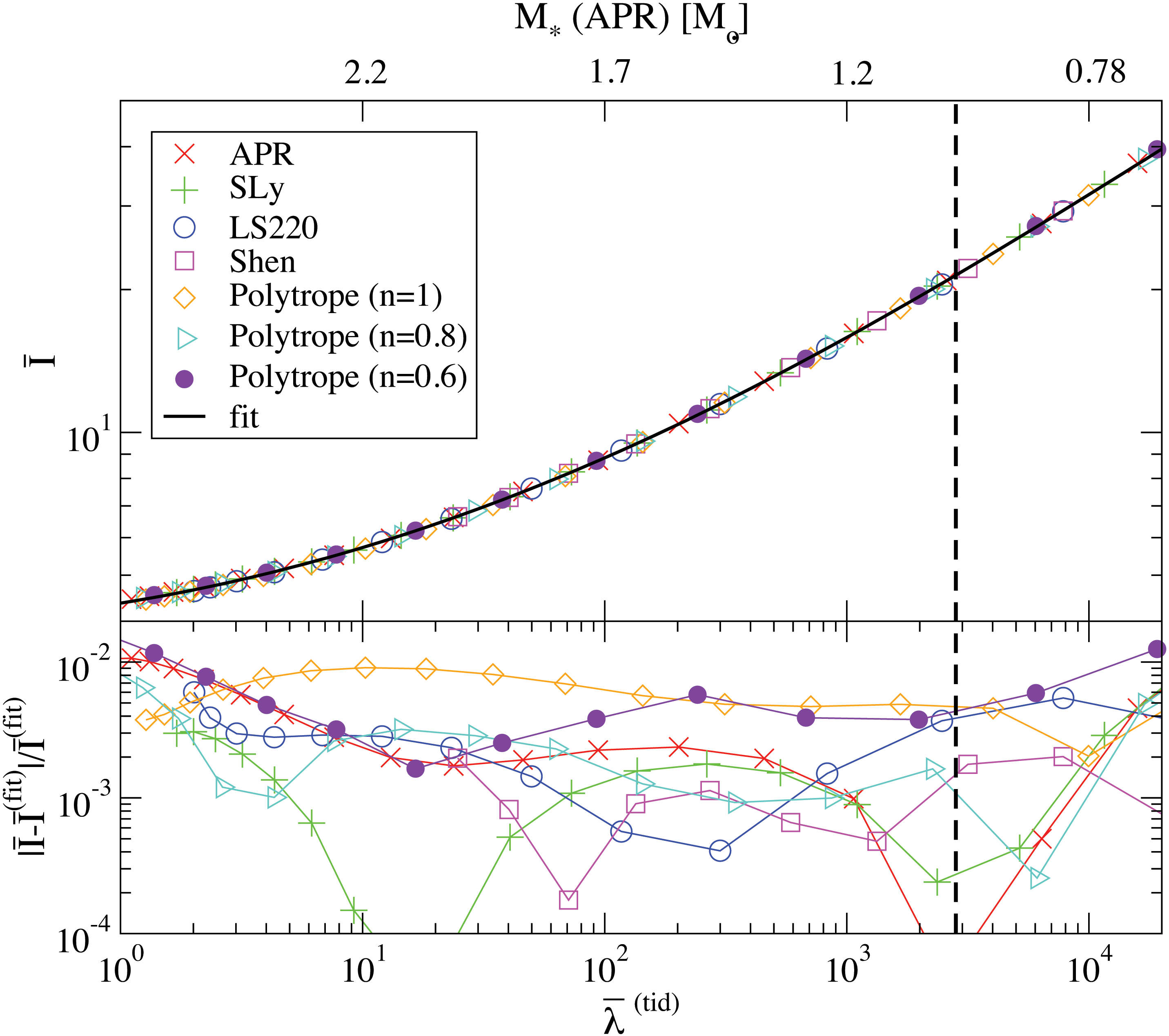}
  \includegraphics[width=7.25cm,clip=true]{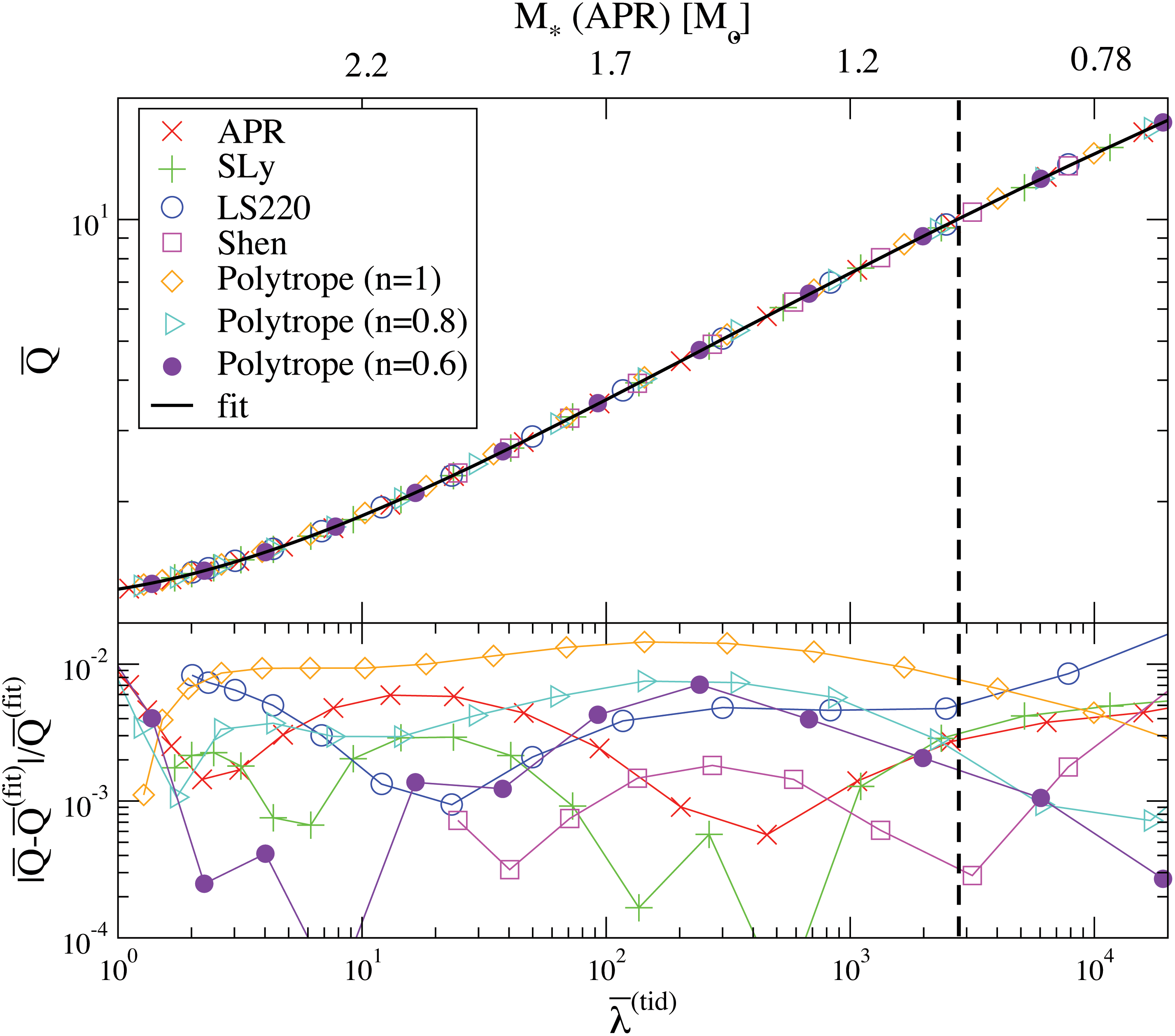}
}
\caption{\emph{Top:} Fitting curves (solid curve) and numerical
  results (points) of the universal I-Love (left) and Q-Love (right)
  relations for various equations of state, normalized as $\bar{I} =
  I/M_{\mathrm{NS}}^{3}$, $\bar{\lambda}^{(\mathrm{tid})}
  = \lambda^{(\mathrm{tid})}/M_{\mathrm{NS}}^{5}$ and $\bar{Q} =
  -Q^{(\mathrm{rot})}/[M_{\mathrm{NS}}^{3}
  (S/M_{\mathrm{NS}}^{2})^{2}]$, $M_{\mathrm{NS}}$ is the neutron-star
  mass,  $\lambda^\mathrm{(tid)}$ is the tidal Love number,
  $Q^\mathrm{(rot)}$ is the rotation-induced quadrupole moment, and
  $S$ is the magnitude of the neutron-star spin angular momentum. The
  neutron-star central density is the parameter varied along each
  curve, or equivalently the neutron-star compactness. The top axis shows the
  neutron star mass for the APR equation of state, with the vertical
  dashed line showing $M_{\mathrm{NS}} = 1\,M_{\odot}$. \emph{Bottom:}
  Relative fractional errors between the fitting curve and the
  numerical results. Observe that these relations are essentially
  independent of the equation of state, with loss of universality at
  the 1\% level. Image reproduced by permission
  from~\cite{Yagi:2013awa}, copyright by APS.}
\label{fig:I-Love-Q} 
\end{figure}}

Given the independent measurement of any two members of the I-Love-Q trio, Yagi and Yunes proposed that one could carry out a (null) model-independent and equation-of-state-independent test of GR~\cite{Yagi:2013bca,Yagi:2013awa} (see~\cite{Yagi:2016bkt,Doneva:2017jop} for reviews on tests of GR with neutron star universal relations). For example, assume that electromagnetic observations of the binary pulsar J0737--3039 have measured the moment of inertia to 10\% accuracy~\cite{lattimer-schutz,kramer-wex,Kramer:2006nb}. The slow-rotation approximation is perfectly valid for this binary pulsar, due to its relatively long spin period. Assume further that a gravitational-wave observation of a neutron-star--binary inspiral, with individual masses similar to that of the primary in J0737--3039, manages to measure the neutron star tidal Love number to 60\% accuracy~\cite{Yagi:2013bca,Yagi:2013awa}. These observations then lead to an error box in the I-Love plane, which must contain the curve in the left-panel of Figure~\ref{fig:I-Love-Q}. 

A similar test could be carried out by using data from only binary pulsar observations or only gravitational wave detections. In the case of the latter, one would have to simultaneously measure or constrain the value of the quadrupole moment and the Love number, since the moment of inertia is not measurable with gravitational wave observations. In the case of the former, one would have to extract the moment of inertia and the quadrupole moment, the latter of which will be difficult to measure. Therefore, the combination of electromagnetic and gravitational wave observations would be the ideal way to carry out such tests. 

I-Love-Q tests of GR are powerful only as long as modified gravity theories predict I-Love-Q relations that are not degenerate with the general relativistic ones. Yagi and Yunes~\cite{Yagi:2013bca,Yagi:2013awa} investigated such a relation in dynamical Chern--Simons gravity and found that such degeneracy is only present in the limit $\zeta_{\mathrm{CS}} \to 0$ (see also Gupta, et al.~\cite{Gupta:2017vsl} for a related work). That is, for any finite value of $\zeta_{\mathrm{CS}}$, the dynamical Chern--Simons I-Love-Q relation differs from that of GR, with the distance to the GR expectation increasing for larger $\zeta_{\mathrm{CS}}$. Yagi and Yunes~\cite{Yagi:2013bca,Yagi:2013awa} predicted that a test similar to the one described above could constrain dynamical Chern--Simons gravity to roughly $\xi_{\mathrm{CS}}^{1/4} < 10 M_{\mathrm{NS}} \sim 15\mathrm{\ km}$, where recall that $\xi_{\mathrm{CS}} = \alpha_{\mathrm{CS}}^{2}/(\beta \kappa)$. This has been demonstrated in Silva, et al.~\cite{Silva:2020acr}, who combined the tidal deformability measurement of GW170817 through gravitational waves and the compactness measurement through X-ray observations with NICER and converted the latter to that of moment of inertia using the universal relation between the compactness and moment of inertia assuming the GR relation holds. Silva, et al.~\cite{Silva:2020acr} also developed a parameterized I-Love relation that can capture deviations from GR in the relation in a model-independent way and found the mapping between phenomenological parameters and the coupling constant in dCS gravity. Pan, et al.~\cite{Pan:2022gwf} studied the possibility of using fast radio burst emitters (like magnetars) in binary neutron stars to measure the quadrupole moment from radio observations and tidal deformability from gravitational wave observations. The authors then pointed out that one can probe non-GR theories like dCS gravity with the universal Q-Love relation. Pan, et al.~\cite{Pan:2022gwf} also constructed a parameterized Q-Love relation similar to the parameterized I-Love one in~\cite{Silva:2020acr}.
Universal relations for neutron stars have also been studied in Einstein--dilaton--Gauss--Bonnet gravity~\cite{Kleihaus:2014lba,Kleihaus:2016dui,Yagi:2016bkt,Saffer:2021gak}, massless and massive scalar-tensor theories~\cite{Doneva:2014faa,Pani:2014jra,Doneva:2016xmf,Yagi:2016bkt,Hu:2021tyw}, $f(R)$ theories~\cite{Doneva:2015hsa,Staykov:2015cfa,Yagi:2016bkt}, Eddington-inspired Born-Infeld gravity~\cite{Sham:2013cya}, higher-dimensional theories~\cite{Chakravarti:2019aup}, Eistein-\AE ther theory~\cite{Vylet:2023pkp}, and Ho\v arava gravity~\cite{Ajith:2022uaw}. Unlike dynamical Chern--Simons gravity, however,  these theories are very well-constrained from prior Solar System and binary pulsar observations, and thus, the I-Love-Q relations are much closer to the general-relativity ones. 

The tests described above, of course, only hold provided the I-Love-Q relations are valid in general relativity and in modified gravity theories. Establishing this analytically is difficult in general, but possible when making some simplifying assumptions and using approximations. In particular, Yagi and Yunes~\cite{Yagi:2013bca,Yagi:2013awa} assumed that the neutron stars are uniformly and slowly rotating, as well as only slightly tidally deformed by their rotational velocity or companion. These assumptions have by now been relaxed and the I-Love-Q relations have been seen to hold~\cite{Lattimer:2012xj,Yagi:2013awa,Chan:2014tva,Chan:2015iou,Pappas:2013naa,Stein:2014wpa,Yagi:2014bxa,Haskell:2013vha,Doneva:2013rha,Pappas:2013naa,Chakrabarti:2013tca,Yagi:2014bxa,Maselli:2013mva}. However, the relations clearly do not hold for newly-born neutron stars, which are rapidly and differentially rotating~\cite{Martinon:2014uua}, or for magnetars, which have strong magnetic fields that contribute to the quadrupolar deformation~\cite{Haskell:2013vha}. Martinon, et al.~\cite{Martinon:2014uua} have found that the I-Love-Q deviate from universality by roughly $20\%$ in proto-neutron stars, but this non-universality effaces away within 2 seconds of evolution, as the stellar entropy relaxes and the star slowly settles down to a barotropic equilibrium (see~\cite{Lenka:2018ehb,Marques:2017zju,Torres-Forne:2019zwz,Raduta:2020fdn} for related works). Moreover,  the radio waves and gravitational waves emitted by millisecond pulsars and neutron-star inspirals are expected to have binary components with normal magnetic fields, for which the magnetic quadrupolar deformation will be small. Therefore, the universal relations should still hold in the systems one would observe to carry out I-Love-Q tests. 

\subsubsection{Consistency Tests}
\label{sec:consistency}

We now review two consistency tests of GR performed by the LIGO/Virgo Collaboration: residual tests and inspiral-merger-ringdown consistency tests. Let us first focus on the former. One can subtract the most probable binary black hole waveform in GR from the signal and test whether the resulting residual is consistent with noise. For the GW150914 event, the collaboration used the BayesWave algorithm~\cite{Cornish:2014kda,Littenberg:2014oda} to rank three different hypotheses: the data contains (i) only Gaussian noise, (ii) Gaussian noise plus uncorrelated noise transients, and (iii) Gaussian noise and an elliptically polarized gravitational-wave signal~\cite{TheLIGOScientific:2016src}. The collaboration computed the signal-to-noise Bayes factor (a measure of significance for the excess power in the data), and the signal-to-glitch Bayes factor (a measure of the coherence of the excess power between Hanford and Livingston detectors). They found that the GW150914 data prefers the first hypothesis over the second and third ones, leading to the conclusion that all the measured power is consistent with the GR prediction. The 95\% upper bound on the residual signal-to-noise ratio was found to be 7.3. This can be translated to a lower bound on the fitting factor of 0.96, which, in turn, means that GR violations in the GW150914 data, if present, are limited to less than 4\% for effects that cannot be captured by redefinition of physical parameters. 

The collaboration carried out residual tests for other events in the gravitational-wave catalogs~\cite{LIGOScientific:2019fpa,LIGOScientific:2020tif,LIGOScientific:2021sio}. For example, using GWTC-3, they compared the signal-to-noise ratios of the signals and residuals for various events and found no correlation between these two signal-to-noise ratios. This indicates that the data is consistent with the GR templates and the residual signal-to-noise ratios depend purely on the detector noise. They also estimated the $p$-values of residual signal-to-noise ratios for each event. This corresponds to the probability of obtaining a background value of the residual signal-to-noise ratio higher than that of the event. They found no significant deviation in the residual data from the expected noise distribution in the individual interferometers.

The second type of consistency test is the inspiral-merger-ringdown one~\cite{Ghosh:2016qgn,Ghosh:2017gfp}. In this consistency test, one compares inferences on the final mass and spin of the remnant black hole in a binary black hole merger obtained from the inspiral and from the post-inspiral parts of a waveform separately. This can be realized as follows. 
First, one estimates the ``inspiral'' masses and spins of the black hole binary components through a phenomenological inspiral-merger-ringdown waveform. With this in hand, one then infers the final mass/spin of the black hole remnant again, but this time through empirical relations between the initial masses/spins and the final mass/spin, which are obtained through numerical relativity simulations in GR~\cite{Healy:2016lce,Hofmann:2016yih,Jimenez-Forteza:2016oae}. One then repeats this analysis for the ``post-inspiral'' signal to find another set of the final mass and spin measurement. Suppose there is an overlap region in the posterior distributions for the final mass and spin from inspiral and post-inspiral. In that case, this indicates that the procedure described above (including that GR is correct) is consistent with the data. One can go one step further and combine such posterior distributions to find a single posterior distribution in $\Delta M_f/\bar {M}_f$ and $\Delta \chi_f/\bar {\chi}_f$. Here, $\Delta M_f$ and $\Delta \chi_f$ are the difference in the final mass and dimensionless spin of the remnant black hole estimated with the post-inspiral only and the inspiral plus numerical relativity fitting formula. The quantities $\bar M_f$ and $\bar \chi_f$ are either the best-fit value of the final mass and dimensionless spin  obtained through the best-fit inspiral-merger-ringdown template or the average of the final mass and dimensionless spin between inspiral and post-inspiral methods. Madekhar, et al.~\cite{Madekar:2024zdj} extended the original inspiral-merger-ringdown consistency tests to a \emph{meta} inspiral-merger-ringdown consistency tests that checks for the inferred final mass and spin from two independent tests of GR and checks for consistency.

We now show results obtained by applying this second consistency test to existing gravitational-wave events.
Such a result for GW150914~\cite{LIGOScientific:2019fpa,TheLIGOScientific:2016src} is shown in Fig.~\ref{fig:IMR}. Observe that the origin (corresponding to GR) is within the 90\% credible region, indicating that data is consistent with the GR assumption. The two-dimensional GR quantile value (defined as the fraction of the posterior enclosed by the isoprobability contour passing through the origin; smaller values mean better consistency with GR) is 28\%.
The LIGO/Virgo Collaboration also carried out this test for other events in the catalog~\cite{LIGOScientific:2020tif,LIGOScientific:2021sio}. The combined posterior distribution for selected events in GWTC-3 (assuming the deviation is the same for all events) is also shown in Fig.~\ref{fig:IMR}, with $\Delta M_f/\bar M_f = -0.02^{+0.07}_{-0.06}$ and $\Delta \chi_f/\bar \chi_f = -0.06^{+0.10}_{-0.07}$~\cite{LIGOScientific:2021sio}. The two-dimensional GR quantile in this case is 79.6\% (see also Zhong, et al.~\cite{Zhong:2024pwb}, who proposed multidimensional hierarchical tests of GR and applied the framework to a two-dimensional inspiral-merger-ringdown consistency tests with GW events in GWTC-3). Figure~\ref{fig:IMR} also presents results for future forecasts obtained through a Fisher analysis computed in Carson and Yagi~\cite{Carson:2019rda,Carson:2019rda,Carson:2020rea}, who assumed that a signal from a GW150914-like event is detected with Cosmic Explorer alone and with multiband observations of Cosmic Explorer and LISA. All of these works assume binaries are quasi-circular. Bhat, et al.~\cite{Bhat:2022amc} studied systematic errors on the inspiral-merger-ringdown consistency tests due to ignoring the effect of eccentricities. They found that the eccentricity $e$ at 10Hz can bring a significant bias in the inferred final mass and spin when $e \gtrsim 0.1$ for aLIGO and $e \gtrsim 0.015$ for Cosmic Explorer. The effect of eccentricity on consistency tests has been studied in~\cite{Shaikh:2024wyn}.

\begin{figure}[t]
\begin{center}
\includegraphics[width=10cm,clip=true]{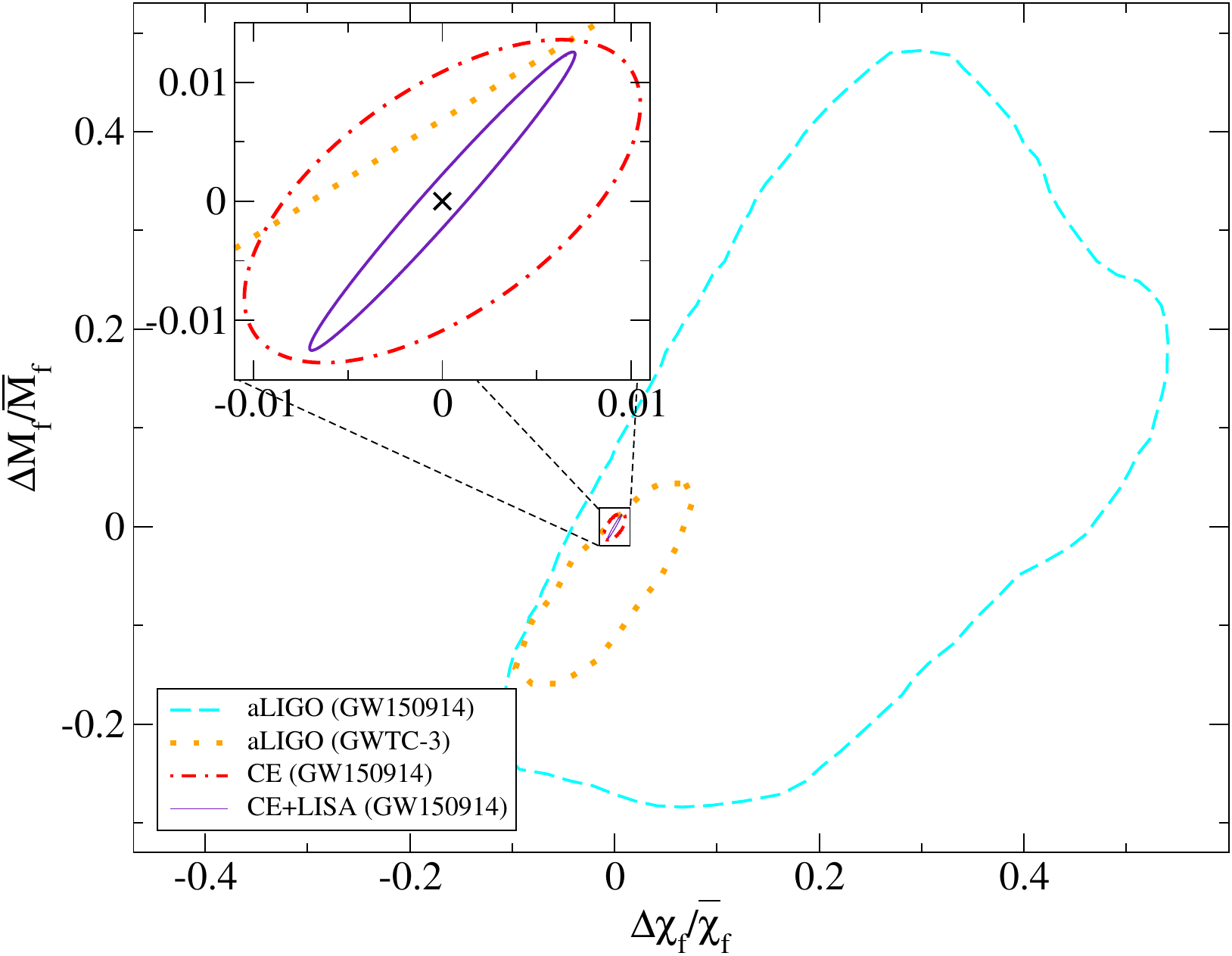} 
\caption{\label{fig:IMR}
Inspiral-merger-ringdown consistency tests of GR in the parameter space of the fractional difference in the estimate of the final mass and spin of the remnant black hole between inspiral and post-inspiral. We present 90\% credible posterior distributions for GW150914~\cite{TheLIGOScientific:2016src} and the combined one for events in GWTC-3~\cite{LIGOScientific:2021sio}. We also present future forecasts with Cosmic Explorer and multiband observations of Cosmic Explorer plus LISA, both assuming they detect signals from GW150914-like events~\cite{Carson:2019rda,Carson:2019rda,Carson:2020rea}.
} 
\end{center}
\end{figure}

Although the inspiral-merger-ringdown tests were designed as consistency tests of GR, one can tweak them to test specific theories and models. This has been investigated by Carson and Yagi for Einstein--dilaton--Gauss--Bonnet gravity~\cite{Carson:2020cqb} and parameterized Kerr black holes~\cite{Carson:2020iik}. They constructed gravitational waveforms including the leading corrections to the inspiral and the ringdown frequency and damping time. They then injected these signals and recovered them with GR templates to estimate systematic and statistical errors on the final mass and spin of the remnant black hole. This allowed them to find upper bounds on the theory and the model if observations are consistent with GR. They found that constraints on Einstein--dilaton--Gauss--Bonnet gravity can be improved by an order of magnitude from current bounds with future multiband observations~\cite{Carson:2020cqb}, while those on the Johannsen--Psaltis metric parameters can be improved by up to three orders of magnitude with future LISA observations~\cite{Carson:2020iik}.

\subsection{Tests of the no-hair theorems}
\label{sec:no-hair}

Another important class of generic tests of GR are those that concern the \emph{no-hair} theorems. Since much work has been done on this area, we have separated this topic from the main generic tests section (\ref{generic-tests}). In what follows, we describe what these theorems are and the possible tests one could carry out with gravitational-wave observations emitted by black-hole--binary systems. 

\subsubsection{The no-hair theorems}

The \emph{no-hair} theorems state that the only stationary, vacuum solution to the Einstein equations that is non-singular outside the event horizon is completely characterized by three quantities: its mass $M$, its spin $S$ and its charge $Q$. This conclusion is arrived at by combining several different theorems. First, Hawking~\cite{Hawking:1971vc,Hawking:1971tu} proved that a stationary black hole must have an event horizon with a spherical topology and that it must be either static or axially symmetric. Israel~\cite{Israel:1967wq,Israel:1967za} then proved that the exterior gravitational field of such static black holes is uniquely determined by $M$ and $Q$ and it must be given by the Schwarzschild or the Reissner--Nordstr\"om metrics. Carter~\cite{Carter:1971zc} constructed a similar proof for uncharged, stationary, axially-symmetric black holes, where this time black holes fall into disjoint families, not deformable into each other and with an exterior gravitational field uniquely determined by $M$ and $S$. Robinson~\cite{Robinson:1975bv} and Mazur~\cite{Mazur:1982db} later proved that such black holes must be described by either the Kerr or the Kerr--Newman metric. See also~\cite{Misner:1973cw,Poisson} for more details. 

The no-hair theorems apply under a restrictive set of conditions~\cite{Cardoso:2016ryw}. First, the theorems only apply in stationary situations. Black-hole horizons can be dynamically deformed in coalescing situations, and if so, Hawking's theorems~\cite{Hawking:1971vc,Hawking:1971tu} about spherical horizon topologies do not apply. This then implies that all other theorems described above also do not apply. Second, the theorems only apply in vacuum. Consider, for example, an axially-symmetric black hole in the presence of a non-symmetrical matter distribution outside the event horizon. One might naively think that this would tidally distort the event horizon, leading to a rotating, stationary black hole that is not axisymmetric. However, Hawking and Hartle~\cite{Hawking:1972hy} showed that in such a case the matter distribution torques the black hole forcing it to spin down, thus leading to a non-stationary scenario. If the black hole is non-stationary, then again the no-hair theorems do not apply by the arguments described at the beginning of this paragraph, and thus non-isolated black holes can have hair. Third, the theorems only apply within GR, i.e., through the use of the Einstein equations. Therefore, it is plausible that black holes in modified gravity theories or in GR with singularities outside any event horizons (naked singularities) will have hair.  

The no-hair theorems imply that the exterior gravitational field of isolated, stationary, uncharged and vacuum black holes (in GR and provided the spacetime is regular outside all event horizons) can be written as an infinite sum of mass and current multipole moments, where only two of them are independent: the mass monopole moment $M$ and the current dipole moment $S$. One can extend these relations to include charge, but astrophysical black holes are expected to be essentially neutral due to charge accretion. If the no-hair theorems hold, all other multipole moments can be determined from~\cite{Geroch:1970cd,Geroch:1970cc,hansen}
\be
M_{\ell}+\mathrm{i}S_{\ell}=M(\mathrm{i}a)^{\ell}\,,
\label{kerrmult}
\ee
where $M_{\ell}$ and $S_{\ell}$ are the $\ell$th mass and current multipole moments. Even if the black-hole progenitor was not stationary or axisymmetric, the no-hair theorems guarantee that any excess multipole moments will be shed-off during gravitational collapse~\cite{Price:1971fb,Price:1972pw}. Eventually, after the black hole has settled down and reached an equilibrium configuration, it will be described purely in terms of $M_{0} = M$ and $S_{1} = S = M a^{2}$, where $a$ is the Kerr spin parameter.

An astrophysical observation of a hairy black hole would not imply that the no-hair theorems are wrong, but rather that one of the assumptions made in deriving these theorems is not appropriate to describe nature. The three main assumptions are stationarity, vacuum and that GR and the regularity condition hold. Astrophysical black holes will generically be hairy due to a violation of the first two assumptions, since they will neither be perfectly stationary, nor exist in a perfect vacuum. Astrophysical black holes will always suffer small perturbations by other stars, electromagnetic fields, other forms of matter, like dust, plasma or dark matter, etc., which will induce non-zero deviations from Eq.~\eqref{kerrmult} and thus evade the no-hair theorems. However, in all cases of interest such perturbations are expected to be too small to be observable, which is why one argues that even astrophysical black holes should obey the no-hair theorems if GR holds. Put another way, an observation of the violation of the no-hair theorems would be more likely to indicate a failure of GR in the extreme gravity regime, than an unreasonably large amount of astrophysical hair. 

Tests of the no-hair theorems come in two flavors: through electromagnetic observations~\cite{Johannsen:2010xs,Johannsen:2010ru,Johannsen:2010bi,Johannsen:2012ng,Qi:2020xzi,Alush:2022juk} and through gravitational wave observations~\cite{Ryan:1995wh,Ryan:1997hg,Collins:2004ex,Glampedakis:2005cf,Babak:2006uv,Barack:2006pq,Li:2007qu,Sopuerta:2009iy,Yunes:2009ry,Vigeland:2009pr,Vigeland:2010xe,Gair:2011ym,Vigeland:2011ji,Rodriguez:2011aa}. The former rely on radiation emitted by accelerating particles in an accretion disk around black holes. However, such tests are not clean as they require the modeling of complicated astrophysics associated with accretion disks, matter and electromagnetic fields. Other electromagnetic tests of the no-hair theorems exist, for example through the observation of close stellar orbits around Sgr~A*~\cite{Merritt:2009ex,Merritt:2011ve,Sadeghian:2011ub,Ajith:2020ydz,Qi:2020xzi,Alush:2022juk} and pulsar--black-hole binaries~\cite{Wex:1998wt}, and through direct images of black holes~\cite{Johannsen:2010ru,Broderick:2013rlq,Psaltis:2015uza,Psaltis:2018xkc,EventHorizonTelescope:2022xqj}. See~\cite{lrr-2008-9} for reviews on these topics. Unlike electromagnetic tests, gravitational wave tests are clean, without much contamination from their astrophysical environment~\cite{Yunes:2010sm,Yunes:2011ws,Kocsis:2011dr,Barausse:2014tra}; little work, however, has gone into quantitatively comparing electromagnetic and gravitational wave tests of the no-hair theorem~\cite{Cardenas-Avendano:2016zml}. Gravitational wave tests of the no-hair theorem can be classified into those carried out with  inspirals and with ringdown waves. We will describe below both of these in some detail below.

\subsubsection{Inspiral tests of the no-hair theorem}
\label{sec:EMRI}

Gravitational wave tests of the no-hair theorems can be performed with the detection of either extreme or comparable mass-ratio inspirals and with the ringdown of comparable-mass black-hole mergers with current ground-based~\cite{Krishnendu:2017shb,Krishnendu:2018nqa,Krishnendu:2019tjp} and future space-borne gravitational-wave detectors~\cite{AmaroSeoane:2012km,AmaroSeoane:2012je}. Extreme mass-ratio inspirals consist of a stellar-mass compact object spiraling into a supermassive black hole in a generic orbit within astronomical units from the event horizon of the supermassive object~\cite{AmaroSeoane:2007aw}. These events outlive the observation time of future detectors, emitting millions of gravitational wave cycles, with the stellar-mass compact object essentially acting as a tracer of the supermassive black hole spacetime~\cite{Sotiriou:2004bm} (see C\'ardenas-Avenda\~no and Sopuerta~\cite{Cardenas-Avendano:2024mqp} for a recent review on tests of GR with extreme mass-ratio inspirals.)
Ringdown gravitational waves are always emitted after black holes merge, as the remnant settles down into its final configuration. During the ringdown, the highly-distorted remnant radiates all excess degrees of freedom and this radiation carries a signature of whether the no-hair theorems hold in its quasi-normal mode spectrum (see, e.g.,~\cite{Berti:2009kk} for a recent review). 

Both electromagnetic and gravitational wave tests need a metric with which to model accretion disks, quasi-periodic oscillations, or extreme mass-ratio inspirals. One can classify these metrics as \emph{direct} or \emph{generic}, paralleling the discussion in Section~\ref{sec:Direct-tests}. Direct metrics are exact solutions to a specific set of field equations, with which one can derive observables. Examples of such metrics are the Manko--Novikov metric~\cite{1992CQGra...9.2477M} and the slowly-spinning black-hole metric in dynamical Chern--Simons gravity~\cite{Yunes:2009hc,Konno:2009kg,Yagi:2012ya,Maselli:2015tta} and in Einstein--dilaton--Gauss--Bonnet gravity~\cite{Yunes:2011we,Pani:2011gy,Maselli:2015tta}. Generic metrics are those that parametrically modify the Kerr spacetime, such that for certain parameter choices one recovers identically the Kerr metric, while for others, one has a deformation of Kerr. Generic metrics can be further classified into two subclasses, Ricci-flat versus non-Ricci-flat, depending on whether they satisfy $R_{\mu \nu} = 0$. 

One might be concerned that such no-hair tests of GR cannot constrain modified gravity theories, because Kerr black holes can also be solutions in the latter~\cite{Psaltis:2007cw}. This is indeed true provided the modified field equations depend only on the Ricci tensor or scalar. In Einstein--dilaton--Gauss--Bonnet or dynamical Chern--Simons gravity, the modified field equations depend on the Riemann tensor, and thus, Ricci-flat metric need not solve these modified set~\cite{Yunes:2011we}. Moreover, just because the metric background is identically Kerr does not imply that inspiral gravitational waves will be identical to those predicted in GR. Most studies carried out to date, be it direct metric tests or generic metric tests, assume that the only quantity that is modified is the metric tensor, or equivalently, the Hamiltonian or binding energy. Inspiral motion, of course, does not depend just on this quantity, but also on the radiation-reaction force that pushes the small object from geodesic to geodesic. Moreover, the gravitational waves generated during such an inspiral depend on the field equations of the theory considered. Therefore, most of metric tests should be considered as partial tests in this sense; in general, modifications in extreme gravity will induce corrections to the Hamiltonian, the radiation-reaction force and wave generation.

\paragraph{Direct metric tests of the no-hair theorem}\mbox{}\\

\noindent
Let us first consider direct metric tests of the no-hair theorem. The most studied direct metric is the Manko--Novikov one, which although an exact, stationary and axisymmetric solution to the vacuum Einstein equations, does not represent a black hole, as the event horizon is broken along the equator by a ring singularity~\cite{1992CQGra...9.2477M}. Just like the Kerr metric, the Manko--Novikov metric possesses an ergoregion, but unlike the former, it also possesses regions of closed time-like curves that overlap the ergoregion. Nonetheless, an appealing property of this metric is that it deviates continuously from the Kerr metric through certain parameters that characterize the higher multiple moments of the solution.

The first geodesic study of Manko--Novikov spacetimes was carried out by Gair, et al.~\cite{Gair:2007kr}. They found that there are two ring-like regions of bound orbits: an outer one where orbits look regular and integrable, as there exist four isolating integrals of the motion; and an inner one where orbits are chaotic and thus ergodic. Gair, et al.~\cite{Gair:2007kr} suggested that orbits that transition from the integrable to the chaotic region would leave a clear observable signature in the frequency spectrum of the emitted gravitational waves. However, they also noted that chaotic regions exist only very close to the central body and are probably not astrophysically accessible. The study of Gair, et al.~\cite{Gair:2007kr} was recently confirmed and followed up by Contopoulos, et al.~\cite{Contopoulos:2011dz}. They studied a wide range of geodesics and found that, in addition to an inner chaotic region and an outer regular region, there are also certain Birkhoff islands of stability. When an extreme mass-ratio inspiral traverses such a region, the ratio of resonant fundamental frequencies would remain constant in time, instead of increasing monotonically. Such a feature would impact the gravitational waves emitted by such a system, and it would signal that the orbit equations are non-integrable and the central object is not a Kerr black hole. Destounis and Kokkotas~\cite{Destounis:2021rko} also derived gravitational waveforms for extreme mass ratio inspirals for Manko--Novikov spacetimes and found that fundamental frequencies undergo sudden jumps when the companion crosses a resonant island.

The study of chaotic motion in geodesics of non-Kerr spacetimes is by no means new. Chaos has also been found in geodesics of Zipoy--Voorhees--Weyl and Curzon spacetimes with multiple singularities~\cite{Sota:1995ms,Sota:1996cv} and in general for Zipoy--Voorhees spacetimes in~\cite{LukesGerakopoulos:2012pq}, of perturbed Schwarzschild spacetimes~\cite{Letelier:1996he}, of Schwarzschild spacetimes with a dipolar halo~\cite{Letelier:1997uv,Letelier:1997fi,Gueron:2001mm} of Erez--Rosen spacetimes~\cite{Gueron:2002jt}, and of deformed generalizations of the Tomimatsy--Sato spacetime ~\cite{Dubeibe:2007hq}. One might worry that such chaotic orbits will depend on the particular spacetime considered, but recently Apostolatos, et al.~\cite{Apostolatos:2009vu} and Lukes--Gerakopoulos, et al.~\cite{LukesGerakopoulos:2010rc} have argued that the Birkhoff islands of stability are a general feature. Although the Kolmogorov, Arnold, and Moser theorem~\cite{Kolmogorov,Arnold,Moser} states that phase orbit tori of an integrable system are only deformed if the Hamiltonian is perturbed, the Poincare--Birkhoff theorem~\cite{Chaos} states that resonant tori of integrable systems actually disintegrate, leaving behind a chain of Birkhoff islands. These islands are only characterized by the ratio of winding frequencies that equals a rational number, and thus, they constitute a distinct and generic feature of non-integrable systems~\cite{Apostolatos:2009vu,LukesGerakopoulos:2010rc}. Given an extreme mass-ratio gravitational-wave detection, one can monitor the ratio of fundamental frequencies and search for plateaus in their evolution, which would signal non-integrability. Of course, whether detectors can resolve such plateaus depends on the initial conditions of the orbits and the physical system under consideration (these determine the thickness of the islands), as well as the mass ratio (this determines the radiation-reaction timescale) and the distance and mass of the central black hole (this determines the SNR).   

Another example of a direct metric test of the no-hair theorem is through the use of the slowly-rotating dynamical Chern--Simons black hole metric~\cite{Yunes:2009hc}. Unlike the Manko--Novikov metric, the dynamical Chern--Simons one does represent a black hole, i.e., it possesses an event horizon, but it evades the no-hair theorems because it is not a solution to the Einstein equations. Sopuerta and Yunes~\cite{Sopuerta:2009iy} carried out the first extreme mass-ratio inspiral analysis when the background supermassive black hole object is taken to be such a Chern--Simons black hole. They used a semi-relativistic model~\cite{Ruffini:1981af} to evolve extreme mass-ratio inspirals and found that the leading-order modification comes from a modification to the geodesic trajectories, induced by the non-Kerr modifications of the background. Because the latter correspond to an extreme-gravity modification to GR, modifications in the trajectories are most prominent for zoom-whirl orbits, as the small compact object zooms around the supermassive black hole in a region of unstable orbits, close to the event horizon.  These modifications were then found to propagate into the gravitational waves emitted, leading to a dephasing that could be observed or ruled out with future gravitational-wave observations to roughly the horizon scale of the supermassive black hole, as has been recently confirmed by Canizares, et al.~\cite{Canizares:2012ji,Canizares:2012is}. However, these studies may be underestimates, given that they treat the black hole background in dynamical Chern--Simons gravity only to first-order in spin and neglect any scalar field charge on the small object. Chaotic orbits in quadratic gravity were studied in~\cite{Cardenas-Avendano:2018ocb,Deich:2022vna}, where the effect of chaos was found to be greatly suppressed relative to that in Manko--Novikov spacetimes. Similar to the dynamical Chern-Simons case, one can also probe other quadratic gravity theories, such as Einstein--dilaton--Gauss--Bonnet gravity, with extreme mass-ratio inspirals. In fact, one can consider the more general setup of a smaller mass black hole endowed with a scalar charge, without choosing a particular theory of gravity and choosing the central black hole to be Kerr~\cite{Maselli:2020zgv,Maselli:2021men,Guo:2022euk,Lestingi:2023ovn,Speri:2024qak}.

A final example of a direct metric test of the no-hair theorems is to consider black holes that are not in vacuum. Barausse, et al.~\cite{Barausse:2006vt} studied extreme--mass-ratio inspirals in a Kerr--black-hole background that is perturbed by a self-gravitating, homogeneous torus that is compact, massive and close to the Kerr black hole. They found that the presence of this torus impacts the gravitational waves emitted during such inspirals, but only weakly, making it difficult to distinguish the presence of matter. Yunes, et al.~\cite{Yunes:2011ws} and Kocsis, et al.~\cite{Kocsis:2011dr} carried out a similar study, where this time they considered a small compact object inspiraling completely within a geometrically thin, radiation-pressure dominated accretion disk. They found that disk-induced migration can modify the radiation-reaction force sufficiently to leave observable signatures in the waveform, provided the accretion disk is sufficiently dense in the radiation-dominated regime and a gap opens up. However, these tests of the no-hair theorem will be rather difficult as most extreme--mass-ratio inspirals are not expected to be in an accretion disk. 

\paragraph{Generic metric tests of the no-hair theorem}\mbox{}\\

\noindent
Let us now consider generic metric tests of the no-hair theorem. Generic Ricci-flat deformed metrics will lead to Laplace-type equations for the deformation functions in the far-field since they must satisfy $R_{\mu \nu} = 0$ to linear order in the perturbations. The solution to such an equation can be expanded in a sum of mass and current multipole moments, when expressed in asymptotically Cartesian and mass-centered coordinates~\cite{Thorne:1980ru}. These multipoles can be expressed via~\cite{Collins:2004ex,Vigeland:2009pr,Vigeland:2010xe}
\begin{equation}
M_{\ell} + \mathrm{i}S_{\ell} = M(\mathrm{i}a)^{{\ell}} + \delta M_{\ell} + \mathrm{i}\delta S_{\ell}\,,
\label{mult}
\end{equation}
where $\delta M_{\ell}$ and $\delta S_{\ell}$ are mass and current multipole deformations. Ryan~\cite{Ryan:1995wh,Ryan:1997hg} showed that the measurement of three or more multipole moments would allow for a test of the no-hair theorem. Generic non-Ricci flat metrics, on the other hand, will not necessarily lead to Laplace-type equations for the deformation functions in the far field, and thus, the far-field solution and Eq.~\eqref{mult} will depend on a sum of $\ell$ and $m$ multipole moments.  Barack and Cutler~\cite{Barack:2006pq} studied how well one can measure the mass, spin and quadrupole moment of a black hole with extreme mass ratio inspirals using LISA. Gravitational waveforms of inspiralling compact objects with generic, non-axisymmetric quadrupole moments with spin-precessing and eccentric orbits were constructed in~\cite{Loutrel:2022ant}.

No-hair tests can also be performed with gravitational waves from stellar-mass binary black holes using ground-based detectors~\cite{Krishnendu:2017shb,Krishnendu:2018nqa,Krishnendu:2019tjp}. In particular, one can constrain deviations of the quadrupole moment $Q$ from Kerr. In order to explain this idea, let us introduce the dimensionless quadrupole moment $\kappa$ (same as $\bar Q$ in Sec.~\ref{sec:I-Love-Q}) as $Q = - \kappa m^3 \chi^2$ for the mass $m$ and dimensionless spin $\chi$ of a black hole, where $\kappa = 1$ for a Kerr black hole. The gravitational waves emitted during compact binary inspirals depend on $\kappa_s$ and $\kappa_a$ which are symmetric and antisymmetric combinations of $\kappa_1$ and $\kappa_2$ for the dimensionless quadrupole moments of binary components. Due to large degeneracies between $\kappa_s$ and $\kappa_a$, it would be extremely challenging to independently measure both quadrupole parameters simultaneously. Thus, one typically assumes $\kappa_a=0$ and measures $\delta \kappa_s \equiv \kappa_s -1$, the deviation in the symmetric combination of the dimensionless quadrupole moment from the Kerr expectation. This assumption corresponds to having compact stars with identical spin-induced deformations in a binary. The LIGO/Virgo Collaboration has derived bounds on $\delta \kappa_s$ with gravitational-wave events in the catalogs collected~\cite{LIGOScientific:2020tif,LIGOScientific:2021sio}. Multiplying the likelihood of $\kappa_s$ of each signal in selected events in GWTC-3, the collaboration found $\delta \kappa_s = -16.0 ^{+13.6}_{-16.7}$~\cite{LIGOScientific:2021sio}. On the other hand, with the hierarchical analysis discussed in Sec.~\ref{subsubsection:ppE}, the combined bound becomes $\delta \kappa_s = -26.3^{+45.8}_{-52.9}$~\cite{LIGOScientific:2021sio}. The bounds on the positive $\delta\kappa_s$ side are stronger because of how $\delta \kappa_s$ is correlated with the effective inspiral spin parameter.
Li, et al.~\cite{Li:2023zbm} carried out a no-hair test with GW150914 and GW200129 and reported a significant deviation in the quadrupole moment from the Kerr case for the latter event, though more events and further analyses are necessary to validate the deviation.
The no-hair test can be extended to include the octupole moment~\cite{Saini:2023gaw}. The importance of precession on no-hair tests for the inspiral is discussed in~\cite{Loutrel:2023boq}. Mahapatra, et al.~\cite{Mahapatra:2023ydi} studied the effect of amplitude corrections to multipolar tests of GR with gravitational waves, as well as carrying out multiparameter tests~\cite{Mahapatra:2023uwd}.

Let us now return to tests of the no-hair relations from extreme mass-ratio inspirals, which require the construction of a parametrically-deformed black hole metric.
The first attempt to construct a generic, Ricci-flat metric was by Collins and Hughes~\cite{Collins:2004ex}: the \emph{bumpy black-hole metric}. In this approach, the metric is assumed to be of the form
\begin{equation}
g_{\mu \nu} = g_{\mu \nu}^{(\mathrm{Kerr})} + \epsilon \; \delta g_{\mu \nu}\,,
\end{equation}
where $\epsilon \ll 1$ is a bookkeeping parameter that enforces that $\delta g_{\mu \nu}$ is a perturbation of the Kerr background. This metric is then required to satisfy the Einstein equations linearized in $\epsilon$, which then leads to differential equations for the metric deformation. Collins and Hughes~\cite{Collins:2004ex} assumed a non-spinning, stationary spacetime, and thus $\delta g_{\mu \nu}$ only possessed two degrees of freedom, both of which were functions of radius only: $\psi_{1}(r)$, which must be a harmonic function and which changes the Newtonian part of the gravitational field at spatial infinity; and $\gamma_{1}(r)$ which is completely determined through the linearized Einstein equations once $\psi_{1}$ is specified. One then has the freedom to choose how to prescribe $\psi_{1}$ and Collins and Hughes investigate~\cite{Collins:2004ex} two choices that correspond physically to point-like and ring-like naked singularities, thus violating cosmic censorship~\cite{Penrose:1969pc}. Vigeland and Hughes~\cite{Vigeland:2009pr} and Vigeland~\cite{Vigeland:2010xe} then extend this analysis to stationary, axisymmetric spacetimes via the Newman--Janis method~\cite{Newman:1965tw,Drake:1998gf}, showing how such metric deformations modify Eq.~\eqref{mult}, and computing how these bumps imprint themselves onto the orbital frequencies and thus the gravitational waves emitted during an extreme--mass-ratio inspiral.  

That the bumps represent unphysical matter should not be a surprise, since by the no-hair theorems, if the bumps are to satisfy the vacuum Einstein equations they must either break stationarity or violate the regularity condition. Naked singularities are an example of the latter. A Lorentz-violating massive field coupled to the Einstein tensor is another example~\cite{Dubovsky:2007zi}. Gravitational wave tests with bumpy black holes must then be understood as {\emph{null tests}}: one assumes the default hypothesis that GR is correct and then sets out to test whether the data rejects or fails to reject this hypothesis (a null hypothesis can never be proven). Unfortunately, however, bumpy black hole metrics cannot parameterize spacetimes in modified gravity theories that lead to corrections in the field equations that are not proportional to the Ricci tensor, such as for example in dynamical Chern--Simons or in Einstein--dilaton--Gauss--Bonnet modified gravity. 

Other bumpy black hole metrics have also been proposed. Glampedakis and Babak~\cite{Glampedakis:2005cf} proposed a different type of stationary and axisymmetric bumpy black hole through the Hartle--Thorne metric~\cite{Hartle:1968si}, with modifications to the quadrupole moment. They then constructed a ``kludge'' extreme mass-ratio inspiral waveform and estimated how well the quadrupole deformation could be measured~\cite{Babak:2006uv}. However, this metric is valid only when the supermassive black hole is slowly-rotating, as it derives from the Hartle--Thorne ansatz.  Johansen and Psaltis~\cite{Johannsen:2011dh} proposed yet another metric to represent bumpy stationary and spherically-symmetric spacetimes. This metric introduces one new degree of freedom, which is a function of radius only and assumed to be a series in $M/r$. Johansen and Psaltis then rotated this metric via the Newman--Janis method~\cite{Newman:1965tw,Drake:1998gf} to obtain a new bumpy metric for axially-symmetric spacetimes. However, such a metric possesses a naked ring singularity on the equator, and naked singularities on the poles. As before, none of these bumpy metrics can be mapped to known modified gravity black hole solutions, in the Glampedakis and Babak case~\cite{Glampedakis:2005cf} because the Einstein equations are assumed to hold to leading order in the spin, while in the Johansen and Psaltis case~\cite{Johannsen:2011dh} because a single degree of freedom is not sufficient to model the three degrees of freedom contained in stationary and axisymmetric spacetimes~\cite{Stephani:2003tm,Vigeland:2011ji}. 

The first generic non-Ricci-flat bumpy black-hole metric so far is that of Vigeland, Yunes and Stein~\cite{Vigeland:2011ji}, and its simplified extension by Johannsen~\cite{Johannsen:2015pca}, which was further extended by Carson and Yagi~\cite{Carson:2020dez,Yagi:2023eap} and a similar black-hole metric with Kerr symmetry found by Papadopoulos and Kokkotas~\cite{Papadopoulos:2018nvd,Papadopoulos:2020kxu}, Chen~\cite{Chen:2020aix} (breaking the $Z_2$ symmetry of the spacetime), and Delaporte, et al.~\cite{Delaporte:2022acp} (breaking the circularity of spacetime). Vigeland, Yunes and Stein~\cite{Vigeland:2011ji} allowed generic deformations in the metric tensor, only requiring that the new metric perturbatively retained the Killing symmetries of the Kerr spacetime: the existence of two Killing vectors associated with stationarity and axisymmetry, as well as the perturbative existence of a Killing tensor (and thus a Carter-like constant), at least to leading order in the metric deformation. Such requirements imply that the geodesic equations in this new background are fully integrable, at least perturbatively in the metric deformation, which then allows one to solve for the orbital motion of extreme--mass-ratio inspirals by adapting previously existing tools. Brink~\cite{Brink:2008xx,Brink:2008xy,Brink:2009mq,Brink:2009mt,Brink:2009rf} studied the existence of such a second-order Killing tensor in generic, vacuum, stationary and axisymmetric spacetimes in Einstein's theory and found that these are difficult to construct exactly. By relaxing this exact requirement, Vigeland, Yunes and Stein~\cite{Vigeland:2011ji} found that the existence of a perturbative Killing tensor poses simple differential conditions on the metric perturbation that can be analytically solved. Moreover, they also showed how this new bumpy metric can reproduce all known modified gravity black hole solutions in the appropriate limits, provided these have an at least approximate Killing tensor; thus, these metrics are still vacuum solutions even though $R \neq 0$, since they satisfy a set of modified field equations. The imposition that the spacetime retains the Kerr Killing symmetries also leads to a bumpy metric that is well-behaved everywhere outside the event horizon (no singularities, no closed-time-like curves, no loss of Lorentz signature). Gair and Yunes~\cite{Gair:2011ym} studied how the geodesic equations are modified for a test-particle in a generic orbit in such a spacetime and showed that the bumps are indeed encoded in the orbital motion, and thus, in the gravitational waves emitted during an extreme-mass-ratio inspiral. This work was extended by Moore, et al.~\cite{Moore:2017lxy} for probing a parameterized black hole spacetime preserving Kerr symmetry with gravitational waves from extreme mass ratio inspirals. Destounis, et al.~\cite{Destounis:2020kss} constructed a new metric that contains a parameter characterizing the violation of the Kerr symmetry and studied how well one can constrain this parameter with future gravitational-wave observations from extreme mass-ratio inspirals. Carson and Yagi~\cite{Carson:2020iik} derived the leading correction to the inspiral waveform for a generic modification to non-Kerr spacetime. The authors also derived corrections to ringdown following the post-Kerr formulation~\cite{Glampedakis:2017dvb}. They then provided a future forecast on tests of non-Kerr spacetime with future gravitational-wave observations by performing parameterized tests and 
inspiral-merger-ringdown consistency tests of GR. Non-Kerr corrections to the waveform found in~\cite{Carson:2020iik} were used by Santos, et al.~\cite{Santos:2024pfa} to place bounds on deviations from Kerr with GWTC-3 (see also~\cite{Das:2024mjq}).
Kumar, et al.~\cite{Kumar:2024utz} constructed gravitational waves from extreme-mass-ratio inspirals with the central black hole represented by the parameterized Kerr spacetime developed by Yagi, et al.~\cite{Yagi:2023eap}, and studied the prospect for probing such a spacetime with LISA.

Another approach to construct generic, non-Ricci-flat bumpy black hole-metrics is that recently pursued by Konoplya, Rezzolla and Zhidenko~\cite{Rezzolla:2014mua,Konoplya:2016jvv,Konoplya:2018arm,Konoplya:2020hyk,Ma:2024kbu}. In this approach, one models the bumpy metric with the most general, stationary and axisymmetric line element, which is typically parametrized in terms of five free metric functions. These functions are expressed in terms of a compact radial coordinate (equivalent to the Schwarzschild factor in general relativity) and an infinite continuous fraction. Rezzolla, Zhidenko and Konoplya have shown that this metric is capable of reproducing the predictions of a large class of modified theories, including Einstein--dilaton--Gauss--Bonnet black holes, dilatonic  black holes and Kerr black holes~\cite{Rezzolla:2014mua,Konoplya:2016jvv}.  Moreover, this parametrized metric has also been used to model the shadows cast by the light-ring of this metric on the light emitted by an accretion disk~\cite{Younsi:2016azx}. C\'ardenas-Avenda\~no, et al.~\cite{Cardenas-Avendano:2019zxd} compared gravitational-wave versus X-ray observations to test the parameterized spacetime of Rezzolla and Zhidenko. The authors derived corrections to the gravitational waveforms and also simulated X-ray observations for the parameterized spacetime. They found that current gravitational-wave observations place stronger bounds on the spacetime than future X-ray observations can. The gravitational-wave bounds will further improve with future gravitational-wave detectors.
The Konoplya--Rezzolla--Zhidenko metric was constrained with GWTC-2 by Shashank and Bambi~\cite{Shashank:2021giy}.

\subsubsection{Ringdown tests of the no-hair theorem}
\label{sec:rd}

Let us now consider tests of the no-hair theorems with gravitational waves emitted by comparable-mass binaries during the ringdown phase. Gravitational waves emitted during ringdown can be described by a superposition of exponentially-damped sinusoids~\cite{Berti:2005ys}:
\begin{equation}
\label{eq:ringdown-waves}
h_{+}(t) + i \; h_{\times}(t) = \frac{M}{r} \sum_{\ell m n} \left\{
{\cal{A}}_{\ell m n} e^{i (\omega_{\ell m n} t + \phi_{\ell m n})} e^{-t/\tau_{\ell m n}} S_{\ell m n}
+
{\cal{A}}_{\ell m n}' e^{i (-\omega_{\ell m n} t + \phi_{\ell m n}')} e^{-t/\tau_{\ell m n}} S_{\ell m n}^{*} \right\}
\,,
\end{equation}
where $r$ is the distance from the source to the detector, the asterisk stands for complex conjugation, the real mode amplitudes ${\cal{A}}_{\ell mn}$ and ${\cal{A}}_{\ell mn}'$ and the real phases $\phi_{n \ell m}$ and $\phi_{n \ell m}'$ depend on the initial conditions, $S_{\ell m n}$ are spheroidal functions evaluated at the complex quasinormal ringdown frequencies $\omega_{n \ell m} = 2 \pi f_{n \ell m} + i/\tau_{n \ell m}$, and the real physical frequency $f_{n \ell m}$ and the real damping times $\tau_{n \ell m}$ are both functions of the mass $M$ and the Kerr spin parameter $a$ only, provided the no-hair theorems hold. These frequencies and damping times can be computed numerically or semi-analytically, given a particular black-hole metric (see~\cite{Berti:2009kk} for a review). The Fourier transform of a given $(\ell,m,n)$ mode is~\cite{Berti:2005ys}
\begin{align}
\tilde{h}_{+}^{(\ell,m,n)}(\omega) &= \frac{M}{r} {\cal{A}}_{\ell m n}^{+} \left[e^{i \phi_{\ell m n}^{+}} S_{\ell m n} b_{+}(\omega) + e^{-i \phi_{\ell m n}^{+}} S_{\ell m n}^{*} b_{-}(\omega) \right]\,, 
\\
\tilde{h}_{\times}^{(\ell,m,n)}(\omega) &= \frac{M}{r} {\cal{A}}_{\ell m n}^{\times} \left[e^{i \phi_{\ell m n}^{\times}} S_{\ell m n} b_{+}(\omega) + e^{-i \phi_{\ell m n}^{\times}} S_{\ell m n}^{*} b_{-}(\omega)\right]\,, 
\end{align}
where we have defined ${\cal{A}}_{\ell m n}^{+,\times} e^{i \phi_{\ell m n}^{+,\times}} \equiv {\cal{A}}_{\ell m n} e^{i \phi_{\ell m n}} \pm {\cal{A}}' e^{-i \phi_{\ell m n}'}$ as well as the Lorentzian functions
\begin{equation}
b_{\pm}(\omega) = \frac{\tau_{\ell m n}}{1 + \tau_{\ell m n}^{2} (\omega \pm \omega_{\ell m n})^{2}} .
\end{equation}
Ringdown gravitational waves will all be of the form of Eq.~\eqref{eq:ringdown-waves} provided that the characteristic nature of the differential equation that controls the evolution of ringdown modes is not modified, i.e., provided that one only modifies the potential in the Teukolsky equation or other subdominant terms, which in turn depend on the modified field equations.

Tests of the no-hair theorems through the observation of black-hole ringdown date back to Detweiler~\cite{Detweiler:1980gk}, and it has been worked out in detail by Dreyer, et al.~\cite{Dreyer:2003bv}. Let us first imagine that a single complex mode is detected ${\omega}_{\ell_{1} m_{1} n_{1}}$  and one measures separately its real and imaginary parts. Of course, from such a measurement, one cannot extract the measured harmonic triplet $(\ell_{1},m_{1},n_{1})$, but instead one only measures the complex frequency ${\omega}_{\ell_{1} m_{1} n_{1}}$. This information is not sufficient to extract the mass and spin angular momentum of the black hole because different quintuplets $(M,a,\ell,m,n)$ can lead to the same complex frequency ${\omega}_{\ell_{1} m_{1} n_{1}}$. The best way to think of this is graphically: a given observation of ${\omega}_{\ell_{1} m_{1} n_{1}}^{(1)}$ traces a line in the complex $\Omega_{\ell_{1} m_{1} n_{1}} = M \omega_{\ell_{1} m_{1} n_{1}}^{(1)}$ plane; a given $(\ell,m,n)$ triplet defines a complex frequency $\omega_{\ell m n}$ that also traces a curve in the complex $\Omega_{\ell m n}$ plane; each intersection of the measured line $\Omega_{\ell_{1} m_{1} n_{1}}$ with $\Omega_{\ell m n}$ defines a possible doublet $(M,a)$; since different $(\ell,m,n)$ triplets lead to different $\omega_{\ell m n}$ curves and thus different intersections, one ends up with a set of doublets $S_{1}$, out of which only one represents the correct black-hole parameters. We thus conclude that a single mode observation of ringdown gravitational waves is not sufficient to test the no-hair theorem~\cite{Dreyer:2003bv,Berti:2005ys}.

Let us then imagine that one has detected two complex modes, ${\omega}_{\ell_{1} m_{1} n_{1}}$ and ${\omega}_{\ell_{2} m_{2} n_{2}}$. Each detection leads to a separate line ${\Omega}_{\ell_{1} m_{1} n_{1}}$ and ${\Omega}_{\ell_{2} m_{2} n_{2}}$ in the complex plane. As before, each $(n,\ell,m)$ triplet leads to separate curves $\Omega_{\ell m n}$ which will intersect with both ${\Omega}_{\ell_{1} m_{1} n_{1}}$ and ${\Omega}_{\ell_{2} m_{2} n_{2}}$ in the complex plane. Each intersection between $\Omega_{\ell m n}$ and ${\Omega}_{\ell_{1} m_{1} n_{1}}$ leads to a set of doublets $S_{1}$, while each intersection between $\Omega_{\ell m n}$ and ${\Omega}_{\ell_{2} m_{2} n_{2}}$ leads to another set of doublets $S_{2}$. However, if the no-hair theorems hold sets $S_{1}$ and $S_{2}$ must have at least one element in common. Therefore, a two-mode detection allows for tests of the no-hair theorem~\cite{Dreyer:2003bv,Berti:2005ys}. However, when dealing with a quasi-circular black-hole--binary inspiral within GR one knows that the dominant mode is $\ell=2=m$. In such a case, the observation of this complex mode by itself allows one to extract the mass and spin angular momentum of the black hole. Then, the detection of the real frequency in an additional mode, such as the $\ell=3=m$, $\ell=4=m$ or $(\ell,m) = (2,1)$ modes that are the next subleading modes~\cite{Bhagwat:2016ntk}, can be used to test the no-hair theorem~\cite{Berti:2005ys,Berti:2007zu,Kamaretsos:2011um}. Generally, detecting higher harmonic modes of $\ell \neq 2$ or $m \neq 2$ is more promising than detecting overtone modes of $n >0$ for unequal-mass binaries, while the latter is easier to detect for nearly equal-mass systems~\cite{JimenezForteza:2020cve,Ota:2021ypb,Ota:2022fds}. One can also perform a no-hair test by studying the consistency between the mode amplitude ratio and the phase difference that can only be in narrow regions in parameter space in GR~\cite{Forteza:2022tgq}. No-hair tests through ringdown observations can probe non-Kerr black holes~\cite{Carson:2020iik,Dey:2022pmv}, as discussed in Sec.~\ref{sec:EMRI}, through e.g.~the post-Kerr formalism~\cite{Glampedakis:2017dvb}, and exotic compact objects~\cite{Westerweck:2021nue} (see Sec.~\ref{sec:hairy-BHs} for more details).

Although the logic behind these tests is clear, one must study them carefully to determine whether all systematic and statistical errors are sufficiently under control so that they are feasible. Berti, et al.~\cite{Berti:2005ys,Berti:2007zu} investigated such tests carefully through a frequentist approach. First, they found that a matched-filtering type analysis with two-mode ringdown templates would increase the volume of the template manifold by roughly three orders of magnitude. A better strategy then is perhaps to carry out a Bayesian analysis, like that of Gossan, et al.~\cite{Gossan:2011ha}; through such a study one can determine whether a given detection is consistent with a two-mode or a one-mode hypothesis. Berti, et al.~\cite{Berti:2005ys,Berti:2007zu} also calculated that a SNR of ${\cal{O}}(10^{2})$ in the ringdown part of the signal would be needed to detect the presence of two ringdown modes in the signal and to resolve their frequencies, so that no-hair tests would be possible. Although this is difficult to imagine with single detections by aLIGO, such tests would be possible with third-generation ground-based detectors, and they should be routine with space-based GW detectors~\cite{Berti:2016lat}. Strong signals are necessary because one must be able to distinguish at least two modes in the signal. Unfortunately, however, whether the ringdown leads to such strong SNRs and whether the sub-dominant ringdown modes are of a sufficiently large amplitude depends on a plethora of conditions: the location of the source in the sky, the mass of the final black hole, which depends on the rest mass fraction that is converted into ringdown gravitational waves (the ringdown efficiency), the mass ratio of the progenitor, the magnitude and direction of the spin angular momentum of the final remnant and probably also of the progenitor and the initial conditions that lead to ringdown. Thus, although such tests are possible, one would have to be quite fortunate to detect a sufficiently loud signal with the right properties so that a two-mode extraction and a test of the no-hair theorems is feasible. See~\cite{Carullo:2018sfu,Baibhav:2018rfk,Bhagwat:2019dtm,Ota:2021ypb,Pacilio:2023mvk} for other works on future prospects for testing GR with ringdown.

Another approach is then to combine multiple events in the hopes to tease out enough information to carry out a ringdown test 
\cite{Yang:2017zxs}. Given a single observation, one can construct the posterior of a parameter that quantifies deformations away from the quasinormal frequencies predicted in GR, given a remnant mass and spin. If the no-hair theorems hold, then this posterior would be peaked at zero, with some width that can be used to compute a confidence region, and thus, a constraint on deviations from black hole baldness. Meidam, et al.~\cite{Meidam:2014jpa} considered adding the posteriors from a catalog of $N$ detections with the third-generation Einstein Telescope (ET) detector to compute the odds-ratio between the hypothesis that the no-hair theorems are satisfied and the hypothesis that they deviate by a constant frequency factor. The authors found that when ${\cal{O}}(10)$ ET posteriors are stacked in this way, then one can test the no-hair theorem to a few percent with the first sub-dominant ringdown modes. Yang, et al.~\cite{Yang:2017zxs} carried out a different analysis: they still considered $N$ inspiral-merger-ringdown detections, but instead of adding the posteriors together, they (i) time- and phase-shifted the events with respect to a fiducial one to align one of the sub-leading modes across the events and (ii) coherently stacked the time- and phase-shifted signals to produce a single ``mega''-signal that would boost the power in the sub-dominant mode. The time- and phase-shifting requires knowledge of how the sub-dominant mode depends on the remnant mass and spin, which Yang, et al.~\cite{Yang:2017zxs} modeled assuming GR, making this a null test. They found that coherently stacking significantly boosts the ability to test the no-hair theorem, allowing a null-test of GR with only ${\cal{O}}(5)$ events using aLIGO at designed sensitivity. They also showed that a significantly smaller number of events are required to test the no-hair relations when coherently stacking because the signal-to-noise ratio of the stacked signal scales as $N^{1/4}$ instead of $N^{1/2}$~\cite{Yang:2017zxs,2009PhRvD..80d2001K} (see Da Silva Costa, et al.~\cite{DaSilvaCosta:2017njq} for a related work).
 
No-hair tests of black holes with ringdown observations have been carried out with the existing gravitational-wave events. Carullo, et al.~\cite{Carullo:2019flw} performed a Bayesian inference on the GW150914 data and found no evidence for more than one quasinormal mode. Isi, et al.~\cite{Isi:2019aib} focused on the $\ell = 2 = m$ modes and found evidence in the GW150914 data for both the fundamental mode $(n=0)$ and at least one overtone mode $(n=1)$ (see~\cite{Isi:2021iql} for a more comprehensive analysis by the same authors). This allowed them to probe the no-hair property of the remnant black hole to a $\sim 10\%$ level. On the other hand, Cotesta, et al.~\cite{Cotesta:2022pci} performed an independent study and found no evidence for the presence of the overtone mode. They pointed out that the analysis for the existence of such a mode is sensitive to the starting time of the ringdown (see Bhagwat, et al.~\cite{Bhagwat:2017tkm} for the importance of the choice of the starting time of the ringdown) and the claim for the detection of the overtone in~\cite{Isi:2019aib} is likely to be dominated by the noise. Their claim was more thoroughly investigated in~\cite{Baibhav:2023clw}. Isi and Farr~\cite{Isi:2022mhy} revisited their analysis and reported that they could not reproduce the results in~\cite{Cotesta:2022pci} and their previous analysis in~\cite{Isi:2019aib} should be robust (see also~\cite{Carullo:2023gtf,Baibhav:2023clw,Isi:2023nif} for further debate by the two groups.) Yet another independent analysis was carried out by Finch, et al.~\cite{Finch:2022ynt}. They took a frequency-domain approach developed in~\cite{Finch:2021qph} by the same authors, as opposed to the time-domain one that had been used previously, and marginalized over the source sky position and the ringdown starting time. The authors found some tentative evidence for the overtone mode but at a much weaker significance than previously reported by Isi, et al.~\cite{Isi:2019aib,Isi:2022mhy}. Ma, et al.~\cite{Ma:2023vvr,Ma:2023cwe} developed a ``rational filter'', a method to clean a particular mode from a ringdown signal. Applying this to GW150914, they removed the fundamental mode and found that the remaining filtered data is consistent with the template that only includes the first overtone.
Bustillo, et al.~\cite{CalderonBustillo:2020rmh} imposed the prior knowledge that the ringdown of GW150914 is sourced by a binary black hole merger and found that the hairy black hole hypothesis is disfavored over the Kerr black hole hypothesis with an odds ratio of 1:600.
Wang, et al.~\cite{Wang:2023xsy} carried out an independent analysis in frequency domain by removing contamination before ringdown. Their results support the existence of the overtone mode and are more consistent with those of Isi, et al.~\cite{Isi:2019aib}. On the other hand, Correia, et al.~\cite{Correia:2023bfn} marginalized over merger time and sky location uncertainties, and found that the evidence for the existence of the overtone mode is rather low.
Regarding no-hair test with just the $n=0$ fundamental mode, Capano, et al.~\cite{Capano:2021etf} found evidence for the existence of the $\ell=3=m$ mode in GW190521 with a Bayes factor of 56 (see~\cite{Capano:2022zqm} for their follow-up analysis). The fractional deviation from GR on the subdominant mode frequency was constrained to $-0.01^{+0.08}_{-0.09}$.

Recently, parameterized ringdown formulations have been developed to test the assumptions of the no-hair theorem~\cite{Cardoso:2019mqo,McManus:2019ulj,Maselli:2019mjd,Carullo:2021dui,Franchini:2022axs}. Cardoso, et al.~\cite{Cardoso:2019mqo} assumed a spherically-symmetric background black hole solution, that all GR deviations are small, and that the perturbed equations decouple from each other. The authors then parameterized deviations in the perturbation potential from GR, where generic parameters are coefficients of the potential deviation expanded about spatial infinity. One can then solve such perturbed equations and find ringdown frequencies perturbatively. McManus, et al.~\cite{McManus:2019ulj} improved~\cite{Cardoso:2019mqo} by allowing the perturbed equations to couple to each other and by going to second order in perturbation parameters. They showed that their framework can be applied to various non-GR theories, including dynamical Chern-Simons gravity, Horndeski theories and effective field theories. The original work by Cardoso, et al.~\cite{Cardoso:2019mqo} was refined by Franchini and V\"olkel~\cite{Franchini:2022axs} to include additional coefficients that have some advantage when doing parameter estimation. The parameterized quasi-normal mode framework was further extended to include overtones~\cite{Hirano:2024fgp} and even to include tidal Love numbers of black holes~\cite{Katagiri:2023umb}. Maselli, et al.~\cite{Maselli:2019mjd,Maselli:2023khq} further extended these previous works by including spin to the background black hole solution. The spin is here treated perturbatively and the framework is known as the parametrized ringdown spin expansion coefficients (PARSPEC). Carullo~\cite{Carullo:2021dui} then applied this framework to existing gravitational-wave events and derived bounds on the correction to the ringdown frequency of the $(\ell,m,n)=(2,2,0)$ mode and the length scales (for various mass dimensions) at which new physics may arise. The parameterized quasi-normal mode framework in~\cite{Cardoso:2019mqo} has recently been extended to include arbitrary spin of black holes by Cano, et al.~\cite{Cano:2024jkd}. The authors introduced a parameterized deviation to the potential in the radial Teukolsky equation for the Kerr background. A parameterized waveform was also constructed using the effective-one-body framework by leaving the complex quasinormal mode frequencies to be free parameters~\cite{Brito:2018rfr}. This is, thus, an inspiral-merger-ringdown waveform, and one can correctly estimate the time at which the signal starts to be dominated by the quasinormal modes, unlike the case where one models the ringdown signal through a superposition of damped sinusoids.

The LIGO/Virgo Collaboration has carried out parameterized tests of ringdown with GWTC-2~\cite{LIGOScientific:2020tif} and GWTC-3~\cite{LIGOScientific:2021sio} adopting two different methods. The first method is the time-domain ringdown analysis pyRing~\cite{Carullo:2019flw,Isi:2019aib}, which is based on damped sinusoids. For example, the collaboration took the Kerr$_{221}$ template (containing the fundamental and first overtone modes for $(\ell,|m|)=(2,2)$) and introduced parameterized deviations from GR, $\delta \hat{f}_{221}$ for the frequency and $\delta \hat{\tau}_{221}$ for the damping time of the overtone mode. Combining 21 gravitational events in GWTC-3 that pass certain criteria, the collaboration found the hierarchically combined bound of $\delta \hat f_{221} = 0.01^{+0.27}_{-0.28}$~\cite{LIGOScientific:2021sio} for the frequency deviation, while the damping time deviation could not be constrained. The combined log odds ratio between non-GR versus GR hypotheses was $\log_{10} \mathcal O^\mathrm{modGR}_\mathrm{GR} = -0.90\pm 0.44$ at 90\% uncertainty~\cite{LIGOScientific:2021sio}, indicating that the GR hypothesis is favored. The second analysis uses the pSEOBNRv4HM waveform model~\cite{Cotesta:2018fcv,Ghosh:2021mrv} (a spinning, non-precessing effective-one-body waveform model with higher modes) in the time domain and the likelihood function in the frequency domain. The frequency and damping time of the fundamental mode with $(\ell,|m|)=(2,2)$ is modified from the GR ones as $f_{220} = f_{220}^\mathrm{GR}(1+\delta \hat f_{220})$ and $\tau_{220} = \tau_{220}^\mathrm{GR}(1+\delta \hat \tau_{220})$. The GR values are predicted from the initial binary's masses and spins through numerical relativity fits~\cite{Taracchini:2013rva,Hofmann:2016yih}. By multiplying posteriors from multiple events in GWTC-3, the collaboration found the bound $\delta \hat f_{220} = 0.02^{+0.03}_{-0.03}$ and $\delta \hat \tau_{220} = 0.13^{+0.11}_{-0.11}$, while hierarchically-combined bounds were $\delta \hat f_{220} = 0.02^{+0.07}_{-0.07}$ and $\delta \hat \tau_{220} = 0.13^{+0.21}_{-0.22}$~\cite{LIGOScientific:2021sio}. Notice that the bound on $\delta \hat \tau_{220}$ from multiplying posteriors does not contain the GR value within the 90\%-credible region. This may be due to the asymmetric prior on $(\delta \hat f_{220},\delta \hat \tau_{220})$, correlations among remnant parameters, imperfect noise modelling, or statistical uncertainties of using only $\sim 10$ events~\cite{LIGOScientific:2021sio}.

Recently, the problem of understanding the quasinormal frequency spectrum of spinning perturbed black holes in theories beyond general relativity has begun to be studied in detail. This problem is extremely difficult because the usual methods that lead to the Teukolsky equation~\cite{Teukolsky:1973ha} break down outside general relativity, either because the black hole background spacetime is not Petrov type D or because the field equations are not Einstein's. For example, spinning black holes in dynamical Chern-Simons gravity~\cite{Yunes:2009hc,Konno:2009kg,Yagi:2012ya,Maselli:2015tta} and Einstein--dilaton--Gauss--Bonnet gravity~\cite{Ayzenberg:2014aka}, computed in the slow-rotational approximation, remain Petrov Type D to linear order in spin~\cite{Sopuerta:2009iy,Yagi:2012ya,Ayzenberg:2014aka}, but become Petrov Type I at second and higher orders in spin~\cite{Yagi:2012ya,Ayzenberg:2014aka,Owen:2021eez}. In order to overcome this problem, several groups have calculated the quasinormal frequency spectrum of spinning perturbed black holes in the slow-rotation approximation~\cite{Wagle:2021tam,Blazquez-Salcedo:2016enn,Pierini:2021jxd,Pierini:2022eim} The problem with these analyses is that they are based on a slow-rotation approximation, while black hole remnant generated in binary coalescence are typically not rotating slowly. For this reason, a new effort has recently been put underway to compute perturbations of arbitrarily-fast rotating black holes outside general relativity. Li, et al.~\cite{Li:2022pcy} and Hussain and Zimmerman~\cite{Hussain:2022ins} have taken the first steps in this direction, deriving a modified Teukolsky equation for a large class of theories outside Einstein's, which holds irrespective of the Petrov type (see also Cano, et al.~\cite{Cano:2023tmv} for related work). The left-hand side of this equation has the same structure as the original Teukolsky equation, but it is now sourced (in the right-hand side) by terms that depend on derivatives of additional fields that may be present in the theory and spin coefficients and other curvature quantities that require metric reconstruction. Work is currently underway to separate this modified equation in a couple of example theories and then to solve the separated equations to find the eigenfrequencies without the slow-rotation approximation. The application of this method to dynamical Chern-Simons gravity for a slowly-spinning black hole has recently been completed by Wagle, et al.~\cite{Wagle:2023fwl}. A different approach was taken by Chung, et al.~\cite{Chung:2023zdq,Chung:2023wkd,Chung:2024ira,Chung:2024vaf} where one can use a spectral decomposition to solve black hole perturbation equations that are coupled and are not separated into radial and angular sectors. This novel framework was first developed for the Schwarzschild background case~\cite{Chung:2023zdq}, and then further extended to the Kerr background~\cite{Chung:2023wkd}, as well as to modified theories of gravity~\cite{Chung:2024ira}, such as scalar Gauss-Bonnet gravity~\cite{Chung:2024vaf}. With the spectral method established, recent work has begun to implement it in other theories of gravity, such as in scalar Gauss-Bonnet with strong coupling~\cite{Blazquez-Salcedo:2024oek}.

\subsubsection{The hairy search for exotica}
\label{sec:hairy-BHs}

Another way to test GR is to modify the matter sector of the theory through the introduction of matter corrections to the Einstein--Hilbert action that violate the assumptions made in the no-hair theorems. More precisely, one can study whether gravitational waves emitted by binaries composed of strange stars, like quark stars, or exotic compact objects that lack a horizon, such as boson stars or gravastars, are different from waves emitted by more traditional neutron-star or black-hole binaries. Horizonless compact objects have been proposed as a way to probe the quantum nature of black holes (see e.g.~\cite{Wang:2019rcf,Oshita:2019sat,Abedi:2020ujo,Addazi:2021xuf,Chakraborty:2022zlq,Chakraborty:2022zlq}). In what follows, we will describe such hairy tests of the existence of compact exotica.  

Boson stars are a classic example of an exotic compact object that is essentially indistinguishable from a black hole in the weak field, but which differs drastically from one in the strong field due to its lack of an event horizon. A boson star is a coherent scalar-field configuration supported against gravitational collapse by its self-interaction. One can construct several Lagrangian densities that would allow for the existence of such an object, including mini-boson stars~\cite{Friedberg:1986tp,Friedberg:1986tq}, axially-symmetric solitons~\cite{Ryan:1996nk}, and nonsolitonic stars supported by a non-canonical scalar potential energy~\cite{Colpi:1986ye}. Boson stars are well-motivated from fundamental theory, since they are the gravitationally-coupled limit of q-balls~\cite{Coleman:1985ki,Kusenko:1998em}, a coherent scalar condensate that can be described classically as a non-topological soliton and that arises unavoidably in viable supersymmetric extensions of the standard model~\cite{Kusenko:1997zq}. In all studies carried out to date, boson stars have been studied within GR, but they are also allowed in scalar-tensor theories~\cite{Balakrishna:1997ek}.

As before, two types of gravitational wave tests for boson stars have been proposed: inspiral tests and ringdown tests. The first studies in the inspiral test class were those that considered extreme--mass-ratio inspirals of a small compact object into a supermassive boson star. Kesden, et al.~\cite{Kesden:2004qx} showed that stable circular orbits exist both outside and inside the surface of the boson star, provided the small compact object interacts with the background only gravitationally. This is because the effective potential for geodesic motion in such a boson-star background lacks the Schwarzschild-like singular behavior at small radius, instead turning over and allowing for a new minimum. Gravitational waves emitted in such a system would then stably continue beyond what one would expect if the background had been a supermassive black hole; in the latter case the small compact object would simply disappear into the horizon. Kesden, et al.~\cite{Kesden:2004qx} found that orbits inside the boson star exhibit strong precession, exciting high frequency harmonics in the waveform, and thus allowing one to easily distinguish between such boson stars from black-hole backgrounds. Palenzuela, Bezares and their collaborators derived 
gravitational waves from binary boson star mergers for compact solitonic boson stars~\cite{Palenzuela:2017kcg} and dark boson stars~\cite{Bezares:2018qwa}. Siemonsen and East~\cite{Siemonsen:2023hko} carried out boson star merger simulations and studied the properties of the remnant. They showed, for the first time, that a remnant rotating boson star can form from a merger of two boson stars.
Destounis, et al.~\cite{Destounis:2023khj} studied non-integrability, chaos and resonances for extreme-mass-ratio inspirals into rotating boson stars.

The late inspiral, merger and ringdown of boson stars is also quite different from that of regular compact objects. Maselli, et al.~and Cardoso, et al.~\cite{Maselli:2016nuu,Cardoso:2017cfl} suggested that one could use the effect of tidal deformability on the gravitational waves to non-Kerr features of spacetime and to distinguish between boson stars and black holes or neutron stars. Tidal effects modify the orbital dynamics, thus imprinting onto the emitted gravitational waves, once the binary components are close enough together that they deform each other and change the gravitational potential. A class of massive boson stars has a minimum tidal deformability that can be much larger than the typical deformability of neutron stars and black holes. Sennett, et al.~\cite{Sennett:2017etc} argued that aLIGO at design sensitivity could distinguish between such massive boson stars and neutron stars or black holes provided an inspiral of a binary of sufficiently large mass or large mass ratio is observed. Moreover, a few studies have found that the merger of boson stars leads to a spinning bar configuration that either fragments or collapses into a Kerr black hole~\cite{Palenzuela:2006wp,Palenzuela:2007dm}. Pacilio, et al.~\cite{Pacilio:2020jza} constructed gravitational waveforms from binary boson star inspirals by including both quadrupole moment and tidal deformability, which was used in~\cite{Vaglio:2023lrd} for a Bayesian parameter estimation study.
Of course, the gravitational waves emitted during such a merger, and the subsequent ringdown will be drastically different from those produced when black holes merge. Indeed, Berti and Cardoso~\cite{Berti:2006qt} calculated the quasi-normal mode spectrum of boson stars and found that it is different from that of a Kerr black hole.  

Another exotic compact object that has been considered in the past are gravastars (see e.g.~\cite{Mottola:2023jxl} for a review). Gravitational vacuum stars, or gravastars for short, are compact objects that consist of a Schwarzschild exterior and a de Sitter interior, separated by an infinitely thin shell with finite tension and anisotropic pressure~\cite{Mazur:2001fv,Chapline:2000en,Cattoen:2005he}. The Lagrangian density for a gravastar is simply the Einstein-Hilbert action, but with a suitable stress-energy tensor that allow for a phase transition at or near where the Schwarzschild event horizon would have been. Mazur and Mottola~\cite{Mazur:2001fv} and later Visser and Wiltshire~\cite{Visser:2003ge} were able to construct appropriate stress-energy tensors such that the resulting object was thermodynamically and dynamically stable (see also~\cite{Chirenti:2007mk}), given a physically reasonable equation of state for the matter degrees of freedom. The current motivation for the gravastar picture comes  from attempting to resolve the information loss problem, for example through quantum phase transitions~\cite{Chapline:2000en,Chapline:2000en}.

As in the boson star case, the gravastar picture can also be tested with gravitational waves emitted in their inspiral and merger. Pani, et al.~\cite{Pani:2009ss,Pani:2010em} calculated the gravitational waves emitted by a small compact object in a quasi-circular orbit around a supermassive gravastar, using a radiative-adiabatic waveform generation model~\cite{Poisson:1993vp,Hughes:1999bq,Hughes:2001jr,Yunes:2009ef,2009GWN.....2....3Y,Yunes:2010zj}, instead of the kludge scheme used by Kesden, et al.~\cite{Barack:2003fp,Babak:2006uv,2009GWN.....2....3Y}. They concluded that the waves emitted during such inspirals are sufficiently different that they could be used to discern between a Kerr black hole and a gravastar. The quasinormal ringdown of gravastars has also been found to be drastically different from that of Kerr black holes, opening the possibility of using such waves for a test of their existence~\cite{Chirenti:2007mk,Pani:2009ss}. Such quasinormal modes of gravastars were tested against the GW150914 event by Chirenti and Rezzolla~\cite{Chirenti:2016hzd}, who concluded it was unlikely that the remnant of GW150914 was a gravastar.

Unfortunately, these exotic compact object alternatives typically encounter theoretical difficulties, for example due to instabilities in their stellar evolution or due to their own rotation. Cardoso, et al.~\cite{Cardoso:2007az,Cardoso:2008kj} and Pani, et al.~\cite{Pani:2010jz} have pointed out that \emph{all} horizonless compact objects with stable circular photon orbits, including boson stars and gravastars, are likely to be unstable to ergoregion instabilities if spinning rapidly, unless their surface is sufficiently absorbing~\cite{Maggio:2017ivp} (see~\cite{Chirenti:2008pf,Hod:2017wks,Maggio:2018ivz,Vicente:2018mxl,Zhong:2022jke} for related works on ergoregion instabilities). On top of these ergoregion instabilities, there are nonlinear trapping instabilities even for non-rotating horizonless compact objects~\cite{Keir:2014oka,Cardoso:2014sna,Cunha:2017qtt}. Moreover, the dynamical and non-linear stability of these objects in coalescence events is not clear. For example, in the case of an extreme mass-ratio inspiral into a supermassive boson star, the small compact object will accrete scalar field once it has entered the boson star's surface, possibly forcing the boson star to collapse into a black hole. Linear stability analysis of boson stars are not sufficient to prove that such objects are also non-linearly stable. Finally, horizonless objects typically encounter observational difficulties when considering accreting systems. Broderick and Narayan~\cite{Broderick:2007ek} have argued that the heating of gravastars due to accretion should be enough to stringently constrain them observationally. 

What is worse, for entire classes of exotic compact objects the scenarios discussed above cannot be even considered because of a lack of a definite Lagrangian density from which to derive their dynamics. For example, the gravastar model is sometimes referred to as a ``cut-and-paste'' spacetime, in that the interior de Sitter metric is glued to an exterior BH metric through a boundary layer of exotic matter. To date, nobody has yet shown that such objects arise naturally in dynamical gravitational collapse in a given theory of gravity. Of course, nobody has studied either how such objects would behave dynamically and what the associated gravitational waves would look like in the highly non-linear regime of merger (special cases of boson stars being the only exception). Efforts have been made to consider such exotic objects as ``straw-men'' to rule out using gravitational wave data. However, the use of such pathological objects as an even ``in-principle'' caveat to the evidence of black holes in gravitational wave data is questionable~\cite{Yunes:2016jcc}.  

When a merger remnant is a horizonless exotic compact object, gravitational waves can get trapped between the potential barrier and the surface, reflecting back and forth many times and producing gravitational-wave echoes~\cite{Cardoso:2016rao,Cardoso:2017cqb,Cardoso:2016oxy}. Abedi, et al.~\cite{Abedi:2016hgu} analyzed the data for GW150914, GW151226 and LVT151012 (now GW151012) and found some evidence for the presence of echoes in the data at a false alarm rate of 1\% (corresponding to a $2.5\sigma$ detection). This analysis was later challenged by Ashton, et al.~\cite{Ashton:2016xff}, Westerweck, et al.~\cite{Westerweck:2021nue} and Nielsen, et al.~\cite{Nielsen:2018lkf}, who reanalyzed the data and did not find such evidence for the echoes. Other groups conducted independent analyses and also found no significant evidence of echoes~\cite{Lo:2018sep,Uchikata:2019frs,Wang:2020ayy}. On the other hand, Abedi, et al.~\cite{Abedi:2018npz} further claimed to find tentative evidence of echoes with GW170817 one second after the merger at a $4.2\sigma$ confidence level and mentioned that the null results by other groups can be consistent with theirs if echoes contribute the most at lower frequencies and/or in binary mergers with more extreme
mass ratio~\cite{Abedi:2020sgg}.

The LIGO/Virgo Collaboration also searched for echoes in signals of gravitational-wave events in GWTC-2~\cite{LIGOScientific:2020tif} and GWTC-3~\cite{LIGOScientific:2021sio}. For GWTC-2, the collaboration used a template-based search~\cite{Lo:2018sep}, using the model described in~\cite{Abedi:2016hgu}. The waveform model takes and repeats the modified ringdown portion of the IMRPhenomPv2 waveform with 5 additional parameters: the relative amplitude of the echoes, the damping factor between each echo, the
start time of ringdown, the time of the first echo with respect
to the merger, and the time delay between each echo. The collaboration computed the Bayes factor comparing the hypotheses with and without echoes and found no statistically significant evidence of echoes. For GWTC-3, they carried out a morphology-independent approach~\cite{Tsang:2018uie,Tsang:2019zra} via BayesWave~\cite{Cornish:2014kda,Littenberg:2014oda} that uses sine Gaussians as basis functions for modeling gravitational waves. Once again, the features of echoes are captured by introducing 5 extra parameters. The Collaboration estimated the $p$-value for the log Bayes factors between hypotheses for signal versus noise and found that the measurement is consistent with no echoes within 90\% credibility.

\newpage
\section{Results of Gravitational Wave Tests with Pulsar Timing Data}
\label{section:PTA-tests}
%

Due to the nature of pulsar timing experiments, PTAs offer advantages over interferometers for detecting new polarizations or constraining the polarization content of GWs. For instance, each line of sight to a pulsar can
be used to construct an independent projection of the
various GW polarizations, and since PTAs typically observe tens of pulsars, linear combinations of the data
can be formed to measure or constrain each of the six
polarization modes many times over. A number of analyses of pulsar timing data have been performed searching for evidence of additional polarization modes which arise in  modified theories of gravity. Additionally, PTAs have an enhanced response to the longitudinal polarization modes~\cite{Chamberlin:2011ev,Cornish:2017oic,OBeirne:2019lwp}. Indeed, the constraint on the energy density of longitudinal modes inferred from recent NANOGrav data is about three orders of magnitude better than the constraint for the transverse modes~\cite{Cornish:2017oic}. Robust searches for evidence of modified gravity via non-Einsteinian modes are complex and, to illustrate this, it is worth summarizing some of the work here.

To date the most comprehensive searches for non-Einsteinian polarization modes in pulsar timing data have been performed by the NANOGrav collaboration on their 12.5-year data set~\cite{NANOGrav:2021ini} as well as using PPTA DR2 data~\cite{Wu_2022}. We summarize the NANOGrav results here.  In~\cite{NANOGrav:2021ini}, the data were analyzed with a suite of Bayesian and frequentist techniques assuming that the observed stochastic common red noise process found in~\cite{NANOGrav:2020bcs} is due to various combinations of the possible modes that can exist in metric theories of gravity. Interestingly, NANOGrav found a monopolar correlation signature which was the origin of some claims of detection of the scalar-transverse (ST) modes in the NANOGrav 12.5-year data set~\cite{Chen:2021wdo}, favored with a Bayesian odds ratio of about 100 to 1 (these authors also analyzed the IPTA Second Data release finding a Bayesian odds ratio in favor of scalar transverse modes of about 30:1~\cite{Chen:2021ncc}).  On theoretical grounds, we expect the presence of ST correlations to be accompanied by the standard quadrupolar $+$- and $\times$-modes of General Relativity: metric theories of gravity have at least the $+$- and $\times$-modes and possibly additional modes.

A detailed analysis of the NANOGrav 12.5-year data set revealed that the significance of non-quadrupolar correlations is reduced significantly (the Bayes factor drops to about 20) when one of the pulsars, J0030+0451, is excluded from the analyses. This pulsar has a history of being problematic in detection searches~\cite{Hazboun_2020}, and the results point
to the possibility of noise modeling issues involving this
MSP. The apparent albeit weak presence of non-Einsteinian modes of gravity is likely unphysical. Having found no statistically significant evidence in favor of any correlations, NANOGrav placed upper limits on the amplitudes of all possible subsets of polarization modes of gravity predicted by metric spacetime
theories. 

More recently~\cite{NANOGrav:2023ygs}, the NANOGrav 15-yr data were searched for evidence of a gravitational-wave background with quadrupolar (HD) and ST correlations, finding that HD correlations are the best fit to the data and no significant evidence in favor of ST
correlations. We discuss the details of this work below.

\subsection{Details of the Analyses of the NANOGrav 12.5-yr dataset}

Here we discuss some of the detailed results of the NANOGrav 12.5-yr dataset search for additional polarization modes~\cite{NANOGrav:2021ini}. We begin by summarizing the naming convention used for the various models in their analyses. 

The two general types of Bayesian models used in this paper are referred to as \emph{M2A} and \emph{M3A} which were also used in~\cite{NANOGrav:2020bcs}. The M2A model includes 
various white noise terms for each pulsar, an intrinsic red noise term for each pulsar, and an uncorrelated so-called common red noise process, which has the same spectral index and amplitude across all pulsars.  The M2A model does not include correlations between pulsars so its full PTA covariance matrix is block-diagonal. The M3A models include identical noise processes as M2A, with the common red noise process correlated across pulsars. Therefore, for the M3A models the full PTA covariance matrix has non-vanishing off-diagonal components. 
If the spectral index of the common process is fixed, it is given in square brackets after either M2A or M3A; and for M3A models, the type of correlations are given in square brackets before M3A. For example, [HD]M3A[13/3] refers to a model where the common process has Helligs--Downs (quadrupolar correlations) with a spectral index of $13/3$. When more than one type of common correlated red noise process is included, it is indicated by more than one type of correlation in the square bracket preceding the term ``M3A'' and additional spectral indices. For instance, [HD,ST]M3A[13/3,5] means that the M3A contains two different correlated common signals: a red noise process with spectral index of 13/3 and quadrupolar Hellings--Downs correlations, and second red noise process with spectral index of 5 with scalar-transverse correlations.

\epubtkImage{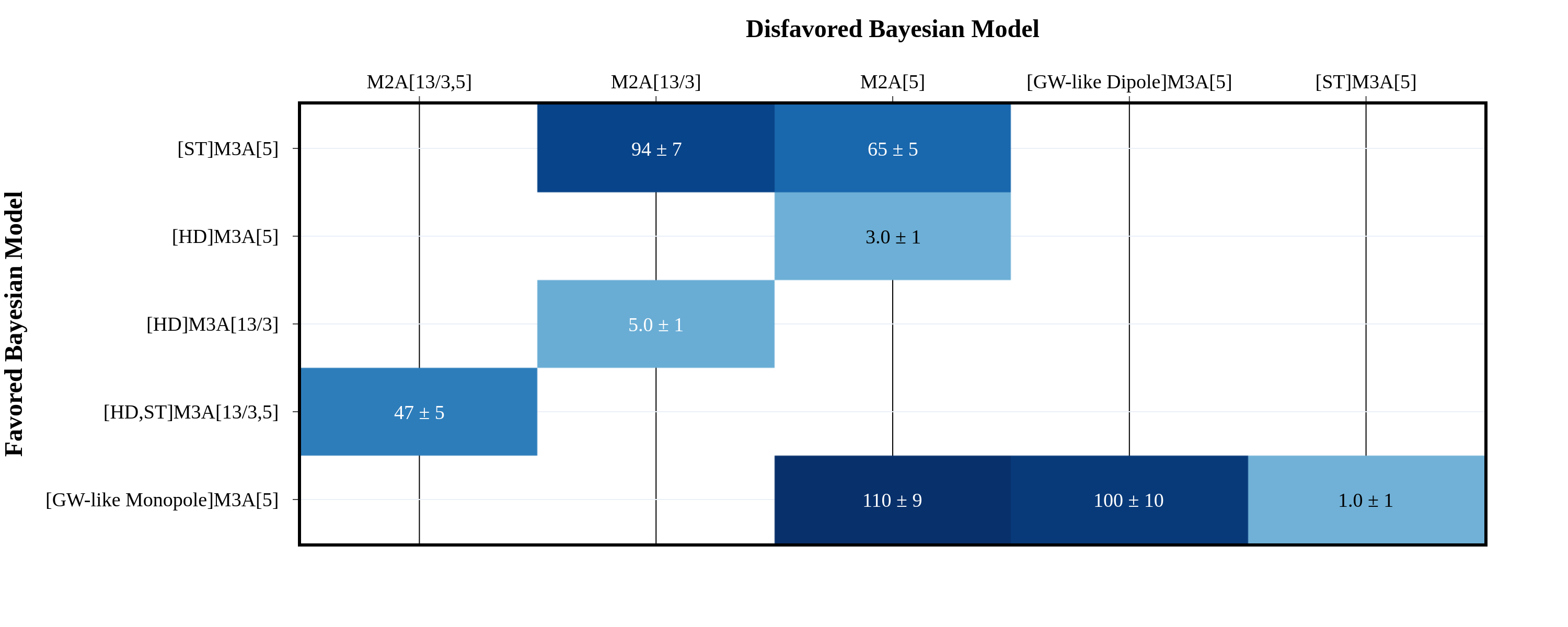}{%
\begin{figure*}[hbt!]
\includegraphics[width=\textwidth]{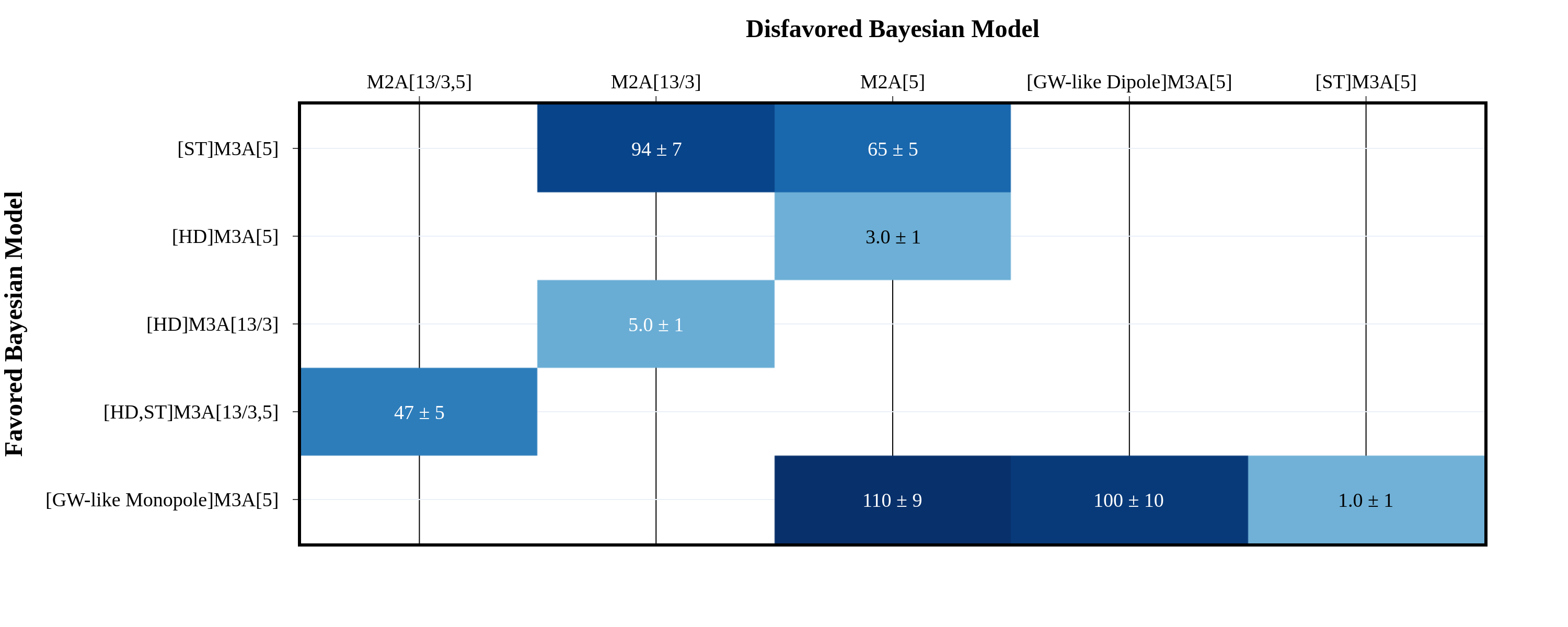}
\caption{The table shows the odds ratios for comparisons of various models. Darker shades of blue correspond to higher odds ratios. The model with the highest odds is the GW-like monopole, where the correlations between pulsars are not angular separation dependent and fixed to $1/2$ (to be compared with a clock-error monopole where the correlations between pulsars are unity). Figure taken from~\cite{NANOGrav:2021ini}.}
\label{BFHeatmap}
\end{figure*}}

NANOGrav calculated the results of several different Bayesian analyses shown in Figs.~\ref{BFHeatmap} (cite NG12.5 here). The model with the highest odds is a GWB with GW-like monopolar correlations (where the correlations between pulsars are not angular separation dependent and fixed to $1/2$, to be compared with a clock-error monopole where the correlations between pulsars are unity). The odds of [GW-like Monopole]M3A[5] are greater than 100 compared with a model with no correlations and the same spectral index M2A[5]. NANOGrav was also able to reproduce the results of [Chen, et al.], where [ST]M3A[5] was compared to a model without correlations and a spectral index of 13/3, M2A[13/3] finding odds of around 94:1 in favor of [ST]M3A[5], which is consistent with the results in [Chen, et al.]. This results in part due to the difference in spectral index between the two models (the NANOGrav 12.5-yr dataset prefer a steeper spectral index), and a degeneracy between ST and HD correlations at low S/N. Finally, as we've mentioned, the result is very sensitive to the inclusion of one MSP, J0030$+$0451, which indicates a problem with noise modeling for that pulsar. NANOGrav also examined the problem using the frequentist optimal statistic and found consistent results.

From a theoretical perspective, some models shown in Fig.~\ref{BFHeatmap} are not viable. Metric theories of gravity always contain the two Einsteinian $+$- and $\times$-modes so that even when a model with ST-only correlations is found to have high odds, this model does not correspond to a metric theory of gravity. However, models with multiple spatial correlations that include the TT modes, such as [HD,ST]M3A[13/3,5] are theoretically viable.

Having found no compelling evidence for gravitational waves, NANOGrav proceeded to present upper limits on all possible polarization content in metric theories of gravity. The naming convention they introduced for this model classification is the letters MG (an abbreviation for metric theory of gravity), followed by 4 digits which can be unity or zero depending on which polarization modes (TT, ST, VL, SL) are present in any particular theory. For example, MG1000 is a metric theory of gravity with only the TT modes present, i.e. Einstein gravity; MG1100 is a theory with TT and ST modes, e.g, Jordan--Fierz--Brans--Dicke gravity, etc. In this classification there are only 8 families of metric theories of gravity, since the TT mode is present in all metric theories of gravity. 

The results for the upper limits are shown in Fig.~\ref{upplim}. It is worth noting the higher sensitivity of pulsar timing arrays to the longitudinal polarizations, VL and SL. This results in upper limits that are about an order of magnitude smaller for VL modes and about two orders of magnitude smaller for the SL mode compared with the TT mode. These upper limits are theory agnostic, and can be used to constrain the parameters of metric theories of gravity that couple to these modes.

\epubtkImage{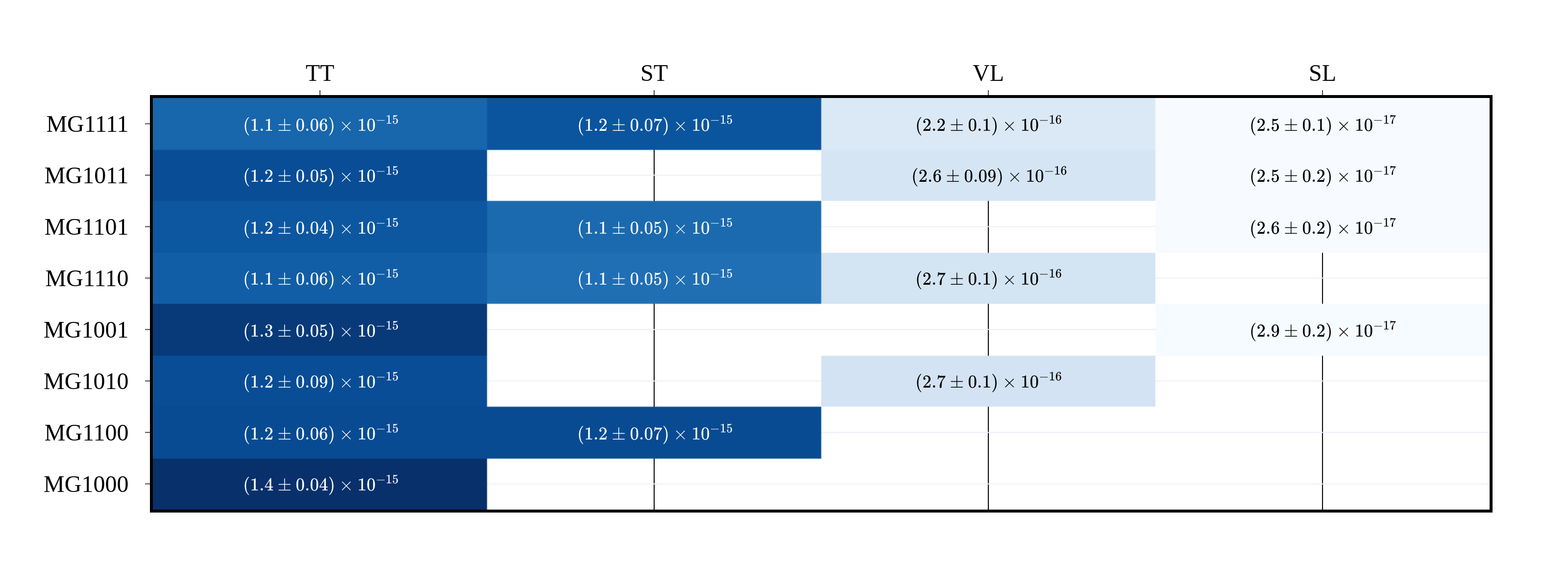}{%
\begin{figure*}
    \centering
    \includegraphics[width=\linewidth]{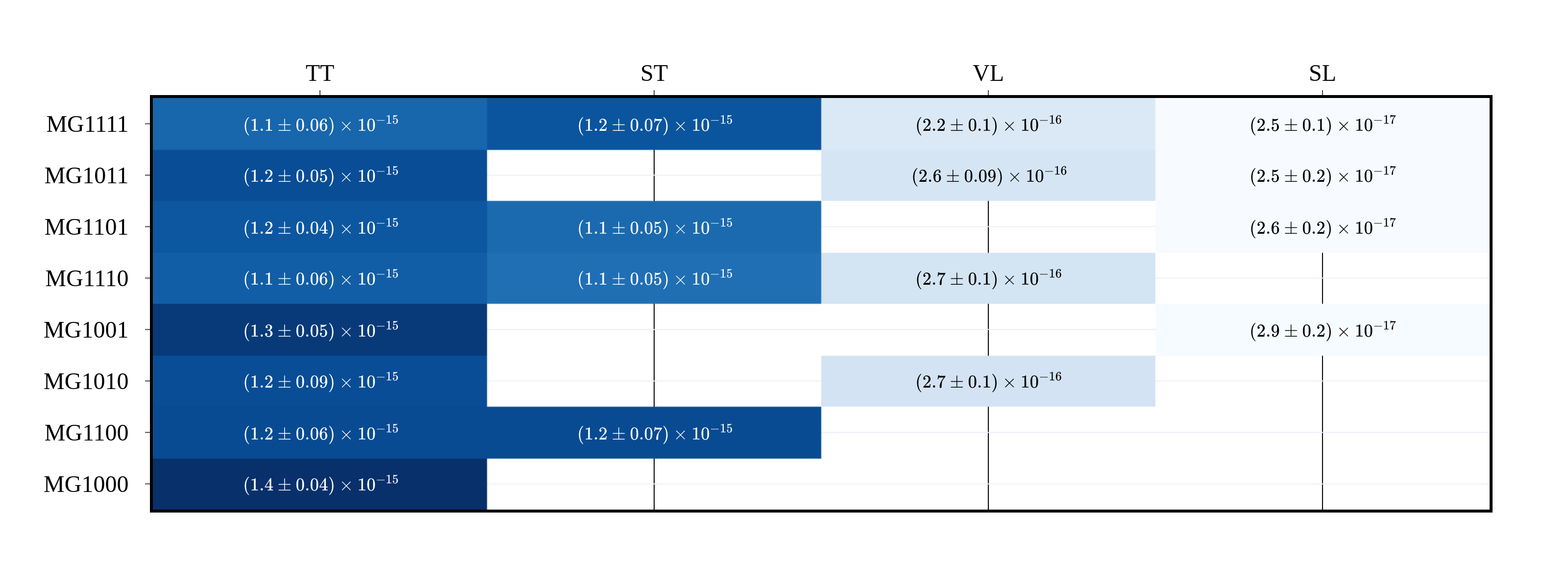}
    \caption{95\%  upper limits on the amplitude of GWs for the eight families of metric theories of gravity. For simplicity, the spectral index for gravitational waves was fixed at $\gamma = 5$, which corresponds to a flat spectrum in $\Omega_{\rm GW}$, the ratio of the energy density in GWs to the critical density. Figure taken from~\cite{NANOGrav:2021ini}}
    \label{upplim}
\end{figure*}}

\subsection{Details of the Searches in the NANOGrav 15-yr dataset}

NANOGrav also recently searched their 15-year data set for evidence of a gravitational wave background with all possible transverse modes: HD and ST correlations~\cite{NANOGrav:2023ygs}(see left panel of Fig.~\ref{fig:ORFs}). This analysis was restricted to the transverse modes because their overlap reduction functions have no frequency dependence and are more simple to model. Analyses that include the remaining 3 modes (SL and VL), with frequency dependent overlap reduction functions, are underway.

\epubtkImage{TTST-TTSTBayesFactors.png}{%
\begin{figure}
\includegraphics[width=0.5\linewidth]{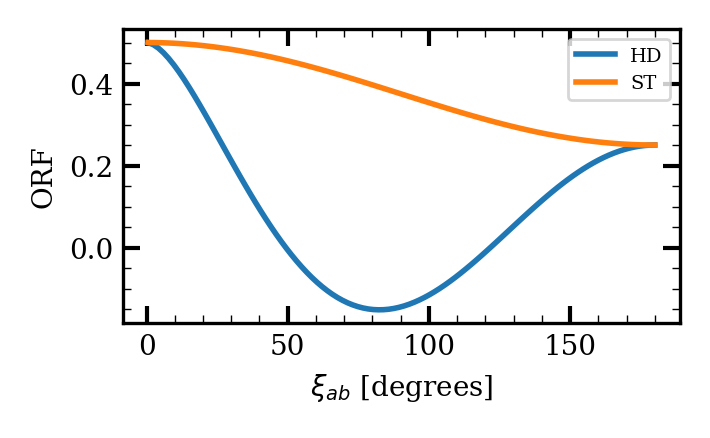}
\includegraphics[width=0.5\linewidth]{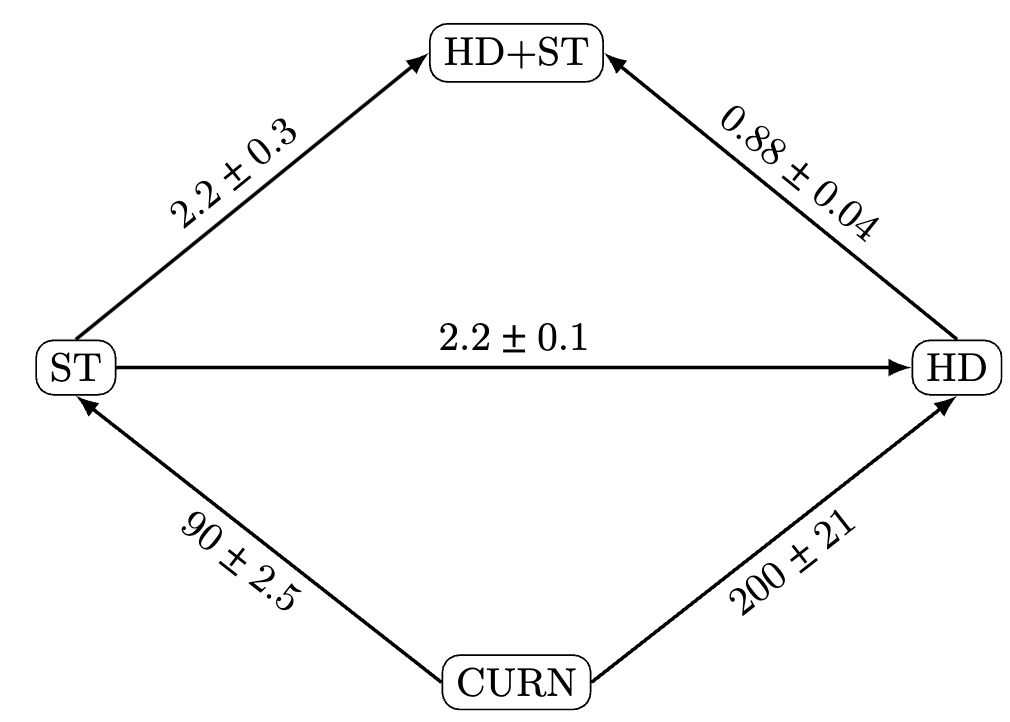}
\caption{{\it Left panel:} Plot of the correlation signatures for the transverse modes as a function of angular separation. The blue curve shows the Hellings and Downs curve, and the orange curve shows the correlations produced by ST gravitational waves. These two types of correlations were searched for in NANOGrav's analysis of the 15-yr dataset in separate and combined analyses with Bayesian and frequentist techniques. {\it Right panel:} Odds ratios for HD and ST models and HD+ST models compared to one another along with an common uncorrelated red noise model (CURN). See the main text for details. Figure credit~\cite{NANOGrav:2023ygs}.}
\label{fig:ORFs}
\end{figure}}

The right panel of Fig.~\ref{fig:ORFs} shows the odds ratios for a series of Bayesian runs on NANOGrav's 15-yr dataset. The base model is a stochastic common uncorrelated red noise process (CURN), i.e. a stochastic process that has the same amplitude and spectral index for all pulsars but no correlations among the different pulsars. The HD model, which is the prediction of General Relativity, has odds of $\sim$200:1 relative to the CURN model, and the ST model odds are $\sim$90:1. Interestingly, the odds ratios of standard Bayesian analyses of the data did not show a strong preference for either correlation signature, with odds ratios $\sim 2$ in favor of HD versus ST correlations, and $\sim 1$ for HD plus ST correlations versus HD correlations alone, and further analyses were required to establish that HD correlations are indeed the best fit to the data, with no significant evidence in favor of ST correlations. 

\epubtkImage{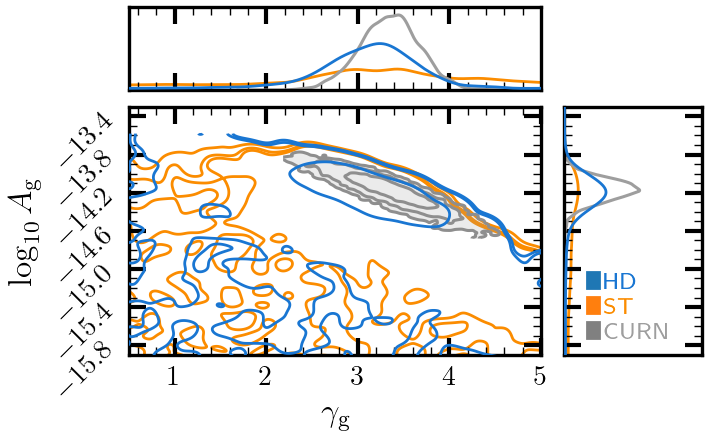}{%
\begin{figure*}[!ht]
\centering
\includegraphics[width = 0.7\linewidth]{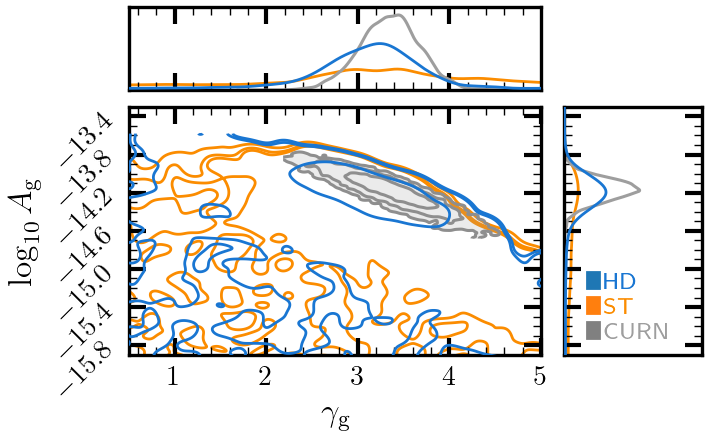}%
\caption{Plot of the posteriors for the amplitudes ($A_g$) and spectral indices ($\gamma_g$) for HD (blue) and ST (orange) from the combined HD+ST model and the CURN model (grey)  (the models at the top and bottom of the right panel of Fig.~\ref{fig:ORFs}). Figure credit~\cite{NANOGrav:2023ygs}.}
\label{fig:bayes_amps}
\end{figure*}}

Looking at the posteriors for the amplitudes and spectral indices of the HD and ST modes when searched for simultaneously, NANOGrav found that the posteriors for the amplitude and spectral index of ST correlations are uninformative, with the HD stochastic process accounting for the majority of the correlated signal. This is shown in Fig.~\ref{fig:bayes_amps} where we plot the posteriors for the amplitudes and spectral indices for the HD and ST stochastic processes as recovered from a combined HD+ST model as well as the CURN model. These two models correspond to the top and bottom of the right panel of Fig.~\ref{fig:ORFs}.  The amplitude and spectral index recovery for ST is poor when modeled alongside HD. The posterior distribution for ST is uninformative and HD adequately describes the total signal as recovered by CURN. 

In addition, using the optimal statistic, a frequentist detection technique that uses only the correlation information~\cite{Anholm:2008wy,Chamberlin:2014ria,Vigeland:2018ipb,Sardesai:2023qsw}, NANOGrav found results consistent with the Bayesian analyses, i.e. similar signal-to-noise-ratios for each of the correlation, but inconsistent parameter estimation recovery for ST. Figure~\ref{fig:OSamps} shows the optimal statistic amplitude recovery for the HD and ST amplitudes when searched for separately (right panel) and jointly (left panel), compared with the CURN model, and in both cases the HD amplitude is significantly more consistent with the CURN process than ST. 

\epubtkImage{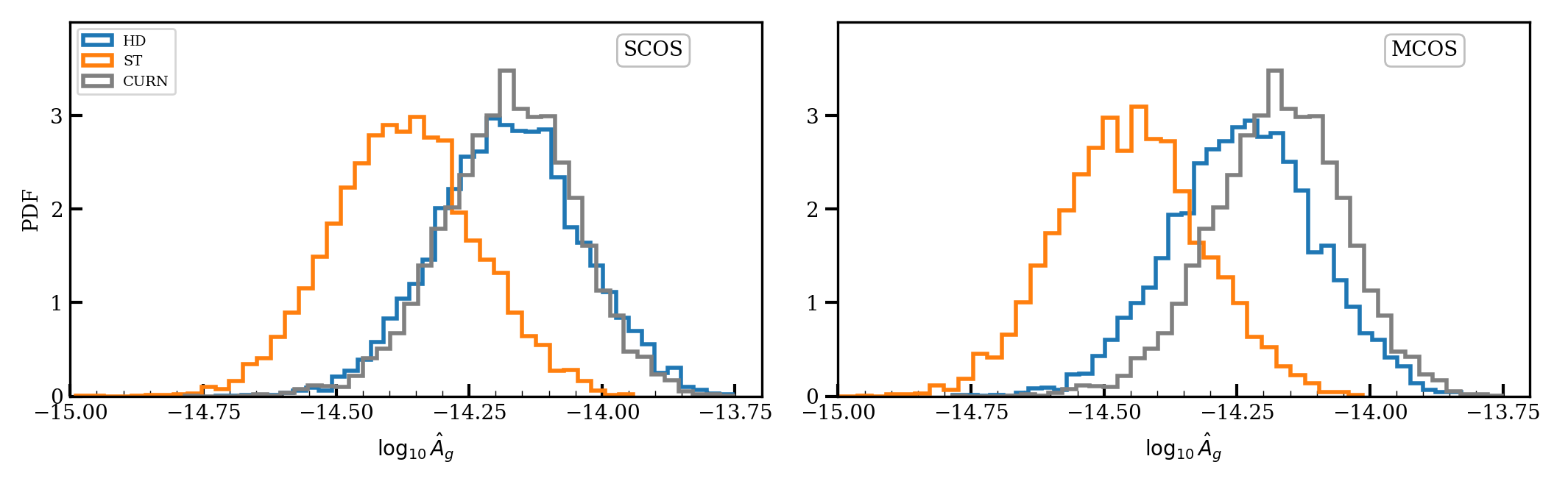}{%
\begin{figure*}[!htp]
    \centering
    \includegraphics[width =\linewidth]{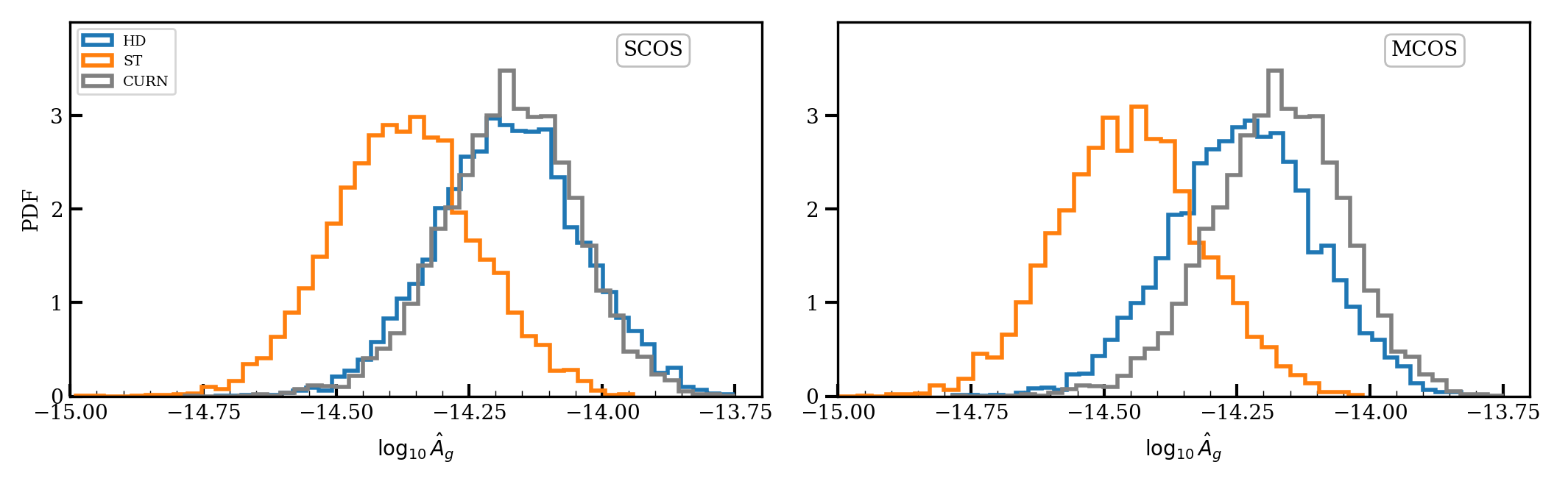}
    \caption{The left panel shows the recovered amplitudes from single-component noise marginalized optimal statistic (SCOS), where the HD (blue) and ST (orange) correlations are searched for separately, and the right panel shows the recovered aplitudes for the multi-component noise marginalized optimal statistic (MCOS)(\emph{right}) where the HD and ST correlations are fit for simultaneously. The CURN (gray) is also shown to determine the consistency of the amplitude recovery of HD and ST with the common red noise process.  Figure credit~\cite{NANOGrav:2023ygs}.}
    \label{fig:OSamps}
\end{figure*}}

\pagebreak

\newpage
\section{Musings About the Future}
\label{section:musings}

Gravitational waves hold the key to testing Einstein's theory of general relativity (GR) to new exciting levels in the previously unexplored extreme gravity regime. Depending on the type of wave that is detected, e.g., compact binary inspirals, mergers, ringdowns, continuous sources, supernovae, etc, different tests will be possible. Irrespective of the type of wave detected, two research trends seem currently to be arising: direct tests and generic tests. These trends aim at answering different questions. With direct tests, one wishes to determine whether a certain modified theory is consistent with the data. Generic tests, on the other hand, ask whether the data is statistically consistent with our canonical beliefs, or put another way, whether there are any statistically-significant deviations present in the data. The approaches, however, are interconnected because once a generic test has established that no statistically significant deviations from GR are present in the data, then this information can be recast to draw inferences from the data on constraints on specific GR deviations. 

Gravitational waves have now been detected with ground-based detectors, opening up an entire new area in experimental relativity. Many concrete efforts are currently underway to develop and extend formalisms and implementation pipelines to test Einstein's theory in extreme gravity. Currently, the research groups separate into two classes: theory and implementation. The theory part of the research load is being carried out at a variety of institutions without a given focal point. The implementation part is being done mostly within the LIGO Scientific Collaboration, the Virgo Scientific collaboration, and the pulsar timing consortia. Cross-communication between the theory and implementation groups has been flourishing in recent years and one expects the interdisciplinary work to continue and expand in the future. 

So many accomplishments have been made in the past 50 years that it is almost impossible to list them all here. From the implementation side, perhaps one of the most important is the actual construction and operation of the initial and advanced ground-based instruments that have given us the first gravitational wave observations. This is a tremendously important engineering and physics accomplishment. Similarly, the construction of impressive pulsar timing arrays, and the timing of these pulses to nanosecond precision is an instrumental and data analysis feat to be admired. Without these observatories, there would be no gravitational wave physics, and of course, no tests of Einstein's theory in extreme gravity. On the theory side, perhaps the most important accomplishment has been the understanding of the inspiral phase to extremely high post-Newtonian order and the merger phase with numerical simulations. The latter, in particular, had been an unsolved problem for over 50 years. It is these accomplishments that then allow us to postulate modified inspiral template families and study mergers in modified gravity, since we understand what the GR model is. This is particularly true if one is considering small deformations away from Einstein's theory, as it would be impossible to perturb about an unknown solution.  

The main questions that are currently at the forefront are the following. On the theory side of things, one would wish to understand the inspiral, merger and ringdown in extreme gravity modifications to GR. We have here discussed only a few of them, such as dynamical Chern--Simons gravity, Einstein--dilaton-Gauss--Bonnet theory and theories with preferred frames, such as Einstein--Aether theory or Ho{\v{r}}ava--Lifshitz gravity. The first step in this direction is to develop higher post-Newtonian order models for the inspiral phase. Such a task, of course, is very difficult, given that the complexity of the calculation in GR alone is already daunting. Simultaneously, the second step is to numerically simulate the merger in modified gravity theories. This task is also very difficult because the characteristic structure of the evolution equations in modified gravity is likely different from that in GR, sometimes requires a reformulation of the standard evolution methods. Once these two steps are complete, one then needs to construct an inspiral-merger-ringdown model that smoothly connects these two phases of coalescence. This, in turn, requires many numerical simulations of modified gravity mergers to properly sample the parameter space, a task that is still not complete today in GR. 

On the implementation side of things, there is also much work that remains to be done. Currently, efforts are ongoing on the implementation and improvement of Bayesian frameworks for hypothesis testing, one of the most promising approaches to testing Einstein's theory with gravitational waves. Present studies concentrate mostly on single-detectors, but by the beginning of the next decade we expect four or five detectors to be online, and thus, one will have to extend these implementations. The use of multiple detectors also opens the door to the extraction of new information, such as multiple polarization modes, a precise location of the source in the sky, etc. Moreover, the evidence for a given model increases dramatically if the event is observed in several detectors. One therefore expects that the strongest tests of GR will come from leveraging the data from all detectors in a multiply-coincident event, perhaps also including information from electromagnetic counterparts. 

Research is moving toward the construction of robust techniques to test Einstein's theory, and in particular, testing the general principles that serve as foundations of GR. This allows one to answer general questions, such as: Does the graviton have a mass? Are compact objects represented by the Kerr metric and the no-hair theorems satisfied? Does the propagating metric perturbation possess only two transverse-traceless polarization modes? What is the rate of change of a binary's binding energy? Do naked singularities exist in nature and are orbits chaotic? Is Lorentz-violation present in the propagation of gravitons? The more questions of this type that are generated and the more robust the methods to answer them are, the more stringent the test of Einstein's theories and the more information we will obtain about the gravitational interaction in a previously unexplored regime.

\newpage
\section*{Acknowledgements}
\label{section:acknowledgements}

We would like to thank Emanuele Berti, Vitor Cardoso, William Nelson,
Bangalore Sathyaprakash, and Leo Stein for many discussions. We would
also like to thank Laura Sampson and Tyson Littenberg for helping us
write parts of the data analysis sections. Finally, we would like to
thank Matt Adams, Katerina Chatziioannou, Tyson Littenberg, and Laura
Sampson for proofreading earlier versions of this
manuscript. 
Nicol\'as Yunes would like to acknowledge support from the Simmons Foundation through Award No. 896696, the NSF through award PHY-2207650, and NASA through Grant No. 80NSSC22K0806.
Xavier Siemens would like to acknowledge support from the NSF CAREER
award number 0955929, the 
PIRE award number 0968126, and award number 0970074. 
K.Y. acknowledges support from NSF Grant PHY-1806776, PHY-2207349, NASA Grant No.
80NSSC20K0523, a Sloan Foundation Research Fellowship, and the Owens Family Foundation.
LIGO Laboratory and Advanced LIGO are funded by the United States National Science Foundation (NSF) as well as the Science and Technology Facilities Council (STFC) of the United Kingdom, the Max-Planck-Society (MPS), and the State of Niedersachsen/Germany for support of the construction of Advanced LIGO and construction and operation of the GEO600 detector. Additional support for Advanced LIGO was provided by the Australian Research Council. Virgo is funded, through the European Gravitational Observatory (EGO), by the French Centre National de Recherche Scientifique (CNRS), the Italian Instituto Nazionale di Fisica Nucleare (INFN) and the Dutch Nikhef, with contributions by institutions from Belgium, Germany, Greece, Hungary, Ireland, Japan, Monaco, Poland, Portugal, Spain. The construction and operation of KAGRA are funded by Ministry of Education, Culture, Sports, Science and Technology (MEXT), and Japan Society for the Promotion of Science (JSPS), National Research Foundation (NRF) and Ministry of Science and ICT (MSIT) in Korea, Academia Sinica (AS) and the Ministry of Science and Technology (MoST) in Taiwan. 

\newpage
\bibliography{refs}

\end{document}